%% file: thesis-arxiv.tex
\documentclass{sotonthesis-arxiv}

\usepackage{graphicx}
\graphicspath{{./img/}}
\usepackage{caption}
\usepackage[figure, table]{totalcount}

\usepackage[british]{babel}

\usepackage{siunitx}
\DeclareSIUnit\solarMass{\mbox{$M_\odot$}}
\DeclareSIUnit\year{yr}
\DeclareSIUnit\gauss{G}
\DeclareSIUnit\erg{erg}
\DeclareSIUnit\electronVolt{eV}
\DeclareSIUnit\dyne{dyn}

\usepackage{graphics}

\title{Gravitational waves from deformed neutron stars: mountains and tides}
\author{Fabian Gittins}

\date{September 2021}

\supervisor{Nils Andersson}
\degree{Doctor of Philosophy}
\university{University of Southampton}
\faculty{Faculty of Social, Human and Mathematical Sciences}
\academicunit{Mathematical Sciences}

\makeatletter
\renewcommand{\@pnumwidth}{2em}
\makeatother

\begin{document}
\pagestyle{empty}
\frontmatter

\maketitle

\pdfbookmark[0]{Abstract}{abstract}
\begin{abstract}
    \input{./frontmatter/abstract}
\end{abstract}

\dedication{\input{./frontmatter/dedication}}

\pagestyle{fancy}

\pdfbookmark[0]{\contentsname}{toc}
\tableofcontents
\cleardoublepage{}

\iftotalfigures{}
    \listoffigures
    \cleardoublepage{}
    \iftotaltables{}
        \listoftables
        \cleardoublepage{}
    \fi
\fi

\pagestyle{empty}

\authorshipdeclaration{\input{./frontmatter/authorship-references}}

\acknowledgements{\input{./frontmatter/acknowledgements}}

\phantomsection{}
\addcontentsline{toc}{chapter}{Notation}
\input{./frontmatter/notation}
\cleardoublepage{}

\pagestyle{fancy}
\setlength{\parskip}{1.3em}

\def\order{{\mathcal{O}}}

\mainmatter{}
\raggedbottom{}

\input{./sections/chapter-1}
\input{./sections/chapter-2}
\input{./sections/chapter-3}
\input{./sections/chapter-4}
\input{./sections/chapter-5}
\input{./sections/chapter-6}
\input{./sections/chapter-7}

\renewcommand{\theHchapter}{A\arabic{chapter}}
\appendix
\input{./sections/appendix-A}
\input{./sections/appendix-B}
\input{./sections/appendix-C}
\input{./sections/appendix-D}

\cleardoublepage{}
\phantomsection{}

\begin{refcontext}[sorting=nyt]
    \printbibliography[heading=bibintoc]  
\end{refcontext}

\end{document}

%% file: frontmatter/abstract.tex
With the remarkable advent of gravitational-wave astronomy, we have shed light 
on previously shrouded events: compact binary coalescences. Neutron stars are 
promising (and confirmed) sources of gravitational radiation and it proves 
timely to consider the ways in which these stars can be deformed. Gravitational 
waves provide a unique window through which to examine neutron-star interiors 
and learn more about the equation of state of ultra-dense nuclear matter. In 
this work, we study two relevant scenarios for gravitational-wave emission: 
neutron stars that host (non-axially symmetric) mountains and neutron stars 
deformed by the tidal field of a binary partner. Although they have yet to be 
seen with gravitational waves, rotating neutron stars have long been considered 
potential sources. By considering the observed spin distribution of accreting 
neutron stars with a phenomenological model for the spin evolution, we find 
evidence for gravitational radiation in these systems. We study how mountains 
are modelled in both Newtonian and relativistic gravity and introduce a new 
scheme to resolve issues with previous approaches to this problem. The crucial 
component of this scheme is the deforming force that gives the star its 
non-spherical shape. We find that the force (which is a proxy for the star's 
formation history), as well as the equation of state, plays a pivotal role in 
supporting the mountains. Considering a scenario that has been observed with 
gravitational waves, we calculate the structure of tidally deformed neutron 
stars, focusing on the impact of the crust. We find that the effect on the 
tidal deformability is negligible, but the crust will remain largely intact up 
until merger.


%% file: frontmatter/dedication.tex
This, I dedicate to my Uncle, 

\vspace{2\baselineskip}
Robert Ekblom (1956--2015), 

\vspace{2\baselineskip}
who suggested I become a quantum physicist...


%% file: frontmatter/authorship-references.tex
\begin{itemize}
\item \citet{2019MNRAS.488...99G},
\item \citet{2020PhRvD.101j3025G},
\item \citet{2021MNRAS.500.5570G},
\item \citet{2021MNRAS.507..116G}.
\end{itemize}


%% file: frontmatter/acknowledgements.tex
All I have accomplished has been made possible through the Lord, ``for it is God 
who works in you to will and to act in order to fulfil his good purpose'' 
(Philippians 2:13). For everything involved in the pursuit of a PhD, I want to 
thank and praise my Lord Jesus Christ, through whom ``all things were created: 
things in heaven and on earth, visible and invisible, whether thrones or powers 
or rulers or authorities; all things have been created through him and for him. 
He is before all things, and in him all things hold together'' 
(Colossians 1:16-17). He is not only the creator and sustainer of the 
Universe, including upholding the rich and intricate physical laws of nature 
that govern the magnificent stars, Jesus is also my personal saviour, who ``was 
pierced for our transgressions, he was crushed for our iniquities; the 
punishment that brought us peace was on him, and by his wounds we are healed'' 
(Isaiah 53:5). This was the ultimate manifestation of his love for us.

It has been a privilege to study at the University of Southampton and I wish to 
acknowledge some of the main blessings I have enjoyed during my time here.

Firstly, I thank my supervisor, Nils Andersson, who has been a superb mentor to 
me. Never has there been a moment when he was not willing to fit in time to give 
advice on my research and I have always looked forward to our discussions. I am 
sincerely grateful for his unwavering assistance and his excellent teaching and 
support.

I would also like to extend my gratitude to the rest of the Gravity Group at the 
University of Southampton, in particular the members focused on neutron-star 
physics. It has been a pleasure to be among such a talented ensemble of 
physicists, who have played an important role in my education.

In addition to the academic support I have benefited from, I would like to 
acknowledge the encouragement and prayer of my dear friends at Christ Church 
Southampton. I am grateful for the family that I have found there and the 
manifold ways in which I have been supported. In particular, I wish to thank 
three of my closest friends that I have made during my time here: Tom Bell, 
Thomas Emerson and Matt Magee. I have truly been blessed to have met such wise 
and thoughtful brothers in Christ, who continually challenge me and bring out my 
best qualities. ``Walk with the wise and become wise, for a companion of fools 
suffers harm'' (Proverbs 13:20).

Many teachers, peers, friends and relatives that I had before moving to 
Southampton have continued to offer advice and support. They are too numerous 
to acknowledge personally, but they should know that I am incredibly grateful.

Finally, the greatest debt of gratitude that I have to offer, under Christ, is 
to my family. My parents, Mark and Caroline Gittins, loved me since before I was 
born and have wonderfully modelled self-sacrificial love to me and my siblings. 
My brother, Adam Gittins, has been my best friend for our entire lives and can 
always make me smile. My sister, Caitlin Gittins, has taught me much about 
kindness, compassion and unwavering loyalty. Without their love and support, I 
would certainly not be the man I am today.


%% file: frontmatter/notation.tex
\chapter*{Notation}

\textbf{Constants and units.}

For relativistic calculations, we work in geometric units, 
\begin{equation*}
    G = \SI{6.67428e-8}{\centi\metre\cubed\per\gram\per\second\squared} 
        = 1
\end{equation*}
and 
\begin{equation*}
    c = \SI{2.99792458e10}{\centi\metre\per\second} = 1,
\end{equation*}
where $G$ is Newton's gravitational constant and $c$ is the speed of light in a 
vacuum. In these units, the mass of the Sun is 
$M_\odot = \SI{1.98892e33}{\gram} = \SI{1.47700}{\kilo\metre}$.

\textbf{Indices, Einstein summation convention, metric signature.}

Spatial indices are denoted with late Latin characters $i, j, k, \ldots$ and 
spacetime indices are denoted with early Latin characters $a, b, c, \ldots$. We 
will reserve the indices $\ell$ and $m$ exclusively for spherical-harmonic 
modes.

The Einstein summation convention will be used, where repeated indices indicate 
a summation, \textit{e.g.}, 
\begin{equation*}
    v^i v_i = \sum_i v^i v_i, \quad u^a u_a = \sum_a u^a u_a,
\end{equation*}
where $v^i$ and $u^a$ are an arbitrary three-vector and four-vector, 
respectively. We do not assume summations over repeated $\ell, m$.

The signature of the spacetime metric $g_{a b}$ is $(-, +, +, +)$. Hence, 
time-like four-vectors have negative lengths.

\textbf{Covariant and partial derivatives.}

The covariant and partial derivatives are denoted by $\nabla_a$ and 
$\partial_a$, respectively. We use the traditional conventions for the covariant 
derivative so that it reduces to the partial derivative for scalars,
\begin{equation*}
    \nabla_a f = \partial_a f,
\end{equation*}
where $f$ is an arbitrary scalar quantity, and it commutes with the 
metric, 
\begin{equation*}
    \nabla_a g_{b c} = 0.
\end{equation*}

We work in a coordinate basis,
\begin{equation*}
    \partial_a f \equiv \frac{\partial f}{\partial x^a},
\end{equation*}
thus, indices will run over the coordinates, \textit{e.g.}, in spherical polar 
coordinates, we have $x^t = t, x^r = r, x^\theta = \theta, x^\phi = \phi$, which 
can be written compactly as $x^a = (t, r, \theta, \phi)$. It should be noted 
that the distinction between raised and lowered indices will be meaningful (as 
opposed to when working with an orthonormal basis).

Covariant derivatives obey the following law for an arbitrary tensor 
$T_{a b \ldots}^{\hphantom{a b \ldots} c d \ldots}$:
\begin{equation*}
\begin{split}
    \nabla_a T_{b c \ldots}^{\hphantom{b c \ldots} d e \ldots} 
        = \partial_a T_{b c \ldots}^{\hphantom{b c \ldots} d e \ldots} 
        &- \Gamma^f_{\hphantom{f} b a} 
        T_{f c \ldots}^{\hphantom{b c \ldots} d e \ldots} 
        - \Gamma^f_{\hphantom{f} c a} 
        T_{b f \ldots}^{\hphantom{b c \ldots} d e \ldots} - \ldots \\ 
        &+ \Gamma^d_{\hphantom{d} f a} 
        T_{b c \ldots}^{\hphantom{b c \ldots} f e \ldots} 
        + \Gamma^e_{\hphantom{d} f a} 
        T_{b c \ldots}^{\hphantom{b c \ldots} d f \ldots} + \ldots,
\end{split}
\end{equation*}
where the connection coefficients (known as 
\textit{Christoffel symbols} for coordinate bases) are given by 
\begin{equation*}
    \Gamma^a_{\hphantom{a} b c} = \frac{1}{2} g^{a d} 
        (\partial_b g_{c d} + \partial_c g_{b d} - \partial_d g_{b c}).
\end{equation*}

\textbf{Riemann, Ricci and Einstein tensors.}

The Riemann tensor is defined as 
\begin{equation*}
    R^a_{\hphantom{a} b c d} = \partial_c \Gamma^a_{\hphantom{a} b d} 
        - \partial_d \Gamma^a_{\hphantom{a} b c} 
        + \Gamma^a_{\hphantom{a} e c} \Gamma^e_{\hphantom{e} b d} 
        - \Gamma^a_{\hphantom{a} e d} \Gamma^e_{\hphantom{e} b c}.
\end{equation*}
The Ricci tensor is $R_{a b} = R^c_{\hphantom{c} a c b}$ and the Ricci 
scalar is $R = R_a^{\hphantom{a} a}$.

The Einstein tensor is given by 
\begin{equation*}
    G_{a b} = R_{a b} - \frac{1}{2} g_{a b} R.
\end{equation*}

\textbf{Eulerian and Lagrangian perturbations, Lie derivatives.}

Generally, we will reserve the symbols $\delta$ and $\Delta$ to denote the 
Eulerian and Lagrangian perturbations, respectively, of a quantity. The 
perturbations are related by 
\begin{equation*}
    \Delta f = \delta f + \mathcal{L}_\xi f,
\end{equation*}
where $\mathcal{L}_\xi$ is the Lie derivative along the Lagrangian displacement 
vector $\xi^a$.

The Lie derivative along a scalar is simply 
\begin{equation*}
    \mathcal{L}_\xi f = \xi^a \nabla_a f.
\end{equation*}
For an arbitrary tensor, the Lie derivative gives 
\begin{equation*}
\begin{split}
    \mathcal{L}_\xi T_{a b \ldots}^{\hphantom{a b \ldots} c d \ldots} 
        = \xi^e \nabla_e T_{a b \ldots}^{\hphantom{a b \ldots} c d \ldots} 
        &+ T_{e b \ldots}^{\hphantom{e b \ldots} c d \ldots} \nabla_a \xi^e 
        + T_{a e \ldots}^{\hphantom{a e \ldots} c d \ldots} \nabla_b \xi^e 
        + \ldots \\ 
        &- T_{a b \ldots}^{\hphantom{a b \ldots} e d \ldots} \nabla_e \xi^c
        - T_{a b \ldots}^{\hphantom{a b \ldots} c e \ldots} \nabla_e \xi^d 
        - \ldots.
\end{split}
\end{equation*}
Because of the symmetry of the connection coefficients, 
$\Gamma^a_{\hphantom{a} b c} = \Gamma^a_{\hphantom{a} c b}$, one can 
equivalently express the Lie derivatives in terms of partial derivatives. This 
is intuitive as the Lie derivative $\mathcal{L}_\xi$ is with respect to a frame 
that is dragged by $\xi^a$ and should not depend on the precise geometry.


%% file: sections/chapter-1.tex
\chapter{Introduction}

Over a century ago, \citet{1916AnP...354..769E} completed his formulation of 
what was to become one of the most rich and elegant theories in the history of 
physics: the \textit{general theory of relativity}. It was a revolutionary 
theory of gravity that replaced the steadfast (but ultimately fundamentally 
flawed) description provided by Newton two hundred years earlier. Remarkably, 
Einstein showed that gravity should not be interpreted as a force, but instead 
as a curved geometry.

General relativity was highly controversial to begin with. In order to assess 
its validity, Einstein proposed three tests: the precession of the perihelion of 
Mercury, the bending of light in gravitational fields and the gravitational 
redshift. The theory of gravity passed these three classical tests with flying 
colours and, indeed, all subsequent tests 
\citep[for a review, see][]{2014LRR....17....4W}!

What was to become the most contentious prediction of general relativity was the 
existence of \textit{gravitational waves}. Einstein himself was the first to 
show that relativity permitted small wave-like solutions that travelled at the 
speed of light on flat, Minkowskian spacetime 
\citep{1916SPAW.......688E, 1918SPAW.......154E}. However, questions were raised 
whether these waves were physical in nature or if they were geometric artefacts 
that could be transformed away. This issue would not be theoretically resolved 
until the Chapel Hill conference in 1957 \citep{1957RvMP...29..352B}. At the 
conference, Pirani argued that if one considered a pair of freely falling 
particles subjected to a gravitational wave, the particles would experience 
genuine motions with respect to one another 
\citep{Pirani1956, PhysRev.105.1089}. Thus, such waves must be physical. This 
insight inspired Bondi's famous ``sticky bead'' argument \citep{Bondi1957} and 
eventually culminated in the development and construction of kilometre-long 
interferometers designed to measure these minute ripples in spacetime. Indeed, 
the very fact that gravitational waves interact so weakly with matter meant that 
a combination of cataclysmic events and ultra-sensitive instruments were 
required in order to have any hope of detecting them. It was a truly gargantuan 
task.

Eventually, history was made on 14th September 2015. After decades of 
experimental and theoretical toil, the two laser interferometers that 
constitute LIGO heard a faint whisper that lasted a fraction of a second 
\citep{2016PhRvL.116f1102A}. This whisper excellently matched the predictions 
from numerical simulations of the merger of two black holes with masses 
\SI{36}{\solarMass} and \SI{29}{\solarMass}. This event was dubbed 
\textit{GW150914} and with it began an entirely new era of astronomy.

Subsequently, gravitational-wave detectors (including the Virgo instrument that 
joined the network during the second observing run in 2017) have detected a 
plethora of binary-black-hole mergers 
\citep{2019PhRvX...9c1040A, 2021PhRvX..11b1053A}. Within this impressive 
catalogue, there exists a particularly special event that occurred on 17th 
August 2017, \textit{GW170817} \citep{2017PhRvL.119p1101A}.

\begin{figure}[h]
    \centering
	\includegraphics[width=0.8\textwidth]{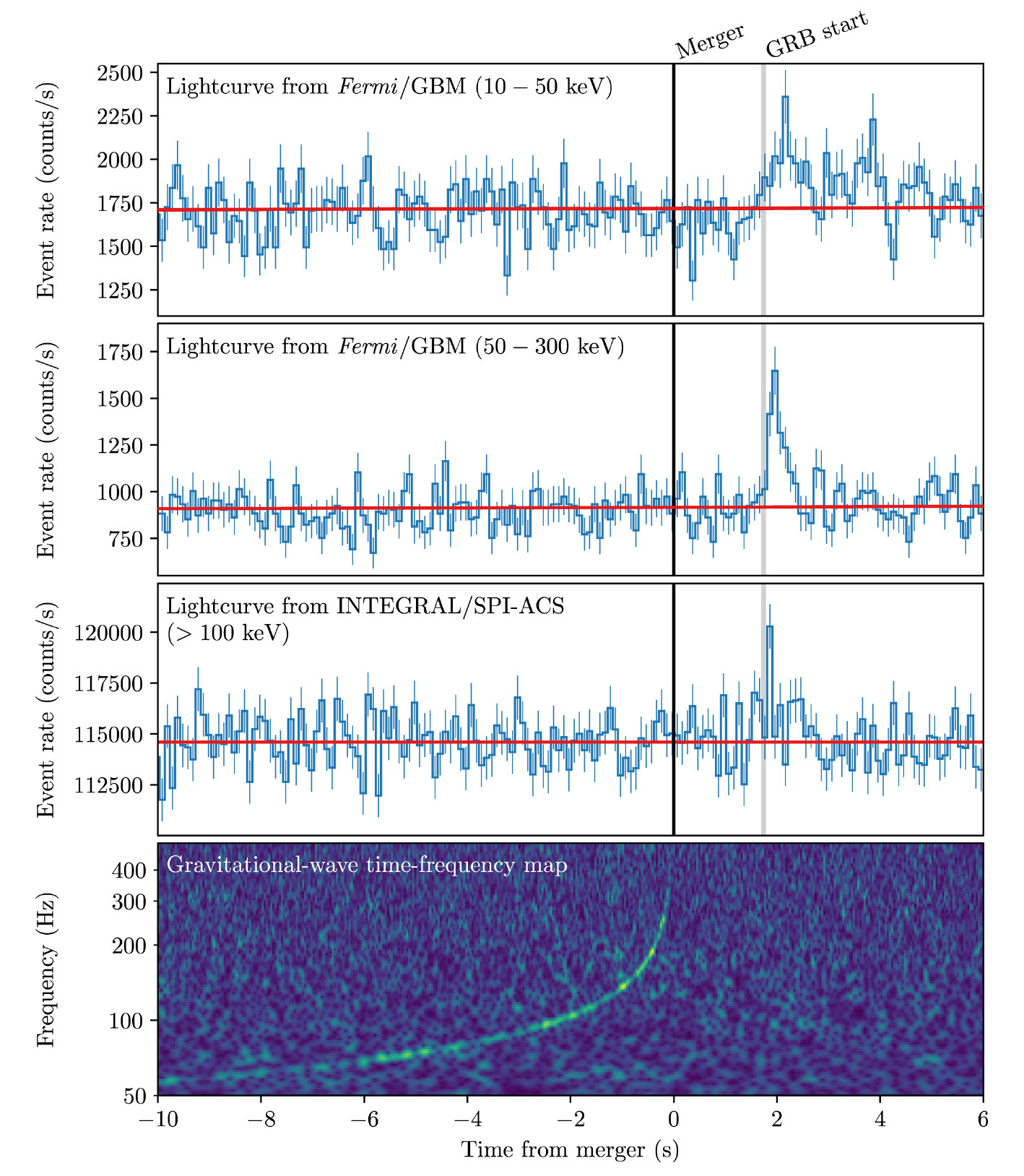}
    \caption[Multi-messenger observation of binary-neutron-star inspiral 
             GW170817]{The joint multi-messenger detection of GW170817 and the 
             short gamma-ray burst GRB 170817A. The top three panels show the 
             gamma-ray observations with different instruments and energy ranges. 
             The bottom panel shows the time-frequency map of the 
             gravitational-wave detection obtained from a coherent combination 
             of the signals in the two LIGO detectors. [Reproduced from 
             \citet{2017ApJ...848L..13A}.]}
	\label{fig:GW170817}
\end{figure}

The signal from GW170817 was quite unlike any of the previous black-hole mergers 
(see Fig.~\ref{fig:GW170817}). The event was observed by the ground-based 
detectors for over a minute and, given how faint the signal was in the Virgo 
instrument, it was reasonably well localised. Soon after this source was 
observed with gravitational waves, a trigger message was circulated around 
electromagnetic observers across the world. What followed was an unprecedented 
symphony of detections all across the electromagnetic spectrum; from 
low-frequency radio waves, all the way up to high-frequency gamma rays. This has 
been heralded as the first \textit{multi-messenger event} and is deservedly the 
most celebrated observation in neutron-star astronomy. One of the many exciting 
results from this multi-messenger event was the confirmation that neutron-star 
mergers are (quite literally) treasure troves: they are the centres of 
production for gold, platinum and uranium 
\citep{2017Natur.551...80K, 2017Natur.551...67P}! In addition, the 
electromagnetic counterpart to the gravitational-wave signal provided a 
\textit{standard siren} that enabled an independent distance measurement to the 
source in order to estimate the Hubble parameter 
\citep{1986Natur.323..310S, 2017Natur.551...85A}. Thus, nicely demonstrating how 
such detections will facilitate precision cosmology. The electromagnetic signals 
also presented further opportunities to test general relativity by investigating 
the speed of gravity, Lorentz invariance and the equivalence principle 
\citep{2017ApJ...848L..13A}, as well as providing constraints on various 
modified theories of gravity 
\citep{2017PhRvL.119y1302C, 2017PhRvL.119y1304E, 2017PhRvL.119y1301B}.

At present, the third LIGO/Virgo observing run has completed. During it, a 
second binary-neutron-star system was observed \citep{2020ApJ...892L...3A}. The 
mass of this system was significantly larger than that of any other known 
neutron-star binary. Additionally, for the first time ever, the 
gravitational-wave instruments detected not one but two neutron star-black hole 
coalescences, just ten days apart from one another \citep{2021ApJ...915L...5A}. 
The question is what more can we learn about these curious objects.

\section{Neutron stars}

Neutron stars are fated to begin their lives in a particularly dramatic fashion: 
in the explosive and cataclysmic furnaces known as supernovae. They are one of 
two possible endpoints of stellar evolution for massive stars (with initial 
masses $\gtrsim \SI{8}{\solarMass}$) -- the other being the remarkably simple 
black holes.%
\footnote{The apparent mathematical simplicity of black holes has culminated in 
the so-called no-hair theorem. The no-hair theorem states (but does not 
rigorously prove) that isolated black holes in equilibrium can be fully 
characterised by three externally observable parameters: their mass $M$, angular 
momentum $J$ and electric charge $Q$. Because there are so few astrophysically 
motivated evolutionary paths that may produce a black hole with a non-zero 
charge, it is common to assume that realistic black holes will have $Q = 0$.}
Massive stars begin their lives on the main sequence and evolve off it as 
subsequent millions of years of nuclear fusion burns the elements in the core 
all the way up to iron. Once the fuel in the core has been exhausted, the star 
starts to collapse, as there is no longer any outward-acting radiative pressure 
and the core is left with just the degeneracy pressure to counteract the inward 
gravitational force. This results in a Type II supernova explosion and leaves a 
compact remnant behind in the form of a neutron star. Assuming the remnant is 
stable (with a mass $\lesssim \SI{3}{\solarMass}$), it will not collapse further 
to form a black hole. 

Neutron stars are extremely compact bodies, with typical masses around 
\SI{1.4}{\solarMass} and radii of the order of \SI{10}{\kilo\metre}, that 
support the star against gravity by neutron-degeneracy pressure. Since they are 
so compact, relativistic effects become important and they are able to spin-up 
to high frequencies ($\sim \SI{700}{\hertz}$). Neutron stars harbour strong 
magnetic fields that can reach magnitudes of $\sim \SI{e15}{\gauss}$. For these 
reasons, neutron stars make excellent, extreme, astrophysical laboratories that 
can probe conditions we could not possibly hope to synthesise in terrestrial 
experiments. They enable us to study physics on a vast range of scales; from 
the very small nuclear interactions that dictate their ultra-dense structure, 
all the way up to the gravitational effects that come into play in neutron-star 
binaries and give rise to gravitational radiation. 

\citet{1934PNAS...20..254B, 1934PNAS...20..259B} were the first to propose the 
existence of neutron stars, motivated by seeking an explanation for the origin 
of supernovae. This came just two years after the neutron had been discovered by 
\citet{Chadwick1932}. Not long after, \citet{1939PhRv...55..364T} and 
\citet{1939PhRv...55..374O} derived the general-relativistic hydrostatic 
equations of equilibrium that could describe non-rotating neutron-star models. 
However, it was not until 1967 that they were originally discovered in the form 
of a radio pulsar with a period of \SI{1.33}{\second} 
\citep{1968Natur.217..709H}. The source of the radio emission initially posed 
quite the puzzle and was humorously designated LGM-1, standing for ``Little 
Green Men''. More discoveries of these \textit{pulsars} (pulsating stars) 
followed in the succeeding years, unveiling the mystery and associating these 
sources with fast rotation rates and strong magnetic fields. These sources were 
identified as neutron stars. 

\begin{figure}[h]
    \centering
	\includegraphics[width=0.8\textwidth]{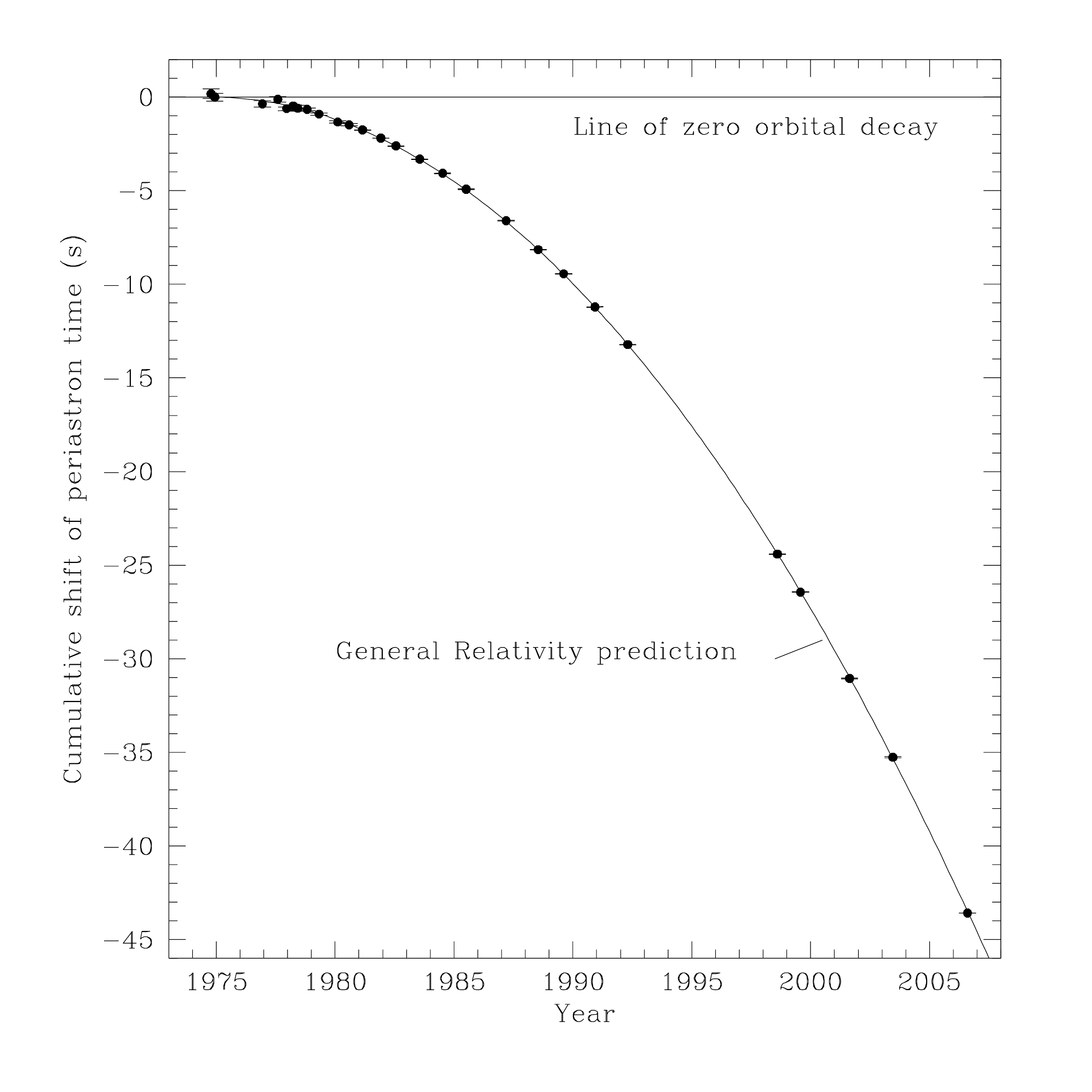}
    \caption[Observations of the Hulse-Taylor binary pulsar PSR B1913+16]{The 
             orbital decay of the Hulse-Taylor binary pulsar PSR B1913+16 
             compared with the prediction of general relativity. The data points record the observed change in the periastron and the solid line 
             indicates the theoretical prediction from general relativity for 
             the binary radiating gravitational waves. [Reproduced from 
             \citet{2010ApJ...722.1030W}.]}
	\label{fig:HulseTaylor}
\end{figure}

Another famous discovery came in 1974. \citet{1975ApJ...195L..51H} observed the 
first pulsar in a neutron-star binary. It was quickly realised that the 
Hulse-Taylor binary could serve as an indirect test for gravitational waves 
\citep{1975ApJ...196L..63W}. The test was a simple one: if the binary was 
emitting gravitational waves, which would take away energy and angular momentum 
from the system, its orbit would shrink. This prediction of general relativity 
was confirmed with excellent accuracy, resulting in Hulse and Taylor being 
awarded the Nobel Prize in 1993, and provided observational support for the 
existence of gravitational radiation. Recent measurements have continued to show 
remarkable agreement with general relativity \citep{2010ApJ...722.1030W}, as 
shown in Fig.~\ref{fig:HulseTaylor}.

Neutron stars are known to exist in accreting binaries. In 1971, the Uhuru 
satellite detected a pulsating, compact X-ray source that was confirmed to be a 
neutron star in a binary with a massive O-type star \citep{1972ApJ...172L..79S}. 
This provided proof of neutron stars existing in close binary systems accreting 
gas from companion stars which gives rise to their characteristic X-ray 
pulsations. Such systems in which the stellar companion has a mass 
$\lesssim M_\odot$ are referred to as low-mass X-ray binaries. Shortly after the 
discovery of the first millisecond pulsar \citep{1982Natur.300..615B}, low-mass 
X-ray binaries became candidate systems for producing these rapidly spinning 
neutron stars. In the same year, \citet{1982Natur.300..728A} and 
\citet{1982CSci...51.1096R} suggested that accretion could transfer angular 
momentum to the neutron star, thus providing a mechanism for the star to spin 
up. This has since been called the \textit{recycling scenario}, owing its name 
to the ability of the mechanism to spin-up and ``revive'' pulsars that have 
stopped their radio emission and entered the \textit{pulsar graveyard}. The 
first accreting millisecond X-ray pulsar to be observed was SAX J1808.4--3658, 
which was detected by the Rossi X-ray Timing Explorer in 1998 
\citep{1998Natur.394..344W}. This provided an exciting confirmation of the 
recycling scenario. 

The biggest unknown in neutron-star astrophysics is the elusive 
\textit{equation of state}. The equation of state encodes the detailed 
microscopic interactions that take place at ultra-high densities. Such a 
description is crucial in order build astrophysically correct neutron-star 
models. A plethora of candidates have been proposed and used to generate stellar 
models. Traditionally, one has constrained the equation of state by observing 
two macroscopic neutron-star parameters: the mass and radius. Promisingly, with 
the advent of gravitational-wave astronomy, we can now examine neutron-star 
interiors using gravitational radiation.

\begin{figure}[h]
    \centering
	\includegraphics[width=0.8\textwidth]{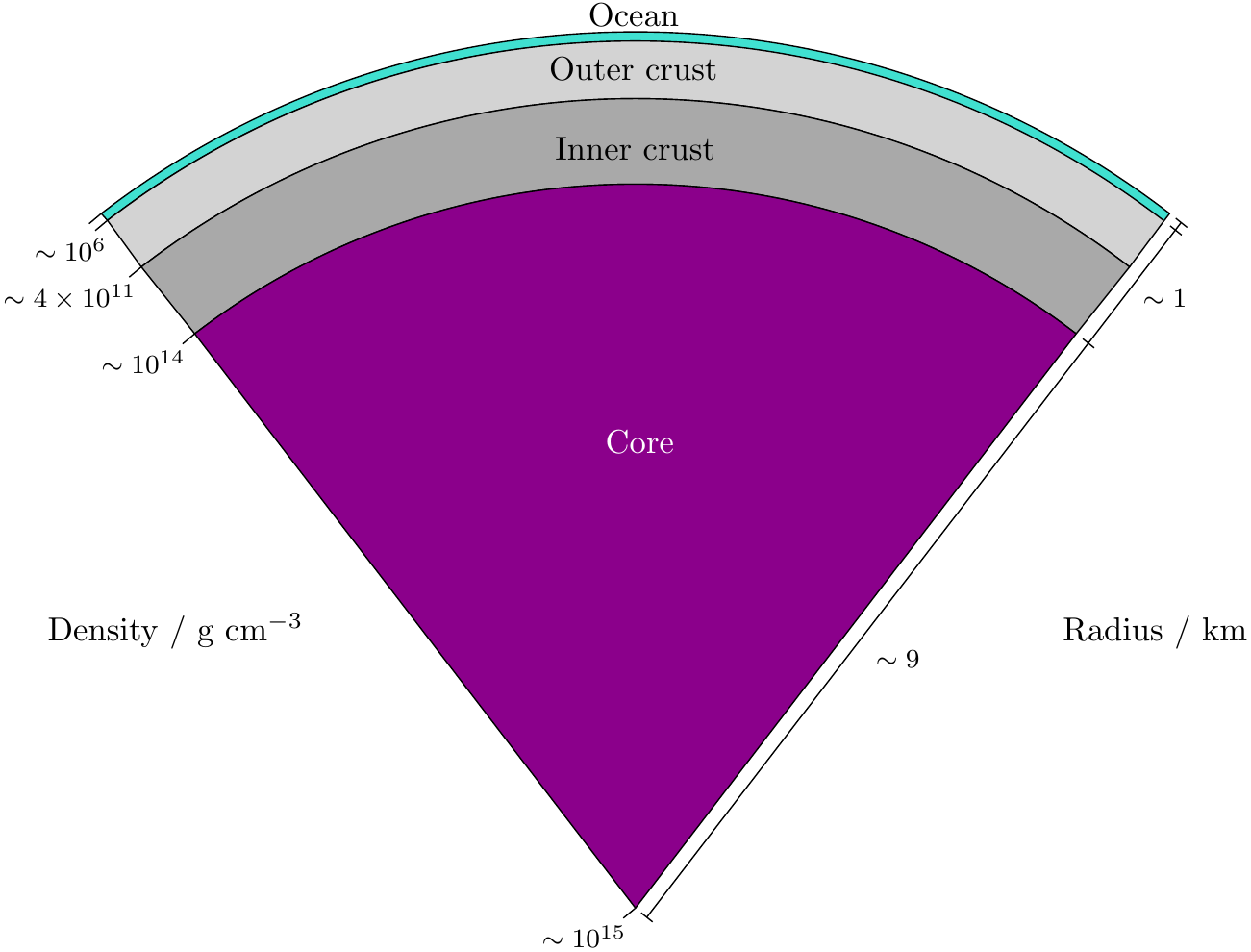}
    \caption[Cross section of a neutron star]{Illustration of the interior 
             layers in a neutron star indicating (from low to high density; from 
             the surface to the centre) the ocean, outer crust, inner crust and 
             core.}
	\label{fig:CrossSection}
\end{figure}

The interior of a neutron star has a series of distinct regions from the surface 
to the centre, as shown in Fig.~\ref{fig:CrossSection}. As one progresses to the 
core of the star, the density increases monotonically and (in some sense) so 
does our uncertainty of the physics. The outermost layer of a mature, isolated 
neutron star is a thin \textit{ocean}; at most a few metres thick. Beneath the 
ocean, it becomes energetically favourable for nuclei to capture electrons and 
undergo inverse beta decay. This region is the solid \textit{outer crust} and is 
composed of nuclei. At a critical density, around 
\SI{4e11}{\gram\per\centi\metre\cubed}, one reaches \textit{neutron drip} -- so 
called because this is the point neutrons begin to drip out of the nuclei to 
form a superfluid that permeates the solid lattice. This marks the top of the 
\textit{inner crust} and is where our understanding becomes increasingly 
challenged.

The inner crust is expected to contain remarkable nuclear structures that occur 
as the matter becomes frustrated and undergoes a series of phase transitions. 
These are known as the \textit{nuclear-pasta phases}. At higher densities still, 
one enters the \textit{core} at $\sim \SI{1.7e14}{\gram\per\centi\metre\cubed}$. 
This region is primarily a neutron superfluid. However, in the very depths of 
the core -- a few times the nuclear-saturation density, 
$\rho_\text{sat} \approx \SI{2.8e14}{\gram\per\centi\metre\cubed}$ -- various 
exotic states of matter may be present, such as deconfined quarks.

\section{Gravitational-wave scenarios}

In order to generate gravitational radiation, massive bodies must be accelerated 
in curved spacetime. For this reason, neutron stars have long been considered 
promising candidate sources of gravitational waves 
\citep{1978MNRAS.184..501P, 1984ApJ...278..345W}. This is owed to their extreme 
compactness (rivalled only by black holes) and their role in some of the most 
cataclysmic events in the Universe.

There are a variety of probable mechanisms through which neutron stars can 
radiate gravitational waves. These include binary inspiral and merger 
\citep{2010CQGra..27q3001A}, various modes of oscillation 
\citep[and their corresponding instabilities;][]
{1998ApJ...502..708A, 1999ApJ...516..307A} and rotating neutron stars that 
are deformed away from perfect sphericity \citep{1998ApJ...501L..89B}. Indeed, 
the exciting, recent gravitational-wave detections of binary neutron stars has 
ignited a revitalised effort in searching for gravitational waves from other 
neutron-star scenarios.

\subsection{Deformed, rotating neutron stars}

Rotating neutron stars with non-axisymmetric deformations are natural 
gravitational-wave emitters. Indeed, provided the deformation is misaligned with 
the rotation axis, such stars will continuously radiate at twice their spin 
frequency.%
\footnote{In general, spinning neutron stars will precess. When precession is 
involved, the gravitational waves will be radiated at a frequency equal to the 
sum of the spin frequency and the frequency of precession, as well as twice this 
value.}
The associated (quadrupole) deformations are commonly known as 
\textit{mountains}. Rapidly rotating neutron stars have enjoyed the attention of 
a large number of searches using gravitational-wave data. These searches have 
been split into two strategies: looking for evidence of gravitational radiation 
in specific pulsars 
\citep{2004PhRvD..69h2004A, 2005PhRvL..94r1103A, 2007PhRvD..76d2001A, 
2008ApJ...683L..45A, 2010ApJ...713..671A, 2017PhRvD..95l2003A, 
2017PhRvD..96l2006A, 2017ApJ...839...12A, 2017ApJ...847...47A, 
2018PhRvL.120c1104A, 2019PhRvD..99l2002A, 2019PhRvD.100l2002A, 
2019ApJ...879...10A, 2020ApJ...902L..21A, 2021ApJ...913L..27A, 
2011PhRvD..83d2001A, 2011ApJ...737...93A, 2014ApJ...785..119A, 
2015PhRvD..91b2004A, 2015PhRvD..91f2008A, 2021ApJ...906L..14Z} 
and wide-parameter surveys for unobserved sources 
\citep{2005PhRvD..72j2004A, 2007PhRvD..76h2001A, 2008PhRvD..77b2001A, 
2009PhRvD..79b2001A, 2016PhRvD..94d2002A, 2017PhRvD..96f2002A, 
2018PhRvD..97j2003A, 2019PhRvD.100b4004A, 2012PhRvD..85b2001A, 
2013PhRvD..87d2001A, 2020PhRvL.125q1101D, 2021PhRvD.103f3019D, 
2021ApJ...909...79S}.

Assuming torque balance, the brightest X-ray sources should be the loudest in 
gravitational-wave emission. Hence, Scorpius X-1 has long been considered a 
potential candidate for gravitational-wave emission and has been the focus of a 
number of targeted gravitational-wave searches 
\citep{2007PhRvD..76h2001A, 2017PhRvD..95l2003A, 2017ApJ...847...47A, 
2015PhRvD..91f2008A, 2019PhRvD.100l2002A}. These searches, while only yielding 
upper limits so far, have helped develop and improve data-analysis techniques 
that can be used in the future, as the sensitivities of gravitational-wave 
detectors are improved. Detecting gravitational waves from rotating neutron 
stars will be a challenge, since only a few of the most rapidly accreting 
neutron stars are expected to be detectable with the current generation of 
gravitational-wave detectors \citep{2008MNRAS.389..839W}. The main limiting 
factors in detecting gravitational waves from these systems are the precision 
with which the spin and the orbital parameters are measured. If these are not 
well known, then it becomes computationally very expensive to run 
gravitational-wave searches, since the searches need to run over a large 
parameter space. 

\subsection{Binary neutron stars}

Binaries are also intuitive sources for gravitational radiation. By virtue of 
orbiting a common centre of mass, the two components of a binary generate a 
time-dependent mass asymmetry with respect to the orbital axis. Both binary 
black holes and binary neutron stars have been successfully detected using 
gravitational waves. Indeed, it has almost become routine for the LIGO/Virgo 
detectors to observe binary-black-hole mergers.

As gravitational waves are radiated from a binary, they carry energy and angular 
momentum away. This means the orbit of the binary shrinks and thus, provided 
enough time within the lifetime of the Universe, the fate of all binaries is to 
coalesce and merge. Because these binaries are expected to be less massive than 
the vast majority of black-hole binaries, the signal associated with the 
inspiral lasts for a longer time in the detectability frequency band of 
ground-based detectors. This signal includes information about the extended 
nature of neutron stars, known as \textit{finite-size effects}, where the 
dominant contribution is due to the tidal deformation that each star's 
gravitational field induces on the other. Promisingly, this information can be 
used to provide model-independent constraints on the equation of state. This is 
quite an appealing prospect since it would seem that there is no current 
model-independent technique to measure the neutron-star radius. Various radius 
estimates have been made using X-ray spectroscopy from quiescent neutron stars 
\citep{2018MNRAS.476..421S}, thermonuclear X-ray bursts 
\citep{2010ApJ...722...33S, 2016ApJ...820...28O, 2017A&A...608A..31N} and 
accretion-powered millisecond pulsars \citep{2018A&A...618A.161S}. However, all 
these methods suffer from being susceptible to systematic errors 
\citep{2013arXiv1312.0029M, 2016EPJA...52...63M}. With the detection of 
GW170817, studies have been able to use the lack of an observed tidal imprint to 
obtain constraints on the equation of state using gravitational waves 
\citep{2017PhRvL.119p1101A, 2017ApJ...850L..34B, 2018PhRvL.120q2703A, 
2018ApJ...857L..23R, 2018PhRvL.120z1103M, 2018PhRvL.121i1102D, 
2018PhRvL.121p1101A, 2019PhRvX...9a1001A}.

\section{Outlook}

This thesis is concerned with the structure and dynamics of deformed neutron 
stars and focuses on two scenarios in which they can emit gravitational 
radiation: rotating neutron stars with mountains and tidally deformed binaries.

In Chap.~\ref{ch:Structure}, we delve into the fundamental physics that will be 
needed for the rest of the thesis. We discuss how the structure of neutron 
stars is formally described in Newtonian and relativistic gravity and how 
non-spherical stellar models are built. We also describe some basic 
gravitational-wave theory. In Chap.~\ref{ch:PopulationSynthesis}, we study the 
spin evolution of accreting neutron stars -- systems which have long been 
considered potential sources of gravitational waves. In particular, we explore 
whether gravitational waves are likely to play an important role in their 
dynamics by considering different torques acting on the stars and evolving the 
systems through their lifetimes. Should such systems host non-axisymmetric 
mountains, they will naturally radiate gravitational waves. Thus, we go on to 
analyse how mountains are modelled in Newtonian gravity in 
Chap.~\ref{ch:Mountains}. We describe how non-spherical stellar models with and 
without elastic crusts can be constructed. By considering previous studies of 
this problem, we show that the deforming force that sustains the mountains is a 
crucial component and consider some simple examples. We take this problem one 
step closer to realism by generalising the calculation to general relativity in 
Chap.~\ref{ch:RelativisticMountains}. We derive the relativistic equations of 
structure for non-spherical stars subjected to a deforming force and consider 
how the equation of state affects the size of the mountains. In 
Chap.~\ref{ch:Tides}, we use our derived relativistic formalism to study the 
relevance of the elastic crust in tidally deformed neutron stars. We calculate 
the tidal deformabilities for realistic stellar models and also consider when 
the crust may fracture during the inspiral. We conclude and provide an outlook 
on future directions in Chap.~\ref{ch:Conclusions}.

Supplemental material is provided in the appendices. In 
Appendix~\ref{app:Numerical}, we discuss solving the equations of stellar 
structure for non-rotating, spherical fluid stars. In 
Appendix~\ref{app:Harmonics}, we summarise the properties of the spherical 
harmonics. Appendix~\ref{app:NumericalScheme} is dedicated to the numerical 
techniques used for solving the perturbation equations. We obtain the interface 
conditions for the perturbation functions in relativity in 
Appendix~\ref{app:Interface}.


%% file: sections/chapter-2.tex
\chapter{The structure of neutron stars}
\label{ch:Structure}

This chapter serves as an introduction to some fundamental aspects of 
neutron-star structure. Throughout this chapter, we will work in both arenas 
of gravitation: Newtonian physics and general relativity. We begin with some 
thermodynamics in Sec.~\ref{sec:EOS}, in order to motivate and introduce the 
infamous neutron-star equation of state. We shall see how the equation of state 
plays an important role in the interior structure of neutron stars as we derive 
the equations of structure for non-rotating, fluid stars in 
Sec.~\ref{sec:StellarStructure}. To construct more complex and realistic stellar 
models later in the thesis, we provide a brief introduction to perturbation 
theory in Sec.~\ref{sec:Perturbations}. Armed with an understanding of 
perturbation theory, in Sec.~\ref{sec:Multipoles}, we go on to derive and 
analyse the gravitational multipole moments that describe the shape of a star 
and are crucial in generating gravitational radiation. At the end of the 
chapter, Sec.~\ref{sec:GravitationalWaves}, we provide a brief introduction to 
how gravitational waves arise in relativity and how they are produced.

\section{The equation of state}
\label{sec:EOS}

It is well established that stellar material is excellently approximated as a 
perfect fluid; neutron stars are no exception to this. (Although this treatment 
needs to be refined when one considers additional physics such as superfluidity 
and magnetic fields. Indeed, we will include an elastic crust from 
Chap.~\ref{ch:Mountains} onwards.) Fluids consist of many fluid elements and 
each fluid element is a collection of particles.%
\footnote{Each fluid element comprises a large number of particles that all 
undergo individual motions on a scale associated with the mean-free path. The 
central assumption behind fluids is that the size of a fluid element is much 
greater than the mean-free path of the particles. At this scale, the fluid is 
a continuum.}
In order to understand the many-particle interactions, it is natural to consider 
the thermodynamics of the system. In this way, one need not concern themselves 
with the behaviour of individual particles, but instead track the parameters of 
the fluid elements. 

The natural starting point is the first law of thermodynamics, which states that 
the change in energy $dE$ of a many-particle-species system is given by 
\begin{equation}
    dE = T dS - p dV + \sum_\text{x} \mu_\text{x} dN_\text{x},
    \label{eq:FirstLawFull}
\end{equation}
where $T$ is the temperature, $dS$ is the change in entropy, $p$ is the 
pressure, $dV$ is the change in volume and $\mu_\text{x}$ and $dN_\text{x}$ are 
the chemical potential%
\footnote{This should not be confused with the magnetic moment $\mu$ in 
Chap.~\ref{ch:PopulationSynthesis}.}
and change in particle number, respectively, of a species $\text{x}$. The first 
term, $T dS$, may be identified as the heat into the system; the second, 
$- p dV$, corresponds to the work done on the system; and the final term, 
$\mu_\text{x} dN_\text{x}$, is associated with the change in particle number of 
species $\text{x}$. Thus, there is an equation of state for the system 
$E(S, V, \{N_\text{x}\})$ from which one can obtain the following: 
\begin{equation}
    T = \left( \frac{\partial E}{\partial S} \right)_{V, \{N_\text{x}\}}, \quad
        p = - \left( \frac{\partial E}{\partial V} \right)_{S, \{N_\text{x}\}}, 
        \quad 
        \mu_\text{x} = \left( \frac{\partial E}{\partial N_\text{x}} 
        \right)_{S, V, \{N_\text{y}\}},
\end{equation}
where $\{N_\text{y}\}$ is the set of particle numbers excluding $N_\text{x}$. 
Here, the subscripted notation next to a bracket makes explicit the variables 
that are being treated as constants in the partial derivative. One describes the 
variables $E$, $S$, $V$ and $N_\text{x}$ as \textit{extensive} variables, since 
any variation in $S$, $V$ and $N_\text{x}$ changes the energy of the system. 
Conversely, variables $T$, $p$ and $\mu_\text{x}$ are called \textit{intensive} 
variables. We may call $E(S, V, \{N_\text{x}\})$ the equation of state since it 
is a thermodynamic relation from which one can obtain the other state variables 
through (\ref{eq:FirstLawFull}).

From (\ref{eq:FirstLawFull}), one can derive the Euler relation 
\citep{2021LRR....24....3A} 
\begin{equation}
    E = T S - p V + \sum_\text{x} \mu_\text{x} N_\text{x}.
    \label{eq:EulerRelationFull}
\end{equation}
The Euler relation (\ref{eq:EulerRelationFull}) is essentially the integrated 
form of the first law (\ref{eq:FirstLawFull}). It is convenient to write 
(\ref{eq:EulerRelationFull}) in terms of densities, 
\begin{equation}
    \varepsilon = n T s - p + n \sum_\text{x} \mu_\text{x} Y_\text{x},
    \label{eq:EulerRelation}
\end{equation}
where $\varepsilon = E / V$ is the total energy density, $s = S / N$ is the 
entropy per particle, $Y_\text{x} = N_\text{x} / N$ is the abundance of species 
$\text{x}$, $n = N / V$ is the total number density and $N$ is the total number 
of particles. We assume that the total number of particles in the system is 
conserved, 
\begin{equation}
    N = \sum_\text{x} N_\text{x} = \text{const},
\end{equation}
so we can write the first law (\ref{eq:FirstLawFull}) as
\begin{equation}
    d\varepsilon = n T ds + \frac{\varepsilon + p}{n} dn 
        + n \sum_\text{x} \mu_\text{x} dY_\text{x},
    \label{eq:FirstLaw}
\end{equation}
with 
\begin{equation}
    n T = \left( \frac{\partial \varepsilon}{\partial s} 
        \right)_{n, \{Y_\text{x}\}}, \quad 
        \frac{\varepsilon + p}{n} = 
        \left( \frac{\partial \varepsilon}{\partial n} 
        \right)_{s, \{Y_\text{x}\}}, \quad 
        n \mu_\text{x} = \left( \frac{\partial \varepsilon}{\partial Y_\text{x}} 
        \right)_{s, n, \{Y_\text{y}\}}.
    \label{eq:Intensive1}
\end{equation}
Therefore, we have $\varepsilon = \varepsilon(s, n, \{Y_\text{x}\})$.%
\footnote{It should be noted that the particle abundances are not independent, 
$\sum_\text{x} Y_\text{x} = 1$. This illustrates a feature of extensive systems: 
the number of variables required to describe the system has been reduced by one 
in working with densities and removing $V$.}
We may identify the enthalpy from (\ref{eq:FirstLaw}) as 
$h = (\varepsilon + p) / n$. Putting this into (\ref{eq:EulerRelation}), we have 
\begin{equation}
    p = - \varepsilon + \left( \frac{\partial \varepsilon}{\partial s} 
        \right)_{n, \{Y_\text{x}\}} s 
        + \sum_\text{x} \left( \frac{\partial \varepsilon}{\partial Y_\text{x}} 
        \right)_{s, n, \{Y_\text{y}\}} Y_\text{x}.
    \label{eq:Intensive2}
\end{equation}
The fact we can obtain all the intensive parameters of the system from 
$\varepsilon(s, n, \{Y_\text{x}\})$, via Eqs.~(\ref{eq:Intensive1}) and 
(\ref{eq:Intensive2}), justifies naming it the equation of state for our system.

If one assumes that the system is in chemical equilibrium with respect to 
microscopic reactions that involve the different particle species, over reaction 
timescales we have 
\begin{equation}
    \sum_\text{x} \mu_\text{x} dY_\text{x} = 0
\end{equation}
and (\ref{eq:FirstLaw}) reduces to 
\begin{equation}
    d\varepsilon = n T ds + \frac{\varepsilon + p}{n} dn.
    \label{eq:FirstLawSimple}
\end{equation}
This assumption applies well to equilibrium neutron stars, where diffusive 
processes may be neglected. Furthermore, since the lepton contribution to the 
total energy is small relative to the baryons, the total number of particles can 
be assumed to simply be the total number of baryons. This means that the energy 
density depends solely on the entropy per baryon $s$ and baryon-number density 
$n$, $\varepsilon = \varepsilon(s, n)$.

Mature neutron stars are cold, so one can assume that $T = 0$. Thus, the 
equation of state is reduced by one parameter, $\varepsilon = \varepsilon(n)$. 
Such an equation of state is termed \textit{barotropic}. However, since the 
thermodynamic quantities are all related via the first law, we are free to work 
with a different quantity. For cold stars, 
\begin{equation}
    p(n) = - \varepsilon(n) + \mu(n) n,
\end{equation}
where $\mu(n) = d\varepsilon / dn = (\varepsilon + p) / n$. It is, therefore, 
common in neutron-star astrophysics to refer to $p = p(n)$ [or, equivalently, 
$p = p(\varepsilon)$] as the cold-matter equation of state.

It should be emphasised that a single fluid pressure-density relation does not 
entirely tell the full story. Neutron stars exhibit different phases of matter. 
For example, the crust of a neutron star is solid and its modelling will require 
an understanding of the shear stresses. In particular, it is common to treat the 
crust as an elastic solid, which requires a shear modulus (as we shall consider 
from Chap.~\ref{ch:Mountains} onwards). We will now see how the equation of 
state is a crucial ingredient in producing spherical, fluid bodies.

\section{Equations of stellar structure}
\label{sec:StellarStructure}

In order to construct models of the stellar interior, one uses the equations of 
stellar structure. In this section, we derive the Newtonian and relativistic 
equations of stellar structure for non-rotating, spherically symmetric, fluid 
stars.

\subsection{Newtonian stars}

Spherical stars are modelled as perfect fluids. A perfect fluid has, by 
definition, no viscosity or heat flow. It is also isotropic, so the magnitude of 
the pressure is the same in all directions. We will neglect temperature (or, at 
the very least, assume our stars are cold), so we assume that the equation of 
state is barotropic. A barotropic perfect-fluid configuration, with mass density 
$\rho$, isotropic pressure $p$ and velocity $v^i$, is a solution 
$(\rho, p, v^i)$ to the following standard fluid equations 
\citep[for a modern reference on fluid mechanics, and other aspects of classical 
physics, the reader is referred to][]{2017mcp..book.....T}: 
\begin{gather}
    \partial_t \rho + \nabla_i (\rho v^i) = 0, 
    \label{eq:Continuity}\\
    \rho (\partial_t + v^j \nabla_j) v_i = - \nabla_i p - \rho \nabla_i \Phi, 
    \label{eq:Euler}\\
    p = p(\rho) 
    \label{eq:EOS}
\end{gather}
and the gravitational potential $\Phi$ is provided by 
\textit{Poisson's equation}, 
\begin{equation}
    \nabla^2 \Phi = 4 \pi G \rho. 
    \label{eq:Poissons}
\end{equation}

Equation~(\ref{eq:Continuity}) is the differential law of mass conservation, 
also known as the \textit{continuity equation}. In the absence of a source term 
on the right-hand side, the continuity equation relates the rate of change 
of mass in a fixed volume to the flux of mass into that volume. The mass density 
is simply related to baryon-number density by $\rho = m_\text{b} n$, where 
$m_\text{b}$ is the baryon mass. Therefore, the continuity equation can also be 
interpreted as the Newtonian law of baryon conservation, 
$\partial_t n + \nabla_i (n v^i) = 0$.

The equation of motion for a perfect fluid is the \textit{Euler equation} 
(\ref{eq:Euler}). This is the differential law of momentum conservation. The 
Euler equation provides the following physical interpretation. The acceleration 
felt by the fluid, given by the convective derivative of the velocity 
$(\partial_t + v^j \nabla_j) v^i$ has two contributions: the gravitational 
field and the pressure gradient. 

Poisson's equation (\ref{eq:Poissons}) is the Newtonian field equation for 
gravity. It accounts for the coupling between the gravitational field and the 
matter that sources it.

Finally, we use the equation of state (\ref{eq:EOS}) to close this system of 
equations. These expressions have been written in a fully covariant way, which 
means that one has the freedom to work in whichever inertial reference frame 
they wish. 

Non-rotating equilibrium stars are self-gravitating fluid spheres. Since they 
are in equilibrium and static, the time derivatives and velocity vanish.%
\footnote{It is natural to adopt a reference frame that is at rest relative to 
the fluid.}
This means the continuity equation~(\ref{eq:Continuity}) is trivially satisfied 
and we obtain the equation of hydrostatic equilibrium from (\ref{eq:Euler}), 
\begin{equation}
    \nabla_i p = - \rho \nabla_i \Phi. 
    \label{eq:HydrostaticEquilibrium}
\end{equation}
This shows how a non-rotating star is supported against gravity by the fluid 
pressure gradient.

Since the star is static, it must be spherically symmetric and all quantities 
will depend solely on the radius from the centre. The continuity equation 
(\ref{eq:Continuity}) indicates the presence of a conserved quantity with 
density $\rho$. This quantity is the total mass 
\begin{equation}
    M \equiv \int_0^R 4 \pi r^2 \rho(r) dr,
    \label{eq:TotalMass}
\end{equation}
where $R$ is the stellar radius. (Equivalently, the total number of baryons is 
conserved.) Therefore, the mass enclosed in radius $r$, $m(r)$, is given by 
\begin{subequations}\label{eqs:NewtonianStellarStructure}
\begin{equation}
    \frac{dm}{dr} = 4 \pi r^2 \rho.
    \label{eq:dm_dr}
\end{equation}
Equation~(\ref{eq:Poissons}) integrates to 
\begin{equation}
    \frac{d\Phi}{dr} = \frac{G m}{r^2}. 
    \label{eq:dPhi_dr}
\end{equation}
Equation~(\ref{eq:HydrostaticEquilibrium}) becomes 
\begin{equation}
    \frac{dp}{dr} = - \rho \frac{d\Phi}{dr} = - \frac{G m \rho}{r^2}. 
    \label{eq:dp_dr}
\end{equation}
\end{subequations}
Equations~(\ref{eqs:NewtonianStellarStructure}) constitute the 
\textit{Newtonian equations of stellar structure}. They describe a non-rotating, 
fluid star with $(\rho, p, \Phi)$, supplemented by an equation of state 
(\ref{eq:EOS}). 

With an equation of state specified, one solves the stellar-structure equations  
by imposing the following boundary conditions: 
\begin{description}
    \item[At the centre,] the enclosed mass must vanish, $m(0) = 0$, and the 
        central density (or, equivalently, the central pressure) may be freely 
        specified, $\rho(0) = \rho_\text{c}$ [$p(0) = p_\text{c}$]. 
    \item[At the surface,] the pressure must go to zero, $p(R) = 0$, and the 
        internal potential needs to match to the exterior, 
        $\Phi(R) = - G M / R$.%
        \footnote{Although, the total mass is calculated as an integral of the 
        density over its total volume (\ref{eq:TotalMass}), we choose to define 
        the total mass from the exterior potential. Conceptually, this is a 
        subtle distinction that enables us to also use $m$ and $M$ to denote the 
        analogous quantities in the relativistic case, as we shall elucidate 
        below. This definition is commonly distinguished as the total 
        \textit{gravitational mass} because it is read off from the 
        gravitational potential.} 
\end{description}

Since the central density may be chosen arbitrarily, the structure equations 
admit a one-parameter family of solutions that depend on the equation of state 
to close the system. For a given equation of state, varying the central density 
provides a variety of equilibrium configurations with different masses and 
radii, $[M(\rho_\text{c}), R(\rho_\text{c})]$. This results in the 
\textit{mass-radius diagram} for that equation of state, $M = M(R)$, which is a 
common diagnostic to differentiate between equations of state (see 
Fig.~\ref{fig:M-Rexample}). We will discuss the mass-radius diagram in more 
detail below for the relativistic structure equations. 

It is convenient to note that (\ref{eq:dPhi_dr}) decouples from the other 
equations. Therefore, one can obtain the quantities $\rho(r)$, $p(r)$, $m(r)$ 
and $R$ without needing to compute the interior gravitational potential 
$\Phi(r)$.

Accurate descriptions of neutron-star interiors will rely on calculations in 
full general relativity. However, Newtonian models can still be very useful. 
Usually, such calculations are easier to conduct and the physics is simpler to 
understand. A convenient simplification in generating spherical Newtonian stars 
is to assume the equation of state is polytropic, that is 
\begin{equation}
    p(\rho) = K \rho^{1 + 1/n},
    \label{eq:Polytropic}
\end{equation}
where $K$ and $n$ are real, positive constants.%
\footnote{One should be careful to not confuse the index $n$ with the 
baryon-number density.}
In fact, because the relativistic corrections to neutron-star models are more 
significant than variations in the description of nuclear matter, polytropic 
equations of state are quite suitable for Newtonian calculations of neutron 
stars. Polytropes are described in Appendix~\ref{appsec:Polytropes}.

\subsection{Relativistic stars}

At this point, we will depart from the above Newtonian equations of structure 
and look for a relativistic description. As we will see, the procedure we need 
to follow in order to derive the relativistic equations is quite similar in 
spirit. We will need to use the relativistic equivalents of Poisson's equation 
(\ref{eq:Poissons}) and the equation of hydrostatic equilibrium 
(\ref{eq:HydrostaticEquilibrium}) in order to construct static relativistic 
stars. For a comprehensive description of general relativity, the reader is 
referred to \citet{1973grav.book.....M}, \citet{1985fcgr.book.....S} and 
\citet{2003gieg.book.....H}.

In general relativity, space and time are unified to build a four-dimensional 
continuum, known as \textit{spacetime}, and gravity manifests itself as a 
geometric property of this continuum, referred to as \textit{curvature}. Thus, 
space need not be flat as in Newtonian gravity and space and time are 
intricately entwined. Indeed, different observers at different points in 
spacetime will no longer agree upon where and when events happen. Any source of 
matter content in a spacetime deforms the fabric and gives rise to curvature. 

How the matter shapes the geometry is encoded in the relativistic field 
equations, known as the \textit{Einstein equations}. These are given by 
\begin{equation}
    G_a^{\hphantom{a} b} = 8 \pi T_a^{\hphantom{a} b},
    \label{eq:EinsteinEquations}
\end{equation}
where $G_a^{\hphantom{a} b}$ is the Einstein tensor and $T_a^{\hphantom{a} b}$ 
is the stress-energy tensor corresponding to the matter content. 
Equation~(\ref{eq:EinsteinEquations}) is the relativistic analogue of the 
Newtonian field equation (\ref{eq:Poissons}). One should not be fooled by the 
apparent simplicity of (\ref{eq:EinsteinEquations}). In its true form, it is a 
highly non-linear, coupled system of ten second-order partial differential 
equations. To solve this system, one must simultaneously calculate the metric, 
$g_{a b}$, and the stress-energy tensor, $T_{a b}$ (that explicitly depends on 
the metric). Without significant symmetries to take advantage of, it can be very 
difficult task. The Bianchi identities imply 
\begin{equation}
    \nabla_b T_a^{\hphantom{a} b} = 0,
    \label{eq:ConservationEnergyMomentum}
\end{equation}
which is the relativistic conservation law for energy and momentum. It should be 
emphasised that Eqs.~(\ref{eq:EinsteinEquations}) and 
(\ref{eq:ConservationEnergyMomentum}) are not independent. Indeed, the 
information in (\ref{eq:ConservationEnergyMomentum}) is also contained in 
(\ref{eq:EinsteinEquations}). However, due to the challenge in evaluating 
(\ref{eq:EinsteinEquations}), it is occasionally convenient to also use 
(\ref{eq:ConservationEnergyMomentum}) in order to simplify the calculation.

Before we concern ourselves with the gravity of the situation, we will begin 
with some relativistic fluid dynamics. For a perfect fluid -- that is, again, 
free of viscosity and heat -- the stress-energy tensor is 
\begin{equation}
    T_a^{\hphantom{a} b} = (\varepsilon + p) u_a u^b + p \, 
        \delta_a^{\hphantom{a} b} 
        = \varepsilon \, u_a u^b + p \bot_a^{\hphantom{a} b},
    \label{eq:StressEnergyTensor}
\end{equation}
where $\varepsilon$ and $p$ correspond to the energy density%
\footnote{This is an important difference to Newtonian gravity, where the 
corresponding quantity is the (rest-)mass density, $\rho$. The energy density 
is made up of the rest mass-energy and internal energy densities. This is a 
manifestation of the famous mass-energy equivalence.}
and isotropic pressure, respectively, as measured by an observer co-moving with 
the fluid with four-velocity $u^a$. Here, we have introduced the projection 
operator orthogonal to the observer, 
$\bot_a^{\hphantom{a} b} \equiv u_a u^b + \delta_a^{\hphantom{a} b}$. By 
considering projections of (\ref{eq:ConservationEnergyMomentum}) along $u^a$ and 
$\bot_b^{\hphantom{b} a}$, one can show that a relativistic, barotropic perfect 
fluid is a solution $(\varepsilon, p, u^a)$ to the following system 
\citep{2021LRR....24....3A}:
\begin{gather}
    \nabla_a (n u^a) = 0, 
    \label{eq:RelativisticContinuity}\\
    (\varepsilon + p) u^b \nabla_b u_a = - \bot_a^{\hphantom{a} b} \nabla_b p,
    \label{eq:RelativisticEuler}\\
    p = p(\varepsilon),
    \label{eq:RelativisticEOS}
\end{gather}
where the geometry of the spacetime is given by the Einstein 
equations~(\ref{eq:EinsteinEquations}).

Equation~(\ref{eq:RelativisticContinuity}) is the conservation law of baryons 
and is the relativistic generalisation of the continuity equation 
(\ref{eq:Continuity}). In both Newtonian and relativistic gravity, the total 
number of baryons must remain constant. For this reason, it is common to 
distinguish $\rho = m_\text{b} n$ as the \textit{baryon-mass density} or 
\textit{rest-mass density} to bridge the two descriptions of gravity.

Equation~(\ref{eq:RelativisticEuler}) is the relativistic Euler equation 
[\textit{cf.} (\ref{eq:Euler})]. We note that $u^b \nabla_b u^a$ corresponds to 
the fluid four-acceleration and takes into account the spacetime geometry. From 
(\ref{eq:RelativisticEuler}) we may identify the inertial mass per unit volume 
in relativity as $(\varepsilon + p)$; this replaces $\rho$ from the Newtonian 
case.

In order to proceed, we must consider the geometry and choose a frame of 
reference. For the interior of a static, spherical star it is appropriate to use 
the line element 
\begin{equation}
    ds^2 = g_{a b} dx^a dx^b = - e^\nu dt^2 + e^\lambda dr^2 + r^2 (d\theta^2 
        + \sin^2 \theta \, d\phi^2),
    \label{eq:StaticMetric}
\end{equation}
where $dx^a$ denotes the infinitesimal coordinate differences and $\nu$ and 
$\lambda$ are functions of $r$ to be determined.%
\footnote{One should be careful to not confuse the metric potential with the 
spin frequency $\nu$ in Chap.~\ref{ch:PopulationSynthesis}.}
It should be noted that the coordinates in relativity are analogous but do not 
perfectly coincide with the coordinates in Newtonian gravity. One can consider 
the two-surface when $t, r = \text{const}$ in (\ref{eq:StaticMetric}),
\begin{equation}
    ds^2 = r^2 (d\theta^2 + \sin^2 \theta \, d\phi^2).
\end{equation}
Thus, the coordinate $r$ corresponds to the two-sphere, centred around $r = 0$, 
with circumference $2 \pi r$ and surface area $4 \pi r^2$. We can, therefore, 
identify $\theta$ and $\phi$ as the usual angles from spherical polar 
coordinates. However, $r$ is not the radial distance from the centre. Instead, 
we have the proper distance, 
\begin{equation}
    l = \int_0^r e^{\lambda(r')/2} dr'.
\end{equation}
Similarly, the proper time is given by 
$\tau = \int_0^t e^{\nu(r)/2} dt' = e^{\nu(r)/2} t$, where we observe the 
famous red-shift factor.

Because the star is static, it is convenient to choose an observer at rest with 
the fluid.%
\footnote{As with the Newtonian case, there is considerable freedom in the frame 
one chooses to work in. As before, the simplest choice is the fluid's rest 
frame.}
The only non-vanishing component of the fluid four-velocity is the $t$ 
component. So, by normalisation, $u^a u_a = -1$, we have 
\begin{equation}
    u^t = e^{-\nu/2}, \qquad u^i = 0.
\end{equation}
Similar to what we saw in the Newtonian case, baryon conservation 
(\ref{eq:RelativisticContinuity}) is, thus, trivially satisfied. We find the 
relativistic equation of hydrostatic equilibrium from 
(\ref{eq:RelativisticEuler}), 
\begin{equation}
    \nabla_i p = - \frac{1}{2} (\varepsilon + p) \nabla_i \nu.
    \label{eq:RelativisticHydrostaticEquilibrium}
\end{equation}
This conveniently provides us with a physical interpretation for $\nu$ by 
comparing this expression with its Newtonian analogue 
(\ref{eq:HydrostaticEquilibrium}): $\nu$ is proportional to the Newtonian 
potential $\Phi$. 

Working through the Einstein equations (\ref{eq:EinsteinEquations}), we have 
\begin{subequations}
\begin{equation}
    G_t^{\hphantom{t} t} = 8 \pi T_t^{\hphantom{t} t} \quad \longrightarrow 
        \quad \frac{1}{r^2} \frac{d}{dr}[r(1 - e^{-\lambda})] 
        = 8 \pi \varepsilon
    \label{eq:EEtt}
\end{equation}
and 
\begin{equation}
    G_r^{\hphantom{r} r} = 8 \pi T_r^{\hphantom{r} r} \quad \longrightarrow 
        \quad - \frac{1}{r^2} (1 - e^{-\lambda}) + \frac{1}{r} e^{-\lambda} 
        \frac{d\nu}{dr} = 8 \pi p.
    \label{eq:EErr}
\end{equation}
\end{subequations}

To obtain the final form of the structure equations we define 
\citep[motivated by the exterior Schwarzschild solution;][]{1916SPAW.......189S} 
\begin{equation}
    e^\lambda \equiv \frac{1}{1 - 2 m / r},
    \label{eq:explambda}
\end{equation}
which turns (\ref{eq:EEtt}) into 
\begin{subequations}\label{eqs:TOV}
\begin{equation}
    \frac{dm}{dr} = 4 \pi r^2 \varepsilon.
    \label{eq:TOV1}
\end{equation}
Equation~(\ref{eq:EErr}) becomes 
\begin{equation}
    \frac{d\nu}{dr} = \frac{2(m + 4 \pi r^3 p)}{r (r - 2 m)}.
    \label{eq:TOV2}
\end{equation}
From (\ref{eq:RelativisticHydrostaticEquilibrium}), we have 
\begin{equation}
    \frac{dp}{dr} = - \frac{1}{2} (\varepsilon + p) \frac{d\nu}{dr}
        = - \frac{(\varepsilon + p) (m + 4 \pi r^3 p)}{r (r - 2 m)}. 
    \label{eq:TOV3}
\end{equation}
\end{subequations}
Equations~(\ref{eqs:TOV}) make up the \textit{relativistic equations of stellar 
structure} -- also known as the \textit{Tolman-Oppenheimer-Volkoff equations} 
\citep{1939PhRv...55..364T, 1939PhRv...55..374O}. As in the Newtonian case, one 
must provide an equation of state~(\ref{eq:RelativisticEOS}). The boundary 
conditions are essentially the same as described above. The only change is that 
the metric potential must match the exterior Schwarzschild solution, 
$e^{\nu(R)} = 1 - 2 M / R$, where $M$ corresponds to the total mass-energy of 
the star.%
\footnote{Because of our choice to define the total mass using the exterior 
potential, we are justified in identifying the quantities $m$ and $M$ with the 
mass(-energy) in radius $r$ and the total mass(-energy), respectively.}
The numerical evaluation of Eqs.~(\ref{eqs:TOV}) is described in 
Appendix~\ref{appsec:NumericalTOV}.

There are a couple of details worth noting at this point, when comparing the 
relativistic equations to the Newtonian ones. We have pointed out that the 
radial coordinate $r$, in relativity, no longer corresponds to the physical 
distance from the centre of the star. Hence, $R$ does not give the proper radius 
of the star. Instead, the proper radius is 
\begin{equation}
    L = \int_0^R e^{\lambda(r)/2} dr,
\end{equation}
which is different to $R$ by $\order(1)$. Thus, $r$ should be interpreted 
strictly as a coordinate and not a physical distance an observer would measure.

The total mass-energy is given by [\textit{cf.} (\ref{eq:TOV1})]
\begin{equation}
    M = \int_0^R 4 \pi r^2 \varepsilon(r) dr.
    \label{eq:TotalMassEnergy}
\end{equation}
Contrary to the Newtonian mass, it is interesting to note that $M$ need not be 
conserved for a relativistic isolated star. This is because it does not solely 
include the rest mass-energy of the star. The simple form of 
Eq.~(\ref{eq:TotalMassEnergy}), which is reminiscent of the Newtonian result 
(\ref{eq:TotalMass}), slightly obfuscates the identification of $M$ as the total 
mass-energy. However, one can convince themselves that 
(\ref{eq:TotalMassEnergy}) does indeed include all the relevant contributions to 
the total mass-energy by considering the proper volume element for the spacetime 
(\ref{eq:StaticMetric}), 
$dV = e^{\lambda/2} r^2 \sin \theta \, dr \, d\theta \, d\phi$, 
and noting that the energy density contains the rest mass-energy density 
$\rho$ and internal energy density $u$, so $\varepsilon = \rho + u$. Thus, 
(\ref{eq:TotalMassEnergy}) can be written as 
\begin{equation}
\begin{split}
    M &= \int_V \rho dV + \int_V u dV 
        - \int_V (1 - e^{-\lambda/2}) \varepsilon dV \\ 
        &= \int_0^R 4 \pi e^{\lambda/2} r^2 \rho dr 
        + \int_0^R 4 \pi e^{\lambda/2} r^2 u dr 
        - \int_0^R 4 \pi (e^{\lambda/2} - 1) r^2 \varepsilon dr.
\end{split}
    \label{eq:TotalMassEnergy2}
\end{equation}
The first and second terms in (\ref{eq:TotalMassEnergy2}) are simply the total 
rest mass-energy and the total internal energy, respectively. The third term in 
(\ref{eq:TotalMassEnergy2}) is the total gravitational potential energy. This 
exercise further justifies our recognition of $m$ in (\ref{eq:explambda}) as the 
mass-energy within $r$.

\begin{figure}[h]
    \centering
	\includegraphics[width=0.7\textwidth]{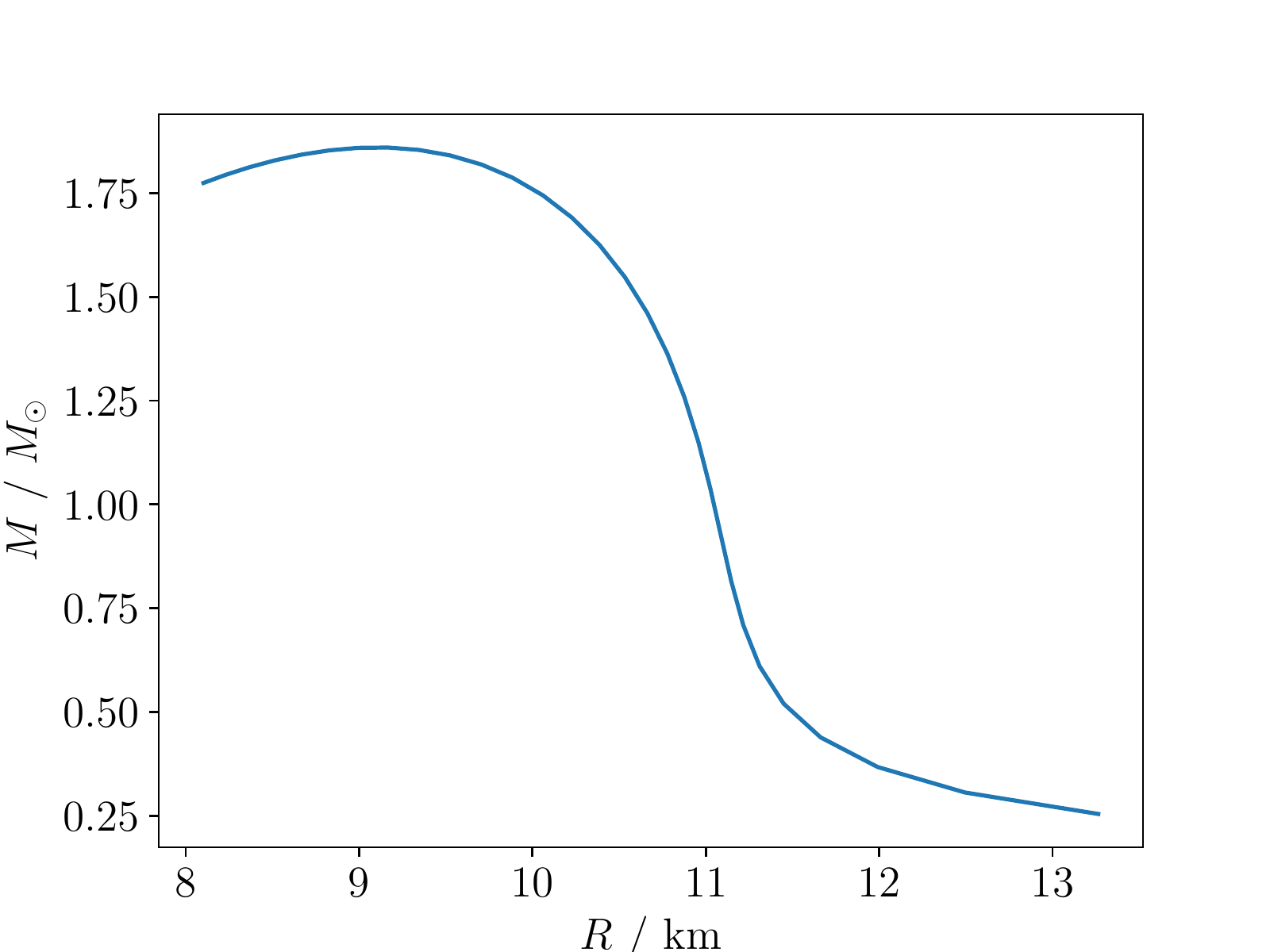}
    \caption[Example mass-radius diagram]{The mass-radius diagram for the 
             BSk19 equation of state 
             \citep{2013A&A...560A..48P}. The $M = M(R)$ curve is obtained by 
             integrating the relativistic equations of stellar structure 
             (\ref{eqs:TOV}) for different values of the central density 
             $\varepsilon_\text{c}$. The range of values for the central density 
             in this diagram are  
             $\SIrange{4.0e14}{6.6e15}{\gram\per\centi\metre\cubed}$.}
	\label{fig:M-Rexample}
\end{figure}

We show an example mass-radius diagram obtained from the relativistic structure 
equations in Fig.~\ref{fig:M-Rexample}. The main features of a neutron-star 
mass-radius diagram are shared among (realistic) equations of state. The larger 
values of the central density correspond to smaller radii and a larger 
compactness. Each equation of state admits a maximum mass that it can support. 
This is a relativistic result and comes from the denominator in (\ref{eq:TOV3}) 
that prevents $m$ from becoming as large as $r / 2$. The maximum mass is 
particularly of interest in excluding certain candidate matter models. It is 
important to note that not all equilibrium configurations correspond to 
\textit{stable} solutions. In general, configurations that lie on the left of 
the maximum mass in the mass-radius diagram are unstable -- small density 
perturbations will grow exponentially in time, leading the star to collapse or 
expand. On the other hand, configurations to the right of the maximum mass are 
stable -- the perturbations away from equilibrium are restored by pressure or 
gravity forces.

It is worth mentioning a common way equation-of-state candidates are 
categorised: equations of state are described as \textit{soft} or 
\textit{stiff}. The stiffness depends on how fast the isotropic pressure $p$ 
increases as a function of the energy density $\varepsilon$ -- the faster, the 
stiffer. Stiff equations of state are so named because the fluid is difficult to 
compress and, therefore, can support larger masses. On the other hand, soft 
equations of state are easier to compress and result in more compact stars.

Although there are presently many questions to be answered concerning what the 
true equation of state must be, there are a number of natural physical 
constraints that realistic equation-of-state candidates must satisfy. A couple 
of key restrictions are: (i) the speed of sound in the fluid $c_\text{s}$, 
where $c_\text{s}^2 \equiv dp / d\varepsilon$, must respect causality, 
$c_\text{s}^2 < 1$, and (ii) the fluid must be thermodynamically stable, that 
is $c_\text{s}^2 > 0$. However, these restrictions are not particularly 
constraining and a variety of popular models violate causality at high 
densities.

\section{Linear perturbations}
\label{sec:Perturbations}

In physics, it is common to approach problems in a perturbative manner. In 
linear perturbation theory, one considers small departures away from an 
equilibrium solution and linearises -- neglecting terms that are of quadratic 
order and higher in the small quantities. Therefore, the problem is split into 
two components: the \textit{equilibrium} (or \textit{background}) solution and 
the \textit{perturbative} part. This technique is particularly powerful when one 
already knows the background solution, \textit{e.g.}, the structure of a 
non-rotating, spherical star. We will make use of perturbation theory in order 
to construct non-spherical stars. In this section, it will be sufficient to 
focus the discussion on Newtonian perturbations. 

In hydrodynamics, there are two methods of describing a fluid. The first 
involves monitoring the behaviour of the fluid at a \textit{given point in 
space}, corresponding to a frame that is at rest. This is known as the 
\textit{Eulerian} description. The other method characterises the behaviour of a
\textit{given fluid element}, which is represented by a frame that is co-moving 
with the fluid flow. This is the \textit{Lagrangian} description. We have, in 
some sense, already encountered these two descriptions. For example, in the 
Euler equation (\ref{eq:Euler}), we see two types of time derivative: the 
ordinary partial derivative $\partial_t$, that keeps the location fixed, and 
the convective derivative $(\partial_t + v^i \nabla_i)$, that measures changes 
from the perspective of a fluid element travelling with velocity $v^i$. In 
relativity, the Lagrangian frame has the same meaning. However, an Eulerian 
frame is a little more complicated \citep{2021LRR....24....3A}.

An Eulerian perturbation, denoted by the $\delta$ symbol, of a quantity $f$ 
(which, in general, can be a scalar, vector or tensor) at a point $x_0^i$ is 
defined by 
\begin{equation}
    f(t, x_0^i) = f_0(x_0^i) + \delta f(t, x_0^i), 
\end{equation}
where $f_0$ is the equilibrium value (which is independent of time). A 
Lagrangian perturbation, denoted by the $\Delta$ symbol, can be expressed by 
\begin{equation}
    f(t, x^i) = f_0(x_0^i) + \Delta f(t, x_0^i), 
    \label{eq:Lagrangian}
\end{equation}
where $x_0^i$ is the position the fluid element was at in the equilibrium 
configuration and $x^i$ is the position it has moved to due to the perturbation. 
Note that, we have defined the Lagrangian perturbation to be a function of the 
equilibrium position.%
\footnote{This is an important detail since, in practice, the perturbation 
equations will tend to be coupled to the background structure. Therefore, when 
one solves the perturbation equations, they do this over the background domain, 
$x_0^i$. Thus, the perturbed quantities $\xi^i$, $\delta f$ and $\Delta f$ 
only have support at $x_0^i$.}
This provides the motivation for the introduction of the Lagrangian displacement 
vector $\xi^i$, which connects fluid elements in the perturbed configuration to 
their positions in the background \citep{1978ApJ...221..937F}, 
\begin{equation}
    x^i = x_0^i + \xi^i(t, x_0^i).
    \label{eq:LagrangianDisplacement}
\end{equation}
To be precise, a fluid element at position $x_0^i$ in the background has moved 
by $\xi^i$ to position $x^i$ in the perturbed configuration. Therefore, one can 
obtain a relation between the Eulerian and Lagrangian perturbations, 
\begin{equation}
\begin{split}
    \Delta f(t, x_0^i) &= f[t, x_0^i + \xi^i(t, x_0^i)] - f_0(x_0^i) \\
        &= f(t, x_0^i) + \xi^j(t, x_0^i) \partial_j f_0(x_0^i) 
        - f_0(x_0^i) + \order(|\xi^i|^2). 
\end{split}
\end{equation}
Note that we have taken the perturbative parameters $\xi^i$, $\delta f$ and 
$\Delta f$ to be small and have assumed $f$ to be a scalar. The perturbative 
terms of second order and above are contained in $\order(|\xi^i|^2)$. Thus, to 
linear order, 
\begin{equation}
    \Delta f = \delta f + \mathcal{L}_\xi f_0, 
    \label{eq:EulerianLagrangian}
\end{equation}
where we have identified the Lie derivative $\mathcal{L}_\xi$ along the vector 
$\xi^i$. Although $f$ has been taken to be a scalar, 
Eq.~(\ref{eq:EulerianLagrangian}) also applies for vectors and tensors.

To obtain the Lagrangian perturbation of the velocity, we must consider the 
velocity in the perturbed configuration, which can be written as 
\begin{equation}
    v^i(t, x^i) = \frac{dx^i}{dt} = \frac{d}{dt} [x_0^i + \xi^i(t, x_0^i)] 
        = v_0^i(x_0^i) + v_0^j(x_0^i) \partial_j \xi^i(t, x_0^i) 
        + \partial_t \xi^i(t, x_0^i).
\end{equation}
We can also expand the velocity as 
\begin{equation}
    v^i(t, x^i) = v^i[t, x_0^i + \xi^i(t, x_0^i)] 
        = v^i(t, x_0^i) + \xi^j(t, x_0^i) \partial_j v_0^i(x_0^i) 
        + \order(|\xi^i|^2).
\end{equation}
Hence, we have 
\begin{equation}
    \delta v^i(t, x_0^i) + \mathcal{L}_\xi v_0^i(x_0^i) 
        = \partial_t \xi^i(t, x_0^i) + \order(|\xi^i|^2).
\end{equation}
By (\ref{eq:EulerianLagrangian}), to leading perturbative order, we find 
\begin{equation}
    \Delta v^i = \partial_t \xi^i. 
    \label{eq:LagrangianVelocity}
\end{equation}

The perturbations also induce a change on the metric. By virtue of the nature of 
the covariant derivative, $\nabla_i g_{j k} = 0$, and the fact the metric in 
Newtonian gravity is flat, $\delta g_{i j} = 0$, we have 
\begin{equation}
    \Delta g_{i j} = \nabla_i \xi_j + \nabla_j \xi_i. 
    \label{eq:LagrangianMetric}
\end{equation}

Another useful expression for fluid perturbations comes from the 
conservation-of-mass principle. Consider an infinitesimal mass $dm_0$ in the 
background configuration. This is given by 
\begin{equation}
    dm_0 = \rho_0(x_0^i) dV_0, 
\end{equation}
where $dV_0$ is an infinitesimal volume in the background configuration. Each 
fluid element undergoes the transformation~(\ref{eq:LagrangianDisplacement}) due 
to the perturbations. The Jacobian of this transformation is 
$[1 + \nabla_i \xi^i + \order(|\xi^i|^2)]$. Therefore, the infinitesimal volume 
in the perturbed configuration, to linear order, is 
\begin{equation}
    dV = [1 + \nabla_j \xi^j(t, x_0^i)] dV_0. 
\end{equation}
Here, we see that the divergence of $\xi^i$ corresponds to the change in volume 
of a fluid element in the perturbation. The variation in the density follows 
from the conservation of mass contained in the volume, 
\begin{equation}
    dm_0 = \rho_0(x_0^i) dV_0 = \rho(t, x^i) dV. 
\end{equation}
The two densities are, thus, related by 
\begin{equation}
    \rho(t, x^i) = [1 - \nabla_j \xi^j(t, x_0^i)] \rho_0(x_0^i) 
        + \order(|\xi^i|^2)
\end{equation}
The Lagrangian perturbation of the density, to first order, is 
\begin{equation}
    \Delta \rho = - \rho_0 \nabla_i \xi^i. 
    \label{eq:MassConservation}
\end{equation}
Note that this expression is completely general and must be satisfied in order 
to conserve the mass in the perturbations. This is equivalently a statement 
about the Newtonian conservation of baryon number, 
$\Delta n = - n_0 \nabla_i \xi^i$.

For relativistic perturbations, we have 
[\textit{cf.} Eqs.~(\ref{eq:LagrangianVelocity}), (\ref{eq:LagrangianMetric}) 
and (\ref{eq:MassConservation}); \citet{2021LRR....24....3A}]
\begin{gather}
    \Delta u^a = \frac{1}{2} u_0^a u_0^b u_0^c \Delta g_{b c}, 
    \label{eq:LagrangianFourVelocity} \\
    \Delta g_{a b} = h_{a b} + \nabla_a \xi_b + \nabla_b \xi_a 
\end{gather}
and 
\begin{equation}
    \Delta n = - \frac{1}{2} n_0 \bot^{a b} \Delta g_{a b},
    \label{eq:BaryonNumberConservation}
\end{equation}
where $h_{a b} = \delta g_{a b}$.

Linear perturbations are useful for a variety of neutron-star problems, 
including asteroseismology and gravitational waves. We will omit the subscript 
denoting background quantities forthwith. 

\section{Multipole moments}
\label{sec:Multipoles}

When characterising the shape of a star, one often considers the star's 
\textit{multipole moments}. Multipole moments quantify the degree of 
non-sphericity of a star and help to determine whether the star in question is a 
promising gravitational-wave emitter. These are most straightforwardly obtained 
from the far-field expansion of the gravitational potential.

\subsection{Newtonian multipole moments}
\label{subsec:NewtonianMultipoles}

Consider a star described by the mass-density distribution $\rho(x^i)$, where 
$x^i$ is a point inside the star. This mass distribution sources a gravitational 
potential field $\Phi(x^i)$ through Poisson's equation (\ref{eq:Poissons}). 
The solution to (\ref{eq:Poissons}) is 
\begin{equation}
    \Phi(X^i) = - \int_V \frac{G \rho(x^i)}{|X^j - x^j|} dV,
    \label{eq:Potential}
\end{equation}
where $X^i$ is the position of an observer measuring the field, outside the 
mass distribution. The volume integration is over the volume of the star. In the 
far-field limit, we have $r \ll \mathcal{R}$, where $r^2 = x^i x_i$ and 
$\mathcal{R}^2 = X^i X_i$. Equation~(\ref{eq:Potential}) may be expanded in 
terms of $\mathcal{R}$ away from the source using Legendre polynomials 
$\mathcal{P}_\ell(x)$, 
\begin{equation}
    \Phi(X^i) = - \sum_{\ell = 0}^\infty \frac{G}{\mathcal{R}^{\ell + 1}} 
        \int_V \rho(x^i) r^\ell \mathcal{P}_\ell(\cos \alpha) dV,
    \label{eq:PotentialLegendre}
\end{equation}
where $\alpha$ is the angle between $X^i$ and $x^i$,
\begin{equation}
    \cos \alpha = \hat{X}^i \hat{x}_i,
\end{equation}
and $\hat{X}^i = X^i / \mathcal{R}$ and $\hat{x}^i = x^i / r$ are the 
corresponding unit vectors. Consequently, we may decompose the potential 
(\ref{eq:PotentialLegendre}) into a series of multipole moments, 
\begin{equation}
    \Phi(X^i) = - \sum_{\ell = 0}^\infty 
        \frac{G Q_\ell(\hat{X}^i)}{\mathcal{R}^{\ell + 1}},
\end{equation}
where 
\begin{equation}
    Q_\ell(\hat{X}^i) = \int_V \rho(x^i) r^\ell \mathcal{P}_\ell(\cos \alpha) dV
    \label{eq:Multipoleell}
\end{equation}
is the $2^\ell$-pole moment of the mass distribution $\rho(x^i)$. Or, more 
precisely, it is the component of the $2^\ell$-pole moment in the $\hat{X}^i$ 
direction. In assuming the far-field limit, each multipole with increasing 
$\ell$ will be successively suppressed by $1 / \mathcal{R}^{\ell + 1}$ and 
provide small contributions. Let us examine the first few multipoles.

\begin{itemize}
    \item[$\ell = 0$:] The monopole moment is simply the net charge of the 
        distribution, 
        \begin{equation}
            Q_0 = \int_V \rho(x^i) dV = M,
        \end{equation} 
        which is independent of the direction $\hat{X}^i$. Hence, the monopole 
        term in the potential is isotropic, 
        \begin{equation}
            \Phi_0(\mathcal{R}) = - \frac{G M}{\mathcal{R}}.
        \end{equation}
    \item[$\ell = 1$:] The dipole moment is the vector 
        \begin{equation}
            p^i = \int_V \rho(x^i) x^i dV
        \end{equation}
        and its component in the $\hat{X}^i$ direction is 
        \begin{equation}
            Q_1(\hat{X}^i) = \int_V \rho(x^i) r \cos \alpha dV = p_i \hat{X}^i.
        \end{equation}
        The dipole potential has the form 
        \begin{equation}
            \Phi_1(X^i) = - \frac{G p_i \hat{X}^i}{\mathcal{R}^2}.
        \end{equation}
        Note that if the star is at rest, one is always free to work in a 
        mass-centred coordinate system where the dipole vanishes.
    \item[$\ell = 2$:] We introduce the quadrupole-moment tensor 
        \begin{equation}
            Q_{i j} = \int_V \rho(x^i) \left( x_i x_j 
                - \frac{1}{3} r^2 g_{i j}\right) dV,
            \label{eq:QuadrupoleTensor}
        \end{equation}
        which is a symmetric and trace-free tensor. Therefore, the quadrupole 
        moment is 
        \begin{equation}
            Q_2(\hat{X}^i) = \int_V \rho(x^i) r^2 
                \left( \frac{3}{2} \cos^2 \alpha - \frac{1}{2} \right) dV 
                = \frac{3}{2} Q_{i j} \hat{X}^i \hat{X}^j.
        \end{equation}
        Thus, the quadrupole potential is 
        \begin{equation}
            \Phi_2(X^i) = - \frac{3}{2} 
                \frac{G Q_{i j} \hat{X}^i \hat{X}^j}{\mathcal{R}^3}.
        \end{equation}
    \item[$\ell \geq 3$:] Similar to the quadrupole-moment tensor, the octupole 
        and higher-order moments are all symmetric and of the form 
        \begin{equation}
            Q_{i j \ldots q}^{\ell} = \int_V \rho(x^i) 
                (\text{homogeneous polynomial of degree } \ell)_{i j \ldots q} 
                dV,
            \label{eq:MultipoleMomentTensor}
        \end{equation}
        where the polynomial follows from the expansion of 
        \begin{equation}
            r^\ell \mathcal{P}_\ell(\cos \alpha) = \frac{(2 \ell - 1)!!}{\ell!} 
                (\text{homogeneous polynomial of degree } \ell)_{i j \ldots q} 
                \hat{X}^i \hat{X}^j \ldots \hat{X}^q.
        \end{equation}
        Note that $i, j, \ldots, q$ constitute a total of $\ell$ indices. The 
        potential due to the $2^\ell$-pole moment is 
        \begin{equation}
            \Phi_\ell(X^i) = - \frac{(2 \ell - 1)!!}{\ell!} 
                \frac{G Q_{i j \ldots q}^{\ell} 
                \hat{X}^i \hat{X}^j \ldots \hat{X}^q}{\mathcal{R}^{\ell + 1}}.
            \label{eq:PotentialMultipolel}
        \end{equation}
\end{itemize}

In general, the $2^\ell$-pole tensor $Q_{i j \ldots q}^{\ell}$ has $3^\ell$ 
components. However, there are symmetries that reduce the number of independent 
components. For example, the quadrupole tensor $Q_{i j}$ has $9$ entries, 
but is symmetric and traceless, 
\begin{equation}
    Q_{i j} = Q_{j i}, \quad Q_i^{\hphantom{i} i} = 0.
\end{equation}
These make up $3 + 1 = 4$ symmetry conditions and so $Q_{i j}$ has $9 - 4 = 5$ 
independent components. For a given $2^\ell$-pole tensor, there are 
$(2 \ell + 1)$ independent components. Likewise, for a given $\ell$ there are 
$(2 \ell + 1)$ spherical harmonics $Y_{\ell m}(\theta, \phi)$.%
\footnote{One should not confuse the harmonic number $m$ with the mass inside a 
star.}
In fact, instead 
of describing the angular dependence of the multipoles' components in the 
direction $\hat{X}^i$ in terms of symmetric tensors, we may expand them in 
spherical harmonics. We discuss some of the properties of the relevant spherical 
harmonics in Appendix~\ref{app:Harmonics}.

For any integer $\ell = 0, 1, 2, \ldots$ and any two unit vectors $\hat{a}^i$ 
and $\hat{b}^i$, the Legendre polynomial of their scalar product may be 
expressed as 
\begin{equation}
    \mathcal{P}_\ell(\hat{a}^i \hat{b}_i) = \frac{4 \pi}{2 \ell + 1} 
        \sum_{m = - \ell}^\ell Y_{\ell m}(\hat{a}^i) Y_{\ell m}^*(\hat{b}^i),
\end{equation}
where the star denotes a complex conjugate. We can apply this result to the 
definition of the multipoles (\ref{eq:Multipoleell}), 
\begin{equation}
    Q_\ell(\hat{X}^i) = \frac{4 \pi}{2 \ell + 1} \sum_{m = - \ell}^\ell 
        Y_{\ell m}(\hat{X}^i) \int_V \rho(x^i) r^\ell Y_{\ell m}^*(\hat{x}^i) 
        dV.
\end{equation}
Hence, we choose to define the spherical-harmonic decomposition of the 
multipoles by%
\footnote{\label{foot:Multipole}Note that this definition of the multipole 
moments is different to that of \citet{1980RvMP...52..299T} 
[\textit{cf.} his Eq.~(5.27a)], which is commonly used in the tidal-deformation literature \citep[see, \textit{e.g.},][]{2008ApJ...677.1216H}.}
\begin{equation}
    Q_{\ell m} \equiv \int_V \rho(x^i) r^\ell Y_{\ell m}^*(\theta, \phi) dV,
    \label{eq:Multipolelm}
\end{equation}
where $(\theta, \phi)$ describe the direction of $\hat{x}^i$. The full 
gravitational potential is then given by 
\begin{equation}
    \Phi(X^i) = - \sum_{\ell = 0}^\infty \sum_{m = - \ell}^\ell 
        \frac{4 \pi G}{2 \ell + 1} 
        \frac{Q_{\ell m} Y_{\ell m}(\vartheta, \varphi)}
        {\mathcal{R}^{\ell + 1}},
    \label{eq:PotentialMultipolelm}
\end{equation}
where $(\vartheta, \varphi)$ are the angles characterising $\hat{X}^i$. We may 
identify two approaches to obtaining the multipole moments. (1) Provided the 
density distribution $\rho(x^i)$, one can integrate through the volume of the 
star, evaluating (\ref{eq:Multipolelm}). (2) If one knows the exterior 
potential at a point $X^i$, then the multipoles can be read off from 
(\ref{eq:PotentialMultipolelm}) (since the potential must be continuous, it is 
often convenient to do this examination at the surface).

It is straightforward to generate spherical background stars using the 
equations of stellar structure (\ref{eqs:NewtonianStellarStructure}). These 
models are fully described by the $\ell = 0$ monopole. If one assumes that the 
star is deformed away from sphericity in a perturbative way, then the multipoles 
that arise due to the corresponding Eulerian density perturbation 
$\delta \rho(x^i)$ are%
\footnote{To formally show this, one must consider the full multipole [given by 
(\ref{eq:Multipolelm})] of the perturbed configuration under the transformation 
(\ref{eq:LagrangianDisplacement}), make use of the Lagrangian relation 
(\ref{eq:Lagrangian}) and linearise.}
\begin{equation}
    Q_{\ell m} = \int_V \delta \rho(x^i) r^\ell Y_{\ell m}^*(\theta, \phi) dV.
    \label{eq:MultipolelmPerturbed}
\end{equation}
Motivated by (\ref{eq:PotentialMultipolelm}), it is useful to decompose the 
rest of the perturbations using spherical harmonics, \textit{e.g.}, 
\begin{equation}
    \delta \rho(x^i) = \sum_{\ell = 0}^\infty \sum_{m = - \ell}^\ell 
        \delta \rho_{\ell m}(r) Y_{\ell m}(\theta, \phi).
    \label{eq:SphericalHarmonicDecomposition}
\end{equation}
However, it is important to note that, should one wish to focus on a specific 
harmonic mode $(\ell, m)$, care must be taken since $Y_{\ell m}$ are, in 
general, complex. One resolves this issue by taking the real part, as physical 
quantities must be real, \textit{e.g.}, 
$\delta \rho (x^i) = \delta \rho_{\ell m}(r) 
\text{Re}[Y_{\ell m}(\theta, \phi)]$. 
With this decomposition, (\ref{eq:MultipolelmPerturbed}) becomes 
\begin{equation}
    Q_{\ell m} = \int_0^R \delta \rho_{\ell m}(r) r^{\ell + 2} dr.
    \label{eq:Multipole}
\end{equation}
The perturbation to the gravitational potential in the exterior can be written 
as 
\begin{equation}
    \delta \Phi(X^i) = \sum_{\ell = 0}^\infty \sum_{m = - \ell}^\ell 
        \delta \Phi_{\ell m}(\mathcal{R}) Y_{\ell m}(\vartheta, \varphi), \quad 
        \delta \Phi_{\ell m}(\mathcal{R}) = - \frac{4 \pi G}{2 \ell + 1} 
        \frac{Q_{\ell m}}{\mathcal{R}^{\ell + 1}}.
    \label{eq:PerturbedPotentialMultipole}
\end{equation}
Note that, for many problems of astrophysical interest, the total mass of the 
star will remain conserved. In these cases, the $\ell = 0$ monopole of the 
perturbation will vanish. [Indeed, this is guaranteed when the constraint 
(\ref{eq:MassConservation}) is applied.] Additionally, one might prefer to work 
in a mass-centred coordinate system, in which there is no $\ell = 1$ dipole 
contribution.

It is important to note that, in the absence of a perturbing force, the star 
will remain spherical. For example, by perturbing the mass density, say, the 
star will simply move to a neighbouring equilibrium configuration along the 
mass-radius curve (Fig.~\ref{fig:M-Rexample}). One must introduce non-sphericity 
into the situation via some deforming force. Simply put: an unforced, fluid 
equilibrium is spherical. Suppose we introduce another potential $\chi$ that is 
related to a deforming force the star is subject to. This potential will adjust 
the total potential by 
\begin{equation}
    \Phi(\mathcal{R}) + \delta \Phi(X^i) + \chi(X^i) = - \frac{G M}{\mathcal{R}} 
        + \sum_{\ell = 0}^\infty \sum_{m = - \ell}^\ell \left[ 
        - \frac{4 \pi G}{2 \ell + 1} 
        \frac{Q_{\ell m}}{\mathcal{R}^{\ell + 1}}
        + \chi_{\ell m}(\mathcal{R}) \right] Y_{\ell m}(\vartheta, \varphi).
\end{equation}
We will make use of this result when we calculate mountains in Newtonian gravity 
in Chap.~\ref{ch:Mountains}.

\subsection{Relativistic multipole moments}

In general relativity, the spacetime metric $g_{a b}$ replaces the Newtonian 
potential $\Phi$. In the far-field exterior, one may write 
\begin{equation}
    - \frac{1 + g_{t t}}{2} = \Phi + \chi.
    \label{eq:MetricPotential}
\end{equation}
Although in Newtonian gravity one is usually able to analyse the gravitational 
and deforming potentials separately, such a neat decoupling does not exist in 
relativity. In relativity, one has the metric that accounts for all the 
potentials acting on the body. For this reason, in some cases it is not possible 
to disentangle the separate potentials from the metric and ambiguities arise.%
\footnote{To be specific, ambiguities occur when different potentials mix in 
powers of $\mathcal{R}$. Much of the work concerning separating multiple 
gravitational effects in relativity has come from studies on tidal deformations 
\citep[see, \textit{e.g.},][]{2018CQGra..35h5002G}.}
However, there are no ambiguities when one considers multipoles of pure $\ell$. 
[Indeed, we will ultimately focus on pure multipoles of order $(\ell, m)$.] 
In this case, we have [\textit{cf.} a variation of (\ref{eq:MetricPotential})]
\begin{equation}
    - \frac{1 + g_{t t} + h_{t t}}{2} = - \frac{M}{\mathcal{R}} 
        + \sum_{m = - \ell}^\ell \left[ 
        - \frac{4 \pi}{2 \ell + 1} B_\ell(\mathcal{R}) 
        \frac{Q_{\ell m}}{\mathcal{R}^{\ell + 1}} 
        + \chi_{\ell m}(\mathcal{R}) \right] Y_{\ell m}(\vartheta, \varphi),
    \label{eq:PerturbedMetricPotential}
\end{equation}
where $h_{a b} = \delta g_{a b}$ is the linearised metric and 
$B_\ell(\mathcal{R})$ is a function that goes to unity in the Newtonian limit 
\citep[this is given as $B_1$ in Table I of][]{2009PhRvD..80h4018B}. 
Equation~(\ref{eq:PerturbedMetricPotential}) is obtained by looking for a 
solution to the Einstein equations in the vacuum around the star (as we shall 
discuss in Chap.~\ref{ch:RelativisticMountains}).

\section{Gravitational waves}
\label{sec:GravitationalWaves}

Having discussed how one formally describes the interior structure and shape of 
a neutron star, we have reached an appropriate point to consider how 
gravitational waves emerge in general relativity. For more detail, the following 
textbooks are recommended: \citet{Maggiore2008} and \citet{2019gwa..book.....A}.

\subsection{The wave equation}

The most straightforward way to see how gravitational waves arise is through the 
linearised theory of gravity. One considers a region of spacetime where the 
metric is close to flat, 
\begin{equation}
    g_{a b} = \eta_{a b} + h_{a b}, \qquad |h_{a b}| \ll 1,
    \label{eq:LinearisedMetric}
\end{equation}
where $\eta_{a b}$ is the Minkowski metric. In Cartesian coordinates, the 
Minkowski metric is given by 
\begin{equation}
    \eta_{a b} dx^a dx^b = - dt^2 + dx^2 + dy^2 + dz^2.
\end{equation}
In linearised gravity, one substitutes (\ref{eq:LinearisedMetric}) into the 
Einstein equations~(\ref{eq:EinsteinEquations}) and then retains terms up to 
linear order in $h_{a b}$. One can show, through the definition 
$\delta_a^{\hphantom{a} b} = g_{a c} g^{c b}$, that the inverse metric is 
\begin{equation}
    g^{a b} = \eta^{a b} - h^{a b} + \order(|h_{a b}|^2),
\end{equation}
where $h^{a b} = \eta^{a c} \eta^{b d} h_{c d}$. Therefore, indices of terms 
of $\order(|h_{a b}|)$ can be raised and lowered with the flat metric 
$\eta_{a b}$.

It should be noted that in writing the metric as in (\ref{eq:LinearisedMetric}), 
we have made a specific choice of frame such that $|h_{a b}| \ll 1$ holds on a 
sufficiently large region of space. In choosing this frame, a residual gauge 
freedom remains. A gauge freedom corresponds to the invariance of a quantity 
under a coordinate transformation. Consider a transformation of coordinates such 
that 
\begin{equation}
    x^a \rightarrow x'^a = x^a + \zeta^a(x^a),
\end{equation}
where $\zeta^a$ is the generator of the transformation and the derivatives 
$|\partial_a \zeta_b|$ are of the same order of smallness as $|h_{a b}|$. By the 
transformation law for tensors under coordinate changes, 
\begin{equation}
    g_{a b}(x^a) \rightarrow g'_{a b}(x'^a) 
        = \frac{\partial x^c}{\partial x'^a} 
        \frac{\partial x^d}{\partial x'^b} g_{c d}(x^a),
\end{equation}
we find 
\begin{equation}
    h_{a b}(x^a) \rightarrow h'_{a b}(x'^a) = h_{a b}(x^a) 
        - [\partial_a \zeta_b(x^a) + \partial_b \zeta_a(x^a)],
    \label{eq:PerturbedTransformation}
\end{equation}
to leading order. If $|\partial_a \zeta_b|$ are small, then the condition 
$|h_{a b}| \ll 1$ is preserved under a transformation.

Since the curvature terms vanish for flat space, one can show that the Riemann 
tensor for the metric (\ref{eq:LinearisedMetric}), to linear order in $h_{a b}$, 
is 
\begin{equation}
    R_{a b c d} = \frac{1}{2} (\partial_b \partial_c h_{a d} 
        + \partial_a \partial_d h_{b c} - \partial_a \partial_c h_{b d} 
        - \partial_b \partial_d h_{a c}).
    \label{eq:PerturbedRiemann}
\end{equation}
Pleasingly, one can show by inserting the 
transformation~(\ref{eq:PerturbedTransformation}) into 
(\ref{eq:PerturbedRiemann}) that the Riemann tensor is invariant under the 
residual gauge transformation, as it should be.

It turns out that the linearised equations of motion can be expressed in a 
more compact form if one introduces the trace-reversed metric perturbation 
\begin{equation}
    \bar{h}_{a b} \equiv h_{a b} - \frac{1}{2} \eta_{a b} h,
\end{equation}
where $h = \eta^{a b} h_{a b}$. One can invert this to show that 
$h_{a b} = \bar{h}_{a b} - \eta_{a b} \bar{h} / 2$, where 
$\bar{h} = \eta^{a b} \bar{h}_{a b} = - h$. Thus, the Einstein 
equations~(\ref{eq:EinsteinEquations}) may be expressed as 
\begin{equation}
    G_{a b} = - \frac{1}{2} (\square \bar{h}_{a b} 
        + \eta_{a b} \partial^c \partial^d \bar{h}_{c d} 
        - \partial^c \partial_b \bar{h}_{a c} 
        - \partial^c \partial_a \bar{h}_{b c}) = 8 \pi T_{a b},
    \label{eq:EinsteinEquations2}
\end{equation}
where $\square \equiv \partial^a \partial_a$ is the flat-space d'Alembertian. 
It is worthwhile noting that the stress-energy tensor in linearised gravity 
enters the problem at $\order(|\bar{h}_{a b}|)$ as the background must be flat. 
Clearly, this will simplify considerably if we make the following gauge choice:
\begin{equation}
    \partial^b \bar{h}_{a b} = 0.
    \label{eq:LorenzGauge}
\end{equation}
This is known as the \textit{Lorenz gauge}. The freedom to impose the gauge 
condition~(\ref{eq:LorenzGauge}) comes from (\ref{eq:PerturbedTransformation}), 
which, in terms of the trace-reversed metric perturbation, is 
\begin{equation}
    \bar{h}_{a b} \rightarrow \bar{h}'_{a b} = \bar{h}_{a b} 
        - (\partial_a \zeta_b + \partial_b \zeta_a 
        - \eta_{a b} \partial_c \zeta^c).
\end{equation}
Thus, 
\begin{equation}
    \partial^b \bar{h}_{a b} \rightarrow (\partial^b \bar{h}_{a b})' 
        = \partial^b \bar{h}_{a b} - \square \zeta_a.
    \label{eq:TraceReversedPerturbedDerivativeTransformation}
\end{equation}
Therefore, we can always ensure (\ref{eq:LorenzGauge}) is satisfied if we 
transform the coordinates such that 
$\square \zeta_a = \partial^b \bar{h}_{a b}$. In this gauge, 
Eq.~(\ref{eq:EinsteinEquations2}) provides the wave equation 
\begin{equation}
    \square \bar{h}_{a b} = - 16 \pi T_{a b}.
    \label{eq:WaveEquation}
\end{equation}
By Eqs.~(\ref{eq:LorenzGauge}) and (\ref{eq:WaveEquation}), the condition 
$\partial^b T_{a b} = 0$ is automatically satisfied.

To summarise, we have seen how general relativity permits small perturbations on 
an otherwise flat background that satisfy the wave 
equation~(\ref{eq:WaveEquation}). These solutions correspond to gravitational 
waves that travel at the speed of light.

We should note the approximations implicit in linearised theory. The bodies that 
generate the gravitational waves are assumed to move in flat spacetime. This 
essentially means the dynamics are described using Newtonian gravity.

\subsection{Interaction with test masses}

To understand how gravitational waves propagate and interact with test masses, 
we consider (\ref{eq:WaveEquation}) outside the source, 
\begin{equation}
    \square \bar{h}_{a b} = 0.
    \label{eq:WaveEquationVacuum}
\end{equation}
We note that the Lorenz gauge is not sufficient to fix the full gauge freedom. 
Further analysis shows that one can introduce an additional coordinate 
transformation $x^a \rightarrow x^a + \zeta^a$, where 
\begin{equation}
    \square \zeta_a = 0,
\end{equation}
that leaves (\ref{eq:TraceReversedPerturbedDerivativeTransformation}) unspoiled. 
Thus, one can impose four further conditions. It is convenient to choose 
$\bar{h} = 0$ so $\bar{h}_{a b} = h_{a b}$ and $h_{t i} = 0$. Therefore, the 
Lorenz condition~(\ref{eq:LorenzGauge}) reads 
\begin{equation}
    \partial^t h_{t t} = 0, \qquad \partial^j h_{i j} = 0.
\end{equation}
One should note that $h_{t t}$ corresponds to the static part of the 
gravitational interaction, which is related to the Newtonian potential of the 
source of the waves. The gravitational wave is the time-dependent part of the 
perturbation and so we are free to set $h_{t t} = 0$. The time-varying aspect of 
the wave is contained in $h_{i j}$. In summary, the gauge is fixed with 
\begin{equation}
    h_{t a} = 0, \qquad h_i^{\hphantom{i} i} = 0, \qquad \partial^j h_{i j} = 0.
\end{equation}
This defines the \textit{transverse-traceless gauge} which we will denote using 
$h^\text{TT}_{i j}$.%
\footnote{One can show by analysing the geodesics of test masses in the 
transverse-traceless gauge that it corresponds to a frame where masses which 
were at rest before the gravitational wave arrives remain at rest after it has 
passed.}
It should be noted that the transverse-traceless gauge does not provide the same 
level of simplification inside the source where $T_{a b} \neq 0$.

In the transverse-traceless gauge, the solution to (\ref{eq:WaveEquationVacuum}) 
is 
\begin{equation}
    h^\text{TT}_{i j} = e_{i j} \exp (i k_a x^a),
    \label{eq:WaveSolution}
\end{equation}
where $k^a = (\omega, k^i)$ is the wave four-vector, $\omega$ is the angular 
frequency of the waves and $e_{i j}$ is the polarisation tensor. Assuming the 
waves propagate along the $z$-axis and ensuring that $h^\text{TT}_{i j}$ is 
symmetric and traceless, we have 
\begin{equation}
    h^\text{TT}_{i j} dx^i dx^j = h_+ \cos [\omega (t - z)] dx^2 
        + 2 h_\times \cos [\omega (t - z)] dx dy 
        - h_+ \cos [\omega (t - z)] dy^2,
\end{equation}
where $h_+$ and $h_\times$ are the two independent wave polarisations 
corresponding to the two degrees of freedom.

Given a plane wave $h_{a b}$ travelling in the direction of the unit vector 
$\hat{n}^i$, already in the Lorenz gauge, one can transform the solution to the 
transverse-traceless gauge using the projection operator 
\begin{equation}
    \Lambda_{i j}^{\hphantom{i j} k l} 
        = P_i^{\hphantom{i} k} P_i^{\hphantom{j} l} 
        - \frac{1}{2} P_{i j} P^{k l}, \qquad 
        P_{i j} = \delta_{i j} - \hat{n}_i \hat{n}_j,
\end{equation}
where $\delta_{i j}$ are the spatial parts of $\eta_{a b}$. Thus, one projects 
the solution into the transverse-traceless gauge by 
$h_{i j}^\text{TT} = \Lambda_{i j}^{\hphantom{i j} k l} h_{k l}$.

With the gauge freedom fully exploited, we can consider how test masses 
move due to a gravitational wave. Both masses are treated as point particles 
freely falling in flat space. They are separated by four-vector 
$\zeta^a(\tau)$ such that one test mass is at $x^a(\tau)$ and the other is at 
$x^a(\tau) + \zeta^a(\tau)$, where $\tau$ is the proper time as measured by a 
clock carried along the trajectory $x^a$. The masses follow geodesics, given by 
the \textit{geodesic equation} 
\begin{equation}
    u^b \nabla_b u^a 
        = \frac{d u^a}{d \tau} + \Gamma^a_{\hphantom{a} b c} u^b u^c = 0,
    \label{eq:Geodesic}
\end{equation}
where $u^a = d x^a / d \tau$. In flat space with Cartesian coordinates, 
$\Gamma^a_{\hphantom{a} b c} = 0$ and we obtain the classical result that 
particles follow straight lines in the absence of a force. Assuming that 
$|\zeta^a|$ is smaller than the variation of the gravitational field, one can 
show by analysing the geodesics followed by both test masses that, to first 
order in $\zeta^a$, 
\begin{equation}
    \frac{D^2 \zeta^a}{D \tau^2} 
        = - R^a_{\hphantom{a} b c d} u^b \zeta^c u^d,
    \label{eq:TidalForce}
\end{equation}
where $D / D \tau \equiv u^a \nabla_a$ is the directional derivative along the 
four-velocity $u^a$. This is the \textit{equation of geodesic deviation}, a 
well-known result in general relativity. It demonstrates that two nearby 
time-like geodesics experience a tidal gravitational force, given by the Riemann 
tensor. When spacetime is flat, $R^a_{\hphantom{a} b c d} = 0$ and the 
separation is fixed or changes at a constant rate. However, in curved spacetime, 
the four-vector $\zeta^a$ will accelerate. Geodesics that are initially 
parallel will eventually converge or diverge.

\begin{figure}[h]
    \centering
	\includegraphics[width=0.95\textwidth]{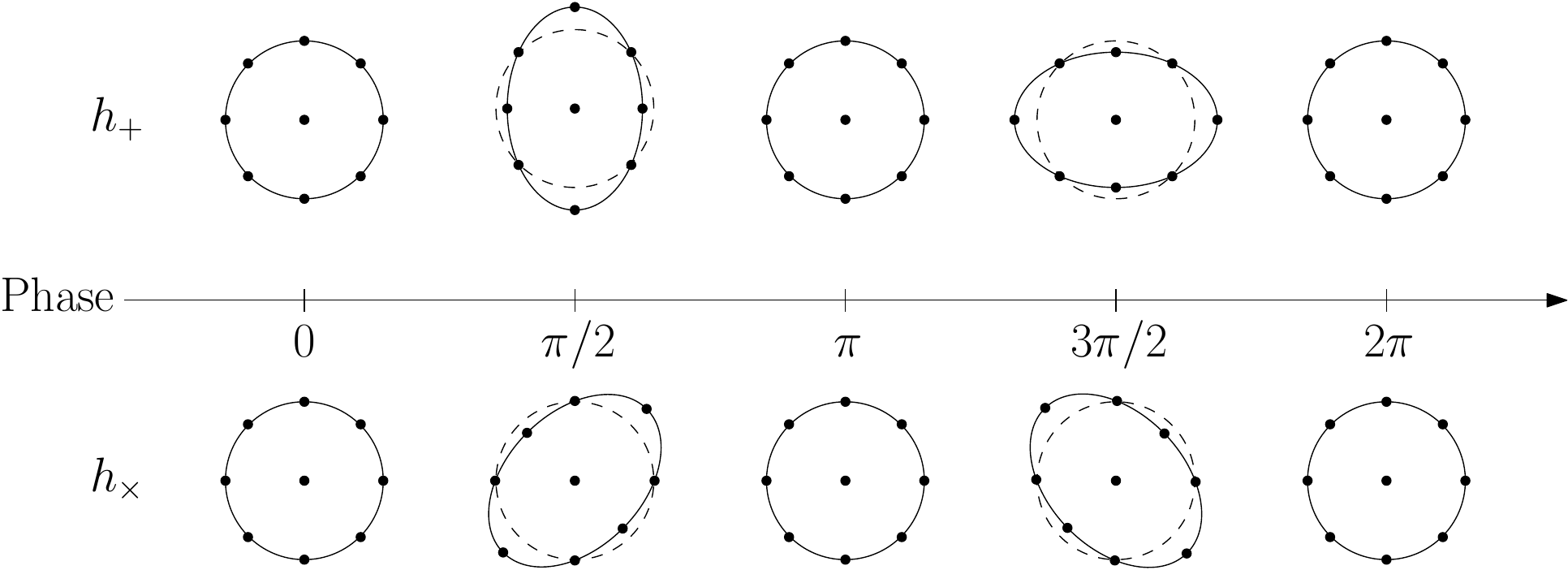}
    \caption[Gravitational wave on a ring of test masses]{The effect of a 
             gravitational wave on a ring of freely falling test masses in the 
             transverse-traceless frame due to the $+$ and $\times$ 
             polarisations. Note that the separation illustrated corresponds to 
             the proper distances between the masses in the transverse-traceless 
             frame.}
	\label{fig:Ring}
\end{figure}

The Riemann tensor is effected by the gravitational wave according to 
(\ref{eq:PerturbedRiemann}) and (\ref{eq:TidalForce}) returns 
\begin{equation}
    \partial_t^2 \zeta^i 
        = \frac{1}{2} \delta^{i j} \partial_t^2 h^\text{TT}_{j k} \zeta^k,
    \label{eq:GeodesicDeviation}
\end{equation}
where we have limited ourselves to linear order in $h_{i j}$ so $t = \tau$. The 
solution to (\ref{eq:GeodesicDeviation}) is simply 
\begin{equation}
    \zeta^i(t) = \left[ \delta^i_{\hphantom{i} k} 
        + \frac{1}{2} \delta^{i j} h^\text{TT}_{j k}(t) \right] \zeta^k(0).
    \label{eq:Separation}
\end{equation}
Equation~(\ref{eq:Separation}) shows how freely falling test masses move 
according to a gravitational wave $h^\text{TT}_{i j}$ in the transverse-traceless 
frame. We show the motion of a ring of test masses due to the two polarisations 
in Fig.~\ref{fig:Ring}. As the wave passes through the ring, the proper distance 
between the masses is stretched and squeezed. This periodic behaviour is the 
basic principle that is exploited by gravitational-wave detectors.

\subsection{Generation of gravitational waves}

We now turn our attention to how the gravitational waves are sourced. Although 
the wave solution~(\ref{eq:WaveSolution}) was known in the 1920s, much 
confusion surrounded the idea of whether the waves were physical in nature 
\citep{2007tste.book.....K}. Many leading physicists weighed into the debate, 
including \citet{Eddington1922} who observed problems with the gauge artefacts 
where they propagated at velocities that depended on the choice of coordinates. 
Einstein himself believed at some point that (\ref{eq:WaveSolution}) was an 
artefact of the linear approximation and no wave solution would exist in the 
full non-linear theory. \citet{EinsteinRosen1937} initially concluded that 
gravitational waves were unphysical upon finding a singularity in a non-linear 
cylindrical wave solution in 1936, only to retract this with the realisation 
that it was merely a coordinate singularity. A recurring theme was the issue of 
coordinate and gauge dependencies, in particular the properties of the Lorenz 
and transverse-traceless gauges. Also problematically, the stress-energy tensor 
of a gravitational field was found to not be invariant under the 
transformation~(\ref{eq:PerturbedTransformation}), casting further doubt on the 
physical nature of gravitational waves.

It turns out that the stress-energy tensor of gravitational waves $t_{a b}$ not 
being gauge invariant is a manifestation of the equivalence principle of 
general relativity: one is free to choose a local inertial frame at any point in 
spacetime where the gravitational wave vanishes. This indicates that we cannot 
localise the effect of the wave. While $t_{a b}$ is not well-defined locally, it 
certainly carries energy. In order to understand the physical effect of the 
wave, one must average over several wavelengths. Indeed, one can show that the 
averaged stress-energy tensor $\langle t_{a b} \rangle$ is gauge invariant. One 
finds that the averaging procedure cancels the gauge terms in 
(\ref{eq:PerturbedTransformation}) and obtains \citep{PhysRev.166.1272}
\begin{equation}
    \langle t_{a b} \rangle = \frac{1}{32 \pi} 
        \langle \partial_a h_{i j}^\text{TT} \partial_b h^{\text{TT} i j} \rangle.
\end{equation}

The solution of (\ref{eq:WaveEquation}) can be obtained using a retarded Green's 
function,
\begin{equation}
    \bar{h}_{a b}(t, X^i) = 4 \int_V \frac{1}{|X^i - x^i|} 
        T_{a b}(t - |X^i - x^i|, x^i) dV,
\end{equation}
where the integration is over the volume of the star. For simplicity, we assume 
that the waves are generated by a weak source and are observed far from the 
source, $r \ll \mathcal{R}$. In this limit, we have 
\begin{equation}
    \bar{h}_{a b}(t, X^i) = \frac{4}{\mathcal{R}} \int_V T_{a b}(t - r, x^i) dV 
        + \order\left( \frac{1}{\mathcal{R}^2} \right).
\end{equation}
What we have done here, in exploiting a Newtonian limit, forms the basis of the 
post-Newtonian expansion. As we have discussed, general relativity is a highly 
non-linear theory, which makes it challenging to evaluate. This contributes to 
the appeal of linear perturbation theory in relativistic problems. The 
post-Newtonian expansion bridges the gap between the linear approximation and 
the full non-linear theory by expanding the relativistic expressions in terms of 
small parameters 
\citep[see, \textit{e.g.},][]{2006LRR.....9....4B, 2014grav.book.....P}.

By using the conservation law $\partial^b T_{a b} = 0$ and noting how the 
divergence terms vanish on the surface, one can derive the relation 
\begin{equation}
    2 \int_V T_{i j} dV = \int_V \partial_t^2 T_{t t} x_i x_j dV.
\end{equation}
Hence, we can write 
\begin{equation}
    \bar{h}_{i j}(t, X^i) = \frac{2}{\mathcal{R}} \partial_t^2 \left[ 
        \int_V  T_{t t}(t - r, x^i) x_i x_j dV \right] 
        \equiv \frac{2}{\mathcal{R}} \partial_t^2 M_{i j}(t - r),
\end{equation}
where we have introduced the (non-traceless) mass-quadrupole moment $M_{i j}$. 
The justification for this identification lies in the fact that, in the 
Newtonian limit, $T_{t t} = \rho$. Transforming to the transverse-traceless 
frame, we arrive at the result 
\begin{equation}
    h^\text{TT}_{i j}(t, X^i) 
        = \frac{2}{\mathcal{R}} \Lambda_{i j}^{\hphantom{i j} k l} 
        \partial_t^2 M_{k l}(t - r)
        = \frac{2}{\mathcal{R}} \partial_t^2 
        Q_{i j}(t - r),
    \label{eq:QuadrupoleFormula}
\end{equation}
where $Q_{i j}$ is the trace-reduced analogue of $M_{i j}$, given by 
(\ref{eq:QuadrupoleTensor}) in this limit. Equation~(\ref{eq:QuadrupoleFormula}) 
is the \textit{quadrupole formula} and it provides the basis for many useful 
gravitational-wave estimates. We see that (i) gravitational waves are generated 
by accelerating sources of matter (entirely analogous to how waves are produced 
in electromagnetism), (ii) the radiation falls off, to leading order, as 
$1 / \mathcal{R}$ and (iii) gravitational waves are quadrupolar. This result is 
true in the full non-linear theory; monopole and dipole radiation are absent in 
general relativity. The lack of monopole radiation turns out to be a consequence 
of Birkhoff's theorem \citep{1921ArMAF..15...18J, 1923rmp..book.....B}, that any 
spherically symmetric spacetime must be static and asymptotically flat. Dipole 
radiation does not exist because there are no opposite charges in gravity.

One can show that the gravitational-wave luminosity is 
\citep{1918SPAW.......154E}
\begin{equation}
    L_\text{GW} = \frac{1}{5} 
        \langle \partial_t^3 Q_{i j}(t - r) \partial_t^3 Q^{i j}(t - r) \rangle
    \label{eq:GravitationalWaveLuminosity}
\end{equation}
This characterises how gravitational waves carry energy away from the source. It 
was the Hulse-Taylor binary that enabled a first observational test for this 
result by studying the binary's orbital decay 
\citep[see Fig.~\ref{fig:HulseTaylor};][]{1975ApJ...195L..51H}.


%% file: sections/chapter-3.tex
\chapter{Gravitational waves from accreting neutron stars}
\label{ch:PopulationSynthesis}

In this chapter, we shall consider the question of whether gravitational waves 
are expected to play a meaningful role in the dynamics of accreting neutron 
stars. This was the subject of \citet{2019MNRAS.488...99G}, which we follow in 
this chapter. In particular, we investigate whether an additional component is 
needed in order to explain the spin evolution of accreting neutron stars. We 
consider gravitational-wave emission as the source of such a component and 
explore what may be the dominant gravitational-wave production mechanism.

To begin with, in Sec.~\ref{sec:AccretingNeutronStars}, we provide a brief 
overview of accreting neutron stars and the puzzle observing their spins has 
presented. In Sec.~\ref{sec:AccretionLMXBs}, we introduce the basic theory 
regarding accretion in low-mass X-ray binaries and discuss how transient 
accretion can affect the spin evolution of accreting neutron stars. In 
Sec.~\ref{sec:GravitationalWavesRotating}, we provide a brief review of 
gravitational radiation in these systems and the different mechanisms that can 
give rise to such emission. We describe our model for the spin evolution of 
accreting neutron stars in Sec.~\ref{sec:SpinModel}. In 
Sec.~\ref{sec:SimulatedPopulations}, we summarise the results of our 
neutron-star population simulations, that include the different 
gravitational-wave-production mechanisms. Finally, we summarise and suggest 
future work in Sec.~\ref{sec:Summary3}. 

\section{Accreting neutron stars}
\label{sec:AccretingNeutronStars}

A promising subset of rotating neutron stars in the context of 
gravitational-wave searches are the accreting neutron stars 
\citep{1978MNRAS.184..501P, 1984ApJ...278..345W}. The reason for this is quite 
intuitive. In order for a star to emit gravitational radiation, it must be 
deformed. Suppose a star is accreting from a companion. As the accreted matter 
gets close to the star, it begins to follow the magnetic-field lines. Provided 
the magnetic poles of the star are misaligned with respect to its axis of 
rotation, then the star quite naturally develops a mass asymmetry as matter 
piles up around the poles. This picture is overly simplified, since there are 
many aspects to this that we do not fully understand. Nevertheless, there is a 
strong case to study these systems more.

The classical picture for the evolution of rapidly spinning neutron stars begins 
with a neutron star accreting gas via a circumstellar accretion disc from their 
companion in a low-mass X-ray binary 
\citep{1982Natur.300..728A, 1982CSci...51.1096R}. This process causes the 
neutron star to spin up and, eventually, the neutron star accretes all the gas 
from its companion such that all that is left of the binary is a radio 
millisecond pulsar. This scenario should, in theory, have no difficulty in 
spinning neutron stars up to their centrifugal break-up frequency 
\citep{1994ApJ...424..823C}, which is generally above $\sim \SI{1}{\kilo\hertz}$ 
for most equations of state \citep{2007PhR...442..109L}. However, the 
fastest-spinning pulsar that has been observed to date is PSR J1748-2446ad, 
which spins at \SI{716}{\hertz} \citep{2006Sci...311.1901H}; well below the 
limit set by the break-up frequency. In fact, the distribution of spins for both 
low-mass X-ray binaries and radio millisecond pulsars has been shown to have a 
statistically significant cut-off at \SI{730}{\hertz} 
\citep{2003Natur.424...42C, 2010ApJ...722..909P}. 

Accreting millisecond X-ray pulsars are a sub-class of low-mass X-ray binaries 
that have been spun up to millisecond periods through accretion 
[see \citet{2021ASSL..461..143P} for a review on these systems]. They are 
characterised by accretion rates $\gtrsim \SI{e-11}{\solarMass\per\year}$ and 
comparatively weak magnetic fields ($\sim \SI{e8}{\gauss}$). Another important 
sub-class of low-mass X-ray binaries are the nuclear-powered X-ray pulsars. 
These pulsars show short-lived burst oscillations during thermonuclear burning 
on their surfaces and are distinct from accreting millisecond X-ray pulsars due 
to being powered by nuclear burning rather than accretion. Nineteen accreting 
millisecond X-ray pulsars and eleven nuclear-powered X-ray pulsars have been 
observed to date. 

\citet{2018A&A...620A..69H} showed that the observed rotation rate limit of 
neutron stars does not correspond to centrifugal break-up and argued that 
additional spin-down torques are required to explain this effect. It is unclear 
what physical process prevents these neutron stars from spinning up to 
sub-millisecond periods. One candidate is the interaction between the magnetic 
field and the accretion disc 
\citep{1978ApJ...223L..83G, 1997ApJ...490L..87W, 2005MNRAS.361.1153A}. 
\citet{2012ApJ...746....9P} demonstrated that a magnetic-field strength at the 
magnetosphere of $\sim \SI{e8}{\gauss}$ could be enough to explain the 
deficiency in accreting neutron stars above $\sim \SI{700}{\hertz}$. More 
recently, it has been shown that transient accretion can have a significant 
impact on the spin evolution of an accreting neutron star 
\citep{2017MNRAS.470.3316D, 2017ApJ...835....4B}. In fact, 
\citet{2017ApJ...835....4B} noted that in the case of transient accretion, a 
magnetosphere of $\sim \SI{e8}{\gauss}$ would no longer be sufficient to explain 
the spin limit. It was suggested by \citet{1998ApJ...501L..89B} and 
\citet{1999ApJ...516..307A} that one would observe a spin-frequency limit to 
accreting neutron stars if they were emitting gravitational waves, thus, 
providing a torque to balance the accretion torques 
\citep{1978MNRAS.184..501P, 1984ApJ...278..345W}. 

\section{Accretion in low-mass X-ray binaries}
\label{sec:AccretionLMXBs}

In a low-mass X-ray binary, the companion star has overfilled its Roche lobe and 
is donating matter to the neutron star through the inner Lagrange point. Given 
this donated gas has some large specific angular momentum, it cannot be 
transferred directly to the surface of the neutron star and instead forms a 
circumstellar accretion disc around it. The gas from the accretion disc is 
channelled onto the magnetic poles of the neutron star along the magnetic-field 
lines. This channelling occurs at what is known as the 
\textit{magnetospheric radius}, $r_\text{m}$, which is the characteristic 
radius where the magnetic field dominates interactions. At this boundary the 
magnetic field can, in turn, be distorted by the accreting gas. This coupling, 
between the field lines and the disc, results in a torque acting on the star 
that can spin it up or down depending on the relative difference between 
$r_\text{m}$ and the \textit{co-rotation radius},
\begin{equation}
	r_\text{c} \equiv \left( \frac{G M}{\Omega^2} \right)^{1 / 3},
\end{equation}
which is the location of a Keplerian disc that rotates with the same frequency 
of the neutron star, where $\Omega = 2 \pi \nu = 2 \pi / P$ is its angular 
frequency, $\nu$ is the spin frequency%
\footnote{We re-emphasise that this is distinct from the metric potential 
$\nu$.}
and $P$ is the spin period. If $r_\text{m} < r_\text{c}$, the neutron star spins 
slower than the accretion disc and so the gas that is channelled onto the 
neutron star has greater specific angular momentum than it, thus, acting to spin 
it up. Conversely, if $r_\text{m} > r_\text{c}$, the neutron star spins faster 
than the disc, which spins it down.

The magnetospheric radius is a somewhat poorly understood quantity. It is 
generally defined as the point where the kinetic energy of in-falling gas 
becomes comparable to the magnetic energy of the magnetosphere. For the 
straightforward case where the gas is radially accreted onto the neutron star, 
one can calculate the Alfv{\'e}n radius, $r_\text{A}$, from:
\begin{equation}
	\frac{1}{2} \rho(r_\text{A}) v(r_\text{A})^2 = \frac{B(r_\text{A})^2}{8 \pi},
\end{equation}
where $\rho(r_\text{A})$, $v(r_\text{A})$ and $B(r_\text{A})$ are the gas 
density, gas speed and magnetic-field strength, respectively, at 
$r_\text{A}$. This calculation gives the standard expression for the 
magnetospheric radius \citep{1972A&A....21....1P}:
\begin{equation}
	r_\text{A} = \left( \frac{\mu^4}{2 G M \dot{M}^2} \right)^{1/7},
\end{equation}
where $\mu = B R^3$ is the magnetic moment of the neutron star%
\footnote{It should be re-emphasised that this is separate from the chemical 
potential $\mu$.}
and $\dot{M}$ is the mass-accretion rate from the disc to the neutron-star 
surface. This picture becomes more complicated when considering accretion from a 
circumstellar disc. A phenomenological factor $\xi$ of order unity is introduced 
to correct for the non-spherical geometry of the problem and also account for 
the extended transition region between where mass is accreted onto the star and 
where mass is ejected in an outflow. This gives the magnetospheric radius as
\begin{equation}
    r_\text{m} = \xi r_\text{A} 
        = \xi \left( \frac{\mu^4}{2 G M \dot{M}^2} \right)^{1/7}.
\end{equation}
Typically, $\xi$ is assumed to fall in the range $0.5 - 1.4$ 
\citep{1996ApJ...465L.111W}. This correction demonstrates that $r_\text{m}$ is 
sensitive to the coupling between the field lines and the accretion disc that 
plays a key role in understanding the magnetospheric radius.

\subsection{Accretion-torque models}
\label{sec:AccretionTorque}

The spin evolution of a neutron star is dictated by the torques acting upon the 
star. By measuring the time derivative of the spin period, $\dot{P}$, (or, 
equivalently, the time derivative of the angular frequency, $\dot{\Omega}$) one 
can gain insight into the physics of accretion processes, as well as other 
aspects such as magnetic-field strengths and gravitational-wave emission. The 
variation in spin is related to the torque exerted onto the neutron star by the 
standard expression
\begin{equation}
	N = I_{z z} \dot{\Omega} = -\frac{2 \pi I_{z z} \dot{P}}{P^2}, 
	\label{eq:TorqueSpinDerivative}
\end{equation}
where $I_{z z}$ is the principal stellar moment of inertia about the rotation 
axis.

For a neutron star accreting from an accretion disc truncated at the 
magnetospheric transition, with $r_\text{m} < r_\text{c}$, the standard spin-up 
torque is 
\begin{equation}
    N_\text{acc} = \dot{M} r_\text{m}^2 \Omega_\text{K}(r_\text{m}) 
        = \dot{M} \sqrt{G M r_\text{m}},
	\label{eq:AccretionDiscN}
\end{equation}
where $\Omega_\text{K}(r_\text{m})$ is the Keplerian angular frequency at 
$r_\text{m}$. For $r_\text{m} > r_\text{c}$, the coupling between the 
magnetosphere and the accretion disc becomes important as the magnetic-field 
lines are threaded through the disc, so extra torques due to magnetic stresses 
come into play. \citet{1979ApJ...234..296G} developed an accretion model based 
on detailed calculations of this coupling. The torque from this model predicts
\begin{equation}
\begin{split}
    \dot{P} &\approx \num{-5.0e-5} \
        \left( \frac{M}{\SI{1.4}{\solarMass}} \right)^{3/7} 
        \left( \frac{I_{z z}}{\SI{e45}{\gram\centi\metre\squared}} \right)^{-1} 
        \left( \frac{\mu}{\SI{e30}{\gauss\centi\metre\cubed}} \right)^{2/7} \\
    &\quad \times \left[ \left( \frac{P}{\SI{1}{\second}} \right) 
        \left( \frac{\dot{M}}{\SI{e-9}{\solarMass\per\year}} \right)^{3/7} 
        \right]^2 n(\omega_\text{s}) \ \si{\second\per\year},
\end{split}
	\label{eq:GhoshLambPdot}
\end{equation}
where $n(\omega_\text{s})$ is the dimensionless torque that accounts for the 
magnetic field-accretion disc coupling and is a function of the fastness 
parameter $\omega_\text{s}$. The fastness parameter is defined as the ratio of 
the neutron-star spin frequency to the Keplerian orbital frequency at the 
magnetospheric boundary,
\begin{equation}
\begin{split}
    \omega_\text{s} \equiv \frac{\Omega}{\Omega_\text{K}(r_\text{m})} 
        &\approx \num{3.1} \ \xi^{3/2} 
        \left( \frac{M}{\SI{1.4}{\solarMass}} \right)^{-5/7} 
        \left( \frac{\mu}{\SI{e30}{\gauss\centi\metre\cubed}} \right)^{6/7} \\
    &\quad \times \left[ \left( \frac{P}{\SI{1}{\second}} \right) 
        \left( \frac{\dot{M}}{\SI{e-9}{\solarMass\per\year}} \right)^{3/7} 
        \right]^{-1}.
\end{split}
\end{equation}
The sign of $n(\omega_\text{s})$ depends on whether the neutron star accretes 
the gas and spins up (the `slow rotator' regime, $\omega_\text{s} < 1$) or 
ejects the gas and spins down 
\citep[the `fast rotator' regime, $\omega_\text{s} > 1$;][]
{1995ApJ...449L.153W}. 
It is interesting to note that a neutron star can still be spun down at long 
spin periods ($P \gg \SI{1}{\second}$) as the magnetic field can be strong 
enough to mean it would still be classified as a fast rotator.

\citet{2014MNRAS.437.3664H} introduced a simple approximation to the 
\citet{1979ApJ...234..296G} model by considering angular momentum changes on the 
neutron star. Matter accreting at the magnetosphere, $r_\text{m}$, has specific 
angular momentum
\begin{equation}
	l_\text{acc} = \pm r_\text{m}^2 \Omega_\text{K}(r_\text{m}),
\end{equation}
where the sign of $l_\text{acc}$ depends on whether there is prograde rotation 
between the accretion disc and the neutron star ($l_\text{acc} > 0$) or 
retrograde rotation ($l_\text{acc} < 0$). Prograde rotation will be assumed 
here. What must also be accounted for is matter that is ejected from the neutron 
star, that will carry specific angular momentum
\begin{equation}
	l_\text{m} = r_\text{m}^2 \Omega.
\end{equation}
Both of these effects produce a torque on the neutron star. The net torque is 
obtained by summing these contributions:
\begin{equation}
    N = \dot{M} (l_\text{acc} - l_\text{m}) 
        = \dot{M} r_\text{m}^2 \Omega_\text{K}(r_\text{m}) (1 - \omega_\text{s}).
	\label{eq:HoN}
\end{equation}
In this relation, one can see the standard spin-up torque due to disc accretion 
(\ref{eq:AccretionDiscN}) that is corrected by the fastness parameter to account 
for interactions spinning down the neutron star. (This model phenomenologically 
accounts for effects such as accretion disc-magnetic field coupling and 
outflows.) This expression can be related to the change in spin period using 
(\ref{eq:TorqueSpinDerivative}) to obtain
\begin{equation}
\begin{split}
    \dot{P} &\approx \num{-8.1e-5} \ \xi^{1/2} 
        \left( \frac{M}{\SI{1.4}{\solarMass}} \right)^{3/7} 
        \left( \frac{I_{z z}}{\SI{e45}{\gram\centi\metre\squared}} \right)^{-1} 
        \left( \frac{\mu}{\SI{e30}{\gauss\centi\metre\cubed}} \right)^{2/7} \\
    &\quad \times \left[ \left( \frac{P}{\SI{1}{\second}} \right) 
        \left( \frac{\dot{M}}{\SI{e-9}{\solarMass\per\year}} \right)^{3/7} 
        \right]^2 (1 - \omega_\text{s}) \ \si{\second\per\year}.
\end{split}
	\label{eq:HoPdot}
\end{equation}
It is clear from (\ref{eq:HoPdot}) that the fastness parameter dictates whether 
the neutron star spins up or down.

A commonly considered aspect of a neutron star's spin evolution is the spin 
equilibrium. This occurs when the spin rate is gradually adjusted until the net 
torque on the star is approximately zero and the accretion flow is truncated at 
the magnetospheric radius, $r_\text{m} \simeq r_\text{c}$. When a neutron star 
reaches spin equilibrium, it is straightforward to estimate its magnetic field, 
assuming that the accretion rate and spin are known. One can estimate the 
spin-equilibrium period, $P_\text{eq}$, from (\ref{eq:HoPdot}) by setting 
$\dot{P} = 0$, when $\omega_\text{s} = 1$:
\begin{equation}
    P_\text{eq} \approx \num{8.2} \ \xi^{3/2} 
        \left( \frac{M}{\SI{1.4}{\solarMass}} \right)^{-5/7} 
        \left(\frac{\mu}{\SI{e26}{\gauss\centi\metre\cubed}}\right)^{6/7} 
        \left(\frac{\dot{M}}{\SI{e-11}{\solarMass\per\year}}\right)^{-3/7} \ 
        \si{\milli\second},
\end{equation}
where we have scaled the period to characteristic accreting millisecond X-ray 
pulsar values. Here, we see that, provided fast accretion rates and small 
magnetic-field strengths (therefore, negligible magnetic-dipole spin-down), it 
should not be difficult for (at least some) accreting neutron stars to reach 
sub-millisecond periods.

The magnetic-field lines rotate with the neutron star. This produces 
magnetic-dipole radiation which causes the neutron star to spin down. The torque 
due to this is well approximated by 
\citep[see, \textit{e.g.},][]{2017MNRAS.470.3316D}
\begin{equation}
	N_\text{EM} = - \frac{2 \mu^2 \Omega^3}{3 c^3}.
\end{equation}
The change in spin due to magnetic-dipole radiation is 
\begin{equation}
    \dot{P}_\text{EM} \approx \num{3.1e-8} \
        \left( \frac{\mu}{\SI{e30}{\gauss\centi\metre\cubed}} \right)^2 
        \left( \frac{I_{z z}}{\SI{e45}{\gram\centi\metre\squared}} \right)^{-1} 
        \left( \frac{P}{\SI{1}{\second}} \right)^{-1} \ \si{\second\per\year}.
	\label{eq:EMPdot}
\end{equation}
The vast majority of pulsars are isolated and their spin evolution can be 
generally described by magnetic-dipole radiation. However, in the case of 
rapidly accreting neutron stars this effect can be essentially negligible. There 
are more accurate numerical models one can use to describe these torques 
\citep[\textit{e.g.}, see][]{2006ApJ...648L..51S}.

\subsection{Transient accretion}
\label{sec:TransientAccretion}

Up until now, the accretion rate has been implicitly assumed to be steady. 
However, many low-mass X-ray binaries exhibit long periods of quiescence, that 
can be of the order of months to years, and short transient outbursts, that can 
last from days to weeks. These outbursts are believed to be caused by 
instabilities in the accretion disc and occur when the mass-accretion rate rises 
above a certain threshold \citep[see, \textit{e.g.},][]{1997ASPC..121..351L}. As 
the companion star donates a steady flow of gas to the accretion disc, the disc 
gets larger and eventually reaches a critical mass to trigger an instability. 
This causes the accretion rate from the disc to the surface of the neutron star 
to increase by several orders of magnitude, giving rise to a transient outburst. 
Once the accretion disc has donated a sufficient amount of gas, the system 
returns to a quiescent state until a new outburst occurs when the disc has 
accumulated enough mass from the companion and the cycle repeats 
\citep{2007A&ARv..15....1D}.

\citet{2017MNRAS.470.3316D} and \citet{2017ApJ...835....4B} have shown that 
transient accretion with a varying accretion rate has a significant impact on 
the spin evolution of a neutron star. Both found that for a given long-term 
average accretion rate, these transients can spin up neutron stars to rates 
several times higher than that of persistent accretors, however, it takes 
approximately an order of magnitude longer to reach these spin-equilibrium 
periods. This demonstrates that for transient systems, like most low-mass X-ray 
binaries, it is not accurate to assume a time-averaged accretion rate but 
instead one must consider the outburst/quiescence phases. 
\citet{2017MNRAS.470.3316D} noted that the two key changes when considering 
transient accretion are: (i) the torque over an outburst is significantly 
smaller than for the persistent case at a given accretion rate and (ii) the 
equilibrium accretion rate is shifted to a lower value. This has the combined 
effect to increase the time it takes for a transient source to reach spin 
equilibrium and decrease its spin-equilibrium period.

\sloppy
\citet{2017MNRAS.470.3316D} and \citet{2017ApJ...835....4B} found that the 
spin-equilibrium period and time to reach spin equilibrium are sensitive to the 
features of the accretion profile. They found that by increasing the duration 
of an outburst by a factor of $10$ the spin-equilibrium period can decrease 
by up to a factor of $2$.

\begin{figure}[h]
    \centering
	\includegraphics[width=0.7\textwidth]{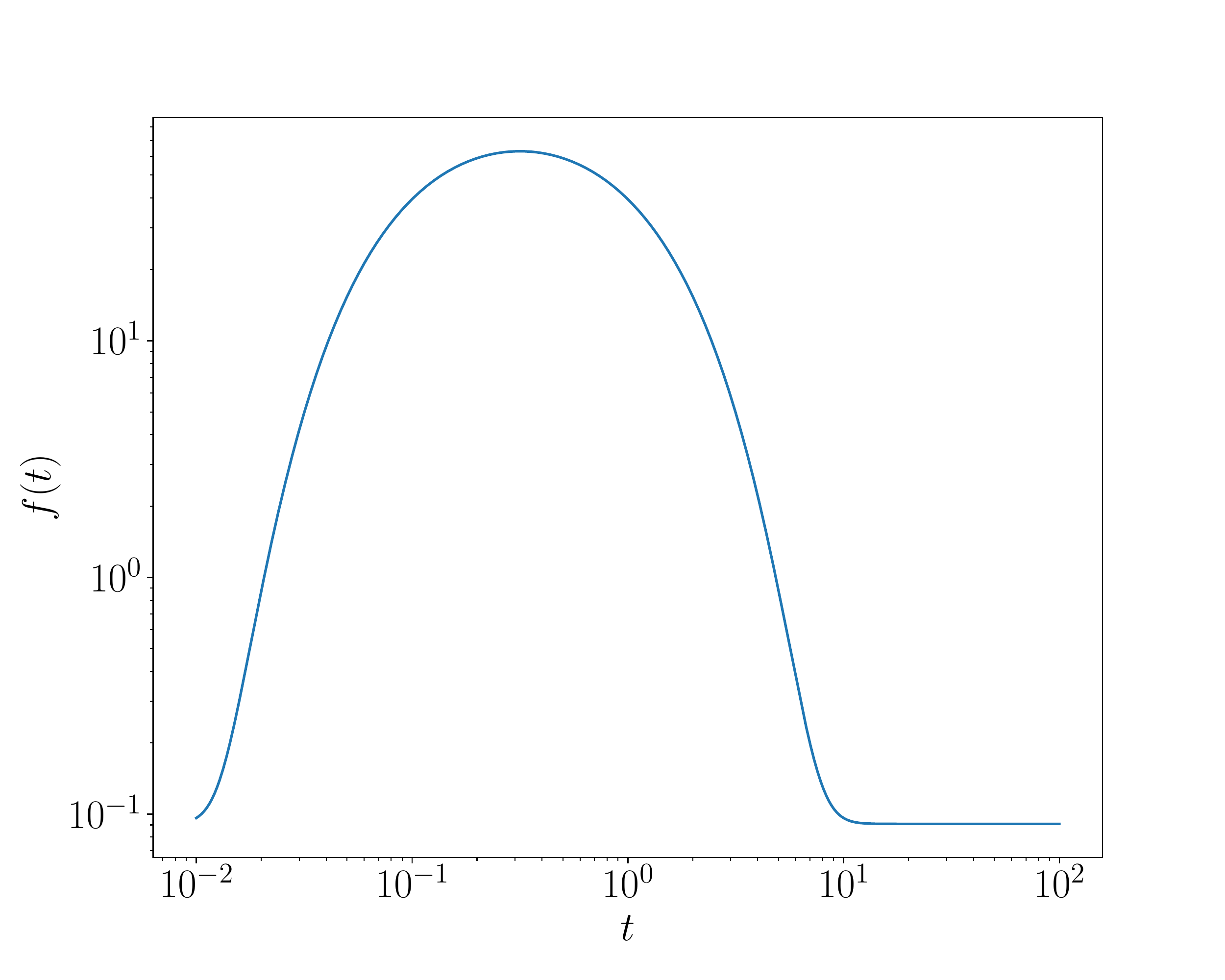}
    \caption[Accretion outburst profile]{Accretion outburst profile, $f(t)$, as 
             a function of time, $t$, 
             where $F_\text{t} = 10$, $f_\text{max} / f_\text{min} = 693.97$ and 
             $T_\text{recurrence} = 100$. The accretion rate and time have 
             arbitrary units. [Recreated from \citet{2017MNRAS.470.3316D}.]}
	\label{fig:AccretionProfile}
\end{figure}

\fussy
For her analysis, \citet{2017MNRAS.470.3316D} used a fast-rise, 
exponential-decay function to model the accretion profile 
[whereas \citet{2017ApJ...835....4B} used a simple sawtooth function]:
\begin{equation}
    f(t) = \exp \left( \sqrt{\frac{2}{F_\text{t}}} \right) 
        \exp \left( - \frac{1}{10 t} - \frac{10 t}{F_\text{t}} \right) 
        + f_\text{min},
	\label{eq:AccretionProfile}
\end{equation}
where $t$ denotes the time from the beginning of the outburst, $F_\text{t}$ is 
an approximate measure of the duration of the outburst and $f_\text{min}$ is 
the minimum. It should be noted that this function models a single 
outburst/quiescence cycle, so in order to model multiple cycles one repeats 
this after a given recurrence time, $T_\text{recurrence}$. Time has arbitrary 
units in this model. The ratio of the maximum to the minimum is
\begin{equation}
    \frac{f_\text{max}}{f_\text{min}} = \frac{1}{f_\text{min}} 
        \exp \left( \frac{\sqrt{2} - 2}{\sqrt{F_\text{t}}} \right) + 1.
	\label{eq:AccretionRatio}
\end{equation}
This accretion profile requires two normalisations. The first normalisation 
chooses $f_\text{max} / f_\text{min}$ to obtain $f_\text{min}$ for a fixed 
$F_\text{t}$ using (\ref{eq:AccretionRatio}). The second normalisation 
is to demand that $\langle f(t) \rangle = 1$. This normalisation depends on 
$T_\text{recurrence}$ and results in $f_\text{min}$ no longer corresponding 
precisely to the minimum value. These normalisations allow one to choose the 
magnitude of the accretion outburst, with respect to the quiescent accretion 
rate, and also mean that one can simply choose an average accretion rate over 
one cycle by multiplying (\ref{eq:AccretionProfile}) by the chosen 
average. Thus, the time-dependent accretion rate is given by
\begin{equation}
	\dot{M}(t) = \langle \dot{M} \rangle f(t),
	\label{eq:AccretionRate}
\end{equation}
where $f(t)$ has been appropriately normalised. The canonical profile used by 
\citet{2017MNRAS.470.3316D} had $F_\text{t} = 10$, 
$f_\text{max} / f_\text{min} = 693.97$ and $T_\text{recurrence} = 100$ and is 
shown in Fig.~\ref{fig:AccretionProfile}.

\section{Gravitational radiation from a rotating star}
\label{sec:GravitationalWavesRotating}

In order for a spinning neutron star to emit gravitational radiation, it must 
host a non-axisymmetric mass deformation. This is made obvious through the 
quadrupole-moment tensor, given by (\ref{eq:QuadrupoleTensor}) in the Newtonian 
limit. The quadrupole moment is particularly relevant for calculating the 
gravitational-wave strain of a given source.

Indeed, it turns out that the dominant multipole moment for gravitational-wave 
emission is the $(\ell, m) = (2, 2)$ moment, $Q_{2 2}$, which is defined by 
(\ref{eq:Multipole}). A rotating source with a quadrupole moment will radiate 
gravitational waves at a frequency that is double its rotation frequency. 
Gravitational waves carry energy and angular momentum away from the source. The 
gravitational-wave luminosity is given by 
(\ref{eq:GravitationalWaveLuminosity}), thus, the torque that spins down the 
star is 
\begin{equation}
    N_\text{GW} = - \frac{L_\text{GW}}{\Omega} 
        = - \frac{G}{c^5} \frac{1}{5 \Omega} 
        \langle \partial_t^3 Q_{i j} \partial_t^3 Q^{i j} \rangle,
\end{equation}
One can show that a uniformly spinning star has 
\begin{equation}
    \langle \partial_t^3 Q_{i j} \partial_t^3 Q^{i j} \rangle 
        = \frac{256 \pi}{15} \Omega^6 Q_{2 2}^2,
\end{equation}
which gives 
\begin{equation}
    N_\text{GW} = - \frac{256 \pi}{75} \frac{G \Omega^5 Q_{2 2}^2}{c^5}.
    \label{eq:TorqueGW}
\end{equation}
Equation~(\ref{eq:TorqueGW}) describes the rate at which the deformed star loses 
angular momentum due to a quadrupole deformation.

In addition to the quadrupole moment, it is fashionable to consider the fiducial 
ellipticity, which is defined as \citep{2005PhRvL..95u1101O} 
\begin{equation}
    \epsilon \equiv \sqrt{\frac{8 \pi}{15}} \frac{Q_{2 2}}{I_{z z}}, 
    \label{eq:Ellipticity}
\end{equation}
where the principal stellar moment of inertia is taken to have the fiducial 
value of $I_{z z} = \SI{e45}{\gram\centi\metre\squared}$. It should be noted 
that this fiducial principal moment of inertia can be different to the star's 
actual principal moment of inertia by a factor of a few. This parameter is often 
reported in observational papers.

The braking torque due to gravitational waves (\ref{eq:TorqueGW}) corresponds to 
a spin-down rate of
\begin{equation}
    \dot{P}_\text{GW} \approx \num{1.4e-19} \
        \left( \frac{I_{z z}}{\SI{e45}{\gram\centi\metre\squared}} \right)^{-1} 
        \left( \frac{Q_{2 2}}{\SI{e37}{\gram\centi\metre\squared}} \right)^2 
        \left( \frac{P}{\SI{1}{\second}} \right)^{-3} \ \si{\second\per\year}.
	\label{eq:GWPdot}
\end{equation}

In order to estimate how strong a quadrupole is needed in order to considerably 
influence the spin evolution of the neutron star, it is useful to balance 
(\ref{eq:TorqueGW}) with the accretion-magnetosphere torque from (\ref{eq:HoN}). 
This leads to
\begin{equation}
\begin{split}
    Q_{2 2} &\approx \num{4.2e37} \ \xi^{1 / 4} 
        \left( \frac{M}{\SI{1.4}{\solarMass}} \right)^{3 / 14} 
        \left( \frac{\mu}{\SI{e30}{\gauss\centi\metre\cubed}} \right)^{1 / 7} 
        \left( \frac{\dot{M}}{\SI{e-9}{\solarMass\per\year}} \right)^{3 / 7} \\
	&\quad \times \left(\frac{f}{\SI{500}{\hertz}}\right)^{- 5 / 2} 
        (1 - \omega_\text{s}) \ \si{\gram\centi\metre\squared}.
\end{split}
    \label{eq:BalanceQ}
\end{equation}
For a typical accreting millisecond X-ray pulsar with $B \sim \SI{e8}{\gauss}$ and 
$\dot{M} \sim \SI{e-11}{\solarMass\per\year}$, this gives a quadrupole moment 
of $Q_{2 2} \sim \SI{e36}{\gram\centi\metre\squared}$ in order to achieve spin 
equilibrium at $f \sim \SI{500}{\hertz}$. One can express a mass quadrupole 
in terms of the fiducial ellipticity (\ref{eq:Ellipticity}). Therefore, in order 
to balance the accretion torque with gravitational-wave spin-down, one requires 
$\epsilon \sim \num{e-9}$. This is far smaller than the maximum deformation a 
neutron-star crust can sustain for most reasonable equations of state 
(see Chaps.~\ref{ch:Mountains} and \ref{ch:RelativisticMountains}). A recent 
population-based analysis has suggested that $\epsilon \approx \num{e-9}$ is the 
minimum ellipticity of millisecond pulsars \citep{2018ApJ...863L..40W}.

\begin{figure}[h]
    \centering
	\includegraphics[width=0.7\textwidth]{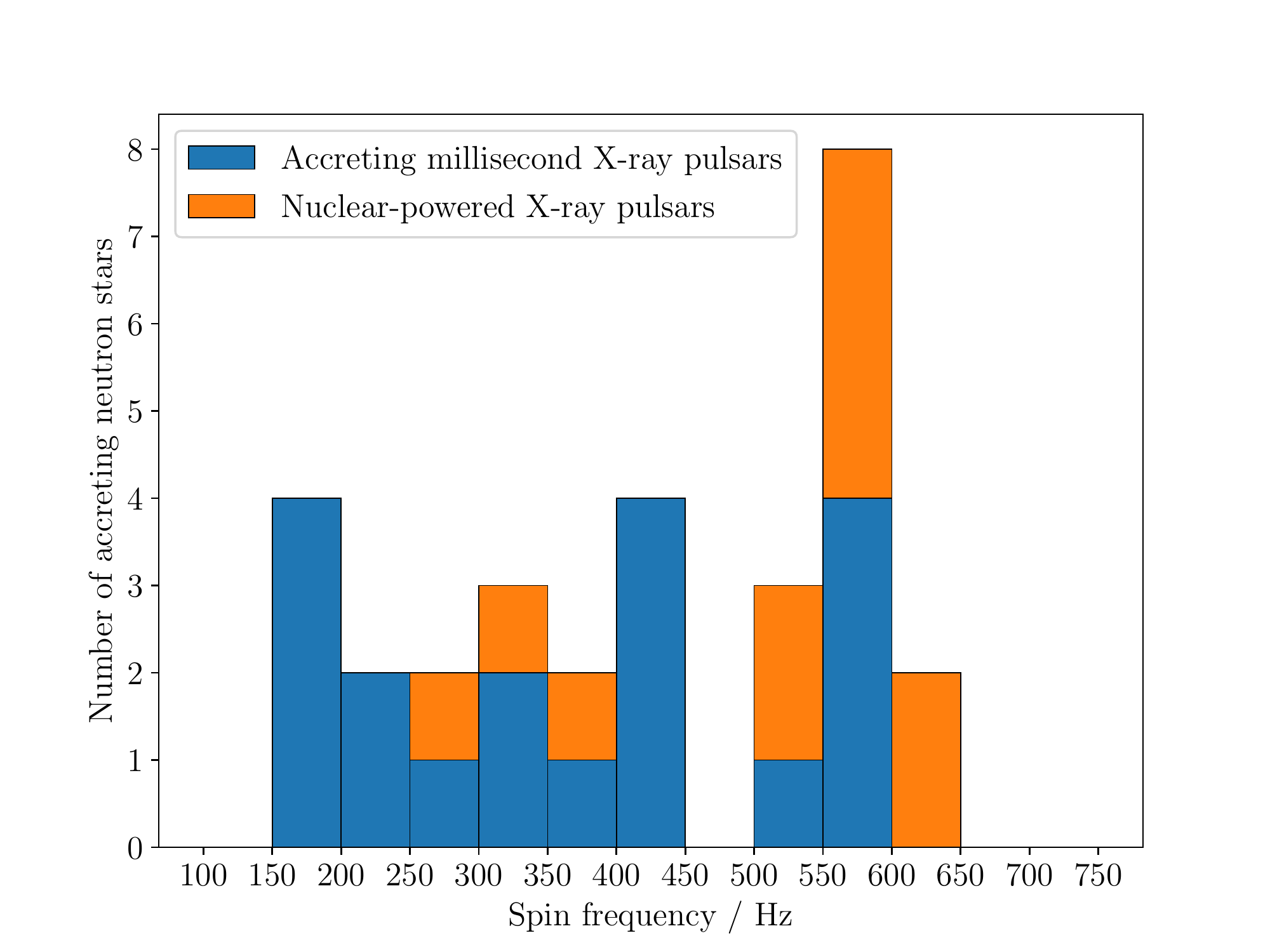}
    \caption[Spin-frequency distribution for accreting millisecond neutron 
             stars]{Distribution of spin frequencies for accreting neutron stars 
             with millisecond periods. The accreting neutron-star population 
             comprises accreting millisecond X-ray pulsars (blue) and 
             nuclear-powered X-ray pulsars (orange).}
	\label{fig:SpinDistribution}
\end{figure}

An outstanding problem in understanding rapidly spinning accreting neutron stars 
is their peculiar spin distribution (see Fig.~\ref{fig:SpinDistribution}). It 
is this unusual shape and, in particular, the sharp cut-off at 
$\sim \SI{600}{\hertz}$ that has motivated the search for gravitational waves 
from these systems. This is an appealing explanation since the braking torque 
due to gravitational waves scales as the fifth power of the spin frequency for 
deformed, rotating neutron stars [see (\ref{eq:TorqueGW})]. 
\citet{2017ApJ...850..106P} have shown that among accreting millisecond X-ray 
pulsars and nuclear-powered X-ray pulsars, there appear to be two 
sub-populations. One sub-population is at a relatively low spin-frequency, with 
a mean spin of $\approx \SI{300}{\hertz}$. The second sub-population has a 
higher peak and a mean of $\approx \SI{575}{\hertz}$. This faster sub-population 
has a very narrow range and is composed of a mixture of accreting millisecond 
X-ray pulsars and nuclear-powered X-ray pulsars. The two sub-populations are 
separated by a transition region around $\approx \SI{540}{\hertz}$. 
\citet{2017ApJ...850..106P} argued that, when considering various accretion 
torque models, no model naturally explains the presence of a fast sub-population 
and postulated that, whatever mechanism that causes this clustering, it must set 
in quickly -- as soon as the pulsars reach a certain spin threshold. It was 
noted by \citet{2017ApJ...850..106P} that this is a subtly different problem to 
the one of accreting neutron stars not spinning close to their break-up 
frequency. These two problems make gravitational waves a promising avenue to 
explore. Gravitational waves can help justify the transition region between the 
two sub-populations and provide a physical meaning to it (the region in which 
gravitational-wave emission starts to become significant), and naturally explain 
the cut-off at $\sim \SI{600}{\hertz}$. [See \citet{2008AIPC.1068..130H} and 
\citet{2014A&A...566A..64P} for additional work on the spin-frequency 
distribution of millisecond pulsars.]

There are a number of different ways a mass asymmetry could arise in an 
accreting neutron star. \citet{1998ApJ...501L..89B} originally proposed that 
interior temperature asymmetries misaligned with respect to the spin-axis of the 
neutron star could produce a significant quadrupole through 
temperature-sensitive electron captures. Hotter regions of the crust would have 
electron captures at lower pressures and so the density drop would occur at 
higher altitudes in the hotter parts of the crust. This is known as a 
\textit{thermal mountain} 
\citep{1998ApJ...501L..89B, 2000MNRAS.319..902U, 2005ApJ...623.1044M, 
2006MNRAS.373.1423H, 2006ApJ...641..471P}. 
Another mechanism through which mass quadrupoles can be built are through 
mountains sustained by magnetic stresses, called \textit{magnetic mountains} 
\citep{2002PhRvD..66h4025C, 2008MNRAS.385..531H}. These can occur when a neutron 
star has a sufficiently large toroidal or poloidal magnetic field that will act 
to distort the neutron star into an oblate or prolate shape and will naturally 
produce a quadrupole if the spin- and magnetic-axes are misaligned. A third way 
through which gravitational waves can arise is through internal \textit{r}-mode 
instabilities 
\citep{1998ApJ...502..708A, 1999ApJ...516..307A, 1999ApJ...517..328L, 
2000ApJ...534L..75A, 2002MNRAS.337.1224A, 2002ApJ...574L..57H, 
2002ApJ...578L..63W, 2006PhRvD..73h4001N, 2007PhRvD..76f4019B}. 
In a perfect fluid, these modes are unstable for all rates of rotation due to 
gravitational-wave emission.

We explored whether gravitational waves could explain the observed distribution 
and, if so, whether there is a preference for any of the 
gravitational-wave-production mechanisms. For our analysis, we did not consider 
mountains solely created by the magnetic field, nor did we consider magnetic 
mountains built through accretion. For these cases, the magnetic fields are not 
strong enough to sustain sufficiently large mountains for the spin evolution of 
these systems to be noticeably affected 
\citep[see][]{2008MNRAS.385..531H, 2011MNRAS.417.2696P, 2017PhRvL.119p1103H}.

\section{Spin-evolution model}
\label{sec:SpinModel}

We constructed a model for the spin evolution of an accreting neutron star. We 
incorporated the accretion-magnetosphere coupling by using the model of 
\citet{2014MNRAS.437.3664H} (\ref{eq:HoPdot}) and included a torque due to 
gravitational-wave spin-down (\ref{eq:GWPdot}). The spin rate is a first-order 
time derivative and so  can be evolved numerically. Our assumed canonical values 
for a low-mass X-ray binary are shown in Table~\ref{tab:CanonicalLMXB}. For 
our canonical neutron star we did not include gravitational-wave effects. We 
assumed our neutron stars to be incompressible (that is, they have a constant 
density), which affected the moment of inertia. For simplicity, we did not model 
the magnetic-field evolution. The time a neutron star is evolved for is denoted 
as the evolution time.

\begin{table}[h]
    \centering
	\caption[Canonical values for a low-mass X-ray binary]{Canonical values for 
             a low-mass X-ray binary.}
    \label{tab:CanonicalLMXB}
    \resizebox{\textwidth}{!}{  
	\begin{tabular}{c c c c c c c}
		\hline\hline
        $M$ / \si{\solarMass} & $R$ / \si{\kilo\metre} & $B$ / \si{\gauss} 
        & Initial spin period / \si{\second} & $\xi$ & 
        $\langle \dot{M} \rangle$ / \si{\solarMass\per\year} 
        & $Q_{2 2}$ / \si{\gram\centi\metre\squared} \\
		\hline
		$1.4$ & $10$ & $10^8$ & $0.1$ & $0.5$ & $\num{5e-11}$ & $0$ \\
		\hline
    \end{tabular}
    }
\end{table}

Our model can evolve both persistent and transient accretors. For transient 
accretors we used a fast-rise, exponential-decay function, described in 
Sec.~\ref{sec:TransientAccretion} by 
Eqs.~(\ref{eq:AccretionProfile})--(\ref{eq:AccretionRate}), and evolved the 
time-averaged spin-derivative, $\langle \dot{P}(P, \dot{M}) \rangle$, which, 
for a given neutron star, is a function of the spin and accretion rate. This 
average was obtained by averaging the spin derivative over one 
outburst/quiescence cycle. The time-average was evolved rather than the 
instantaneous spin rate, $\dot{P}(P, \dot{M})$, to simplify the integration 
procedure. Otherwise the integration procedure would have needed to take into 
account the full fast-rise, exponential-decay features of the accretion profile. 
For persistent accretors this was not a problem and so we could simply evolve 
$\dot{P}(P, \dot{M})$. Unless specified otherwise we used the following values 
for the transient accretion profile: $F_\text{t} = \SI{10}{\year}$, 
$T_\text{recurrence} = \SI{100}{\year}$ and 
$f_\text{max} / f_\text{min} = \num{e4}$. This was chosen for simplicity and to 
limit the explorable parameter space. Most of our simulations turned out to be 
relatively insensitive to the exact values of these parameters. Of course, 
should one be interested in modelling individual systems with this profile, then 
particular care would need to be taken when tuning these parameters.

\begin{figure}[h]
    \centering
	\includegraphics[width=0.7\textwidth]{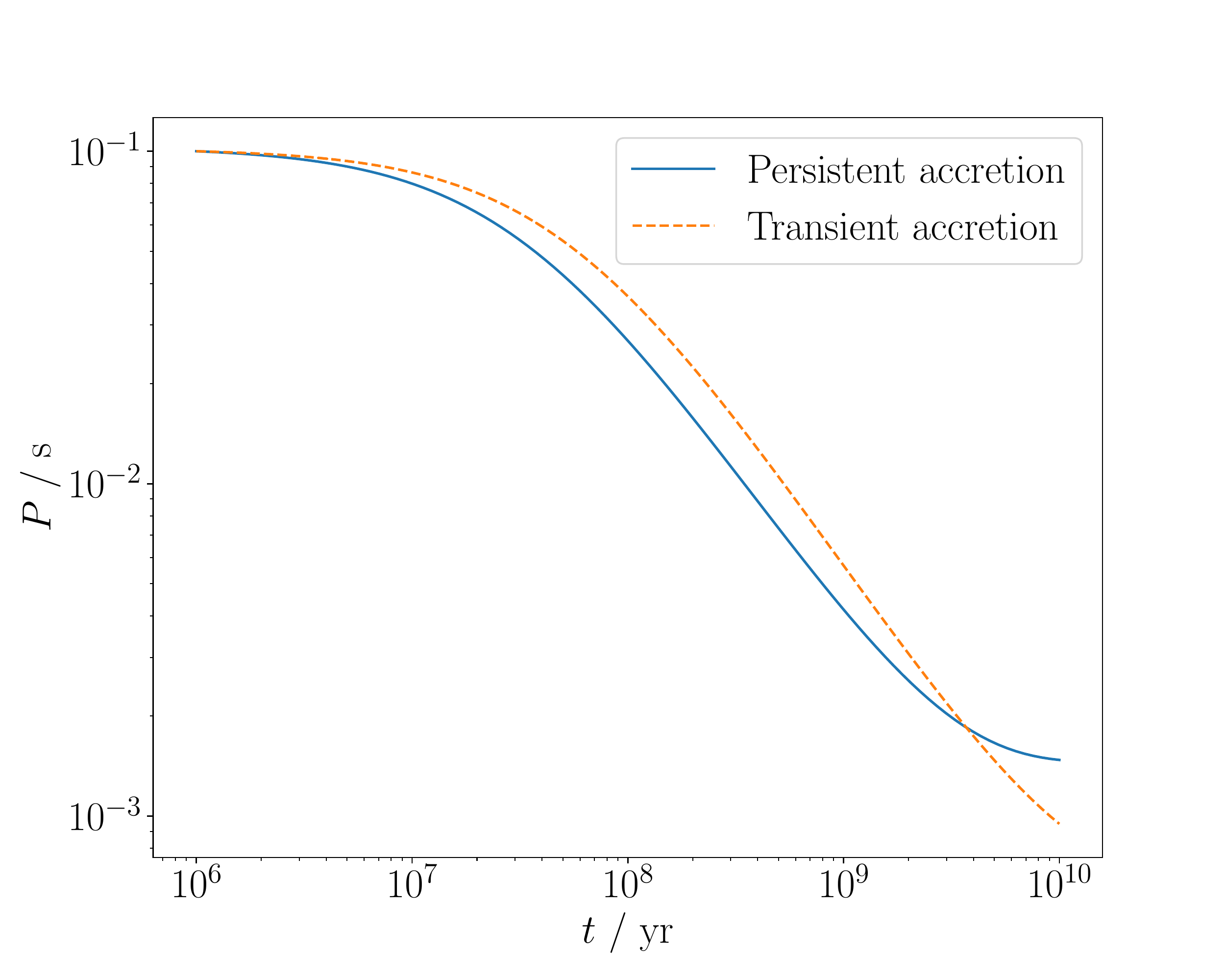}
    \caption[Spin evolution of an accreting neutron star]{The spin evolution of 
             the canonical accreting neutron star with persistent accretion 
             (blue solid line) and transient accretion (orange dashed line) with 
             initial values from Table~\ref{tab:CanonicalLMXB}. The persistent 
             accretor initially spins up faster than the transient accretor. 
             However, towards the end of its evolution the persistent accretor 
             begins to slow down and the transient accretor overtakes and 
             reaches a faster final spin.}
	\label{fig:CanonicalEvolution}
\end{figure}

Figure~\ref{fig:CanonicalEvolution} shows the spin evolution of the canonical 
accreting neutron star with persistent and transient accretion. As was found by 
\citet{2017MNRAS.470.3316D} and \citet{2017ApJ...835....4B}, we see that the 
persistently accreting neutron star initially spins up faster and reaches a 
final spin of $\nu = \SI{678}{\hertz}$. The transient system spins up slower but 
obtains a faster final spin of $\nu = \SI{1055}{\hertz}$. However, neither of 
the systems were evolved long enough to reach spin equilibrium. The upper limit 
of $\SI{e10}{\year}$ for the evolution time was chosen since no system can 
evolve for longer than the age of the Universe.

\section{Simulated populations}
\label{sec:SimulatedPopulations}

In order to obtain a distribution of spins with which to compare to the 
observed distribution, we used a Monte Carlo population-synthesis method to 
draw the initial parameters from a given set of distributions and evolve each 
neutron star [see \citet{1998ApJ...497L..97P} for another neutron-star 
population-synthesis study]. Each neutron star was assigned a mass $M$, radius 
$R$, magnetic-field strength $B$, initial spin period, average accretion rate 
$\dot{M}$, mass-quadrupole moment $Q_{2 2}$ and evolution time. We evolved 
$1000$ neutron stars in each simulation.

\begin{table}[h]
    \centering
	\caption[Initial values and evolution parameters for population synthesis]
             {Initial values and evolution parameters for population synthesis.}
	\label{tab:InitialValues}
	\begin{tabular}{ l c l }
		\hline\hline
        \multicolumn{1}{c}{Parameter} & \multicolumn{1}{c}{Distribution} 
        & \multicolumn{1}{c}{Values} \\ 
		\hline
		$M$ / \si{\solarMass} & Single-value & $1.4$ \\ 
		$R$ / \si{\kilo\metre} & Single-value & $10$ \\ 
		$\log_{10} (B \, / \, \si{\gauss})$	& Gaussian	& $\mu = 8.0$, $\sigma = 0.1$ \\
		Initial spin period / \si{\second}	& Flat	& $0.01-0.1$ \\
		$\xi$	& Single-value & $0.5$ \\
        $\log_{10}(\langle \dot{M} \rangle \, / \, \si{\solarMass\per\year})$ 
        & Gaussian & $\mu = -11.0 + \log_{10}(5)$, $\sigma = 0.1$ \\
		$Q_{2 2}$ / \si{\gram\centi\metre\squared} & Single-value & $0$ \\
		Evolution time / \si{\year} & Flat-in-the-log & $10^9 - 10^{10}$ \\
		\hline
	\end{tabular}
\end{table}

The first simulations were evolved using the distributions shown in 
Table~\ref{tab:InitialValues}. We fixed the masses and radii at 
$\SI{1.4}{\solarMass}$ and $\SI{10}{\kilo\metre}$, respectively, to match the 
canonical values for neutron stars. Typically, accreting millisecond X-ray 
pulsars are measured to have magnetic fields of $\sim \SI{e8}{\gauss}$ and so 
the field strength was taken from a log-Gaussian distribution with mean 
$\mu = 8.0$ and standard deviation $\sigma = 0.1$. The initial spin period was 
drawn from a flat distribution between $\SIrange{0.01}{0.1}{\second}$, which our 
simulations turned out to be relatively insensitive to. The correction factor 
$\xi$ was chosen to be $0.5$. The average accretion rate was motivated by 
observations of low-mass X-ray binaries and was given by a log-Gaussian with 
$\mu = -11.0 + \log_{10}(5)$ and $\sigma = 0.1$. For the initial simulations we 
assumed there was no gravitational-wave component. We found that for evolution 
times much less than $\SI{e9}{\year}$, the neutron stars would not have enough 
time to spin up to frequencies above $\SI{100}{\hertz}$ and so the evolution 
time was taken from a flat-in-the-log distribution between 
$\SIrange{e9}{e10}{\year}$. The distribution was chosen to be flat-in-the-log 
in order for it to be scale-invariant 
\citep[as was used by][]{1998ApJ...497L..97P}, thus, parametrising our 
uncertainty in the value of the evolution time.

\begin{figure}[h]
    \includegraphics[width=0.49\textwidth]{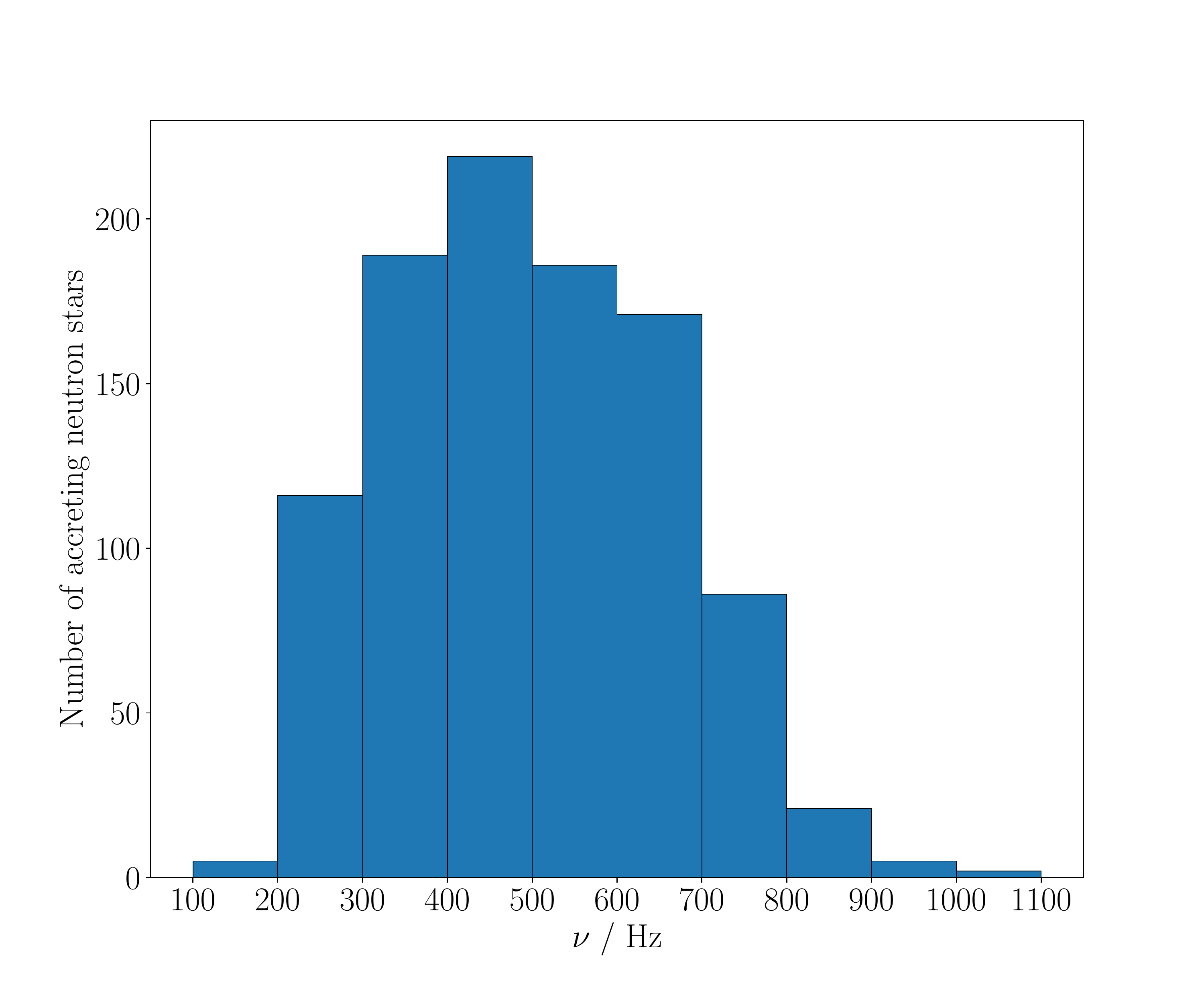}
    \includegraphics[width=0.49\textwidth]{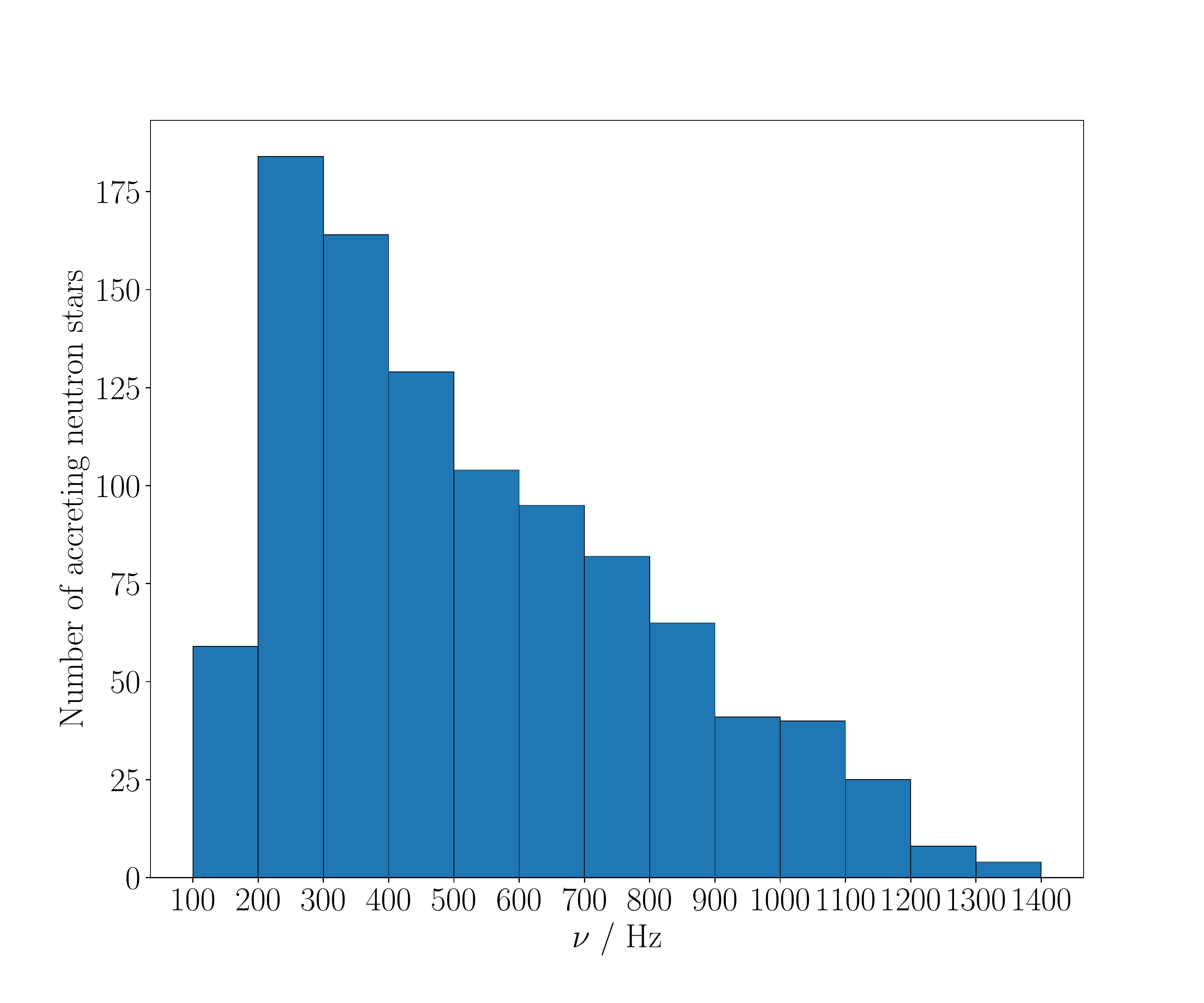}
    \caption[Spin-frequency distributions for simulated accreting neutron stars]
             {Distributions of spin frequencies for simulated persistently 
             accreting neutron stars (left panel) and transiently accreting 
             neutron stars (right panel) with initial distributions from 
             Table~\ref{tab:InitialValues}.}
	\label{fig:NoQSpins}
\end{figure}

The resultant spin-frequency distributions for persistent and transient 
accretors are shown in Fig.~\ref{fig:NoQSpins}. One can see for this simple 
case that for both persistent and transient accretion, we do indeed obtain 
neutron stars that spin in excess of $\SI{1}{\kilo\hertz}$. More generally, we 
observe that we get many neutron stars that spin faster than the observed 
spin-frequency limit of $\sim \SI{600}{\hertz}$ and, as one might expect, we 
find more high-frequency neutron stars for the transient case. This is because 
transient accretion enables these stars to spin to higher frequencies than with 
persistent accretion, provided they evolve for long enough. For both 
simulations, we have not obtained the characteristic behaviour of the observed 
distribution since there is no evidence for a pile-up of neutron stars at high 
frequencies.

In order to quantify how different our simulated populations are to the observed 
population, we applied a Kolmogorov-Smirnov test to the distributions. This test 
allows one to compare two distributions by testing the null hypothesis that the 
two distributions are the same. We chose a significance level of \num{0.10} 
for this work, which meant that should we have found $p$-values less than this 
value we could reject the null hypothesis with 90\% certainty for that case.%
\footnote{Note that $p$-values should not be confused with the isotropic fluid 
pressure $p$.}
For the persistent accretors, we obtained a $p$-value of $p = \num{7.7e-2}$ and 
for the transient accretors, we obtained $p = \num{9.9e-4}$. This meant that we 
could reject the null hypothesis at the 10\% significance level that the 
observed distribution is drawn from the persistent population or the transient 
population. 

\begin{figure}[h]
    \centering
	\includegraphics[width=0.7\textwidth]{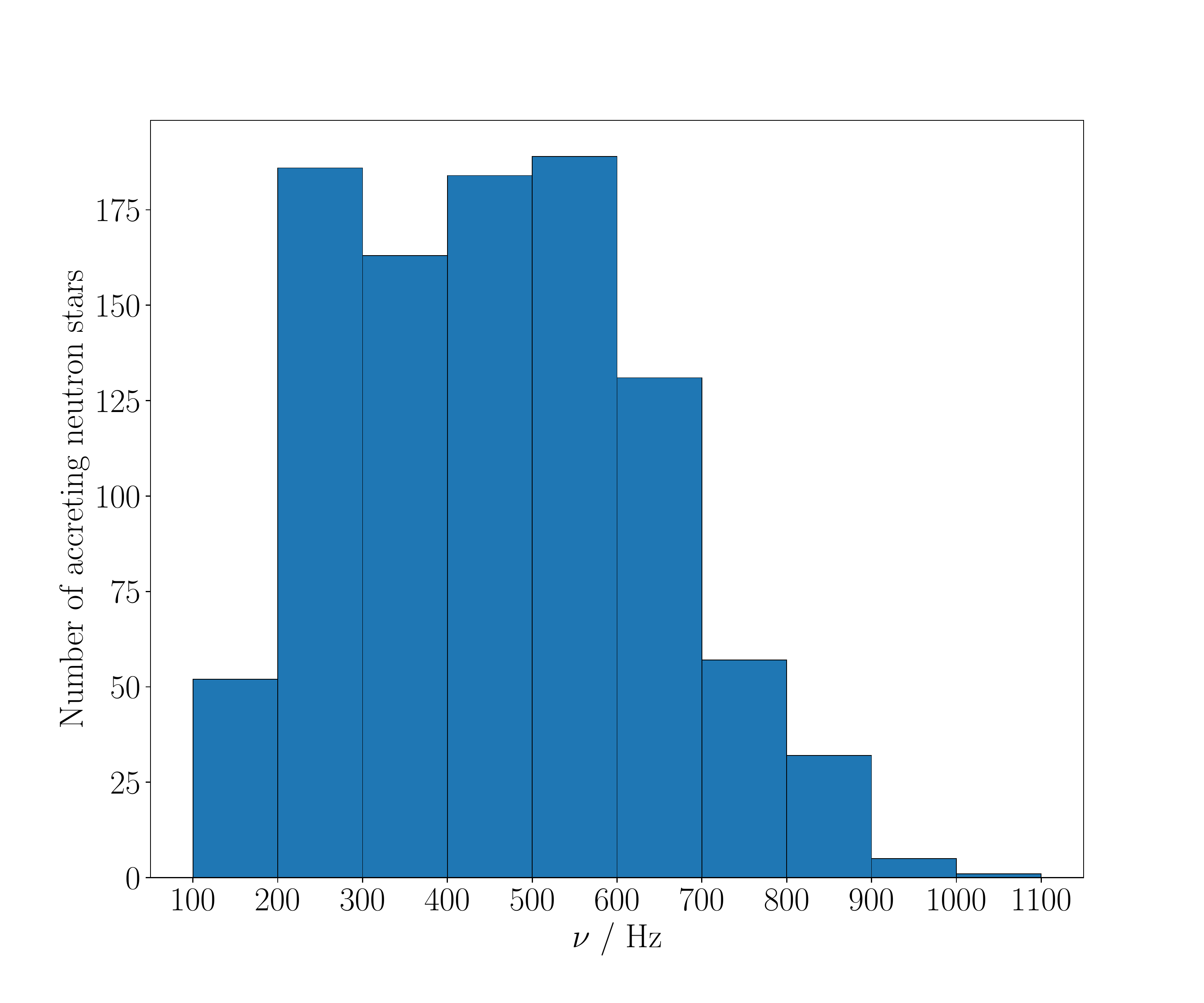}
    \caption[Spin-frequency distribution for simulated transiently accreting 
             neutron stars with the magnetic-dipole torque]{Distribution of spin 
             frequencies for simulated transiently accreting neutron stars with 
             initial distributions from Table~\ref{tab:InitialValues} including 
             the magnetic-dipole torque.}
	\label{fig:NoQSpinsEM}
\end{figure}

We explored the effect that magnetic-dipole radiation (\ref{eq:EMPdot}) has on 
transient accretors using the initial distributions in 
Table~\ref{tab:InitialValues}. The results are displayed in 
Fig.~\ref{fig:NoQSpinsEM}. The inclusion of this additional torque stops many 
of the systems from spinning up to sub-millisecond periods. We obtain 
$p = \num{0.20}$ so we cannot rule out the null hypothesis with any statistical 
certainty. However, in regards to the shape of the distribution, we do not 
obtain a sharp peak at the observed spin-frequency limit. Instead, we find a 
broad peak in the range $\SIrange{200}{600}{\hertz}$.

This demonstrates that a simple model for accretion is not sufficient to explain 
the observations of accreting neutron stars and also suggests that the inclusion 
of magnetic-dipole torques does not resolve this tension either. Therefore, an 
additional component needs to be included into the model.

\subsection{Including gravitational-wave torques}
\label{sec:IncludingGWs}

\begin{figure}[h]
    \includegraphics[width=0.49\textwidth]{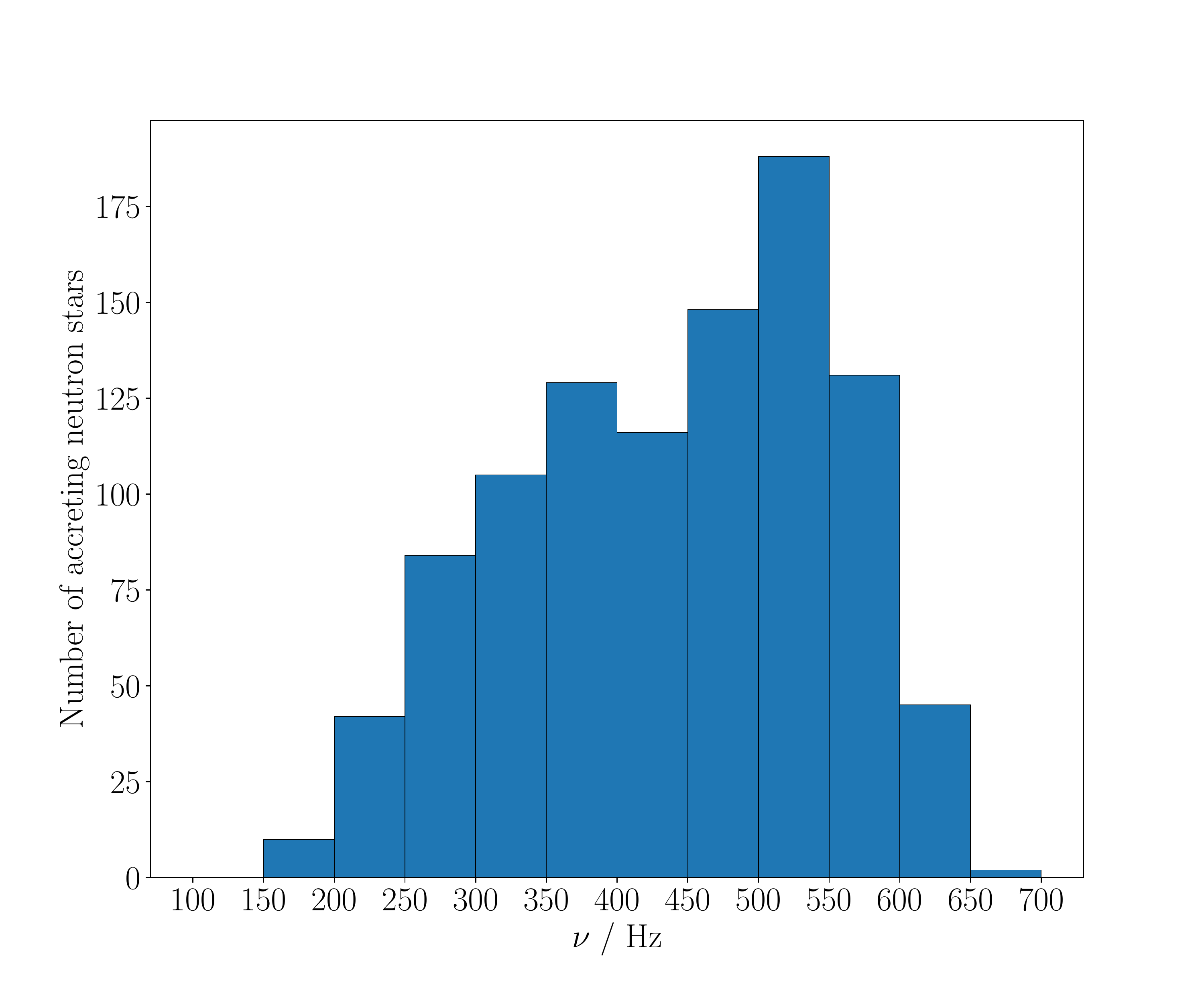}
    \includegraphics[width=0.49\textwidth]{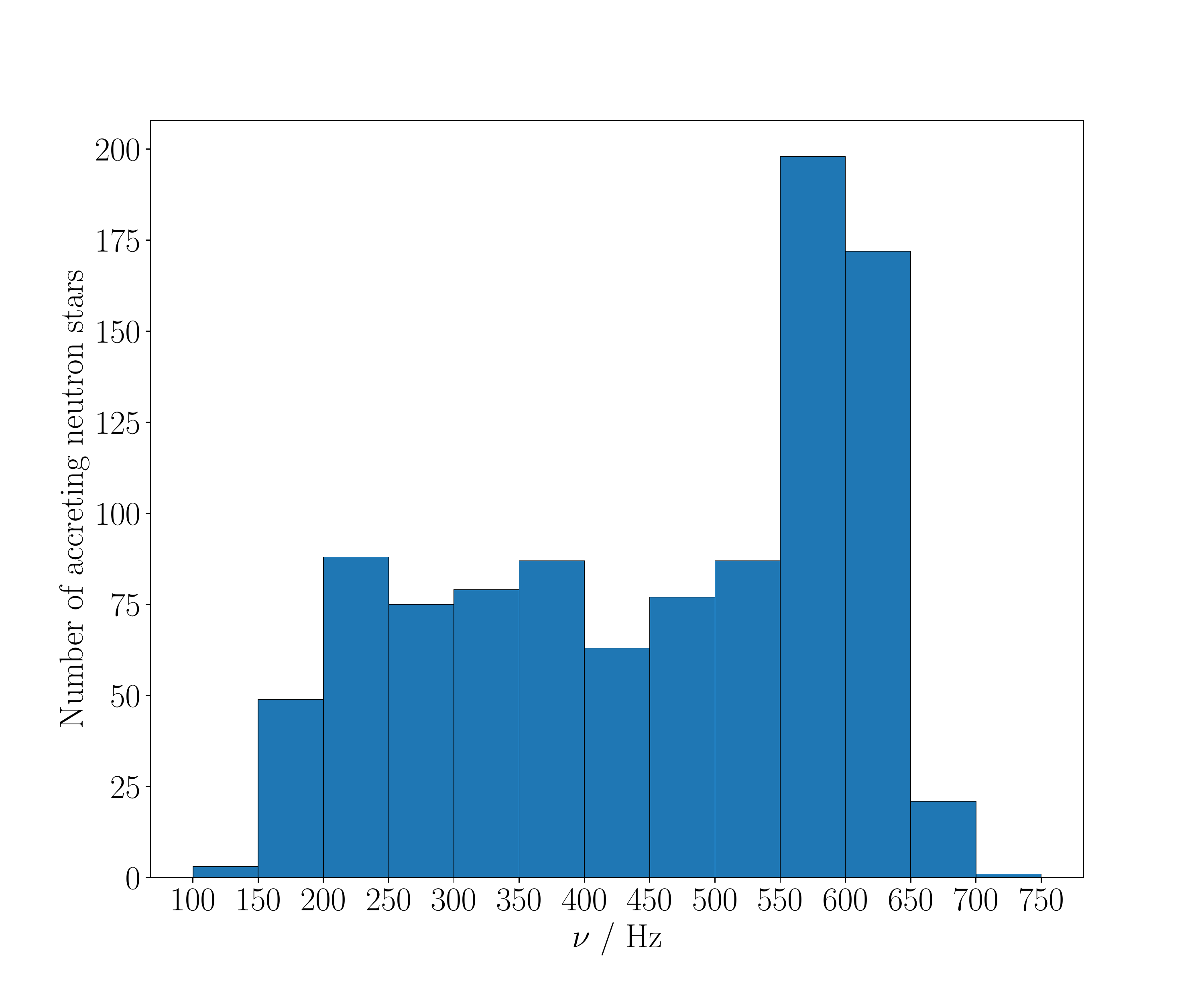}
    \caption[Spin-frequency distributions for simulated accreting neutron stars 
             with a fixed quadrupole]{Distributions of spin frequencies for 
             simulated persistently accreting neutron stars (left panel) and 
             transiently accreting neutron stars (right panel) with initial 
             distributions from Table~\ref{tab:InitialValues} and a fixed 
             quadrupole of $Q_{2 2} = \SI{e36}{\gram\centi\metre\squared}$.}
	\label{fig:FixedQSpins}
\end{figure}

We explored whether including a gravitational-wave component to the spin 
evolution of accreting neutron stars could give us the observed spin 
distribution. Motivated by the necessary quadrupole in order to achieve torque 
balance (\ref{eq:BalanceQ}), we repeated the same simulations but with a fixed 
$Q_{2 2} = \SI{e36}{\gram\centi\metre\squared}$ for all neutron stars. 
(Physically, this can be interpreted as a permanent crustal mountain.) 
Figure~\ref{fig:FixedQSpins} shows the final-spin distributions for these 
simulations. This quadrupole has notably stopped the neutron stars from spinning 
up to sub-millisecond periods and has resulted in a pile-up centred on the 
$\SIrange{500}{550}{\hertz}$ bin for the persistent accretors and at 
$\SIrange{550}{600}{\hertz}$ for the transient accretors. This has appeared 
since the gravitational-wave torque imposes a spin-frequency limit on the 
neutron stars. The peak for transient accretors is promising as this is where 
the peak lies for the spin distribution that we observe (\textit{cf.} 
Fig.~\ref{fig:SpinDistribution}). Interestingly, there is also a broader peak at 
lower frequencies. We found that $\approx 19\%$ of persistently accreting 
neutron stars and $\approx 14\%$ of transiently accreting neutron stars reached 
spin equilibrium by the end of the simulation. The systems that had reached spin 
equilibrium were clustered around the high-frequency peaks. We obtained 
$p = 0.19$ and $p = 0.80$ for the persistent and transient cases, respectively, 
and, therefore, were unable to reject the null hypothesis for both populations.

\begin{figure}[h]
    \centering
	\includegraphics[width=0.7\textwidth]{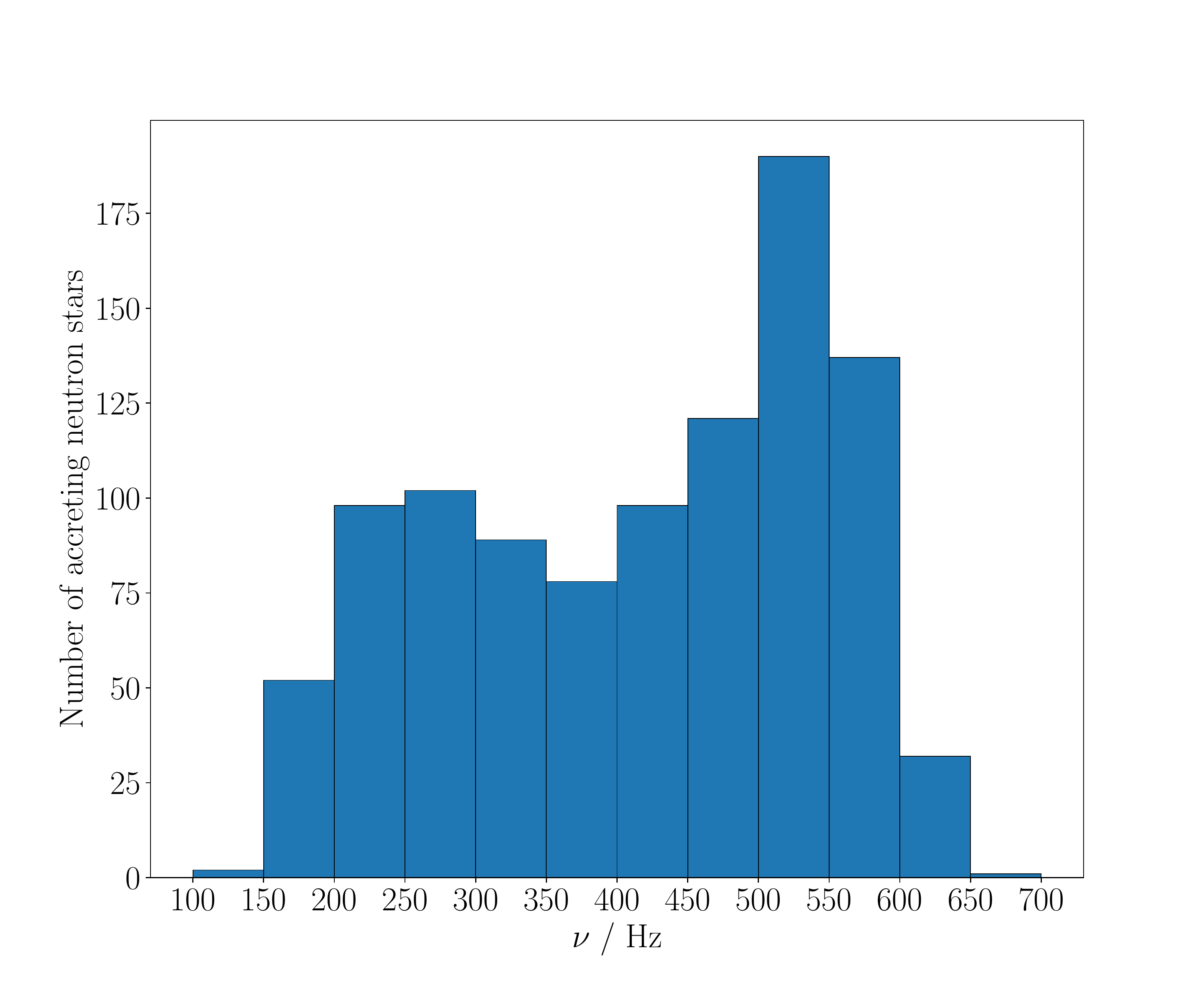}
    \caption[Spin-frequency distribution for simulated transiently accreting 
             neutron stars with a fixed quadrupole and the magnetic-dipole 
             torque]{Distribution of spin frequencies for simulated 
             transiently accreting neutron stars with initial distributions from 
             Table~\ref{tab:InitialValues} with a fixed quadrupole of 
             $Q_{2 2} = \SI{e36}{\gram\centi\metre\squared}$ and including the 
             magnetic-dipole torque.}
	\label{fig:QSpinsEM}
\end{figure}

We considered how magnetic-dipole radiation affects this picture for systems 
undergoing transient accretion. We used the same quadrupole and obtained the 
results shown in Fig.~\ref{fig:QSpinsEM}. Interestingly, this distribution is 
qualitatively similar to the results without magnetic-dipole torques (right 
panel of Fig.~\ref{fig:FixedQSpins}). We recover a pronounced peak at higher 
frequencies $\SIrange{500}{550}{\hertz}$, which by comparison has shifted down 
by $\SI{50}{\hertz}$. Since the features of the distribution remain the same we 
argue that one could obtain a distribution with a peak that matches the observed 
distribution through slight adjustment of the initial values, \textit{e.g.}, the 
quadrupole. Such an adjustment would be justifiable since there is significant 
uncertainty in many of these parameters. We could not reject the null hypothesis 
for these results with $p = 0.39$.

\begin{figure}[h]
    \centering
	\includegraphics[width=0.7\textwidth]{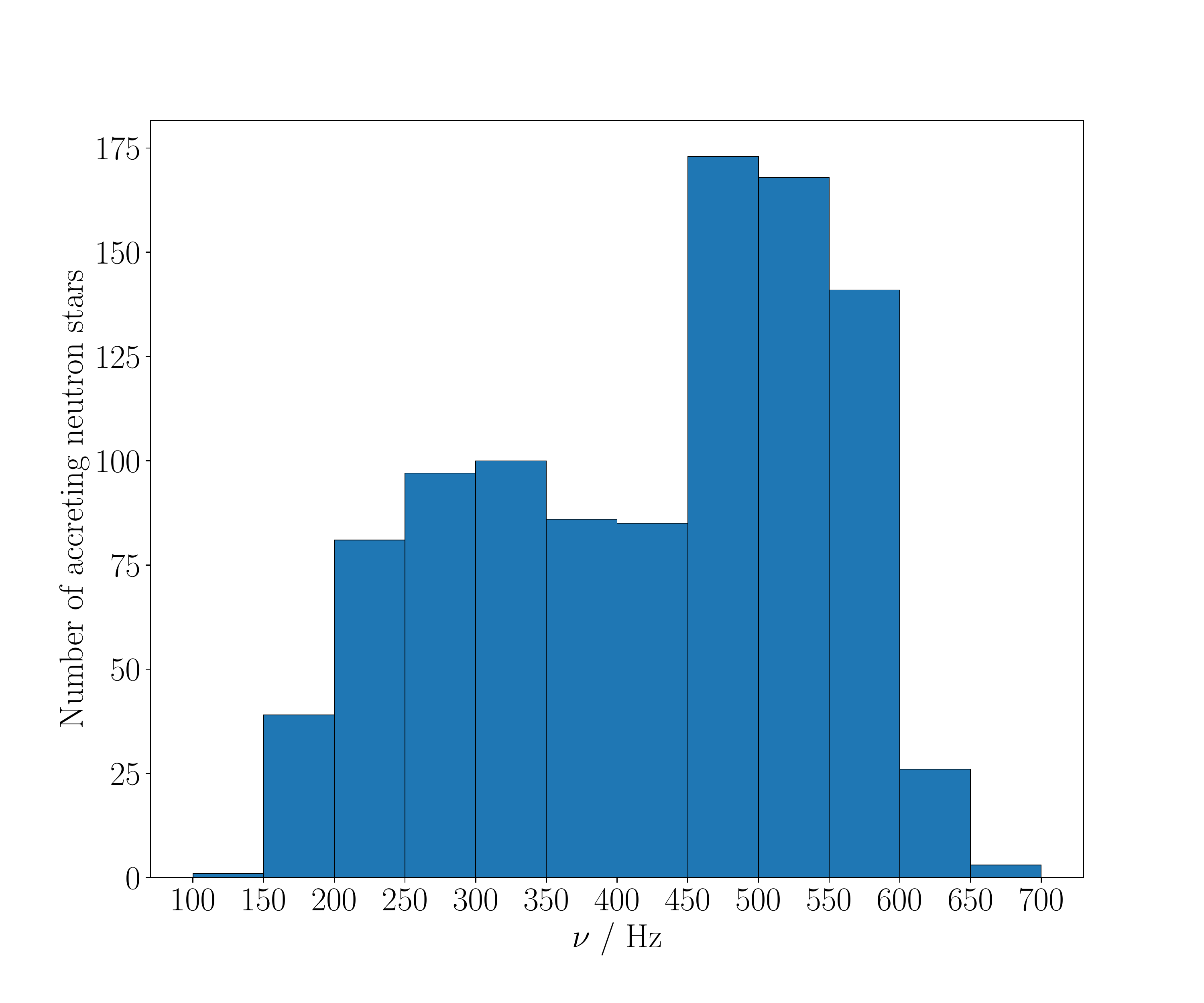}
    \caption[Spin-frequency distribution for simulated transiently accreting 
             neutron stars with the magnetic-dipole torque, a fixed quadrupole 
             and varied outburst durations]{Distribution of spin frequencies for 
             simulated transiently accreting neutron stars with initial 
             distributions from Table~\ref{tab:InitialValues} with a fixed 
             quadrupole of $Q_{2 2} = \SI{e36}{\gram\centi\metre\squared}$, 
             including the magnetic-dipole torque and $F_\text{t}$ distributed 
             flat between $\SIrange{1}{100}{\year}$.}
	\label{fig:QSpinsEMF_t}
\end{figure}

For accreting neutron-star systems the magnetic-dipole torque is expected to be 
negligible during outbursts, but it could play an important role during the 
quiescent phases. We explored a range of outburst durations, $F_\text{t}$ 
distributed uniformly between $\SIrange{1}{100}{\year}$, to assess the impact 
this made on the resultant spin distribution (Fig.~\ref{fig:QSpinsEMF_t}). 
For this wide range of outburst lengths we find a broad peak between 
$\SIrange{450}{600}{\hertz}$. This contrasts the narrow peak in 
Fig.~\ref{fig:SpinDistribution}. In this case, we obtain a \textit{p}-value of 
$p = 0.38$.

\begin{figure}[h]
    \centering
	\includegraphics[width=0.7\textwidth]{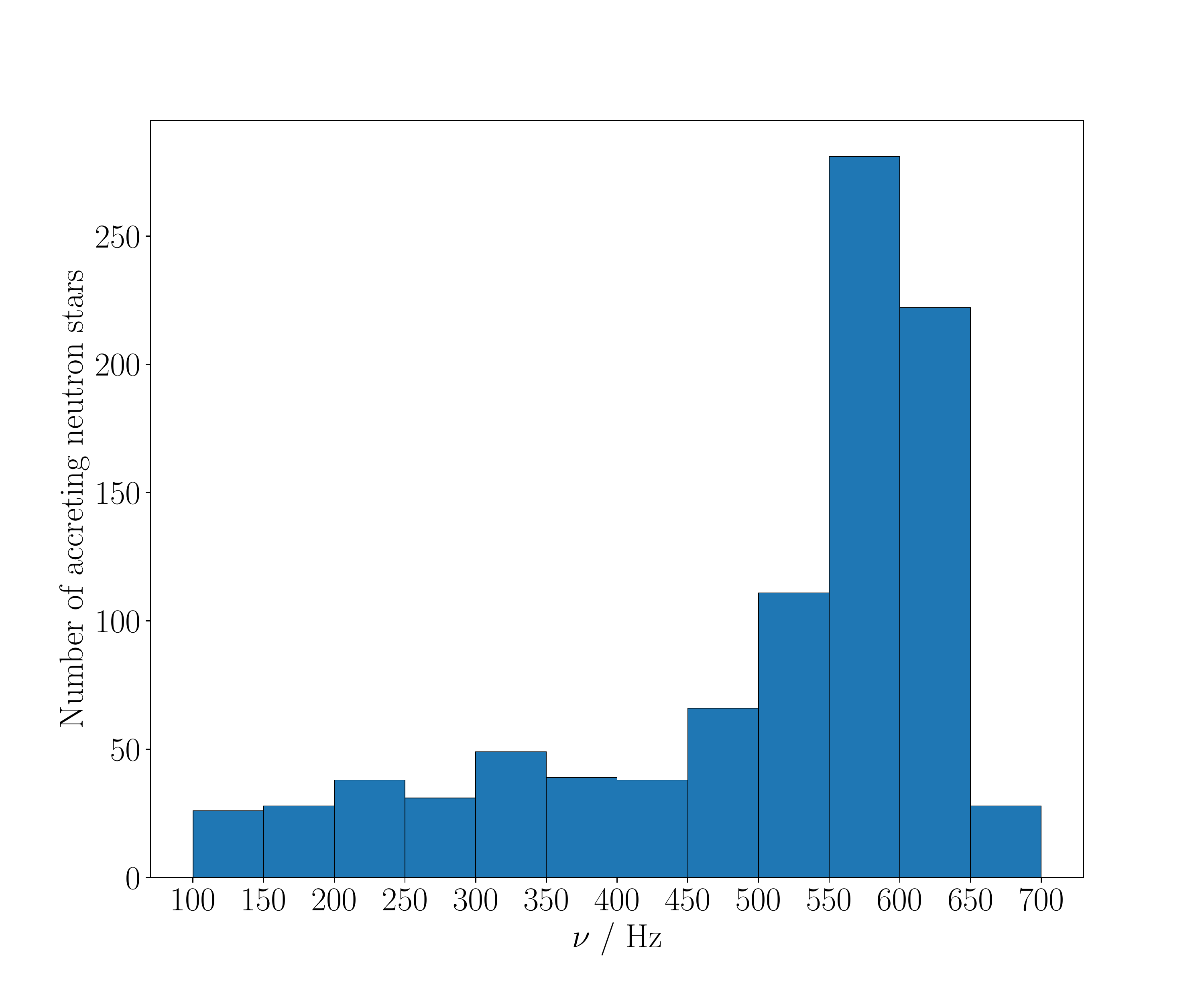}
    \caption[Spin-frequency distribution for simulated transiently accreting 
             neutron stars with a fixed quadrupole and varied evolution 
             times]{Distribution of spin frequencies for simulated transiently 
             accreting neutron stars with initial distributions from 
             Table~\ref{tab:InitialValues} with a fixed quadrupole of 
             $Q_{2 2} = \SI{e36}{\gram\centi\metre\squared}$ and an evolution 
             time distributed flat between $\SIrange{e8}{e10}{\year}$.}
	\label{fig:Ouch}
\end{figure}

We investigated how sensitive the simulated populations were on the 
distribution of the evolution time. We ran a simulation with the same 
quadrupole and the evolution time distributed flat between 
$\SIrange{e8}{e10}{\year}$ for transient accretors. This was motivated by 
assuming that neutron stars are born at a uniform rate, which is intuitively 
what one might expect, and that there are no selection effects to suggest that 
we are more likely to observe younger systems. The resultant spin-distribution 
is shown in Fig.~\ref{fig:Ouch}. As was observed in the case when the time was 
distributed flat-in-the-log, there exists a pronounced peak towards the higher 
spin-frequencies. However, there are far fewer systems spinning at frequencies 
below this peak. For this simulation we obtained a $p$-value of 
$p = \num{1.2e-2}$ and, thus, could reject the null hypothesis. This 
distribution does not match what we observe.

\subsection{Thermal mountains}
\label{sec:Thermal}

One of the most promising avenues for producing a mass quadrupole on a 
fast-spinning, accreting neutron star is through thermal mountains built during 
accretion phases through asymmetries in pycnonuclear reaction rates 
\citep{2017PhRvL.119p1103H}. As a neutron star accretes matter composed of light 
elements, the matter becomes buried by accretion and is then compressed to 
higher densities. This causes the matter to undergo nuclear reactions such as 
electron captures, neutron emission and pycnonuclear reactions 
\citep{1990A&A...227..431H}. If the accretion flow is asymmetric, this can cause 
asymmetries in density and heating that can produce a quadrupole moment. The 
quadrupole moment due to asymmetric crustal heating from nuclear reactions is 
approximated by \citep{2000MNRAS.319..902U}
\begin{equation}
    Q_{2 2} \approx \num{1.3e37} \ 
        \left( \frac{R}{\SI{10}{\kilo\metre}} \right)^4 
        \left( \frac{\delta T_\text{q}}{\SI{e7}{\kelvin}} \right) 
        \left( \frac{E_\text{th}}{\SI{30}{\mega\electronVolt}} \right)^3 
        \ \si{\gram\centi\metre\squared},
	\label{eq:ThermalQ}
\end{equation}
where $\delta T_\text{q}$ is the quadrupolar temperature increase due to the 
nuclear reactions and $E_\text{th}$ is the threshold energy for the reactions 
to occur. The value $\delta T_\text{q}$ will be a fraction of the total heating 
\citep{2001MNRAS.325.1157U}
\begin{equation}
    \delta T \approx \num{2e5} \
        \left( \frac{\mathcal{C}}{k_\text{B}} \right)^{-1} 
        \left( \frac{p_\text{d}}{\SI{e30}{\dyne\per\centi\metre\squared}} 
        \right)^{-1}
        \left( \frac{Q}{\SI{1}{\mega\electronVolt}} \right) 
        \left( \frac{\Delta M}{\SI{e-9}{\solarMass}} \right) \ \si{\kelvin},
	\label{eq:ThermaldeltaT}
\end{equation}
where $k_\text{B}$ is the Boltzmann constant, $\mathcal{C}$ is the heat capacity 
per baryon, $p_\text{d}$ is the pressure at which the reaction occurs, $Q$ is 
the heat released locally due the reactions and $\Delta M$ is the accreted mass. 
Some of this heating will be converted into the quadrupolar temperature 
increase, however, it is unclear quite how much will be converted. 
\citet{2000MNRAS.319..902U} estimate that 
$\delta T_\text{q} / \delta T \lesssim 0.1$, but in reality this ratio is 
poorly understood.

These thermal mountains are built during accretion outbursts. During quiescence 
phases, the deformations are washed away on a thermal timescale 
\citep{1998ApJ...504L..95B}
\begin{equation}
    \tau_\text{th} \approx 0.2 \
        \left( \frac{p_\text{d}}{\SI{e30}{\dyne\per\centi\metre\squared}} 
        \right)^{3 / 4} \ \si{\year}.
\end{equation}
If the system is in quiescence for longer than this timescale then the thermal 
mountain will be washed away and a new mountain will be built during the next 
outburst.

We implemented the expression for a quadrupole moment due to these reactions 
from (\ref{eq:ThermalQ}) and assumed the following values for our 
neutron stars 
[as estimated by \citet{2017PhRvL.119p1103H} for the pulsar J1023+0038]: 
$\mathcal{C} \approx \num{e-6} k_\text{B}$, 
$E_\text{th} = \SI{30}{\mega\electronVolt}$, 
$p_\text{d} = \SI{e30}{\erg\per\centi\metre\cubed}$ and 
$Q = \SI{0.5}{\mega\electronVolt}$. This meant that the quadrupole due to these 
reactions was dependent only on the accreted mass $\Delta M$ and the fraction 
$\delta T_\text{q} / \delta T$. For this mechanism, we only considered 
transient accretion since in persistent accretion these mountains will not wash 
away, but instead will get progressively larger until the crust can no longer 
sustain them. This is effectively modelled through a fixed quadrupole that 
represents the largest mountain that can be built. 

\begin{figure}[h]
    \includegraphics[width=0.49\textwidth]{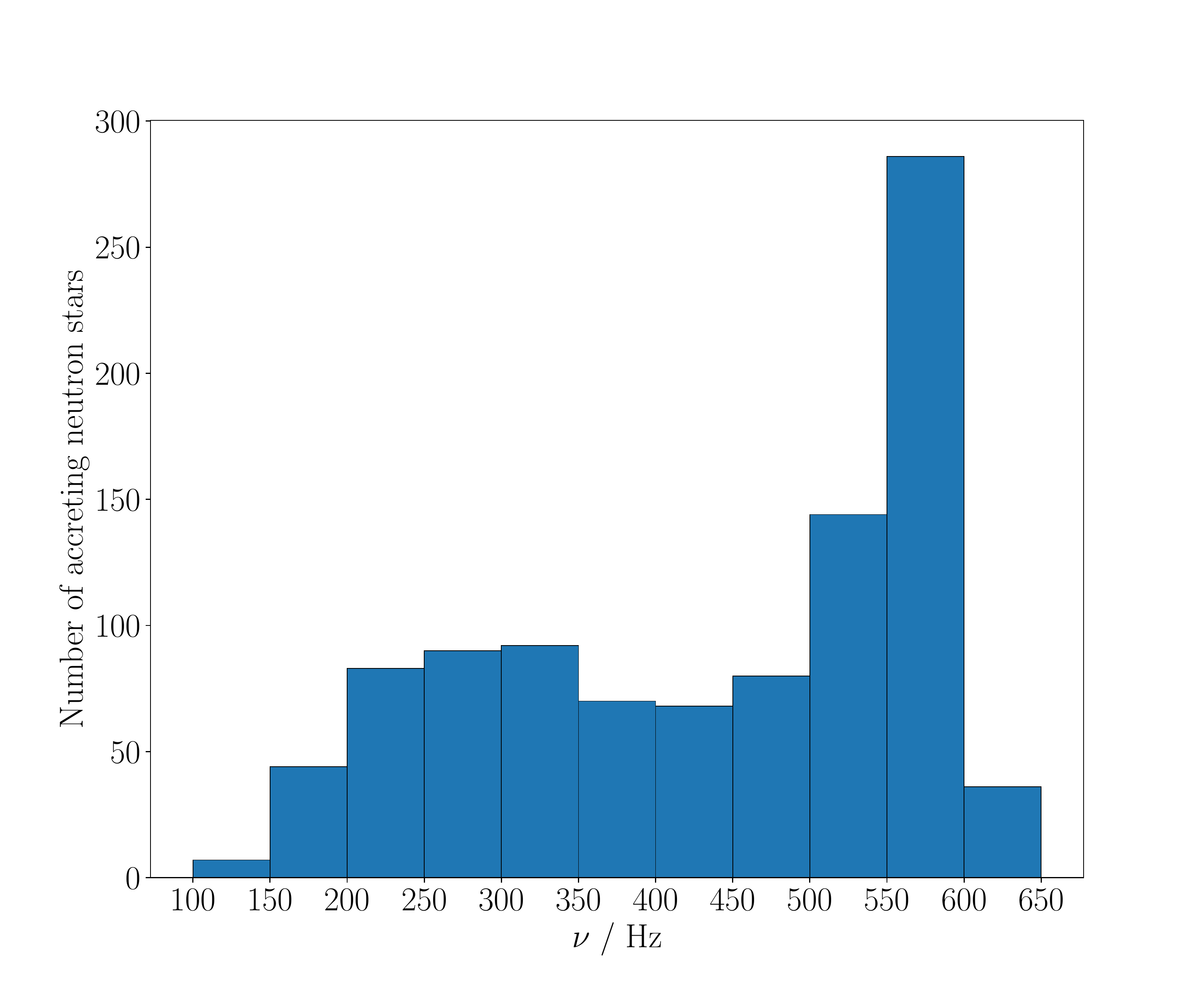}
    \includegraphics[width=0.49\textwidth]{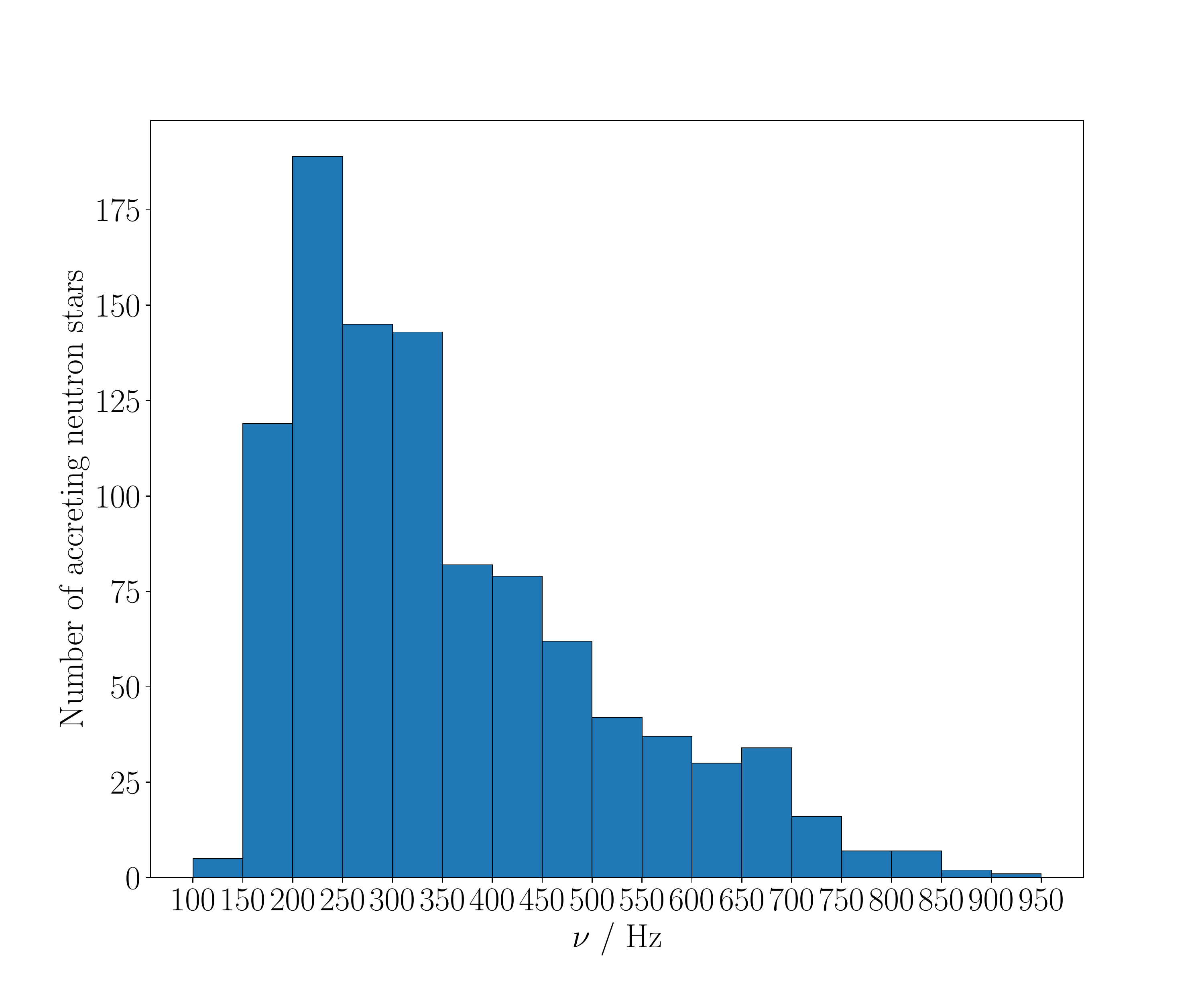}
    \caption[Spin-frequency distributions for simulated transiently accreting 
             neutron stars with thermal mountains]{Distributions of spin 
             frequencies for simulated transiently accreting neutron stars that 
             built thermal mountains during outburst phases with initial 
             distributions from Table~\ref{tab:InitialValues}. The left panel 
             has a fixed $\delta T_\text{q} / \delta T = \num{4e-4}$ and the 
             right panel has $\delta T_\text{q} / \delta T$ distributed 
             flat-in-the-log between $10^{-4}-10^{-2}$.}
	\label{fig:ThermalSpins}
\end{figure}

In our model, we calculated $\Delta M$ by numerically integrating the accretion 
profile from the beginning of the outburst up to $F_\text{t}$. Our quiescence 
phases were long enough for the mountain to wash away during them. Unlike for 
our other prescriptions, we found that for thermal mountains, the specific 
values that parametrise the outburst features were very important in dictating 
the final-spin distribution. Based on observations of X-ray transients, we chose 
$F_\text{t} = \SI{0.1}{\year}$, $T_\text{recurrence} = \SI{2.0}{\year}$ and 
$f_\text{max} / f_\text{min} = \num{e4}$. We simulated neutron stars that built 
thermal mountains with $\delta T_\text{q} / \delta T = \num{4e-4}$. The left 
panel of Fig.~\ref{fig:ThermalSpins} shows the resultant distribution of 
final spins. We can see qualitatively that this distribution has similar 
features to the fixed quadrupole case (see the right panel of 
Fig.~\ref{fig:FixedQSpins}). Another promising aspect of the final-spin 
distribution is the prominence of the high-frequency peak. Like what is 
observed, this peak is narrow and much larger than the values in other 
frequency bins. We found a $p$-value of $p = 0.60$ for this distribution.

The only constraint on the ratio of quadrupolar to total heating comes from the 
non-detection of X-ray emission of quadrupolar flux perturbations during 
quiescence phases in low-mass X-ray binaries, which gives 
$\delta T_\text{q} / \delta T \lesssim 0.1$ \citep{2000MNRAS.319..902U}. 
Currently, there is no reason to believe that this fraction should be constant 
for all neutron stars. To account for our uncertainty in this fraction, we 
distributed $\delta T_\text{q} / \delta T$ flat-in-the-log between 
$10^{-4}-10^{-2}$. The result of this simulation is shown in the right panel of 
Fig.~\ref{fig:ThermalSpins}. The distribution peaks at low frequencies and then 
falls off towards higher frequencies. We obtained $p = \num{7.0e-2}$ for this 
case which meant that we could reject the null hypothesis. This shows how this 
prescription favours $\delta T_\text{q} / \delta T$ being a fixed value. Seeking 
a physical explanation for this preference of $\delta T_\text{q} / \delta T$ 
being a fixed value as opposed to being distributed is beyond the scope of this 
study and has been left for future work.

\subsection{Unstable \textit{r}-modes}
\label{sec:rModes}

An \textit{r}-mode is a fluid mode of oscillation for which the restoring force 
is the Coriolis force. \citet{1998ApJ...502..708A} demonstrated that 
gravitational radiation destabilises the \textit{r}-modes of rotating stars. 
These modes are generically unstable to gravitational-wave emission 
\citep{1998ApJ...502..714F} due to the Chandrasekhar-Friedman-Schutz 
instability, that facilitates the star finding lower energy and angular 
momentum configurations that allow the mode amplitude to grow 
\citep{1970PhRvL..24..611C, 1978ApJ...222..281F}.

The \textit{r}-mode instability has long been considered a potential mechanism 
for imposing a spin limit on neutron stars in low-mass X-ray binaries 
\citep{1999ApJ...516..307A}. The typical picture involves a neutron star being 
spun up through accretion until it enters the \textit{r}-mode instability 
window. This instability region depends primarily on the spin of the neutron 
star and its core temperature. Once a neutron star has entered this region, it 
will emit gravitational radiation and begin to spin down until it reaches 
stability. This is expected to occur on a timescale much shorter than the age of 
the system and should result in most low-mass X-ray binaries being stable. 
However, theoretical models for the \textit{r}-mode instability demonstrate 
that many of the observed accreting neutron stars, in fact, lie inside the 
instability window \citep{2011PhRvL.107j1101H}. This result would be consistent 
if the saturation amplitude for these systems was small, 
$\alpha \approx 10^{-8} - 10^{-7}$, but this is at odds with predictions that 
suggest that the amplitude should be several orders of magnitude higher than 
this \citep{2007PhRvD..76f4019B}.

\citet{1998PhRvD..58h4020O} described a phenomenological model for the 
evolution of \textit{r}-modes and the spin of the star. In this model the 
quadrupole moment for an incompressible neutron star that is unstable due to 
\textit{r}-modes is given by
\begin{equation}
    Q_{2 2} \approx \num{1.67e33} \ \left( \frac{\alpha}{\num{e-7}} \right) 
        \left( \frac{M}{\SI{1.4}{\solarMass}} \right) 
        \left( \frac{R}{\SI{10}{\kilo\metre}} \right)^3 
        \left(\frac{P}{\SI{1}{\second}}\right)^{-1} \ 
        \si{\gram\centi\metre\squared}.
	\label{eq:RModeQ}
\end{equation}
An interesting feature of this expression is its dependence on the spin of the 
neutron star. As a neutron star spins faster the quadrupole moment grows. This 
is different to what is expected from mountains. In fact, \textit{r}-modes and 
mountains could be differentiated from one another through the scaling of the 
associated quadrupoles as well as the frequency of the emitted gravitational 
waves; for mountains the gravitational-wave frequency is $2 \nu$, whereas for 
\textit{r}-modes the frequency is $4 \nu / 3$.%
\footnote{This frequency comes from Newtonian calculations of \textit{r}-modes 
with $\ell = 2$. In going to relativity, the expectation is that there will be 
some low-order correction to this value.}

\begin{figure}[h]
    \includegraphics[width=0.49\textwidth]{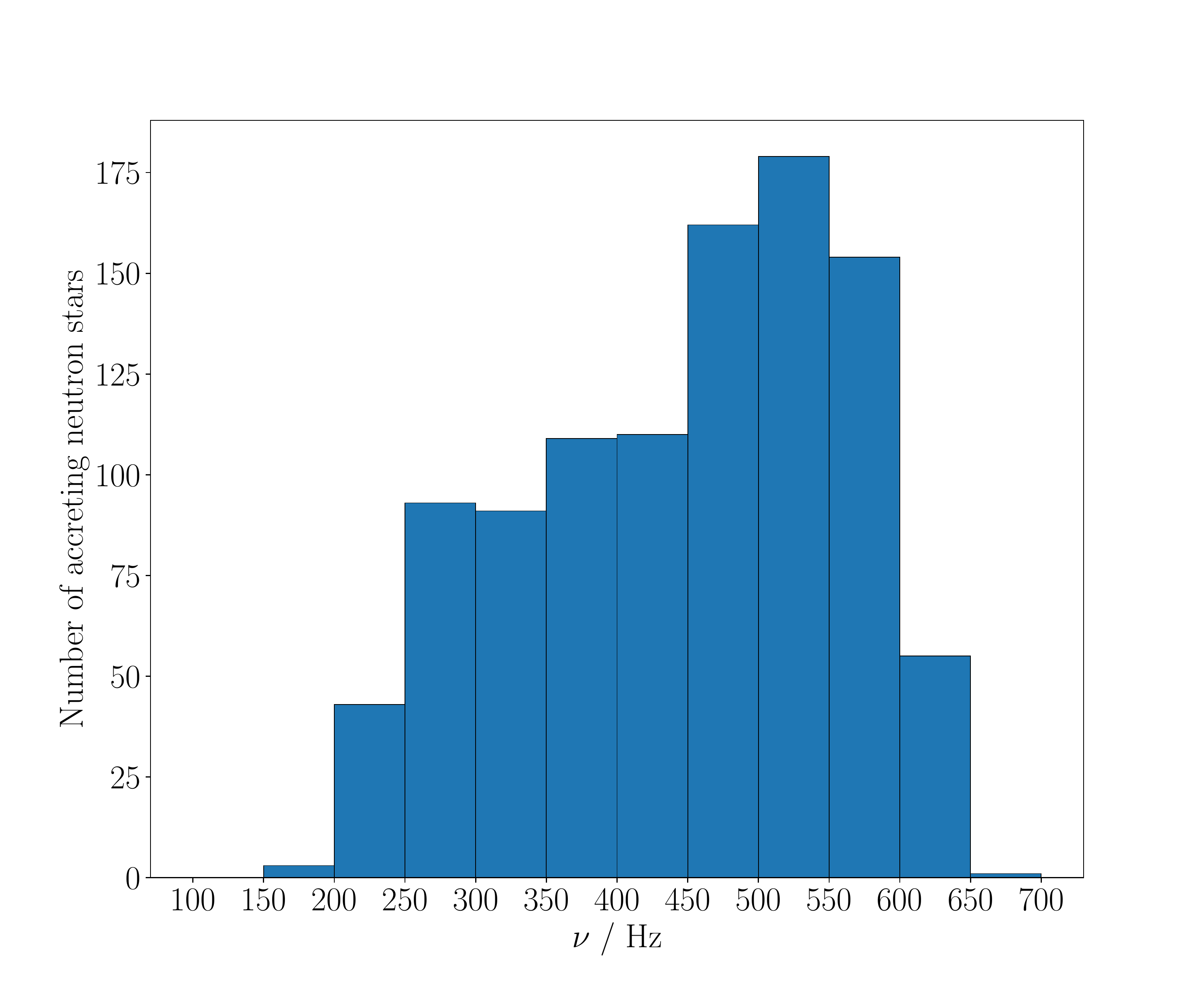}
    \includegraphics[width=0.49\textwidth]{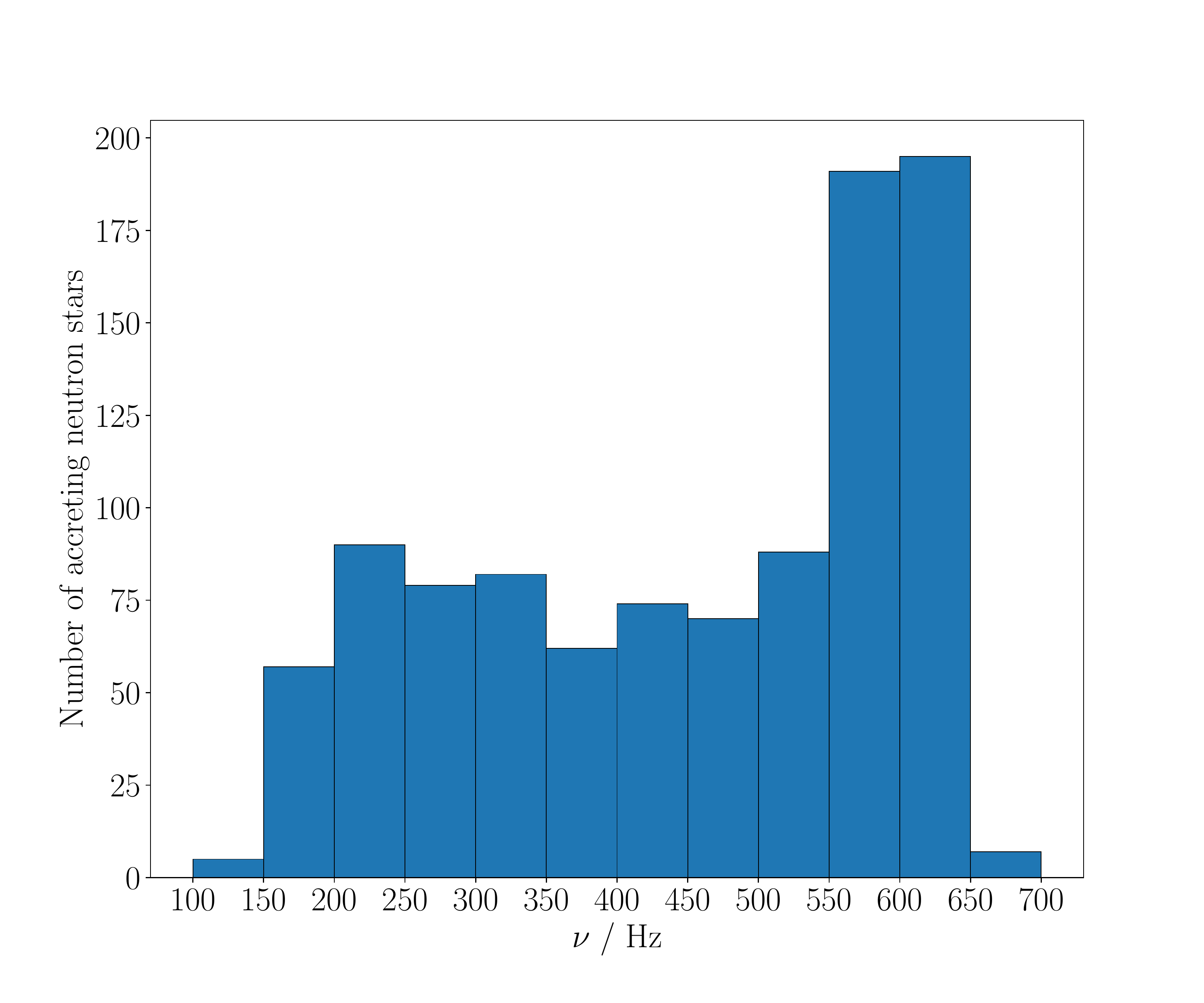}
    \caption[Spin-frequency distributions for simulated accreting neutron stars 
             with unstable \textit{r}-modes]{Distributions of spin frequencies 
             for simulated persistently accreting neutron stars (left panel) and 
             transiently accreting neutron stars (right panel) with unstable 
             \textit{r}-modes with initial distributions from 
             Table~\ref{tab:InitialValues} and $\alpha = \num{e-7}$.}
	\label{fig:RModeSpins1}
\end{figure}

In order to simulate accreting neutron stars with unstable \textit{r}-modes, we 
implemented (\ref{eq:RModeQ}) into our model. We assumed that the 
mode amplitude $\alpha$ remained constant for each neutron star. We repeated the 
previous simulations for persistent and transient accretors with unstable 
\textit{r}-modes and $\alpha = \num{e-7}$. Figure~\ref{fig:RModeSpins1} shows 
the final-spin distributions for those simulations. The unstable 
\textit{r}-modes were sufficient in both cases to give a peak at high 
spin-frequencies. For the persistently accreting neutron stars, the peak was in the 
$\SIrange{500}{550}{\hertz}$ frequency bin and for the transient accretors, 
the peak was in the $\SIrange{550}{600}{\hertz}$ bin. These distributions are 
similar to the case of a permanent quadrupole 
$Q_{2 2} = \SI{e36}{\gram\centi\metre\squared}$ (see 
Fig.~\ref{fig:FixedQSpins}). For transient accretion with unstable 
\textit{r}-modes the peak is narrower and more pronounced indicating that the 
magnitude of the gravitational-wave torque sets in quickly. This is due to the 
scaling of the quadrupole in (\ref{eq:RModeQ}), since it depends linearly on the 
spin. For the persistent case we found a $p$-value of $p = 0.28$ and for the 
transient case $p = 0.57$.

\begin{figure}[h]
    \includegraphics[width=0.49\textwidth]{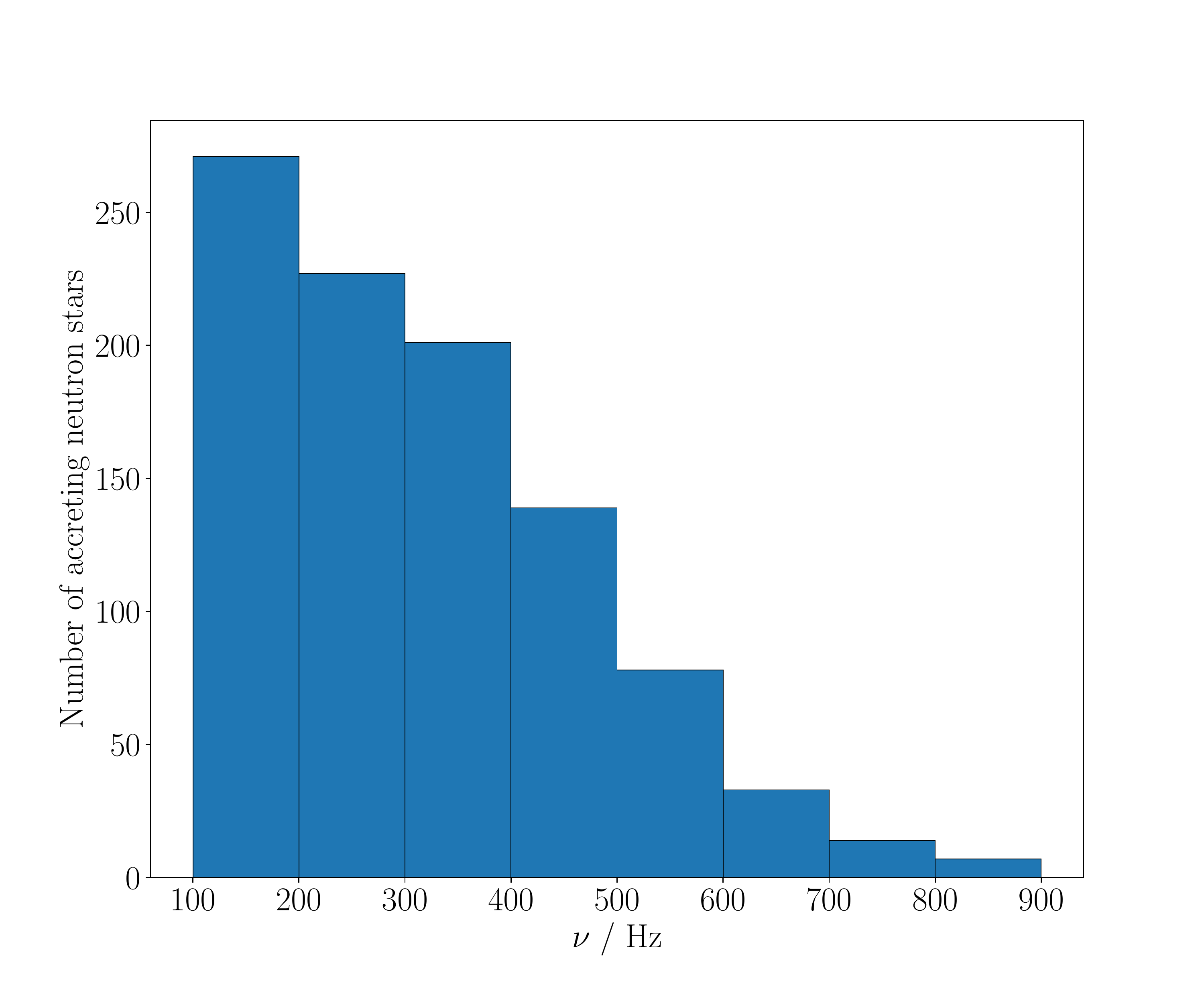}
    \includegraphics[width=0.49\textwidth]{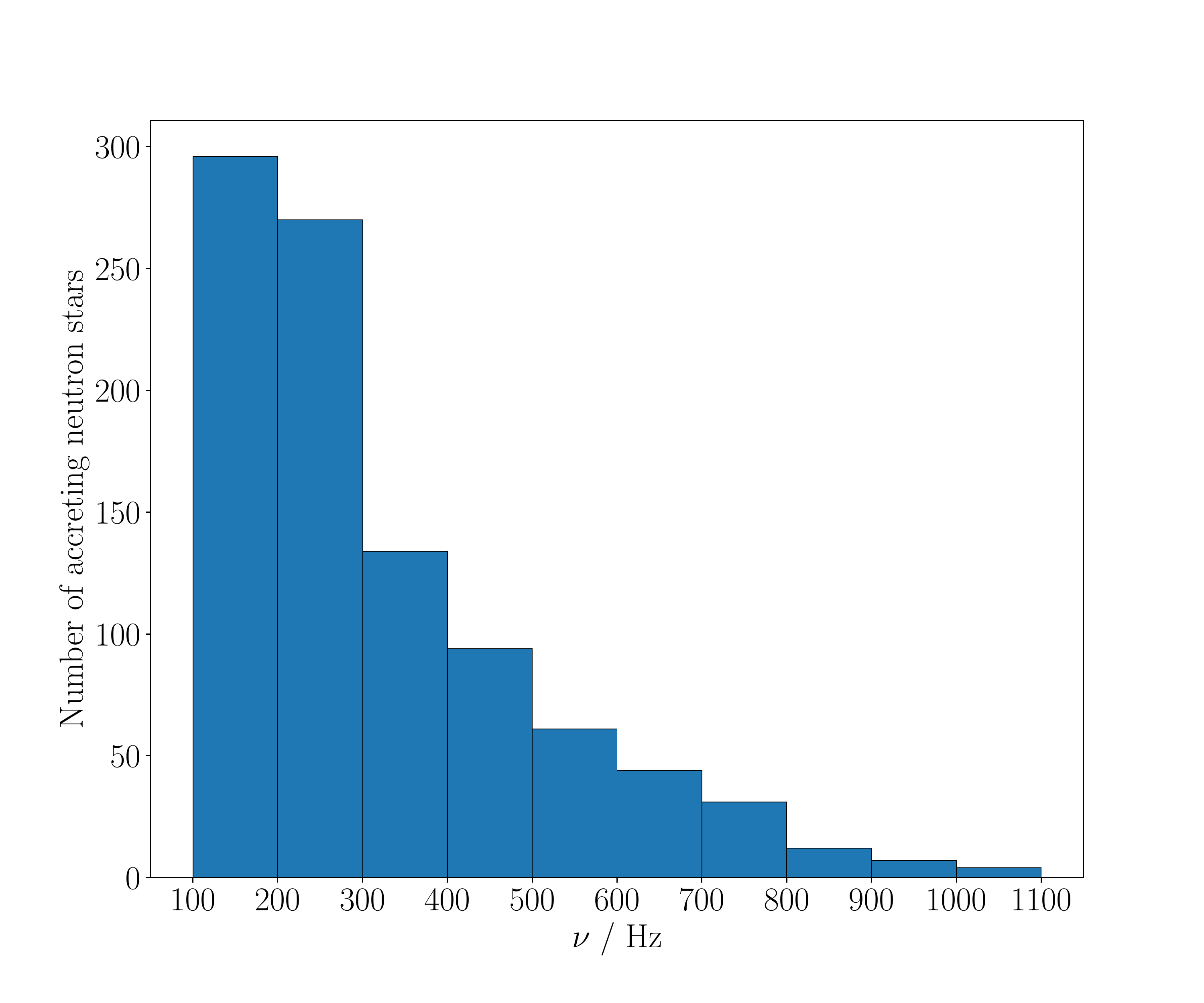}
    \caption[Spin-frequency distributions for simulated accreting neutron stars 
             with unstable \textit{r}-modes and varied amplitudes]{Distributions 
             of spin frequencies for simulated persistently accreting neutron 
             stars (left panel) and transiently accreting neutron stars 
             (right panel) with unstable \textit{r}-modes with initial 
             distributions from Table~\ref{tab:InitialValues} and $\alpha$ 
             distributed flat-in-the-log between $10^{-8} - 10^{-4}$.}
	\label{fig:RModeSpins2}
\end{figure}

We also conducted a simulation where $\alpha$ was distributed flat-in-the-log 
between $10^{-8} - 10^{-4}$. The result is shown in 
Fig.~\ref{fig:RModeSpins2}. We can see, similar to the thermal mountain 
distribution, that both distributions follow an exponentially decreasing 
behaviour. From these distributions we found $p = \num{1.2e-2}$ when the neutron 
stars were persistently accreting and $p = \num{1.5e-3}$ when they were 
transiently accreting. From these $p$-values we can reject the null hypothesis 
and note that the unstable-\textit{r}-modes prescription produces more promising 
results when $\alpha$ is fixed, which is in agreement with current theoretical 
expectations \citep{2003ApJ...591.1129A, 2007PhRvD..76f4019B}.

\section{Summary}
\label{sec:Summary3}

An unresolved problem in the study of low-mass X-ray binaries is the unusual 
spin distribution of rapidly accreting neutron stars and, in particular, why no 
neutron star has been observed to spin close to the centrifugal break-up 
frequency. A potential explanation to this problem comes from gravitational 
waves. Theoretically, gravitational waves could be able to spin down these 
systems away from the break-up frequency. However, there are a number of 
different mechanisms that could give rise to gravitational radiation and it is 
unclear which are the most probable. It is also unclear whether gravitational 
waves are the only way to explain the observed distribution of accreting neutron 
stars. For example, it was recently suggested by \citet{2016ApJ...822...33P} 
that spin-down torques from an enhanced pulsar wind due to a disc-induced 
opening of the magnetic field could have a meaningful effect on the spin 
evolution of an accreting neutron star. Such a torque is not phenomenologically 
accounted for in our method and could be a direction for future work.

In this chapter, we have explored, within the context of our current 
understanding of accretion torques, whether an additional component is required 
in order to describe the spin evolution of accreting neutron stars. We 
investigated whether gravitational-wave emission could be one such explanation 
and have compared competing gravitational-wave mechanisms. We presented our 
model for the spin evolution of an accreting neutron star that accounts for 
accretion and magnetic-field effects, and also includes a gravitational-wave 
spin-down component. Our model is able to simulate persistent and transient 
accretors.

In our simulations with no gravitational-wave torques we obtained neutron stars 
with much higher spins than what is observed. We did not obtain any of the 
characteristic behaviour of the observed spin distribution. In particular, there 
was no evidence of a pile-up at high frequencies. However, by adding a permanent 
quadrupole moment, motivated by torque balance, of 
$Q_{2 2} = \SI{e36}{\gram\centi\metre\squared}$ we obtained qualitatively 
similar behaviour to the observed distribution for the transiently accreting 
neutron-star population. This quadrupole is below the maximum neutron-star 
crusts can support, as we will see in Chaps.~\ref{ch:Mountains} and 
\ref{ch:RelativisticMountains}.

We considered the impact of magnetic-dipole radiation on the results. We found 
that in the case of no gravitational-wave emission one does not obtain the 
observed distribution. With the inclusion of gravitational-wave emission the 
resultant distribution is qualitatively similar to the case of no 
magnetic-dipole radiation. By varying the outburst duration with 
gravitational-wave and magnetic-dipole torques we obtained a distribution with a 
broad high-frequency peak.

We investigated two gravitational-wave-production prescriptions. For thermal 
mountains produced by asymmetric nuclear reactions in the crust, our model was 
sensitive to the precise features of the outburst profile, as well as the ratio 
of quadrupolar to total heating, $\delta T_\text{q} / \delta T$. We found that a 
value of $\delta T_\text{q} / \delta T = \num{4e-4}$ produced a similar 
distribution to what is observed. Promisingly, this gave the characteristic 
pile-up at high frequencies with a narrow, pronounced peak. We examined whether 
the distributions had a preference for $\delta T_\text{q} / \delta T$ being a 
single value or being distributed and found strong evidence arguing that it 
should be a fixed value. Accreting neutron stars with unstable \textit{r}-modes 
and $\alpha = \num{e-7}$ produced similar results to the case with a fixed 
quadrupole moment and the thermal mountain prescription. This prescription 
favoured $\alpha$ being fixed as opposed to being distributed.

The three cases that produced distributions that were qualitatively similar to 
the observed spin distribution -- permanent quadrupole, thermal mountains and 
unstable \textit{r}-modes -- are almost indistinguishable from one another. 
Although, the \textit{r}-mode-instability case could, in theory, be 
differentiated from the other prescriptions. This distinction could come from 
the fact that the quadrupole moment due to unstable \textit{r}-modes scales 
linearly with the spin frequency of the star. Another key difference comes from 
the frequency of the gravitational waves that are emitted through this channel. 
Unstable \textit{r}-modes emit gravitational waves with a frequency of 
$4 \nu / 3$, whereas, gravitational waves due to deformations on a neutron star 
have a frequency of $2 \nu$. Therefore, a gravitational-wave detection would 
indicate which mechanism is active, motivating further searches.

For the \textit{r}-mode scenario, the value for the saturation amplitude that 
we found to agree well with observation ($\alpha = \num{e-7}$) is many orders 
of magnitude below what is currently predicted. Theory would need to explain 
why this is so, or why the instability window is smaller than what is usually 
assumed.

We have not addressed the spin distribution of radio millisecond pulsars in this 
work. Future work could explore how the low-mass X-ray binary population evolves 
into the radio millisecond pulsar population and consider whether gravitational 
waves are relevant in this process and can explain the observed distribution.

In our modelling of transient accretion we considered a simple fast-rise, 
exponential-decay function with a constant average accretion rate. However, in 
these systems it is expected that binary evolution will play a key role in the 
accretion rates and result in a long-term modulation of the accretion rate. 
This, of course, could have a significant effect on the resultant spin 
distribution. Such long-term variations could be explored in a future study.


%% file: sections/chapter-4.tex
\chapter{Neutron-star mountains}
\label{ch:Mountains}

In Chap.~\ref{ch:PopulationSynthesis}, we explored whether gravitational waves 
could describe the spin evolution of rapidly rotating neutron stars found in 
accreting binaries. Promisingly, we found that by including a gravitational-wave 
component we could indeed recreate the spin distribution of these systems, thus, 
justifying further study into the mechanisms that could give rise to 
gravitational waves. In this chapter, we examine neutron-star mountains in a 
Newtonian framework, following \citet{2021MNRAS.500.5570G}.

We start, in Sec.~\ref{sec:Context}, with an introduction to static 
perturbations of neutron stars and a review of prior efforts on estimating the 
maximum mountain. We summarise previous approaches and the important 
assumptions that provide the motivation for this work. In 
Sec.~\ref{sec:Mountains}, we consider the necessary components of a neutron-star 
mountain calculation. We provide a detailed discussion on the usual method of 
calculating mountains and introduce our own scheme, demonstrating the validity 
and equivalence of both approaches. We detail the Newtonian perturbation 
formalism for our mountain scheme in Sec.~\ref{sec:NewtonianPerturbations} and 
pay particular attention to the boundary conditions of the problem. We consider 
three sources for the deformations in Sec.~\ref{sec:Sources} and provide the 
maximum quadrupoles for each scenario. In Sec.~\ref{sec:Summary4}, we end with a 
summary and discuss future directions.

\section{Context}
\label{sec:Context}

As, we have already noted, a rotating neutron star perturbed away from axial 
symmetry will emit gravitational waves. A fluid is continuously deformable. 
However, realistic neutron stars will have elastic crusts that are formed in the 
very early stages of their lives as they begin to cool. The crust can sustain 
shear stresses which, by definition, do not exist in a perfect fluid. It is the 
crust that sets the limit of how much a neutron star can be stretched and 
strained until it breaks. For this reason, it is a physically interesting 
question to ask how much can a neutron star be deformed by.

There have been a number of studies of the maximum quadrupole deformation of a 
neutron star. The earliest of these was conducted by 
\citet{2000MNRAS.319..902U}, who used the Cowling approximation in Newtonian 
gravity to derive an integral expression for the quadrupole moment (we describe 
their approach in detail in Sec.~\ref{subsec:UCB}). They introduced the argument 
that the body will obtain its maximum mountain when the entire crust is strained 
to its elastic yield point. This argument enabled them to straightforwardly find 
the strain tensor that ensures that every point in the crust is maximally 
strained. \citet{2006MNRAS.373.1423H} observed that the approach of 
\citet{2000MNRAS.319..902U} did not respect the required boundary conditions at 
the base and top of the crust and that the Cowling approximation could have a 
large impact on the results. Therefore, they presented a perturbation formalism 
that relaxed the Cowling approximation and enabled them to treat the phase 
transitions appropriately. However, there are inconsistencies in their analysis 
that we explain later. More recent estimates of the maximum elastic deformation 
have been provided by \citet{2013PhRvD..88d4004J}, who carried out their 
calculation in full relativity using a Green's function method. However, since 
they used the covariant analogue to the strain tensor from 
\citet{2000MNRAS.319..902U}, their calculation also ignored the boundary 
conditions on the crust. An important aspect of past studies is the fact that 
the maximum mountains they calculate are independent of the precise mechanisms 
that sourced them. They do not consider the deforming forces or evolutionary 
scenarios that lead to the formation of the mountains.

In \citet{2021MNRAS.500.5570G}, we return to this problem to address some of the 
assumptions of the previous work and detail a formalism that enables one to 
accurately compute the quadrupole deformation throughout the star. As we show, 
in order to satisfy the necessary boundary conditions of the problem, it is 
extremely helpful to characterise the source of the perturbations. This has not 
been done in past calculations. In addition, it is not clear whether strain 
configurations where the majority of the crust is maximally strained can 
actually be reached in a real neutron star. The largest realistic mountain may 
be significantly smaller. These points suggest that future progress on this 
subject will rely on evolutionary calculations that consider the complete 
formation of the mountains 
\citep{1998ApJ...501L..89B, 2000MNRAS.319..902U, 2020MNRAS.494.2839O, 
2020MNRAS.493.3866S}. 

We will decompose our perturbation variables using spherical harmonics as shown 
in (\ref{eq:SphericalHarmonicDecomposition}). For this analysis, it will be 
sufficient to focus on the $(\ell, m) = (2, 2)$ mode relevant for gravitational 
radiation. For this reason, we will drop the mode subscript on our perturbation 
variables.%
\footnote{One should note that, although we restrict ourselves to the 
$(\ell, m) = (2, 2)$ mode, other modes will contribute to the total strain, 
pushing the crustal lattice closer to the breaking strain, while not adding to 
the quadrupole.} 
Because we restrict this analysis to Newtonian gravity, it is inappropriate to 
consider realistic equations of state and we assume a simple polytropic equation 
of state, along with a simple model for the crust elasticity (we refine this 
approach in our relativistic calculation in 
Chap.~\ref{ch:RelativisticMountains}). 

We consider perturbations of a non-rotating, equilibrium, fluid star with mass 
density $\rho$, isotropic pressure $p$ and gravitational potential $\Phi$, 
described by the Newtonian equations of 
structure~(\ref{eqs:NewtonianStellarStructure}). As we demonstrated in 
Sec.~\ref{sec:Perturbations}, we have the following fundamental expressions in 
perturbation theory: the Lagrangian variations of the velocity 
(\ref{eq:LagrangianVelocity}),
\begin{equation}
    \Delta v^i = \partial_t \xi^i,
\end{equation}
and the mass density (\ref{eq:MassConservation}), 
\begin{equation}
    \Delta \rho = - \rho \nabla_i \xi^i.
\end{equation}
By considering perturbations of the Euler equation~(\ref{eq:Euler}) and 
Poisson's equation~(\ref{eq:Poissons}) and making use of our static background, 
$\delta v^i = \Delta v^i = \partial_t \xi^i$, we obtain the following equations 
that govern the perturbations: 
\begin{gather}
    \delta \rho + \nabla_i (\rho \xi^i) = 0, 
    \label{eq:PerturbedContinuity}\\
    \rho \partial_t^2 \xi_i = - \nabla_i \delta p - \delta \rho \nabla_i \Phi 
        - \rho \nabla_i \delta \Phi, 
    \label{eq:PerturbedEuler}\\
    \delta p = c_\text{s}^2 \delta \rho 
    \label{eq:PerturbedEOS}
\end{gather}
and 
\begin{equation}
    \nabla^2 \delta \Phi = 4 \pi G \delta \rho. 
    \label{eq:PerturbedPoissons}
\end{equation}
Since we focus on static perturbations, we amend the perturbed Euler equation 
(\ref{eq:PerturbedEuler}) by 
\begin{equation}
    0 = - \nabla_i \delta p - \delta \rho \nabla_i \Phi 
        - \rho \nabla_i \delta \Phi + f_i, 
    \label{eq:PerturbedEulerStatic}
\end{equation}
where $f_i$ is the density of a force that sustains the perturbations. The 
inclusion of this force enables us to produce non-spherical models and will 
prove to be an important component of our analysis, since it enables one to 
satisfy all the boundary conditions of the problem. We note that $f_i$ does not 
correspond to a physical force acting on the star. This force is a proxy for the 
(possibly quite complicated) formation history that results in its 
non-spherical shape. To study neutron stars with an elastic crust, we must 
modify (\ref{eq:PerturbedEulerStatic}) to include the shear stresses, 
\begin{equation}
    0 = - \nabla_i \delta p - \delta \rho \nabla_i \Phi 
        - \rho \nabla_i \delta \Phi + \nabla^j t_{i j} + f_i, 
    \label{eq:PerturbedEulerElastic}
\end{equation}
where $t_{i j}$ is the symmetric and trace-free, shear-stress tensor, assumed to 
enter at the perturbative level. Here, we have used the same sign for the 
shear-stress tensor as in \citet{2000MNRAS.319..902U}. 

As we discuss in detail in Sec.~\ref{subsec:NewtonianInterface}, in order to 
connect the elastic crust of the star with the fluid regions, one needs to 
consider the traction vector. The traction vector is defined using the stress 
tensor. We can identify the perturbed traction from the perturbed Euler 
equation~(\ref{eq:PerturbedEulerElastic}),
\begin{equation}
    T^i = (\delta p g^{i j} - t^{i j}) \nabla_j r.
    \label{eq:PerturbedTraction}
\end{equation}
The traction must be continuous throughout the star. 

We turn our attention to past work on estimating the maximum mountain, that 
we now summarise. We do this to critique some of the assumptions made and set 
the stage for our new calculation. A convenient simplification that this body of 
work makes is to not (explicitly) consider the perturbing force. This is the 
main conceptual difference in our approach. We show how the force enters the 
problem in Sec.~\ref{sec:Mountains} and demonstrate that the formulation is 
consistent. 

\subsection{Ushomirsky, Cutler and Bildsten}
\label{subsec:UCB}

The first (and perhaps most well known) maximum-mountain calculation was 
performed by \citet{2000MNRAS.319..902U}, which we will review here. They 
tackled the problem in Newtonian gravity and adopted the Cowling approximation 
-- neglecting perturbations of the star's gravitational potential, 
$\delta \Phi = 0$. The starting point for this computation is the perturbed 
Euler equation for the elastic crust [(\ref{eq:PerturbedEulerElastic}) with 
$f_i = 0$], 
\begin{equation}
    0 = - \nabla_i \delta p - \delta \rho \nabla_i \Phi 
        - \rho \nabla_i \delta \Phi + \nabla^j t_{i j}.
    \label{eq:PerturbedEulerUCBfull}
\end{equation}
As we shall elaborate upon in Sec.~\ref{sec:Mountains}, one must carefully 
define what the perturbations are with respect to. The perturbation quantities 
in (\ref{eq:PerturbedEulerUCBfull}) connect the spherical, fluid star with the 
non-spherical, strained shape that hosts a mountain and the shear stresses come 
from the crust wishing to have a separate non-spherical, relaxed shape 
[\textit{cf.} (\ref{eq:PerturbedEulerUCB}) and see Fig.~\ref{fig:StarsUCB}]. 
There is no deforming force in (\ref{eq:PerturbedEulerUCBfull}) as the mountains 
are supported solely by the elastic stresses. With the Cowling approximation, 
one can ignore perturbations in the fluid regions of the star, since the absence 
of shear stresses means there is no support for the pressure perturbations by 
the fluid 
[see (\ref{eq:PerturbedEulerUCBfull}) with $\delta \Phi = 0$ and $t_{i j} = 0$]. 
Therefore, only perturbations in the crust contribute to the quadrupole moment. 

It is convenient to decompose the shear-stress tensor using tensor spherical 
harmonics,%
\footnote{We have neglected pieces of axial parity in $t_{i j}$ which are 
proportional to 
$(\nabla_i r \epsilon_{j k n} + \nabla_j r \epsilon_{i k n}) 
\nabla^n r \nabla^k Y_{\ell m}$ and 
$(\nabla_i Y_{\ell m} \epsilon_{j k n} + \nabla_j Y_{\ell m} \epsilon_{i k n}) 
\nabla^n r \nabla^k Y_{\ell m}$.} 
\begin{equation}
    t_{i j} = t_{r r} \left( \nabla_i r \nabla_j r - \frac{1}{2} e_{i j} 
        \right) Y_{\ell m} 
        + t_{r \bot} f_{i j} 
        + t_\Lambda \left( \Lambda_{i j} + \frac{1}{2} e_{i j} Y_{\ell m} 
        \right),
    \label{eq:ShearStressTensorSphericalHarmonics}
\end{equation}
where $t_{r r}$, $t_{r \bot}$ and $t_\Lambda$ are functions of $r$, 
$\beta = \sqrt{\ell (\ell + 1)}$, 
\begin{subequations}\label{eqs:TensorSphericalHarmonicBasis}
\begin{gather}
    e_{i j} = g_{i j} - \nabla_i r \nabla_j r, \\
    f_{i j} = \frac{r}{\beta} (\nabla_i r \nabla_j Y_{\ell m} 
        + \nabla_j r \nabla_i Y_{\ell m}), \\
    \Lambda_{i j} = \left( \frac{r}{\beta} \right)^2 
        \nabla_i \nabla_j Y_{\ell m} 
        + \frac{1}{\beta} f_{i j}.
\end{gather}
\end{subequations}

We obtain an expression for the multipole moment by isolating the mode part of 
$\delta \rho$ from the perturbed Euler 
equation~(\ref{eq:PerturbedEulerUCBfull}). We project the free index of 
(\ref{eq:PerturbedEulerUCBfull}) with $\nabla^i r$ to yield 
\begin{equation}
    \delta \rho + \frac{\rho}{d \Phi / dr} \frac{d \delta \Phi}{dr} 
        = \frac{1}{d \Phi / dr} \left( - \frac{d \delta p}{dr} 
        + \frac{d t_{r r}}{dr} + \frac{3}{r} t_{r r} 
        - \frac{\beta}{r} t_{r \bot} \right).
    \label{eq:PerturbedEulerProjectr}
\end{equation}
We find an expression for $\delta p$ in terms of the shear stresses by 
projecting (\ref{eq:PerturbedEulerUCBfull}) with $\nabla^i Y_{\ell m}$, 
\begin{equation}
    \delta p + \rho \delta \Phi = - \frac{1}{2} t_{r r} 
        + \frac{r}{\beta} \frac{d t_{r \bot}}{dr} + \frac{3}{\beta} t_{r \bot} 
        + \left( \frac{1}{\beta^2} - \frac{1}{2} \right) t_\Lambda.
    \label{eq:PerturbedEulerProjectYlm}
\end{equation}
We combine Eqs.~(\ref{eq:PerturbedEulerProjectr}) and 
(\ref{eq:PerturbedEulerProjectYlm}) and insert them into (\ref{eq:Multipole}) to 
obtain 
\begin{equation}
\begin{split}
    Q_{\ell m} = \int_0^R \frac{r^4}{d \Phi / dr} \Bigg[ 
        &\frac{3}{2} \frac{d t_{r r}}{dr} 
        + \frac{3}{r} t_{r r} 
        - \frac{r}{\beta} \frac{d^2 t_{r \bot}}{dr^2} 
        - \frac{4}{\beta} \frac{d t_{r \bot}}{dr} 
        - \frac{\beta}{r} t_{r \bot} \\ 
        &+ \left( \frac{1}{2} - \frac{1}{\beta^2} \right) 
        \frac{d t_\Lambda}{dr} 
        + \frac{d \rho}{dr} \delta \Phi 
        \Bigg] dr.
\end{split}
\end{equation}
At this point, the Cowling approximation can be invoked, which means the 
multipole is given purely in terms of the shear stresses of the crust. Since the 
shear stresses vanish above and below the crust, by integrating by parts and 
discarding boundary terms, \citet{2000MNRAS.319..902U} found 
\begin{equation}
\begin{split}
    Q_{2 2} = - \int_{r_\text{base}}^{r_\text{top}} \frac{r^3}{d \Phi / dr} 
        \Bigg[ 
        &\frac{3}{2} (4 - U) t_{r r} 
        + \sqrt{\frac{3}{2}} \left( 8 - 3 U + \frac{1}{3} U^2 
        - \frac{r}{3} \frac{d U}{dr} \right) t_{r \bot} \\
        &+ \frac{1}{3} (6 - U) t_\Lambda 
        \Bigg] dr,
\end{split}
\end{equation}
where $r_\text{base}$ and $r_\text{top}$ are the positions of the base and top 
of the crust, respectively, and $U = d \ln (d \Phi / dr) / d \ln r + 2$. In 
discarding the boundary terms, it has been assumed that the shear modulus, 
$\check{\mu}$, is zero at the base and top of the crust.

To obtain the maximum quadrupole, \citet{2000MNRAS.319..902U} imposed their 
neutron-star crust to be in a shape such that it was strained everywhere to the 
breaking point. To define the elastic yield limit, \citet{2000MNRAS.319..902U} 
used the von Mises criterion. This criterion states that the crust will yield 
when the von Mises strain $\bar{\sigma}$, defined by 
\begin{equation}
    \bar{\sigma}^2 \equiv \frac{1}{2} \sigma_{i j} \sigma^{i j}, 
    \label{eq:vonMisesStrain}
\end{equation}
where $\sigma_{i j} = t_{i j} / (2 \check{\mu})$ is the strain tensor,%
\footnote{This is a factor of two different to the expressions in 
\citet{2000MNRAS.319..902U} and \citet{2006MNRAS.373.1423H} but the same as used 
in \citet{2013PhRvD..88d4004J}.}
exceeds the breaking strain, 
\begin{equation}
    \bar{\sigma} \geq \bar{\sigma}_\text{max},
    \label{eq:vonMisesCriterion}
\end{equation}
with $\bar{\sigma}_\text{max}$ being the breaking strain of the crust. One can 
show that 
\begin{equation}
    \sigma_{i j} \sigma^{i j} 
        = \frac{3}{2} \sigma_{r r}^2 \text{Re} (Y_{\ell m})^2 
        + \sigma_{r \bot}^2 \text{Re} (f_{i j})^2 
        + \sigma_\Lambda^2 \text{Re} \left( \Lambda_{i j} 
        + \frac{1}{2} e_{i j} Y_{\ell m} \right)^2,
\end{equation}
where $\sigma_{r r}$, $\sigma_{r \bot}$ and $\sigma_\Lambda$ are the components 
of the strain tensor decomposed as in 
(\ref{eq:ShearStressTensorSphericalHarmonics}). Notice that this expression is a 
consequence of the orthogonality of the 
basis~(\ref{eqs:TensorSphericalHarmonicBasis}). For the $(\ell, m) = (2, 2)$ 
mode, there exists the identity 
\begin{equation}
    \frac{3}{4} \text{Re} (Y_{\ell m})^2 + \frac{3}{4} \text{Re} (f_{i j})^2 
        + \frac{9}{2} \text{Re} \left( \Lambda_{i j} 
        + \frac{1}{2} e_{i j} Y_{\ell m} \right)^2 
        = \frac{15}{32 \pi}.
\end{equation}
Hence, the crust will be strained at every point to the maximum when%
\footnote{It is interesting to note that, although 
Eqs.~(\ref{eqs:MaximumStrain}) certainly constitute a strain configuration that 
is maximally strained at every point in the crust, there are conceivably many 
other solutions that satisfy the von Mises criterion.}
\begin{subequations}\label{eqs:MaximumStrain}
\begin{align}
    \sigma_{r r} &= \sqrt{\frac{32 \pi}{15}} \bar{\sigma}_\text{max}, \\ 
    \sigma_{r \bot} &= \sqrt{\frac{16 \pi}{5}} \bar{\sigma}_\text{max}, \\ 
    \sigma_\Lambda &= \sqrt{\frac{96 \pi}{5}} \bar{\sigma}_\text{max}.
\end{align}
\end{subequations}

For a star with mass $M = \SI{1.4}{\solarMass}$ and radius 
$R = \SI{10}{\kilo\metre}$ strained according to Eqs.~(\ref{eqs:MaximumStrain}), 
\citet{2000MNRAS.319..902U} reported a maximum quadrupole moment of 
\begin{equation}
    Q_{2 2}^\text{max} \approx \num{1.2e39} \ 
        \left( \frac{\bar{\sigma}_\text{max}}{10^{-1}} \right) \ 
        \si{\gram\centi\metre\squared}, 
    \label{eq:Q_22UCB}
\end{equation}
where $\bar{\sigma}_\text{max}$ is the breaking strain of the crust, that we 
take to have the canonical value $\bar{\sigma}_\text{max} = 10^{-1}$ 
\citep{2009PhRvL.102s1102H}. In terms of the fiducial ellipticity, this result 
corresponds to $\epsilon^\text{max} \approx \num{1.6e-6} \ 
(\bar{\sigma}_\text{max} / 10^{-1})$. 

This approach, while elegant, does not enforce the continuity of the traction 
vector at the boundaries of the crust. At the base of the crust, there is a  
transition between the fluid core and the elastic crust. At the top, there is a 
transition between the elastic region and the fluid ocean. At these interfaces, 
there is expected to be a first-order phase transition where the crust sharply 
obtains a non-zero shear modulus. Since the fluid has a vanishing shear modulus, 
the traction can only be continuous if the appropriate strain components go to 
zero at these boundaries. However, due to the fact that 
\citet{2000MNRAS.319..902U} demanded that the crust be maximally strained at 
every point, the strain components have finite values at the interfaces and, 
therefore, one cannot ensure continuity of the traction. 

In defence of the \citet{2000MNRAS.319..902U} approach, one might argue that the 
shear modulus may be assumed to smoothly go to zero at the phase transitions. 
However, this is still problematic. As we show in 
Sec.~\ref{subsec:NewtonianInterface}, such an assumption means that one does not 
have enough equations to uniquely determine the displacement, in the case where 
one does not know the strain. A more realistic assumption might be to take 
almost the entire crust to be at breaking strain, with the exception of an 
infinitesimally small region at the boundaries where the displacement is 
adjusted to satisfy the continuity of the traction. 

Ultimately, the estimate (\ref{eq:Q_22UCB}) may give us an idea of the likely 
maximum mountain, but the calculation is not completely consistent. 

\subsection{Haskell, Jones and Andersson}

\citet{2006MNRAS.373.1423H} set out to relax some of the assumptions made by 
\citet{2000MNRAS.319..902U}. This included dropping the Cowling approximation 
and ensuring the traction is continuous at the appropriate boundaries. They also 
noted that, by insisting the star is strained to the maximum throughout the 
crust, one loses the freedom to impose the boundary conditions of the problem. 

\citet{2006MNRAS.373.1423H} derived a system of coupled ordinary differential 
equations that describe the perturbations in the elastic crust and the fluid 
core relative to a spherically symmetric background star. They numerically 
integrated the perturbation equations and fixed the perturbation amplitude to 
the maximum necessary to begin to break the crust at a point, according the 
von Mises criterion. In their study, \citet{2006MNRAS.373.1423H} obtained the 
largest mountain when they assumed the core to be unperturbed, thus, allowing 
them to use a fully relativistic core combined with Newtonian perturbations in 
the crust. They reported a maximum quadrupole for a star with 
$M = \SI{1.4}{\solarMass}$, $R = \SI{10}{\kilo\metre}$ of 
\begin{equation}
    Q_{2 2}^\text{max} \approx \num{3.1e40} \ 
        \left( \frac{\bar{\sigma}_\text{max}}{10^{-1}} \right) \ 
        \si{\gram\centi\metre\squared}, 
    \label{eq:Q_22HJA}
\end{equation}
that corresponds to an ellipticity of 
$\epsilon^\text{max} \approx \num{4.0e-5} \ 
(\bar{\sigma}_\text{max} / 10^{-1})$. This result is approximately an order of 
magnitude above that of \citet{2000MNRAS.319..902U}. 

The calculation of \citet{2006MNRAS.373.1423H} correctly treated the boundary 
condition at the crust-core interface by demanding that the traction was 
continuous. However, their calculation assumed the relaxed shape -- which the 
strain is taken with respect to -- to be spherical. In general, the relaxed 
shape must be non-spherical, to give an equilibrium solution with a non-zero 
mountain. They did however stipulate that the surface shape of the star was 
deformed in an $(\ell, m) = (2,2)$ way. This effectively meant using an outer 
boundary condition where a traction-like force (\textit{i.e.}, a force per unit 
area) acts at the very surface of the star. Because of this, the maximum 
quadrupoles calculated in this framework turn out to be insensitive to the shear 
modulus of the crust, as they are sustained by this applied surface force. The 
lack of inclusion of a body force (\textit{i.e.}, a force per unit volume) in 
building the mountain meant that their formalism did not have the necessary 
freedom to ensure that the perturbed potential in the interior matches to the 
exterior solution. We discuss this particular subtlety further in 
Sec.~\ref{subsec:NewtonianFluid}. 

By comparing with our new analysis we also note a number of typographical errors 
in their elastic perturbation equations. These errors turn out to have a 
surprisingly dramatic effect. Once they are corrected the maximum quadrupole 
increases by  three orders of magnitude, in sharp contrast with other estimates. 
This, in turn, highlights the conceptual problem with the formulation.

\subsection{Johnson-McDaniel and Owen}

The most recent estimates for the largest possible mountain on a neutron star 
were provided by \citet{2013PhRvD..88d4004J}. They generalised the 
\citet{2000MNRAS.319..902U} argument to relativistic gravity while 
relaxing the Cowling approximation. They evaluated the required integral by 
employing a Green's function. For a $\SI{1.4}{\solarMass}$ star, described by 
the SLy equation of state \citep{2001A&A...380..151D}, they obtained the result 
\begin{equation}
    Q_{2 2}^\text{max} \approx \num{2e39} \ 
        \left( \frac{\bar{\sigma}_\text{max}}{10^{-1}} \right) \ 
        \si{\gram\centi\metre\squared}, 
    \label{eq:Q_22JMO}
\end{equation}
corresponding to $\epsilon^\text{max} \approx \num{3e-6} \ 
(\bar{\sigma}_\text{max} / 10^{-1})$. 

In following the \citet{2000MNRAS.319..902U} approach, the crust was taken to be 
strained to the maximum at every point, which means that the traction vector 
cannot be continuous at the crust boundaries. Furthermore,  they do not use the 
correct expression for the perturbed stress-energy tensor, since it does not 
include variations of the four-velocity. This may be a minor detail, but it 
should still be noted. 

In summary, although some of the above points may have a negligible impact on 
the maximum quadrupole estimates, there are issues with all previous studies 
of the maximum-mountain problem. 

\section{Building mountains}
\label{sec:Mountains}

In this section, we examine what must go into a consistent mountain calculation 
and discuss two methods for modelling mountains on neutron stars. The first 
approach, introduced in \citet{2000MNRAS.319..902U}, involves specifying the 
strain field associated with the mountain. We present a second method that, 
instead of starting with the strain, starts with a description of the perturbing 
force. Both approaches are valid and we demonstrate how they are equivalent. 

To help develop intuition, we will start by briefly discussing the case of 
strains built up in a spinning-down star. We will therefore be considering the 
case of $(\ell, m) = (2, 0)$ perturbations relevant for rotational deformations, 
not the $(\ell, m) = (2, 2)$ relevant to the mountain case. Suppose a young 
neutron star with a molten crust spins  at an angular frequency $\Omega$. At 
this rotation rate, the star cools and the crust solidifies. The star then 
begins to spin down to frequency $\tilde{\Omega} < \Omega$.%
\footnote{This spin-down could be due to the usual radio emission that pulsars 
are well known for.}
Because the star has spun down, it changes shape according to the difference in 
the centrifugal force, $\propto (\Omega^2 - \tilde{\Omega}^2)$. This builds up 
strain in the crust as the shear stresses resist the change in shape. Should the 
star spin down sufficiently, the crust may fracture as stresses get too large. 
In fact, it has been suggested that the elastic yield of the crust in this 
process may be associated with the glitch phenomenon observed in some rotating 
pulsars \citep{1971AnPhy..66..816B, 2015MNRAS.446..865K}. 

Motivated by this example, that does not represent a neutron-star mountain, we 
consider neutron-star models forced away from sphericity by a perturbing force 
$f_i$, which we will choose to give mountain-like $(\ell, m) = (2, 2)$ 
perturbations. The elastic Euler equation (\ref{eq:PerturbedEulerElastic}) then 
becomes 
\begin{equation}
    0 = - \nabla_i p - \rho \nabla_i \Phi + \nabla^j t_{i j} + f_i.
    \label{eq:EulerElasticSource}
\end{equation}
For this exercise, we regard Eq.~(\ref{eq:EulerElasticSource}) as exact and 
consider perturbations of it below. In the fluid regions of the star, which 
cannot support shear stresses, the shear modulus goes to zero so the 
shear-stress tensor vanishes. To condense the notation, we define 
\begin{equation}
    H_i \equiv \nabla_i p + \rho \nabla_i \Phi, 
\end{equation}
which captures the familiar equation of hydrostatic 
equilibrium~(\ref{eq:HydrostaticEquilibrium}) when $H_i = 0$. Therefore, the 
Euler equation (\ref{eq:EulerElasticSource}) can be expressed as 
\begin{equation}
    H_i = f_i + \nabla^j t_{i j}. 
\end{equation}
By considering a variation of $H_i$, we may write 
\begin{equation}
    \delta H_i = \nabla_i \delta p + \delta \rho \nabla_i \Phi 
        + \rho \nabla_i \delta \Phi, 
    \label{eq:PerturbedH_i}
\end{equation}
where the perturbed quantities will need to be carefully defined in what 
follows. 

\begin{figure}[h]
    \centering
	\includegraphics[width=0.7\textwidth]{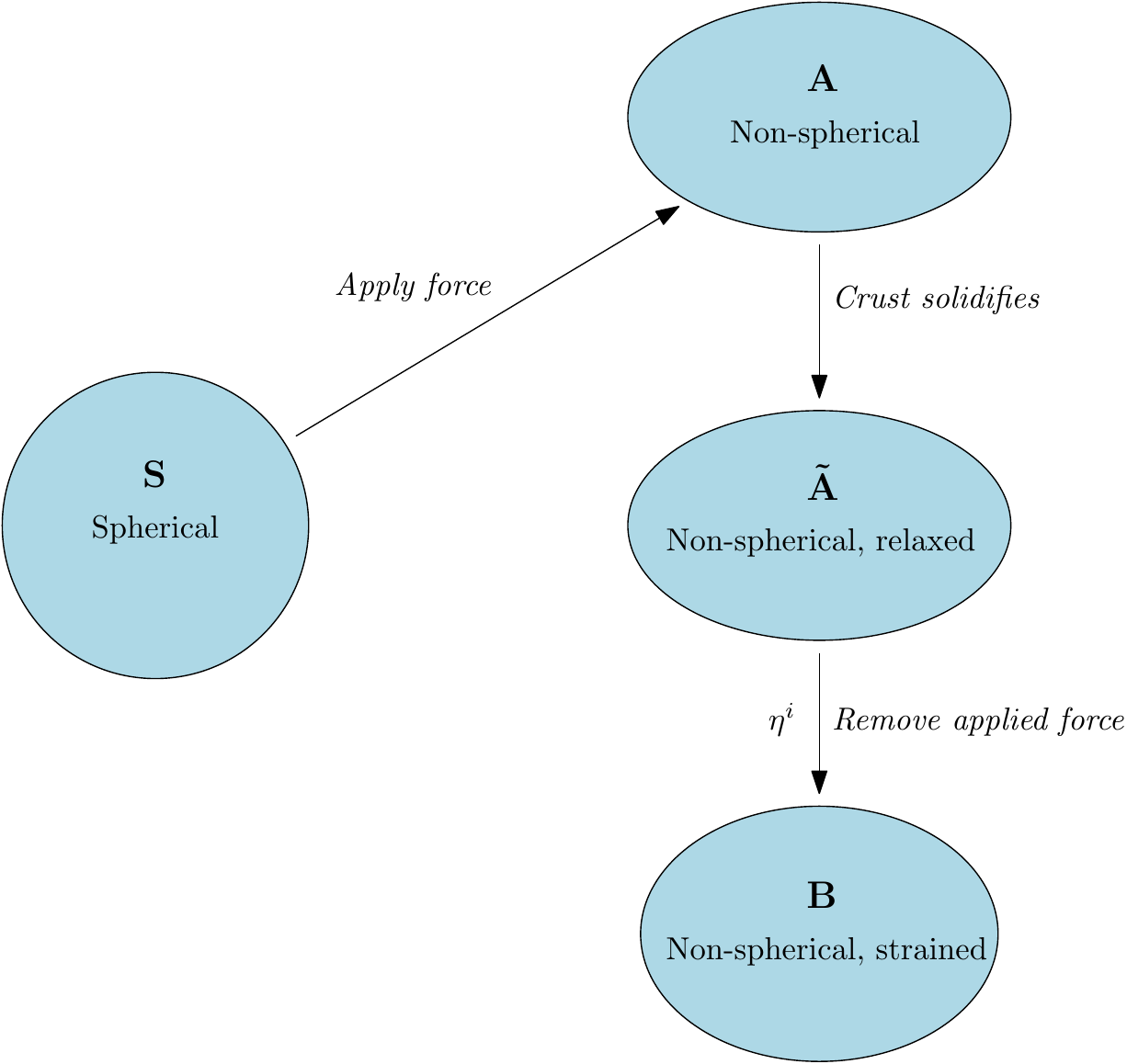}
    \caption[Schematic mountain illustration]{A schematic illustration showing 
             the configurations involved in mountain calculations. Note that 
             previous calculations have typically considered stars S and B, but 
             not (explicitly) A or \~{A}.}
    \label{fig:StarsUCB}
\end{figure}

We now consider a family of four closely-related, equilibrium stars, 
illustrated in Fig.~\ref{fig:StarsUCB}: 

\begin{description}
    \item[Star S] -- A spherical, fluid star with 
        $(\rho_\text{S}, p_\text{S}, \Phi_\text{S})$: 
        \begin{equation}
            H_i^\text{S} = 0. 
            \label{eq:EulerS}
        \end{equation}
    \item[Star A] -- A force is applied to star S, which produces a 
        non-spherical, fluid star with 
        $(\rho_\text{A}, p_\text{A}, \Phi_\text{A})$: 
        \begin{equation}
            H_i^\text{A} = f_i. 
            \label{eq:EulerA}
        \end{equation}
    \item[Star \~{A}] -- The crust of star A solidifies while the force 
        is maintained. This gives rise to a non-spherical, relaxed star with the 
        same structure as star A, although (formally) with a non-zero shear 
        modulus. The star has $(\rho_\text{\~{A}} = \rho_\text{A}, 
        p_\text{\~{A}} = p_\text{A}, \Phi_\text{\~{A}} = \Phi_\text{A})$: 
        \begin{equation}
            H_i^\text{\~{A}} = H_i^\text{A} = f_i. 
        \end{equation}
        Note that, because star A and star \~{A} have the same shape, in 
        general, we only need to refer to star A in the following discussion 
        when specifying the values of perturbed quantities. 
    \item[Star B] -- The force on star \~{A} is removed, which builds up 
        strain in the crust. The associated deformation between these two stars 
        is described by the Lagrangian displacement vector field $\eta^i$. The 
        star is non-spherical and strained with 
        $(\rho_\text{B}, p_\text{B}, \Phi_\text{B})$: 
        \begin{equation}
            H_i^\text{B} = \nabla^j t_{i j}(\eta). 
            \label{eq:EulerB}
        \end{equation}
        Note that it is this star, star B, that we are ultimately interested in: 
        this is the star with a mountain supported in a self-consistent way by 
        elastic strains, with no external force acting.
\end{description}

Note that the force $f_i$ has a simple physical interpretation: it is the force 
that, when applied to our equilibrium star with the mountain (star B), takes us 
to the corresponding unstrained star (star A or, equivalently, \~{A}). Note, 
however, that there is no requirement whatsoever that, in reality, this force 
ever acted upon our star. For a realistic situation, the elastic strains that 
support the deformation of star B will likely have evolved through some complex 
process of plastic flow and cracking, possibly combined with whatever agent 
that caused the asymmetry to develop. The usefulness of $f_i$ is two-fold. 
Firstly, it allows us to explicitly identify the unstrained configuration. 
Secondly, the explicit introduction of the force into the Euler equation 
provides the necessary freedom to determine the displacement vector and satisfy 
all the boundary conditions. 

It is instructive to consider the differences between the stellar models 
described above. Thus, we introduce the notation 
\begin{equation}
    \delta H_i^\text{AB} = H_i^\text{B} - H_i^\text{A}, 
\end{equation}
\textit{i.e.}, $\delta H_i^\text{AB}$ is the quantity that must be added to 
$H_i^\text{A}$ to obtain $H_i^\text{B}$. 

The difference between star B (\ref{eq:EulerB}) and star S (\ref{eq:EulerS}) is 
\begin{equation}
    \delta H_i^\text{SB} = \nabla^j t_{i j}(\eta). 
    \label{eq:PerturbedEulerUCB}
\end{equation}
Expression (\ref{eq:PerturbedEulerUCB}) relates perturbations between the 
strained star -- with a mountain -- and the spherical, reference star to the 
shear stresses induced when the relaxed star is deformed according to the 
displacement $\eta^i$. This is the standard picture for understanding 
neutron-star mountains and, indeed, it is this expression that is used to 
estimate the maximum quadrupole in \citet{2000MNRAS.319..902U} and 
\citet{2013PhRvD..88d4004J}. It is important to note that in these calculations 
one does not have to determine the relaxed shape and, indeed, stars A and \~{A} 
did not appear explicitly in previous calculations. However, we demonstrate that 
the relaxed shape is, in principle, calculable in Sec.~\ref{subsec:Relaxed}. 

\begin{figure}[h]
    \centering
	\includegraphics[width=0.7\textwidth]{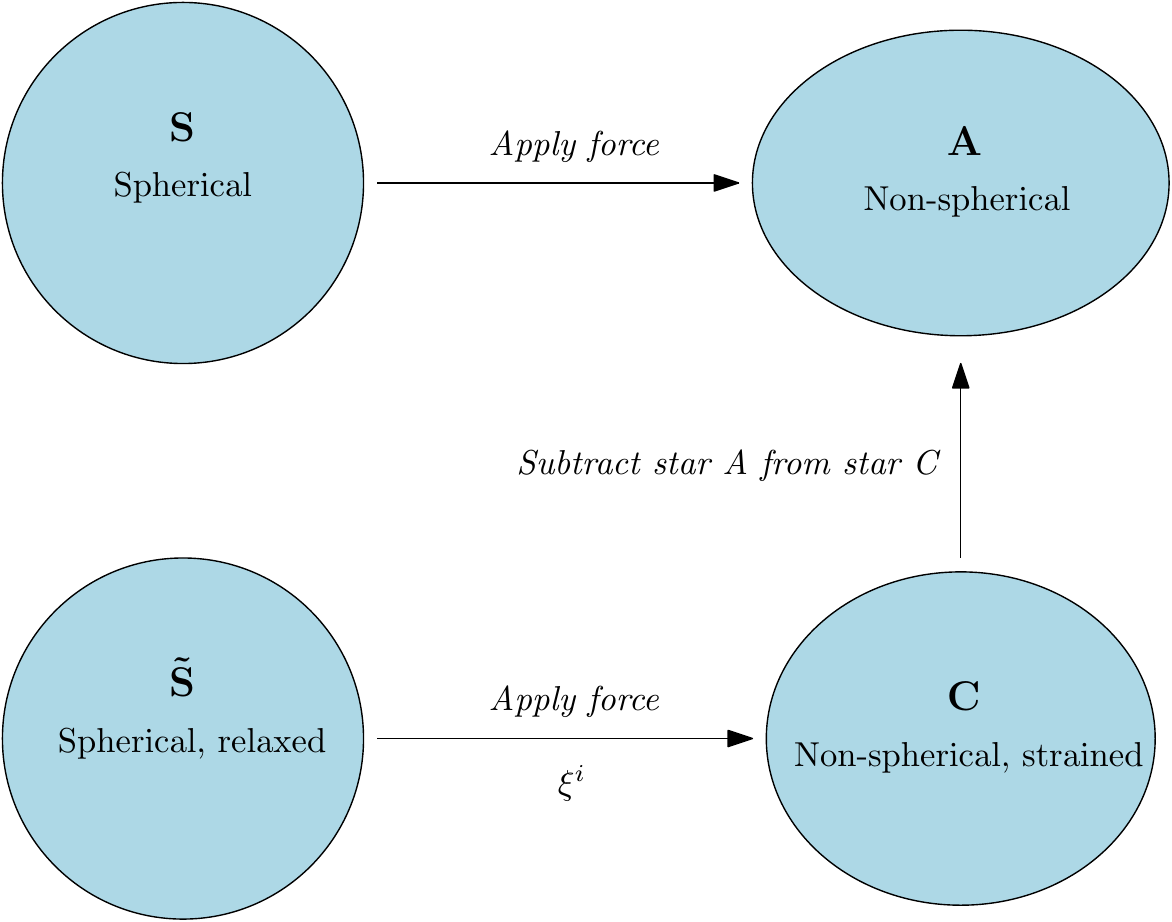}
    \caption[Schematic mountain illustration with the force-based approach]{A 
             schematic illustration showing the configurations in the 
             force-based mountain scheme.}
    \label{fig:StarsScheme}
\end{figure}

As we discuss in more detail below, for a fully consistent calculation that 
satisfies all the boundary conditions of the problem it is not convenient to 
use (\ref{eq:PerturbedEulerUCB}) alone. Rather, we present an alternative 
strategy that makes explicit use of the deforming force. To this end, we 
introduce two additional stars shown in Fig.~\ref{fig:StarsScheme}: 

\begin{description}
    \item[Star \~{S}] -- The crust of star S solidifies. This star has 
        the same shape as star S with a non-zero shear modulus and 
        $(\rho_\text{\~{S}} = \rho_\text{S}, p_\text{\~{S}} = p_\text{S}, 
        \Phi_\text{\~{S}} = \Phi_\text{S})$: 
        \begin{equation}
            H_i^\text{\~{S}} = H_i^\text{S} = 0. 
        \end{equation}
    \item[Star C] -- A force is applied to star \~{S}. This induces 
        stress in the crust, described by the Lagrangian displacement $\xi^i$ 
        and produces a non-spherical, strained star with 
        $(\rho_\text{C}, p_\text{C}, \Phi_\text{C})$: 
        \begin{equation}
            H_i^\text{C} = f_i + \nabla^j t_{i j}(\xi). 
            \label{eq:EulerC}
        \end{equation}
\end{description}

We can then consider the difference between stars A and C. By using 
(\ref{eq:EulerA}) and (\ref{eq:EulerC}), we obtain 
\begin{equation}
    \delta H_i^\text{AC} = \nabla^j t_{i j}(\xi). 
    \label{eq:PerturbedEulerScheme}
\end{equation}

We can note the similarity of (\ref{eq:PerturbedEulerScheme}) to 
(\ref{eq:PerturbedEulerUCB}). Indeed, comparing Figs.~\ref{fig:StarsUCB} 
and \ref{fig:StarsScheme}, we note the following. In Fig.~\ref{fig:StarsUCB}, 
the addition of force $f_i$ maps star B to star A, generating a displacement 
$-\eta^i$, while in Fig.~\ref{fig:StarsScheme}, the addition of the force $f_i$ 
maps star \~{S} to star C, generating a displacement field $\xi^i$. It follows 
that, to a good approximation, these vector fields are related by 
\begin{equation}
    \eta^i =  - \xi^i.
\end{equation}
Comparing (\ref{eq:PerturbedEulerScheme}) to (\ref{eq:PerturbedEulerUCB}) then 
gives the corresponding relation between the associated scalar perturbations, 
\begin{equation}
    \delta H_i^\text{SB} = - \delta H_i^\text{AC}.
    \label{eq:SB_AC_relation}
\end{equation}
These relations are not exact, as in Fig.~\ref{fig:StarsUCB} the force $f_i$ 
acts upon star B, while in Fig.~\ref{fig:StarsScheme} it acts upon star \~{S}, 
but these two stars themselves differ from one another only in a perturbative 
way, so the difference in the action of $f_i$ on the two must be of second 
order.

This immediately suggests a strategy for computing the deformation of star B 
(\textit{i.e.}, $\delta H_i^\text{SB}$ and other perturbed quantities). We can 
easily compute the perturbations linking S and A (\textit{i.e.}, 
$\delta H_i^\text{SA}$ \textit{etc.}), as this is just the perturbation of a 
spherical fluid star by the force $f_i$. We can, with only a little more effort, 
compute the perturbations linking star \~{S} and C (\textit{i.e.}, 
$\delta H_i^\text{SC}$ \textit{etc.}), as this is just the perturbation of a 
spherical elastic star by $f_i$. Then, we can take the difference between these 
two configurations to give the difference between star A and C (\textit{i.e.}, 
$\delta H_i^\text{AC}$ \textit{etc.}), which is, up to an overall sign, the 
deformation of star B relative to star S that we require, as per 
(\ref{eq:SB_AC_relation}).

As we are interested in computing maximum mountains, we will choose the force 
$f_i$ such that the breaking strain is reached at some point in the crust of 
star C. Note, however, that in this force-based approach, we will not be able 
to follow \citet{2000MNRAS.319..902U} and find the solution where the strain is 
reached at \textit{all} points (simultaneously) in the crust. This is a price 
one pays in adopting the force-based approach. (It is possible, at least in 
principle, that one could invent a force that takes the entire crust to breaking 
strain. However, this is beyond the scope of this work.) We have, however, 
reduced the calculation of the mountain to two simpler calculations, both taking 
place on a spherical background and with readily implementable boundary 
conditions.

\subsection{Calculating the relaxed shape}
\label{subsec:Relaxed}

We now briefly examine the relaxed configuration that is implied, but not 
calculated, in maximum-mountain calculations \citet{2000MNRAS.319..902U} and 
\citet{2013PhRvD..88d4004J} to show that it is calculable.

Suppose one knows the strain of star B, $\sigma_{i j}(\eta)$ (see 
Fig.~\ref{fig:StarsUCB}). [This is the case in \citet{2000MNRAS.319..902U} and 
\citet{2013PhRvD..88d4004J}.] From the strain tensor it is possible to obtain 
the displacement vector $\eta^i$ that sources the strain. 

We note the following relations: $\delta \rho$ and $\delta p$ are related 
through the equation of state (\ref{eq:PerturbedEOS}), the perturbed Poisson's 
equation (\ref{eq:PerturbedPoissons}) couples $\delta \rho$ and $\delta \Phi$ 
and (\ref{eq:PerturbedH_i}) links $\delta \rho$, $\delta p$, $\delta \Phi$ and 
$\delta H_i$. Therefore, it follows that if any one of 
$(\delta \rho, \delta p, \delta \Phi, \delta H_i)$ are known, the other 
quantities can, in principle, be calculated. 

We begin with (\ref{eq:PerturbedEulerUCB}). Since we know the strain tensor that 
takes one from star A to star B, we also know $\delta H_i^\text{SB}$. This means 
we have $(\delta \rho_\text{SB}, \delta p_\text{SB}, \delta \Phi_\text{SB})$. It 
is this logic, that enables \citet{2000MNRAS.319..902U} and 
\citet{2013PhRvD..88d4004J} to compute the quadrupole moment from just the 
strain tensor. 

By considering variations between star A (\ref{eq:EulerA}) and star B 
(\ref{eq:EulerB}), we find 
\begin{equation}
    \delta H_i^\text{AB} = - f_i + \nabla^j t_{i j}(\eta). 
\label{eq:PerturbedEulerAB}
\end{equation}
We know $t_{i j}(\eta)$, but not $f_i$ or $\delta H_i^\text{AB}$. However, we 
can obtain $\delta H_i^\text{AB}$. The quantity $\delta H_i^\text{AB}$ is 
generated by the change in shape from star A to star B. This is described by the 
displacement $\eta^i$. In particular, the two density fields $\rho_\text{A}$ 
and $\rho_\text{B}$ are linked through the perturbed continuity equation 
(\ref{eq:PerturbedContinuity}). It, therefore, follows that 
\begin{equation}
    \delta H_i^\text{AB} = \delta H_i^\text{AB}(\eta). 
\end{equation}
We rearrange (\ref{eq:PerturbedEulerAB}) to obtain an expression for the force, 
\begin{equation}
    f_i = - \delta H_i^\text{AB}(\eta) + \nabla^j t_{i j}(\eta). 
\end{equation}
Provided $\eta^i$, we can calculate the force that takes the star from a 
spherical shape (star S) to the relaxed shape (star A). 

Using (\ref{eq:EulerS}) and (\ref{eq:EulerA}), we have 
\begin{equation}
    \delta H_i^\text{SA} = f_i. 
\end{equation}
This determines $\delta H_i^\text{SA}$ and, therefore, also 
$(\delta \rho_\text{SA}, \delta p_\text{SA}, \delta \Phi_\text{SA})$. This means 
one can obtain the shape of the relaxed star, supported by a force, $f_i$, with 
the property that when the force is removed the star obtains a strained 
configuration, according to the displacement vector $\eta^i$. 

\section{Newtonian perturbations}
\label{sec:NewtonianPerturbations}

In order to develop the second strategy of calculating mountains in detail, we 
take the background star to be non-rotating and construct static perturbations 
on top of the background. The background is, thus, spherical and described by 
the Newtonian equations~(\ref{eqs:NewtonianStellarStructure}).

Star C is separated into three layers: a fluid core, an elastic crust and a 
fluid ocean, whereas star A is purely fluid. We choose to include a fluid layer 
outside the crust since at low densities the crustal lattice begins to melt and 
it also simplifies the matching to the exterior gravitational potential 
\citep{2020PhRvD.101j3025G}. The crust only comes into the structure equations 
at the perturbative level.

\subsection{The fluid}
\label{subsec:NewtonianFluid}

In order to calculate the relaxed configuration (star A), we need to introduce 
the force $f_i$. For practical purposes, it is convenient to write the force as 
the gradient of a potential, $\chi$, 
\begin{equation}
    f_i = - \rho \nabla_i \chi. 
    \label{eq:Force}
\end{equation}
This is not the most general expression for the force but it allows us to 
combine $\chi$ with the gravitational potential, which simplifies the analysis. 
To make the notation more compact, we introduce the total perturbed potential 
$U = \delta \Phi + \chi$. 

At this point, we note that this is where our calculation differs from previous 
work \citep{2000MNRAS.319..902U, 2006MNRAS.373.1423H, 2013PhRvD..88d4004J}. 
Previous calculations set out to evaluate the perturbed Euler equation 
(\ref{eq:PerturbedEulerUCB}) where the strain is taken with respect to the 
relaxed shape the crust wants to have (see Fig.~\ref{fig:StarsUCB}). In our 
method, we start with the deforming force and evaluate 
(\ref{eq:PerturbedEulerScheme}), using the subtraction scheme (taking the 
difference between stars C and A) set out in Sec.~\ref{sec:Mountains}. The 
use of this force is a subtle, but important, detail since without it one does 
not have the necessary freedom to impose all the boundary conditions of the 
problem.  We emphasise this point since this issue was somewhat confused in the 
analysis of \citet{2006MNRAS.373.1423H} who calculate perturbations of a 
spherical star but do not explicitly consider the force that sources them. It 
is for this reason that they were unable to satisfy the boundary condition on 
the potential at the surface. This point is elucidated below. 

Recall that, as we discussed earlier, we assume all perturbed quantities to 
be expanded in spherical harmonics, but it will be sufficient for our 
discussion to focus on the $(\ell, m) = (2, 2)$ mode. The system of equations 
that describes fluid perturbations then simplifies to a single second-order 
differential equation for the perturbed potential. From the perturbed Poisson's 
equation (\ref{eq:PerturbedPoissons}), we get 
\begin{subequations}\label{eqs:NewtonianFluidPerturbations}
\begin{equation}
    \frac{d^2\delta \Phi}{dr^2} + \frac{2}{r} \frac{d\delta \Phi}{dr} 
        - \frac{\beta^2}{r^2} \delta \Phi = 4 \pi G \delta \rho. 
    \label{eq:PerturbedPoissonsFluid}
\end{equation}
The perturbed Euler equation (\ref{eq:PerturbedEulerStatic}) returns 
\begin{equation}
    \delta \rho = - \frac{\rho}{c_\text{s}^2} U. 
\end{equation}
\end{subequations}
Therefore, provided a description of the perturbing force, 
Eqs.~(\ref{eqs:NewtonianFluidPerturbations}) give a second-order equation 
that describes the perturbations in the fluid. 

The perturbed potential must satisfy two boundary conditions. At the centre of 
the star the solution must be regular and at the surface it must match to the 
external solution. Therefore, in addition to $\chi$ being regular at the centre 
of the star and continuous at all interfaces, we must have 
\begin{subequations}\label{eqs:PerturbedPotentialBoundaries}
\begin{equation}
    \delta \Phi(0) = 0 
    \label{eq:PerturbedPotentialCentre}
\end{equation}
and 
\begin{equation}
    R \frac{d\delta \Phi}{dr}(R) = - (\ell + 1) \delta \Phi(R). 
    \label{eq:PerturbedPotentialSurface}
\end{equation}
\end{subequations}
Equation~(\ref{eq:PerturbedPotentialCentre}) is simply the statement of 
regularity. Equation~(\ref{eq:PerturbedPotentialSurface}) is mostly readily seen 
by examining the potential in the exterior, $r \geq R$, which must satisfy 
Laplace's equation, $\nabla^2 \delta \Phi = 0$. As discussed in 
Appendix~\ref{app:Harmonics}, this admits two solutions, $c_1 / r^{\ell + 1}$ 
and $c_2 r^\ell$, with constants $c_1$ and $c_2$. Since the exterior 
solution must fall off as $r \rightarrow \infty$, we retain the decaying 
solution. It is simple to show that (\ref{eq:PerturbedPotentialSurface}) is 
satisfied if $\delta \Phi \propto 1 / r^{\ell + 1}$. 

Using the regularity condition~(\ref{eq:PerturbedPotentialCentre}), we consider 
a power-series expansion in small $r$ and input this into 
(\ref{eqs:NewtonianFluidPerturbations}) (as described in 
Appendix~\ref{appsec:NumericalTOV}) to obtain an initial condition, 
\begin{equation}
    \delta \Phi(r) = a_0 r^\ell [1 + \mathcal{O}(r^2)], 
\end{equation}
where $a_0$ is a constant that parametrises the amplitude of the perturbations. 
In the case when $\chi = 0$ and there is no driving force, this initial 
condition provides sufficient information to calculate the perturbations up to 
the surface. At the surface, however, there is no freedom left to impose the 
surface boundary condition (\ref{eq:PerturbedPotentialSurface}) -- except in the 
special case of $a_0 = 0$ where there are no perturbations. [This is the issue 
that the formalism of \citet{2006MNRAS.373.1423H} suffers from, and why in that 
analysis a surface force had to be effectively introduced via a boundary 
condition.] This serves as a simple demonstration of the fact that an unforced, 
fluid equilibrium is a spherical star. 
Equations~(\ref{eqs:NewtonianFluidPerturbations}) with the boundary conditions 
(\ref{eqs:PerturbedPotentialBoundaries}) provide the necessary information to 
calculate perturbations in the fluid regions of the star sourced by a perturbing 
force. A simple example of this is the tidal problem. 

\subsubsection{Aside: the tidal potential}
\label{subsubsec:Tidal}

An external tidal potential field is a solution of Laplace's equation, 
$\nabla^2 \chi = 0$ \citep[see, \textit{e.g.},][]{2020PhRvD.101h3001A}. Because 
of the constraint of regularity at the centre of the tidally deformed star, the 
tidal potential must have the form 
\begin{equation}
    \chi(r) = b_0 r^\ell, 
    \label{eq:InitialChi}
\end{equation}
where $b_0$ parametrises the strength of the tidal field. Therefore, the total 
perturbed potential, for small $r$, must be 
\begin{equation}
    U(r) = a_0 r^\ell[1 + \order(r^2)] + b_0 r^\ell.
\end{equation}
At the surface of the star, the boundary condition 
(\ref{eq:PerturbedPotentialSurface}) must be satisfied. In some sense, this is 
achieved through $a_0$. The constant $a_0$ parametrises the amplitude of 
$\delta \Phi$, which is sourced by the external field. The response of a body 
due to an external tidal field is governed by the Love numbers $k_\ell$. These 
are defined by the ratio of the perturbed potential to the tidal potential at 
the surface, 
\begin{equation}
    k_\ell \equiv \frac{1}{2} \frac{\delta \Phi(R)}{\chi(R)}, 
    \label{eq:LoveNumbers}
\end{equation}
where we recall that the perturbations $\delta \Phi$ are in a specific 
$(\ell, m)$ mode. Since the surface boundary condition must be physically 
satisfied, one can use this information to constrain $k_\ell$ and, therefore, 
$a_0$ for a given tidal field. Using (\ref{eq:PerturbedPotentialSurface}), we 
find 
\begin{equation}
    R \frac{dU}{dr}(R) + (\ell + 1) U(R) = (2 \ell + 1) \chi(R). 
\end{equation}
Thus, defining $y \equiv R [dU(R)/dr] / U(R)$, (\ref{eq:LoveNumbers}) becomes 
\begin{equation}
    k_\ell = \frac{1}{2} \frac{\ell - y}{y + \ell + 1}, 
    \label{eq:k_lNewtonian}
\end{equation}
which is the standard result. This serves as a simple demonstration of how the 
introduction of an external force can enable one to satisfy the boundary 
conditions of the problem. 

\subsection{The elastic crust}

In order to calculate the strained star (star C) in our scheme outlined in 
Sec.~\ref{sec:Mountains} (Fig.~\ref{fig:StarsScheme}), we must consider the 
role of the elastic crust. We reiterate that we consider perturbations with 
respect to a spherical, reference star. 

The elastic material is characterised by the shear-stress tensor 
\begin{equation}
    t_{i j} = \check{\mu} \left( \nabla_i \xi_j + \nabla_j \xi_i 
        - \frac{2}{3} g_{i j} \nabla_k \xi^k \right), 
    \label{eq:ShearStressTensor}
\end{equation}
where $g_{i j}$ is the flat three-metric. This describes an elastic solid that 
obeys Hooke's law: the stress is linearly proportional to the strain. We use the 
static displacement vector appropriate for polar perturbations 
\citep{2000MNRAS.319..902U}%
\footnote{Because we focus on polar deformations, we are able to ignore the 
axial part of the vector spherical harmonics, 
$\epsilon^{i j k} \nabla_j r \nabla_k Y_{\ell m}$, where $\epsilon^{i j k}$ is 
the antisymmetric Levi-Civita symbol. Just as the scalar spherical harmonics, 
$Y_{\ell m}$, form an orthogonal basis, so do the vector spherical harmonics, 
$\nabla^i r Y_{\ell m}$, $\nabla^i Y_{\ell m}$ and 
$\epsilon^{i j k} \nabla_j r \nabla_k Y_{\ell m}$.} 
\begin{equation}
    \xi^i = \xi_r(r) \nabla^i r Y_{\ell m} 
        + \frac{r}{\beta} \xi_\bot(r) \nabla^i Y_{\ell m},
    \label{eq:NewtonianDisplacement}
\end{equation}
where $\xi_r(r)$ and $\xi_\bot(r)$ account for the radial and tangential 
components of the displacement.

To make the application of the boundary conditions straightforward, we consider 
the perturbed traction vector~(\ref{eq:PerturbedTraction}), 
\begin{equation}
    T^i = [\delta p(r) - T_1(r)] \nabla^i r Y_{\ell m} 
        - r T_2(r) \nabla^i Y_{\ell m}, 
    \label{eq:PerturbedTraction2}
\end{equation}
where we have defined the following two variables related to the radial and 
tangential components of the traction: 
\begin{subequations}\label{eqs:TractionVariables}
\begin{equation}
    T_1(r) Y_{\ell m} \equiv t_{r r} = \frac{2 \check{\mu}}{3 r} 
        \left( - 2 \xi_r + \beta \xi_\bot + 2 r \frac{d\xi_r}{dr} \right) 
        Y_{\ell m} 
\end{equation}
and 
\begin{equation}
    T_2(r) \partial_\theta Y_{\ell m} \equiv \frac{t_{r \theta}}{r} 
        = \frac{\check{\mu}}{\beta r} \left( \beta \xi_r - \xi_\bot 
        + r \frac{d\xi_\bot}{dr} \right) \partial_\theta Y_{\ell m}. 
    \label{eq:T_2}
\end{equation}
\end{subequations}

From the perturbed continuity equation (\ref{eq:PerturbedContinuity}), we then 
obtain 
\begin{equation}
\begin{split}
    \delta \rho &= - \rho \frac{d\xi_r}{dr} 
        - \left( \frac{2 \rho}{r} + \frac{d\rho}{dr} \right) \xi_r 
        + \frac{\beta \rho}{r} \xi_\bot \\
        &= - \left( \frac{3 \rho}{r} + \frac{d\rho}{dr} \right) \xi_r 
        + \frac{3 \beta \rho}{2 r} \xi_\bot - \frac{3 \rho}{4 \check{\mu}} T_1.
\end{split}
    \label{eq:PerturbedContinuityElastic} 
\end{equation}

From the definitions of the traction variables (\ref{eqs:TractionVariables}), we 
have the following differential equations that describe the displacement 
vector: 
\begin{subequations}\label{eqs:NewtonianElasticPerturbations}
\begin{equation}
    \frac{d\xi_r}{dr} = \frac{1}{r} \xi_r - \frac{\beta}{2 r} \xi_\bot 
        + \frac{3}{4 \check{\mu}} T_1 
\end{equation}
and 
\begin{equation}
    \frac{d\xi_\bot}{dr} = - \frac{\beta}{r} \xi_r + \frac{1}{r} \xi_\bot 
        + \frac{\beta}{\check{\mu}} T_2. 
\end{equation}
From the radial part of the perturbed Euler equation 
(\ref{eq:PerturbedEulerElastic}) combined with the perturbed continuity equation 
(\ref{eq:PerturbedContinuityElastic}), 
\begin{equation}
\begin{split}
    \bigg( 1 + &\frac{3 c_\text{s}^2 \rho}{4 \check{\mu}} \bigg) \frac{dT_1}{dr} 
        = \rho \frac{dU}{dr} \\
        - &\left\{ \frac{d}{dr}(c_\text{s}^2) \left( 3 \rho 
        + r \frac{d\rho}{dr} \right) 
        + c_\text{s}^2 \left[ \frac{3 \beta^2 \rho}{2 r} + \frac{d\rho}{dr} 
        - \frac{r}{\rho} \left( \frac{d\rho}{dr} \right)^2
        + r \frac{d^2\rho}{dr^2} \right] \right\} \frac{1}{r} \xi_r \\
        + &\left[ \frac{d}{dr}(c_\text{s}^2) 3 \rho + c_\text{s}^2 
        \left( \frac{3 \rho}{r} + \frac{d\rho}{dr} \right) \right] 
        \frac{\beta}{2 r} \xi_\bot \\
        - &\left[ \frac{3}{r} 
        + \frac{d}{dr}(c_\text{s}^2) \frac{3 \rho}{4 \check{\mu}} 
        + c_\text{s}^2 \left( \frac{3 \rho}{r} 
        - \frac{\rho}{\check{\mu}} \frac{d\check{\mu}}{dr}
        + \frac{d\rho}{dr} \right) \frac{3}{4 \check{\mu}} \right] T_1 \\
        + &\left( 1 + \frac{3 c_\text{s}^2 \rho}{2 \check{\mu}}\right) 
        \frac{\beta^2}{r} T_2. 
\end{split}
\end{equation}
Then, from the tangential piece of (\ref{eq:PerturbedEulerElastic}) we find 
\begin{equation}
\begin{split}
    \frac{dT_2}{dr} = \frac{\rho}{r} U - &c_\text{s}^2 \left( 3 \rho 
        + r \frac{d\rho}{dr} \right) \frac{1}{r^2} \xi_r \\
        + &\left[ \frac{3 c_\text{s}^2 \rho}{2} 
        + \left( 1 
        - \frac{2}{\beta^2} \right) \check{\mu} \right] \frac{\beta}{r^2} 
        \xi_\bot \\
        + &\left( \frac{1}{2} - \frac{3 c_\text{s}^2 \rho}{4 \check{\mu}} \right) 
        \frac{1}{r} T_1 - \frac{3}{r} T_2. 
\end{split}
\end{equation}
We also have the perturbed Poisson's equation (\ref{eq:PerturbedPoissonsFluid}), 
that combines with the perturbed continuity equation 
(\ref{eq:PerturbedContinuityElastic}) to give 
\begin{equation}
\begin{split}
    \frac{d^2\delta \Phi}{dr^2} + \frac{2}{r} \frac{d\delta \Phi}{dr} 
        - \frac{\beta^2}{r^2} \delta \Phi 
        = - &4 \pi G \left( \frac{3 \rho}{r} + \frac{d\rho}{dr} \right) \xi_r \\
        + &6 \pi G \frac{\beta \rho}{r} \xi_\bot 
        - 3 \pi G \frac{\rho}{\check{\mu}} T_1.
\end{split}
\end{equation}
\end{subequations}
Equations~(\ref{eqs:NewtonianElasticPerturbations}) form a coupled system of 
ordinary differential equations to describe the perturbations in the elastic 
material. We have compared our perturbation equations with that of 
\citet{2006MNRAS.373.1423H} (in the limit of $\chi = 0$) and noted several 
discrepancies. We find that these mistakes increase the maximum quadrupole 
estimates of \citet{2006MNRAS.373.1423H} by three orders of magnitude. 

\subsection{Interface conditions}
\label{subsec:NewtonianInterface}

At this point, we address the boundary conditions at the fluid-elastic 
interfaces since we wish to connect perturbations in the fluid core and ocean 
with the elastic crust for star C. Provided the density is smooth (which we 
assume), the perturbed potential $\delta \Phi$ and its derivative 
$d\delta \Phi/dr$ must be continuous at an interface. To see how the other 
perturbed quantities behave at an interface, we must consider the perturbed 
traction (\ref{eq:PerturbedTraction2}). 

This admits two quantities that must be continuous: the radial and tangential 
components. Since the shear modulus vanishes in the fluid, continuity of the 
radial traction $(\delta p - T_1)$ provides an algebraic relation that must 
hold true at an interface, 
\begin{equation}
\begin{split}
    \rho U_\text{F} = 
        &\left( 1 + \frac{3 c_\text{s}^2 \rho}{4 \check{\mu}} \right) T_{1 \text{E}} \\
        &+ c_\text{s}^2 \left[ \left( \frac{3 \rho}{r} 
        + \frac{d\rho}{dr} \right) \xi_{r \text{E}} 
        - \frac{3 \beta \rho}{2 r} \xi_{\bot \text{E}} \right], 
\end{split}
    \label{eq:RadialTraction}
\end{equation}
where the subscripts F and E denote the fluid and elastic sides of the 
interface, respectively. We note that the radial displacement $\xi_r$ must be 
continuous at a boundary, however, this does not necessarily have to be the case 
for the tangential piece $\xi_\bot$. From the tangential part of the traction, 
we have $T_2 = 0$ at a fluid-elastic interface. 

In reference to the maximally strained approach of 
\citet{2000MNRAS.319..902U} and \citet{2013PhRvD..88d4004J}, we note that, if 
one assumes the shear modulus smoothly goes to zero at a fluid-elastic 
interface, then the tangential traction condition is trivially satisfied [see 
(\ref{eq:T_2})]. This would effectively result in the displacement vector in the 
crust being arbitrary since there are not enough boundary conditions to 
constrain it. It is not clear how to resolve this issue. 

In the fluid regions of the star, the perturbations are governed by 
Eqs.~(\ref{eqs:NewtonianFluidPerturbations}) and so are described by the 
variables $(d\delta \Phi/dr, \delta \Phi)$. In the crust, we have a more 
complex structure with Eqs.~(\ref{eqs:NewtonianElasticPerturbations}) and 
quantities $(d\delta \Phi/dr, \delta \Phi, \xi_r, \xi_\bot, T_1, T_2)$. We 
assume the force is known. The perturbations in the elastic crust present a 
boundary-value problem. For the six variables, we have six boundary conditions: 
continuity of $d\delta \Phi/dr$ and $\delta \Phi$ at the core-crust transition 
and the two traction conditions -- (\ref{eq:RadialTraction}) and $T_2 = 0$ -- at 
both interfaces. Therefore, the problem is well posed. 

Additionally, it is straightforward to show that the boundary condition on the 
Lagrangian variation of the pressure $\Delta p(R, \theta, \phi) = 0$ is 
trivially satisfied by the background structure.%
\footnote{This boundary condition is intuitive. In the background configuration, 
the surface is defined by $p(R) = 0$. During the perturbation, the individual 
elements of the star are moved according to the Lagrangian displacement 
$\xi^i$ and, therefore, the surface will have moved to 
$R \rightarrow R + \xi^r(R, \theta, \phi)$. This corresponds to 
$\Delta p(R, \theta, \phi) = 0$. One can also convince themselves that this 
holds true by examining the definition of the Lagrangian 
variation~(\ref{eq:Lagrangian}) for $p$ at the surface of the perturbed 
configuration.}

\section{The deforming force}
\label{sec:Sources}

The formalism we detail above requires a description of the deforming force 
that causes the star to have a non-spherical shape. Because of the abstract 
nature of this force, it is difficult to prescribe without a detailed 
evolutionary calculation of the history of the star. As a proof-of-principle 
calculation, we examine three example sources. 

\sloppy
We use a polytropic equation of state (\ref{eq:Polytropic}). 
(See Appendix~\ref{appsec:Polytropes} on how to generate polytropic models.) 
We work with $n = 1$ and generate background models with 
$M = \SI{1.4}{\solarMass}$, $R = \SI{10}{\kilo\metre}$. For the shear-modulus 
profile in the crust, we consider a simple linear model 
\citep{2006MNRAS.373.1423H}, 
\begin{equation}
    \check{\mu}(\rho) = \kappa \rho, 
\end{equation}
where $\kappa = \SI{e16}{\centi\metre\squared\per\second\squared}$. We assume 
the core-crust transition to occur at 
$\rho_\text{base} = \SI{2e14}{\gram\per\centi\metre\cubed}$ 
\citep[which is the same as][]{2000MNRAS.319..902U}, while the crust-ocean 
transition is at 
$\rho_\text{top} = \SI{e6}{\gram\per\centi\metre\cubed}$ 
\citep{2020PhRvD.101j3025G}. 

\fussy
We consider three sources for the perturbations: (i) a potential that satisfies 
Laplace's equation, (ii) a potential that satisfies Laplace's equation but does 
not act in the core and (iii) a thermal pressure perturbation. In reality, the 
fiducial force will depend on the evolutionary history of the star. Since this 
is a complex problem (beyond the scope of this work), the examples we consider 
are indicative in nature and should serve as illustrations of how one can 
calculate mountains using this scheme. The forces we use should not be 
interpreted as having an explicit link with the neutron-star physics. For each 
prescription, we generate two stars -- a relaxed star, that experiences purely 
fluid perturbations (star A in Fig.~\ref{fig:StarsScheme}), and a strained star, 
that experiences elastic perturbations in the crust (star C in 
Fig.~\ref{fig:StarsScheme}). We normalise the perturbations by ensuring the 
strained star reaches breaking strain at a point in the crust, subject to the 
von Mises criterion, and that the relaxed star experiences the same force. This 
allows us to work out the quadrupole moment of each star. Our results for the 
three sources are summarised in Table~\ref{tab:ellipticities}. 

\begin{table}[h]
    \caption[Maximum quadrupoles and ellipticities]{The maximum quadrupoles and 
             ellipticities from the different models. For each case, we show the 
             quadrupole $Q_{2 2}^\text{A}$ and ellipticity 
             $\epsilon^\text{A}$ for the relaxed star (star A) and the 
             difference relative to the strained star (star C) with quadrupole 
             $Q_{2 2}^\text{C}$ and ellipticity 
             $\epsilon^\text{C}$.}
    \label{tab:ellipticities}
    \resizebox{\textwidth}{!}{
    \begin{tabular}{ l c c c c }
        \hline\hline
        Source & $|Q_{2 2}^\text{A}|$ / \si{\gram\centi\metre\squared} 
        & $|\epsilon^\text{A}|$ 
        & $|Q_{2 2}^\text{C} - Q_{2 2}^\text{A}|$ 
        / \si{\gram\centi\metre\squared} 
        & $|\epsilon^\text{C} - \epsilon^\text{A}|$ \\ 
        \hline
        Solution of Laplace's equation & \num{2.4e43} & \num{3.1e-2} 
        & \num{1.7e37} & \num{2.2e-8} \\ 
        Solution of Laplace's equation (outside core) & \num{1.4e41} 
        & \num{1.8e-4} & \num{4.4e38} & \num{5.7e-7} \\ 
        Thermal pressure perturbation & \num{9.2e38} & \num{1.2e-6} 
        & \num{4.0e38} & \num{5.2e-7} \\
        \hline
    \end{tabular}
    }
\end{table}

The structure in the fluid core of both stars may be straightforwardly 
calculated using Eqs.~(\ref{eqs:NewtonianFluidPerturbations}) with boundary 
condition~(\ref{eq:PerturbedPotentialCentre}). The crust of star C presents a 
boundary-value problem with Eqs.~(\ref{eqs:NewtonianElasticPerturbations}) and 
the interface conditions described in Sec.~\ref{subsec:NewtonianInterface}. Both 
star A and C have fluid oceans, where one can integrate 
Eqs.~(\ref{eqs:NewtonianFluidPerturbations}) through to the surface. At this 
point, one can verify that the boundary condition at the 
surface~(\ref{eq:PerturbedPotentialSurface}) is satisfied. The numerical scheme 
we follow for solving the perturbations is described in 
Appendix~\ref{app:NumericalScheme}.

\subsection{A solution of Laplace's equation}

The first example we consider is based on the form of the deforming potential 
for tidal deformations (Sec.~\ref{subsubsec:Tidal}). The source potential is 
taken to be a solution of Laplace's equation, 
\begin{equation}
    \nabla^2 \chi = 0. 
    \label{eq:Laplaces}
\end{equation}
This example is particularly convenient since the perturbed Poisson's equation 
(\ref{eq:PerturbedPoissons}) is simply modified by $\delta \Phi \rightarrow U$. 
Therefore, we may write 
\begin{equation}
    \nabla^2 U = 4 \pi G \delta \rho. 
\end{equation}
The total perturbed potential must be regular at the origin, $U(0) = 0$. 

From (\ref{eq:PerturbedPotentialMultipole}), we can obtain the multipole 
moments from the exterior potential, 
\begin{equation}
    Q_{\ell m} = - \frac{(2 \ell + 1) R^{\ell + 1}}{4 \pi G} \delta \Phi(R). 
    \label{eq:Multipole2}
\end{equation}
By making use of the boundary 
conditions~(\ref{eqs:PerturbedPotentialBoundaries}), one can also write the 
multipole in terms of the total perturbed potential, 
\begin{equation}
    Q_{\ell m} = \frac{R^{\ell + 1}}{4 \pi G} \left[ R \frac{dU}{dr}(R) 
        - \ell U(R) \right]. 
    \label{eq:Multipole3}
\end{equation}
The advantage of writing the multipole in this way is that one does not need to 
disentangle the two potentials ($\chi$ and $\delta \Phi$) from $U$. 

The source potential must be of the form of (\ref{eq:InitialChi}). Its value 
will be chosen to ensure the star is maximally strained at some point in the 
crust. The source potential at the surface is given by 
\begin{equation}
    \chi(R) = \frac{1}{2 \ell + 1} \left[ R \frac{dU}{dr}(R) 
        + (\ell + 1) U(R) \right]. 
    \label{eq:SourcePotentialSurface}
\end{equation}
It is this quantity that we use to ensure that the relaxed and strained stars 
experience the same force. 

To make sure the star is maximally strained we calculate the von Mises 
strain~(\ref{eq:vonMisesStrain}) and use the von Mises 
criterion~(\ref{eq:vonMisesCriterion}). For $(\ell, m) = (2, 2)$ perturbations, 
we have%
\footnote{Since we focus on a specific harmonic, we take the real parts of 
$\sigma_{i j}$ and then contract using (\ref{eq:vonMisesStrain}).}
\begin{equation}
\begin{split}
    \bar{\sigma}^2 = \frac{5}{256 \pi} \Bigg\{ &6 \sin^2 \theta 
        \Bigg[ 3 \sin^2 \theta \cos^2 2 \phi \left( \frac{T_1}{\check{\mu}} \right)^2 \\
        &\qquad\quad+ 4 (3 + \cos 2 \theta - 2 \sin^2 \theta \cos 4 \phi) 
        \left( \frac{T_2}{\check{\mu}} \right)^2 \Bigg] \\
        &+ (35 + 28 \cos 2 \theta + \cos 4 \theta + 8 \sin^4 \theta \cos 4 \phi) 
        \left( \frac{\xi_\bot}{r} \right)^2 \Bigg\}. 
\end{split}
    \label{eq:vonMises}
\end{equation}
Since the von Mises strain is a function of position, we can identify where the 
strain is highest (and, thus, the crust will break first) and take that point to 
be at breaking strain, that we assume to be $\bar{\sigma}_\text{max} = 10^{-1}$ 
\citep{2009PhRvL.102s1102H}. Other estimates for the magnitude of the breaking 
strain include \citet{2018MNRAS.480.5511B} who obtained the smaller value of 
$\bar{\sigma}_\text{max} = 0.04$. However, as was the case for previous 
maximum-mountain calculations, our results are linear in the breaking strain. 
Thus, a smaller breaking strain would result in less pronounced mountains.

\begin{figure}[h]
	\includegraphics[width=0.49\textwidth]{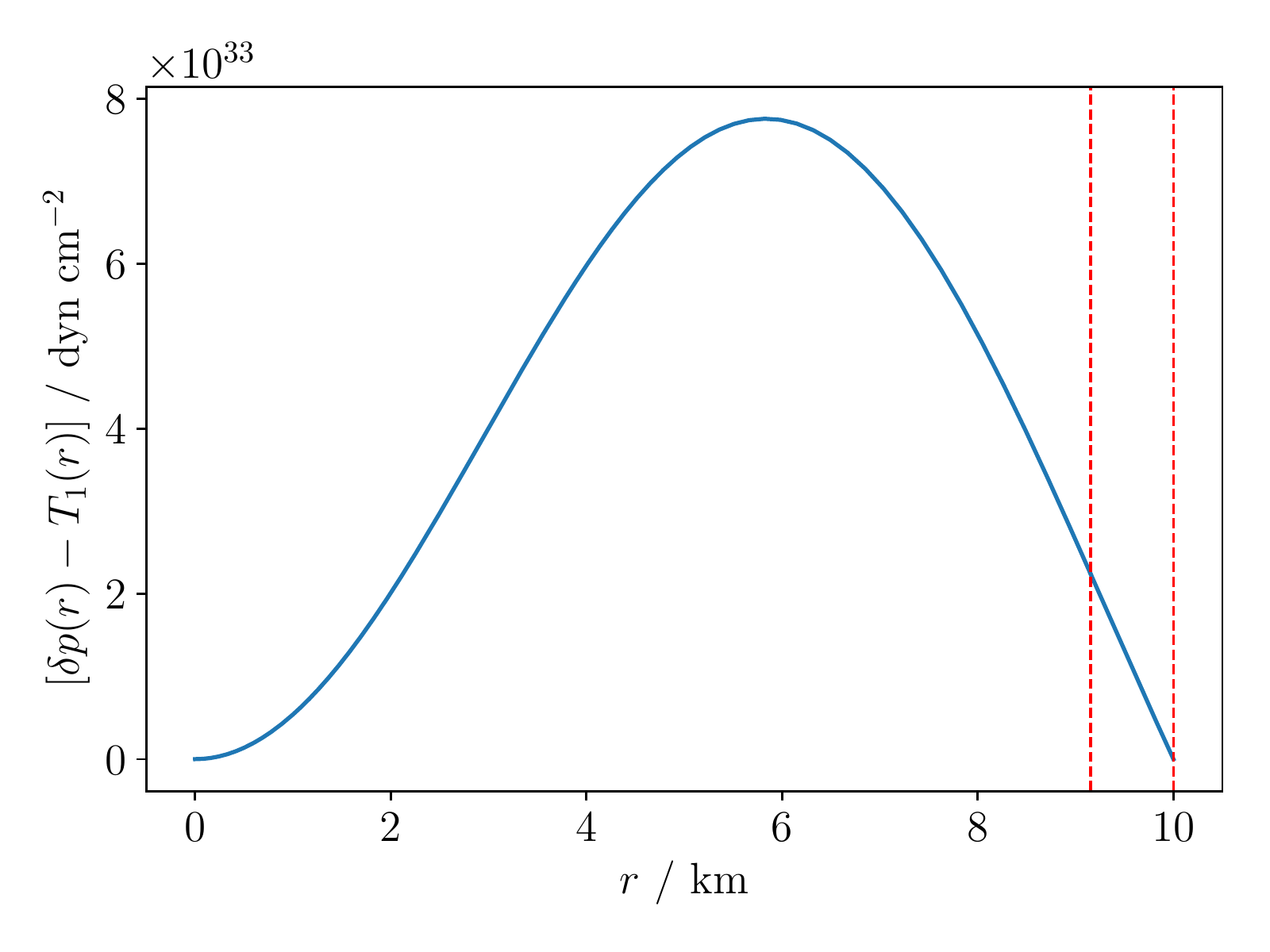}
    \includegraphics[width=0.49\textwidth]{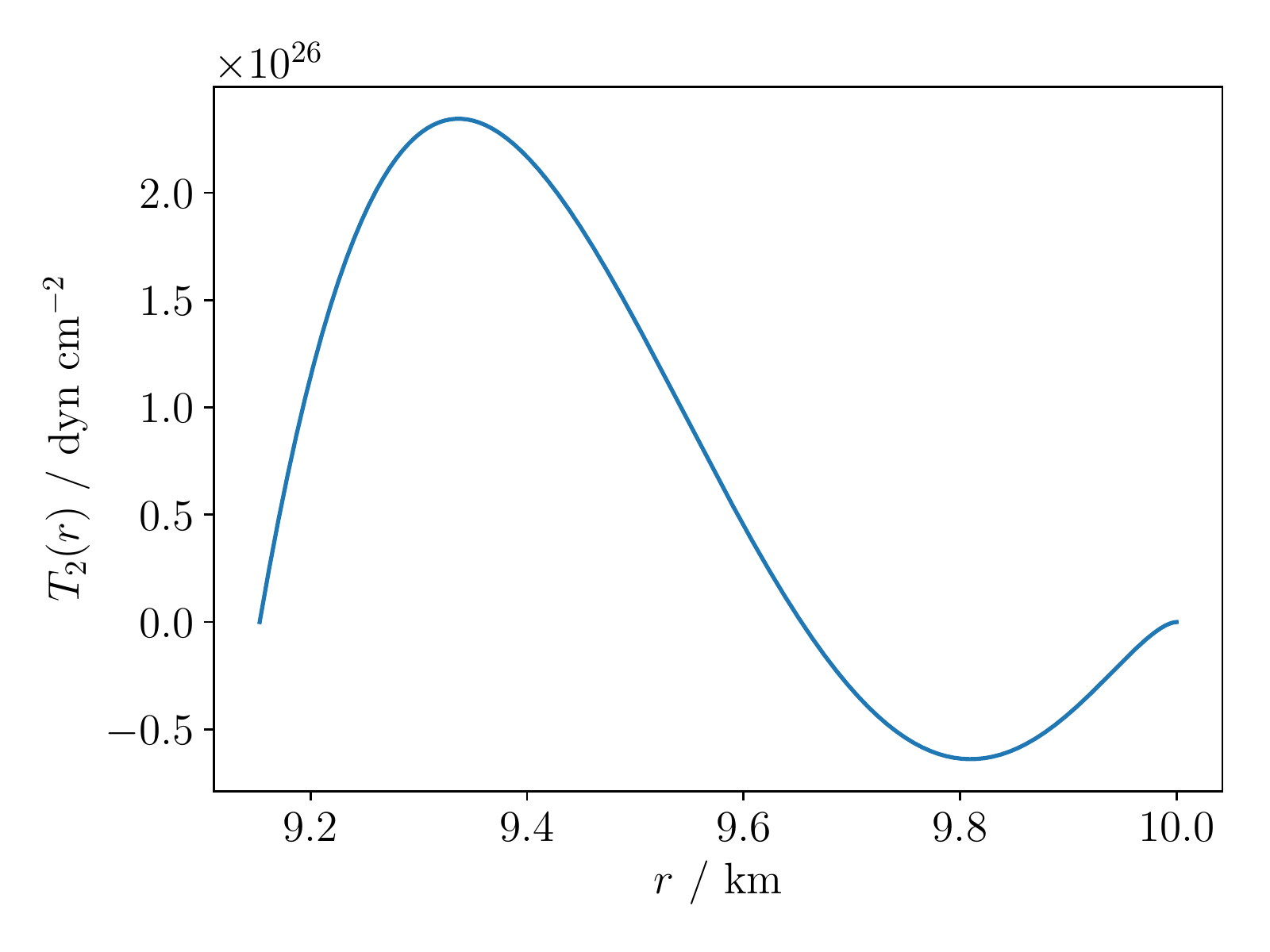}
    \caption[Traction for potential solution to Laplace's equation]{The radial 
             (left panel) and tangential (right panel) components of the 
             perturbed traction as functions of radius 
             for the potential solution to Laplace's equation. 
             The vertical red dashed lines in the left panel indicate the base 
             and the top of the crust. Regarding the horizontal range in the 
             right panel, recall that $T_2$ only has a finite value in the 
             crust.}
    \label{fig:Traction}
\end{figure}

\begin{figure}
    \centering
	\includegraphics[width=0.7\textwidth]{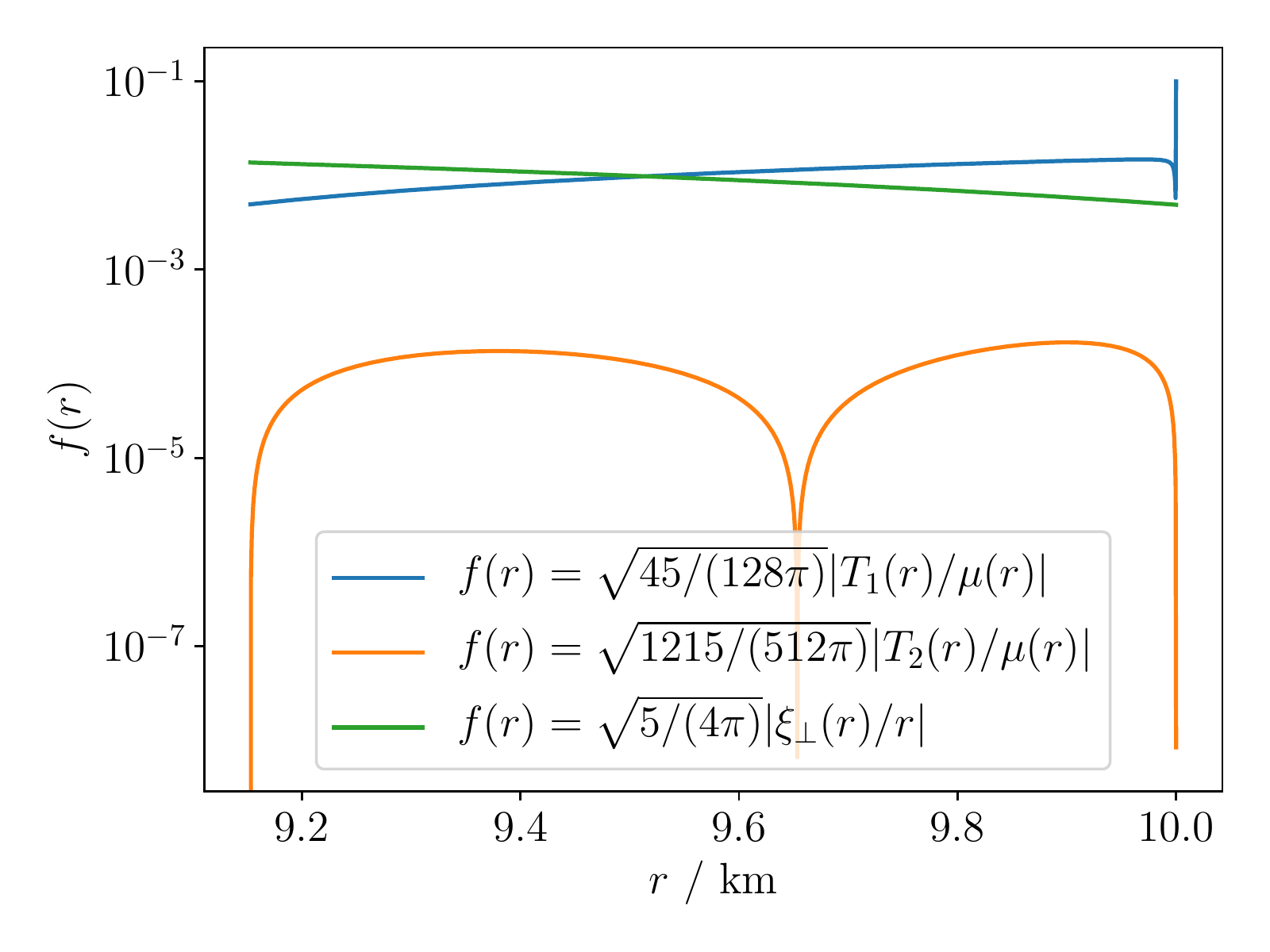}
    \caption[Von Mises strain for potential solution to Laplace's equation]{The 
             strain components in (\ref{eq:vonMises}) maximised over 
             $(\theta, \phi)$ against radius for the potential solution to 
             Laplace's equation.}
    \label{fig:Strain}
\end{figure}

Thus, for the strained star (star C) we integrate 
Eqs.~(\ref{eqs:NewtonianFluidPerturbations}) for the core and ocean and 
integrate Eqs.~(\ref{eqs:NewtonianElasticPerturbations}) in the elastic crust. 
The relaxed star (star A) is generated using 
Eqs.~(\ref{eqs:NewtonianFluidPerturbations}) for the entire star. The 
perturbations are normalised by ensuring that the point in the crust where the 
strain is highest reaches breaking strain, according to (\ref{eq:vonMises}). 
The force associated with this deformation (\ref{eq:SourcePotentialSurface}) is 
then taken to be the same for the relaxed star. Figures~\ref{fig:Traction} and 
\ref{fig:Strain} show the results for the strained star. In 
Fig.~\ref{fig:Traction} we show how the perturbed traction is continuous at the 
fluid-elastic interfaces. We note that Fig.~\ref{fig:Strain} shows how the 
dominant contribution to the von Mises strain comes from the radial traction 
component. This is also true for the other forces we consider. It is at the top 
of the crust that the star is the weakest in the $(\ell, m) = (2, 2)$ mode. The 
quadrupoles are calculated using (\ref{eq:Multipole3}). The relaxed star attains 
a quadrupole of
$|Q_{2 2}^\text{A}| = \SI{2.4e43}{\gram\centi\metre\squared}$, that 
corresponds to an ellipticity of 
$|\epsilon^\text{A}| = \num{3.1e-2}$. The difference between the strained 
and relaxed star is $|Q_{2 2}^\text{C} - Q_{2 2}^\text{A}| 
= \SI{1.7e37}{\gram\centi\metre\squared}$, 
$|\epsilon^\text{C} - \epsilon^\text{A}| = \num{2.2e-8}$. 

The very different sizes of $|\epsilon^\text{A}|$ and 
$|\epsilon^\text{C} - \epsilon^\text{A}|$ reported in 
Table~\ref{tab:ellipticities} have a natural interpretation. The large 
ellipticity represented by $|\epsilon^\text{A}|$ corresponds to a star 
whose deformation is supported by the external force $f_i$, with a size limited 
only by the crustal breaking strain. In this case, the (non-zero) shear modulus 
of the crust plays little role. [It is this sort of configuration that was 
effectively considered in \citet{2006MNRAS.373.1423H}, where in that case the 
force that was implicitly introduced was a force per unit area, applied at the 
surface.] In contrast, the ellipticity represented by 
$|\epsilon^\text{C} - \epsilon^\text{A}|$ is that supported by the 
shear strains of the crust when the external force is removed and, therefore, is 
sensitive to the crust's shear modulus. As is readily captured by simple 
back-of-the-envelope estimates, the relative sizes of these two ellipticities 
are related to the fact that the gravitational binding energy of the star is 
orders of magnitude larger than the Coulomb binding energy of the crustal 
lattice \citep[see, \textit{e.g.},][]{2002CQGra..19.1255J}.

We observe that the ellipticity
$|\epsilon^\text{C} - \epsilon^\text{A}| = \num{2.2e-8}$ is notably 
smaller than what has been found in previous work 
[Eqs.~(\ref{eq:Q_22UCB})--(\ref{eq:Q_22JMO})]. This is not surprising, as 
these previous studies considered strain fields that were maximal everywhere, as 
opposed to at a single point. With a view to producing larger ellipticities, we 
will, therefore, consider some different choices of external force field.

\subsection{A solution of Laplace's equation outside the core}

We consider a special case of the above source: a source potential that does not 
act in the core --  instead, it only manifests itself in the crust and ocean. The 
motivation for considering this special case is regularity at the centre will 
not be a necessary condition on the source potential since it does not exist at 
the origin. Also, we note that \citet{2006MNRAS.373.1423H} found that a similar 
example produced their largest quadrupole moment. We then have the general 
solution to Laplace's equation (\ref{eq:Laplaces}), 
\begin{equation}
    \chi(r) = c_1 / r^{\ell + 1} + c_2 r^\ell. 
    \label{eq:SourcePotential}
\end{equation}
This expression is taken to be true for the base of the crust and above. 

As this  model is somewhat artificial, we have to make a number of 
assumptions with regards to its prescription. We take the core to be 
unperturbed and have $\delta \Phi = \xi_r = 0$ in the core. With the 
introduction of the source potential in the crust, there will be a discontinuity 
in $U$ at the core-crust interface. However, we insist that $\delta \Phi$ must 
be continuous. This discontinuity is relevant for the radial traction 
condition~(\ref{eq:RadialTraction}) where $U_\text{F} = 0$, but has a finite 
value in the crust due to the source potential. 

The quadrupole may be calculated from (\ref{eq:Multipole2}). The matching with 
the total perturbed potential needs to be adjusted to take into account the 
additional $1 / r^{\ell + 1}$ term from the external field. Therefore, we have 
\begin{equation}
    Q_{\ell m} = \frac{R^{\ell + 1}}{4 \pi G} \left[ R \frac{dU}{dr}(R) 
        - \ell U(R) \right] + \frac{2 \ell + 1}{4 \pi G} B. 
\end{equation}

\begin{figure}[h]
	\includegraphics[width=0.49\textwidth]{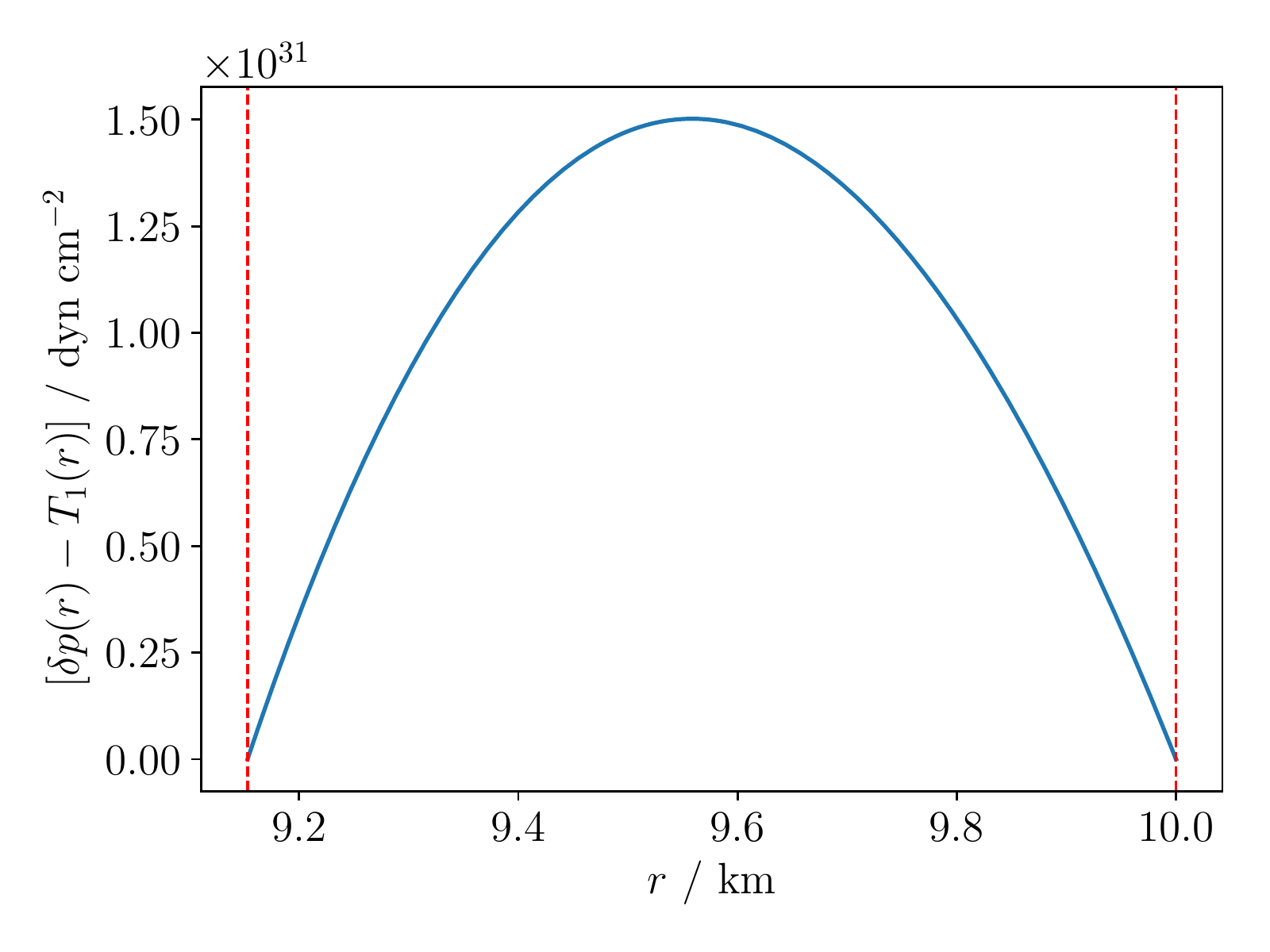}
    \includegraphics[width=0.49\textwidth]{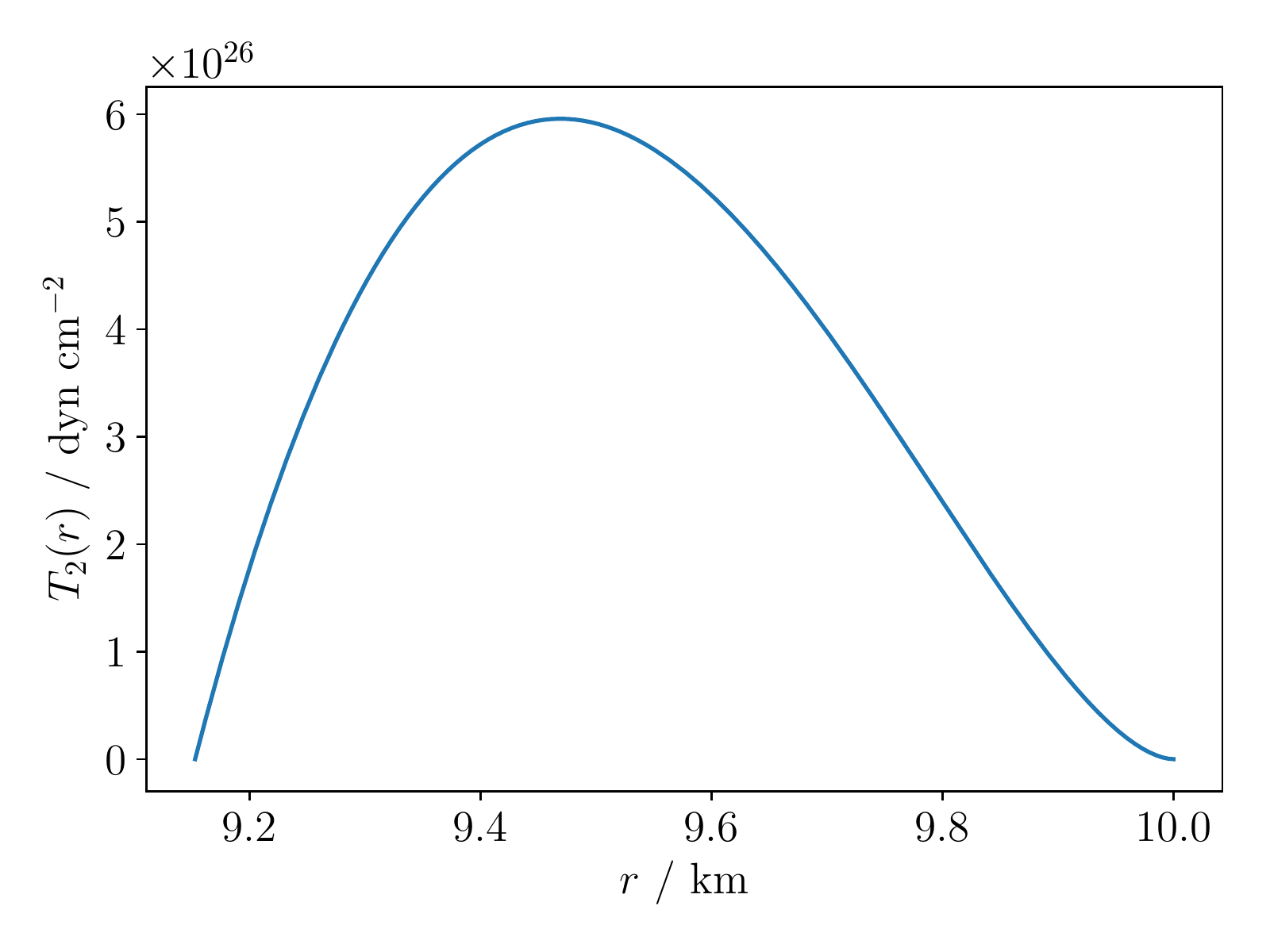}
    \caption[Traction for potential solution to Laplace's equation outside the 
             core]{The radial (left panel) and tangential 
             (right panel) components of the perturbed traction as functions of 
             radius for the potential solution to Laplace's equation outside the 
             core. The vertical red dashed lines in the left panel indicate the 
             base and the top of the crust. Regarding the horizontal range in 
             the right panel, recall that $T_2$ only has a finite value in the 
             crust.}
    \label{fig:TractionSpecial}
\end{figure}

\begin{figure}[h]
    \centering
	\includegraphics[width=0.7\textwidth]{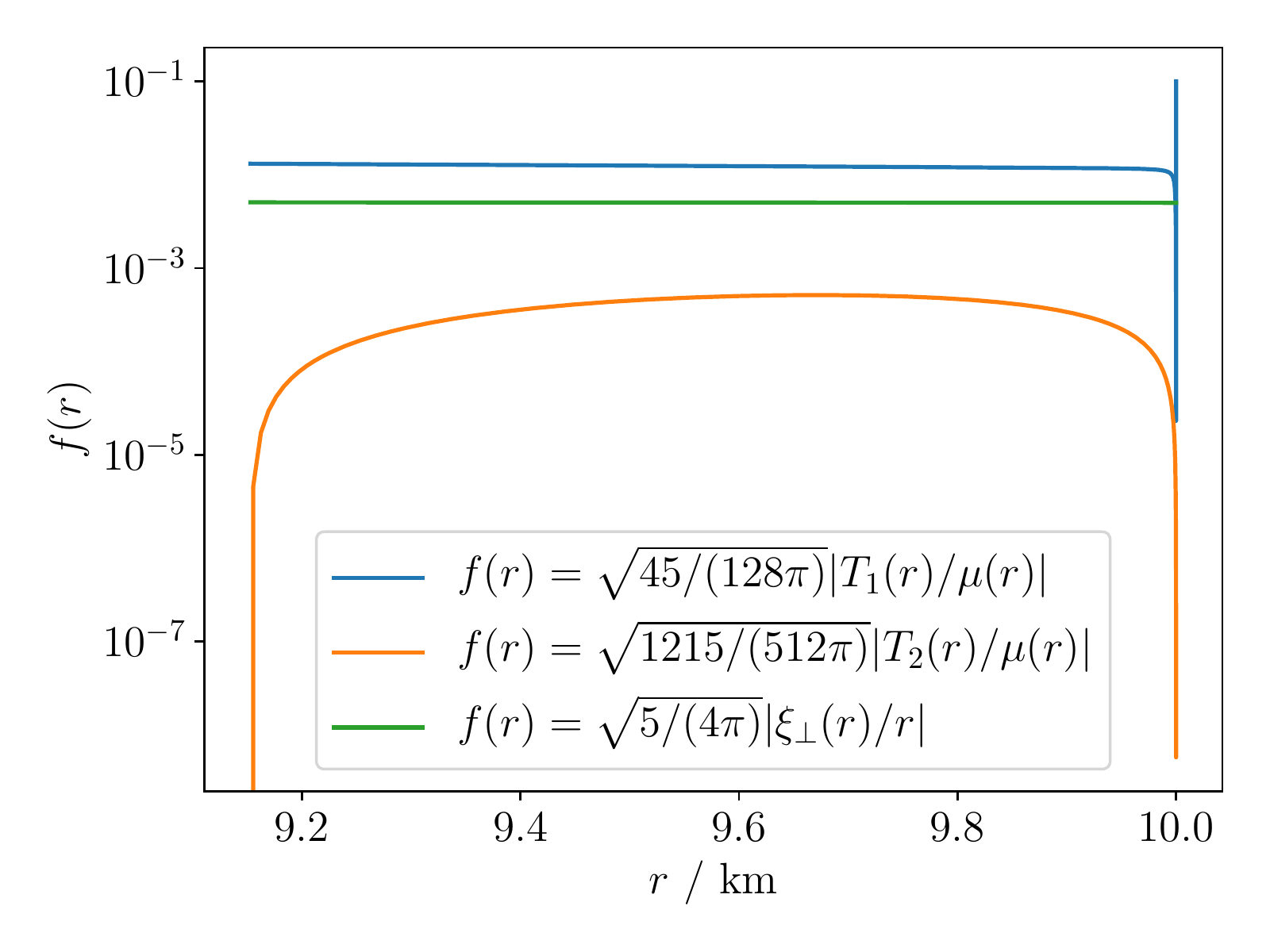}
    \caption[Von Mises strain for potential solution to Laplace's equation 
             outside the core]{The strain components in 
             (\ref{eq:vonMises}) maximised over $(\theta, \phi)$ against 
             radius for the potential solution to Laplace's equation outside the 
             core.}
    \label{fig:StrainSpecial}
\end{figure}

As in the previous case, we generate a relaxed star and a maximally strained 
star. One must vary either $c_1$ or $c_2$ to ensure the surface boundary 
condition~(\ref{eq:PerturbedPotentialSurface}) is satisfied. We normalise the 
relaxed star so that it experiences the same source 
potential~(\ref{eq:SourcePotential}). The 
results for this case are shown in Figs.~\ref{fig:TractionSpecial} and 
\ref{fig:StrainSpecial}. As in the above example, the $T_1$ component dominates 
the von Mises strain and the crust breaks at the top. We find 
$|Q_{2 2}^\text{A}| = \SI{1.4e41}{\gram\centi\metre\squared}$, 
$|\epsilon^\text{A}| = \num{1.8e-4}$ and 
$|Q_{2 2}^\text{C} - Q_{2 2}^\text{A}| 
= \SI{4.4e38}{\gram\centi\metre\squared}$, 
$|\epsilon^\text{C} - \epsilon^\text{A}| = \num{5.7e-7}$. 

Compared to the previous result, the quadrupole difference between the 
relaxed and strained stars  has increased by an order of magnitude. This is 
within a factor of a few of previous maximum-mountain calculations, and 
illustrates the dependence on the force prescription.

\subsection{A thermal pressure perturbation}
\label{subsec:ThermalPressure}

\begin{figure}[h]
	\includegraphics[width=0.49\textwidth]{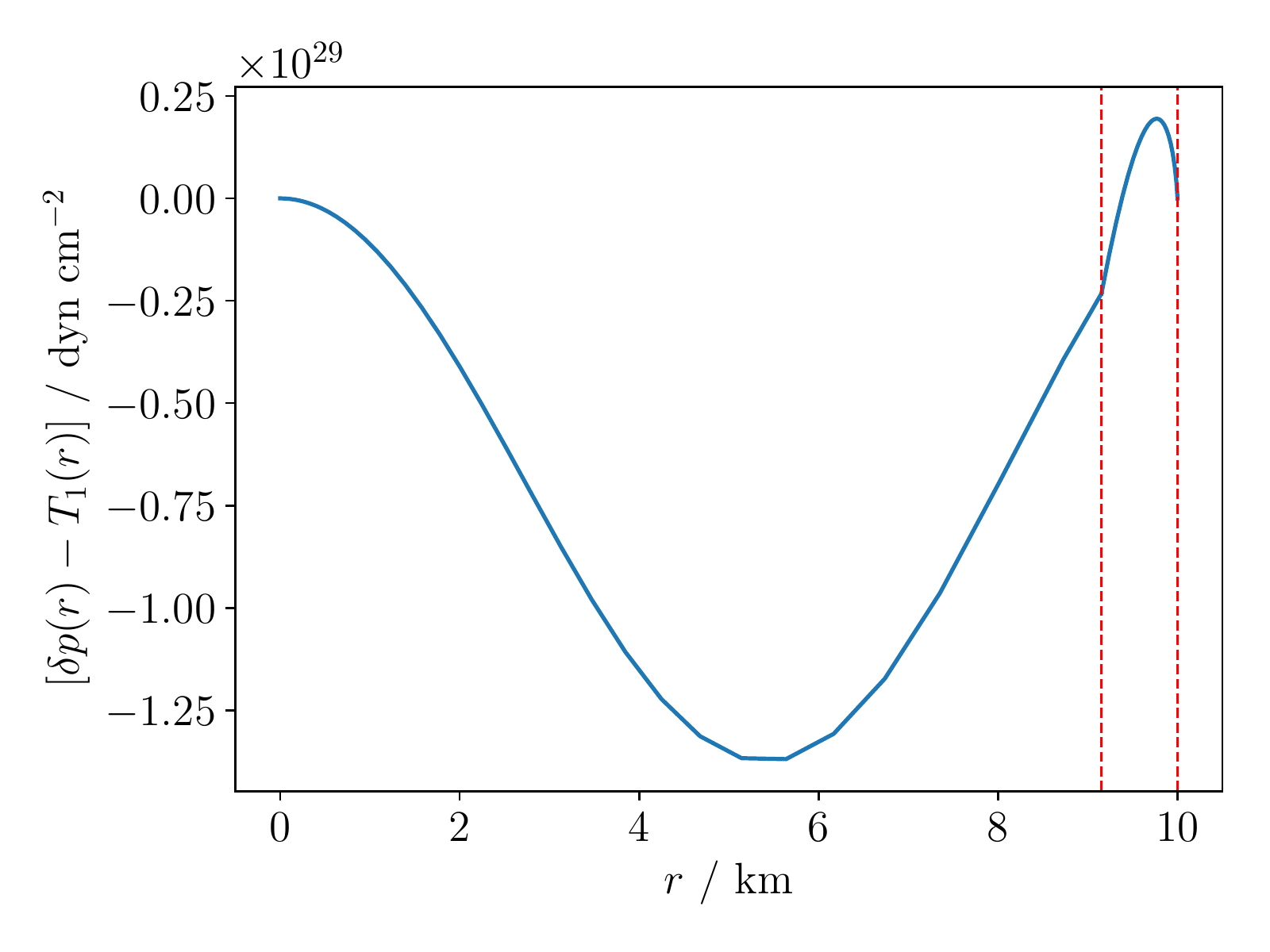}
    \includegraphics[width=0.49\textwidth]{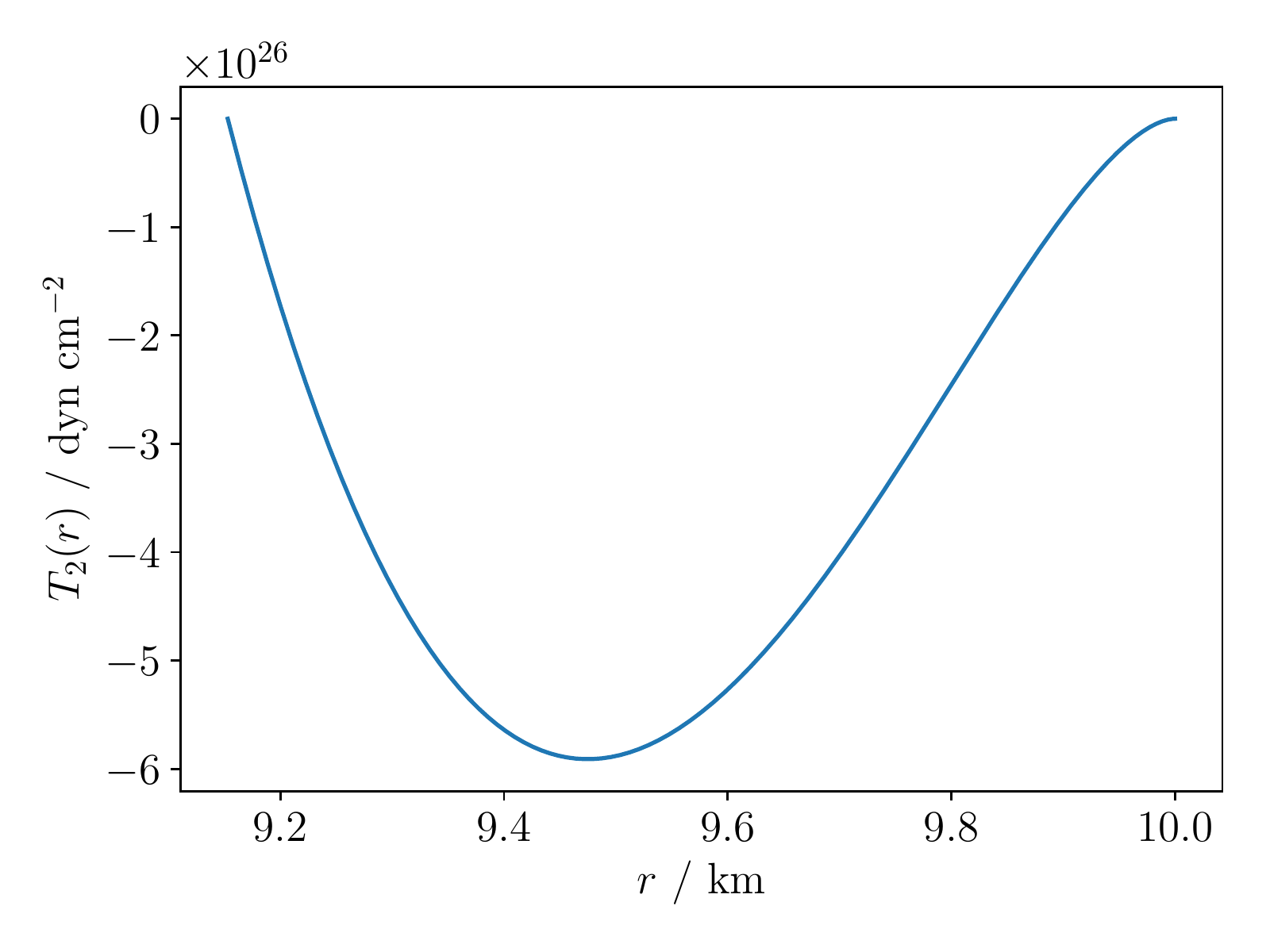}
    \caption[Traction for thermal pressure]{The radial (left panel) and 
             tangential (right panel) components of the perturbed traction as 
             functions of radius for the temperature perturbation. The vertical 
             red dashed lines in the left panel indicate the base and the top of 
             the crust. Regarding the horizontal range in the right panel, 
             recall that $T_2$ only has a finite value in the crust.}
    \label{fig:TractionTemperature}
\end{figure}

\begin{figure}[h]
    \centering
	\includegraphics[width=0.7\textwidth]{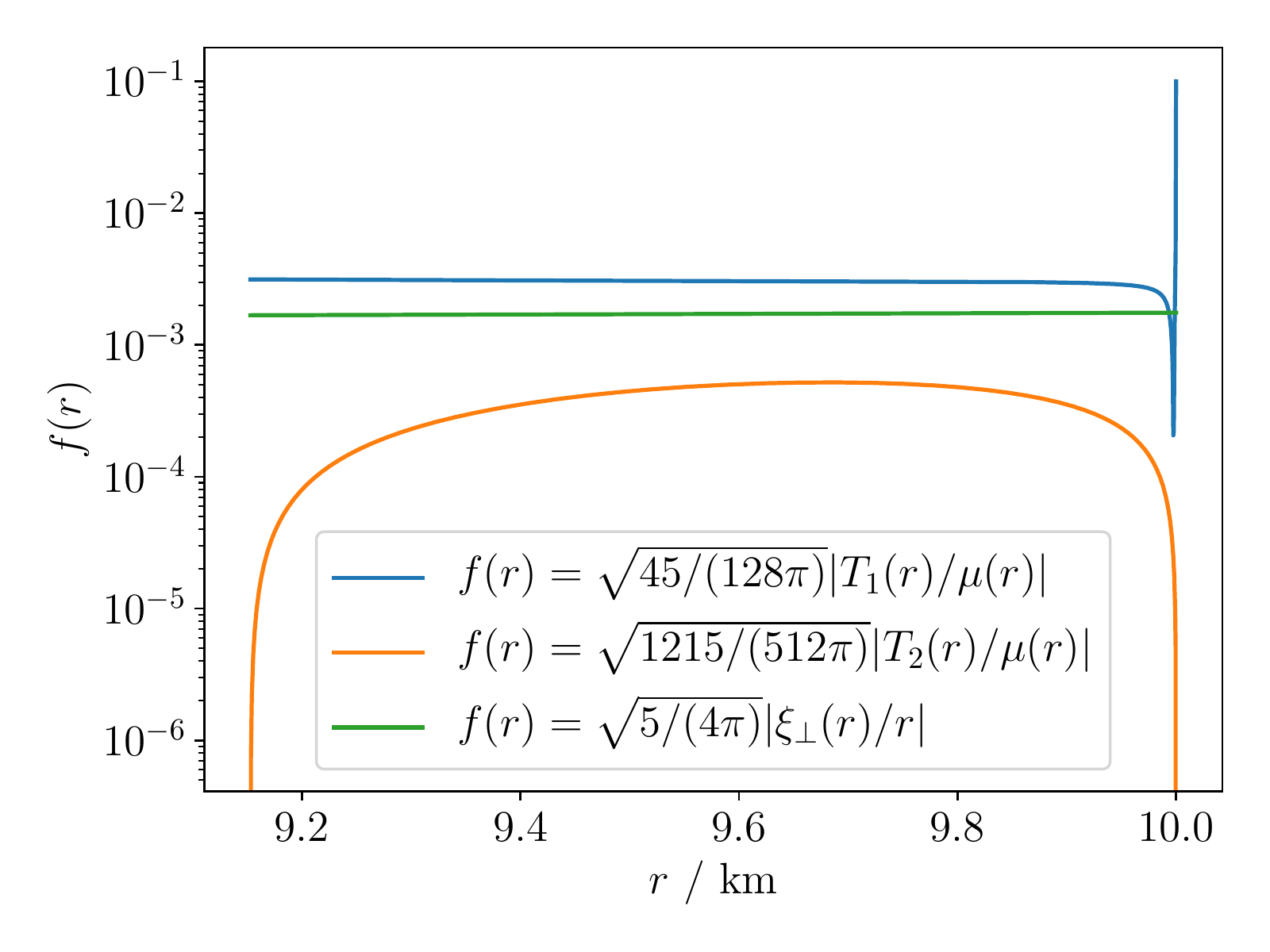}
    \caption[Von Mises strain for thermal pressure]{The strain components in 
             (\ref{eq:vonMises}) maximised over $(\theta, \phi)$ against 
             radius for the temperature perturbation.}
    \label{fig:StrainTemperature}
\end{figure}

The third source for the perturbations we examine is motivated by a thermal 
pressure perturbation. Note that the approach we use for this example 
could be applied more generally to consider non-barotropic matter where the 
pressure is adjusted, relative to the barotropic case, at the perturbative 
level. 
We assume the thermal pressure to be of the ideal-gas form, 
\begin{equation}
    \delta p_\text{th} = \frac{k_\text{B} \rho}{m_\text{b}} \delta T, 
    \label{eq:ThermalPressure}
\end{equation}
where $\delta T$ is the temperature perturbation. To interpret this thermal 
pressure as a force, we identify 
\begin{equation}
    \rho \nabla_i \chi = \nabla_i \delta p_\text{th} 
        = \frac{k_\text{B}}{m_\text{b}} \nabla_i (\rho \delta T). 
\end{equation}
The temperature perturbation must be regular at the origin. For simplicity we 
assume it to be quadratic, 
\begin{equation}
    \delta T(r) = \left( \frac{r}{R} \right)^2 \delta T(R), 
    \label{eq:Temperature}
\end{equation}
where $\delta T(R)$ corresponds to the perturbation of the temperature at the 
surface. Both the relaxed and strained configurations experience the same 
temperature perturbation. We show the results in 
Figs.~\ref{fig:TractionTemperature} and \ref{fig:StrainTemperature}.%
\footnote{We found the crust breaks when $\delta T(R) = \SI{3.5e6}{\kelvin}$. 
The temperature reported here is not a physical temperature perturbation the 
star is subjected to, noting that the background is at zero temperature. It is 
simply a source term for the pressure perturbation~(\ref{eq:ThermalPressure}).}
We obtain the  results 
$|Q_{2 2}^\text{A}| = \SI{9.2e38}{\gram\centi\metre\squared}$, 
$|\epsilon^\text{A}| = \num{1.2e-6}$ and 
$|Q_{2 2}^\text{C} - Q_{2 2}^\text{A}| 
= \SI{4.0e38}{\gram\centi\metre\squared}$, 
$|\epsilon^\text{C} - \epsilon^\text{A}| = \num{5.2e-7}$. This 
result is of the same order of magnitude as the potential outside the core.

It is interesting to note that while the values of the ellipticities 
$|\epsilon^\text{C}|$ and $|\epsilon^\text{A}|$ vary by about four 
orders of magnitude for the three deforming forces we consider, the variation in 
the actual ellipticity of the mountain, 
$|\epsilon^\text{C} - \epsilon^\text{A}|$, is relatively modest, 
about one order of magnitude (see Table~\ref{tab:ellipticities}). This is 
presumably a reflection of the fact that in all three cases we consider the same 
star with the same crustal breaking strain and shear modulus, so all stars have 
a similar ability to support deformations.

\section{Summary}
\label{sec:Summary4}

The question of the maximum mountain a neutron-star crust can support is an 
interesting problem. Such an estimate provides upper limits on the strength of 
gravitational-wave emission from rotating neutron stars, as well as having 
implications for the maximum spin-frequency limit that these systems can attain. 
In Chap.~\ref{ch:PopulationSynthesis}, we found that a fixed quadrupole of 
$Q_{2 2} = \SI{e36}{\gram\centi\metre\squared}$ was sufficient to obtain the 
observed spin distribution of accreting neutron stars.

We returned to this problem to tackle some of the pertinent assumptions made in 
previous work. We have discussed how previous estimates have not dealt 
appropriately with boundary conditions that must be satisfied for realistic 
neutron-star models. The calculations of \citet{2000MNRAS.319..902U} and 
\citet{2013PhRvD..88d4004J} both assumed a specific form for the strain that 
takes the star away from its relaxed shape and ensures the crust is maximally 
strained at every point. However, such a strain is somewhat unphysical since it 
does not respect the continuity of the traction vector.  Additionally, the 
approach of \citet{2006MNRAS.373.1423H}, while satisfying the traction 
conditions at the crust-core boundary, did not obey the boundary condition on 
the potential at the surface. This was due to the calculation assuming the 
relaxed configuration is spherical and implicitly using a surface force to 
deform the star. There were also errors present in the perturbation equations of 
\citet{2006MNRAS.373.1423H} that change their results by several orders of 
magnitude. 

An important simplification of the previous studies was to not explicitly 
calculate the non-spherical, relaxed shape that the strain is taken with respect 
to. As we have shown, such a description requires the introduction of a 
perturbing force that takes the star away from sphericity. We found such a 
discussion was missing in prior studies and, hence, have provided a 
demonstration that shows, provided one has a description of the strain, how the 
relaxed shape can calculated.

We found that including this force is crucial in enabling one to satisfy all the 
boundary conditions. Therefore, we have introduced a novel scheme for 
calculating the maximum quadrupole deformation that a neutron star can sustain 
and have demonstrated how our scheme is entirely equivalent to the approach of 
preceding calculations. Crucially, the formalism satisfies all the boundary 
conditions of the problem. One of the key advantages of our approach is that one 
computes all relevant quantities, including the shape of the relaxed star. 
However, one must provide a prescription for the deforming force. 

There is obviously significant freedom in what one may choose for the form of 
this force and, indeed, the formalism we have presented can be used for 
any deforming force that has the form (\ref{eq:Force}). Furthermore, 
it would not be difficult to adjust this formalism for other forces. 
However, evolutionary calculations will be necessary to fully motivate 
the form of the force. Thus, we surveyed three simple examples for the 
source of the mountains. We obtained the largest quadrupole for the (somewhat 
artificial) case where the perturbing potential is a solution to Laplace's 
equation, but leaves the core unperturbed. All of our results are between a 
factor of a few to two orders of magnitude below that of prior estimates for the 
maximum mountain a neutron star may support. That our results were smaller is 
not surprising, as our maximum mountains were constructed so that the breaking 
strain was reached at only a single point. An immediate question would be if 
there is a reasonable scenario that bridges the gap between relaxed 
configurations associated with a specific force and configurations following 
from specifying the strain. It seems inevitable that the answer will rely on 
evolutionary scenarios, leading to mountain formation, a problem that has not 
yet attracted the attention it deserves. 

An example of a promising scenario through which a rotating neutron star 
may radiate gravitational waves is accretion from a binary companion. As the gas 
is accreted onto the surface of the star, chemical reactions take place that 
change the composition. Such changes in the composition can, in turn, result in 
the star attaining a non-trivial quadrupole moment 
\citep{1998ApJ...501L..89B, 2000MNRAS.319..902U}. Additionally, there has been 
some effort towards calculating mountains on accreting neutron stars that are 
sustained by the magnetic field 
\citep{2005ApJ...623.1044M, 2006ApJ...641..471P, 2011MNRAS.417.2696P}. 
We note that, in our calculation, we only consider barotropic matter. This is 
appropriate to describe equilibrium stellar models. Indeed, if the star is in 
equilibrium, accreted and non-accreted matter may be described using barotropic 
equations of state \citep{2006MNRAS.373.1423H}. However, for evolutionary 
calculations, like those described above, one may need to consider 
non-barotropic features and, as we noted in Sec.~\ref{subsec:ThermalPressure}, 
the formalism we have presented could be used with such aspects at the 
perturbative level.

As an (admittedly phenomenological) indication of a possible solution, it may be 
worth pointing out that our approach to elasticity is somewhat simplistic. We 
have followed the usual assumption that the crust can be well described as an 
elastic solid (represented by a linear stress-strain relation) until it reaches 
the breaking strain, at which point the crust fails and all the strain is 
released. This model accords well with the molecular-dynamics simulations of 
\citet{2009PhRvL.102s1102H}, but it is worth noting that laboratory materials 
tends to behave slightly differently \citep{75da23d2ccdd4ab1b1b32564cc0df76b}. 
In particular, one typically finds that material deforms plastically for some 
level of strain before the ultimate failure. This introduces the yield strain as 
the point above which the stress-strain relationship is no longer linear and 
raises (difficult) questions regarding the plastic behaviour (the matter may 
harden, allowing stresses to continue building, or soften, leading to reduced 
stress as the strain increases). State-of-the-art simulations suggest a narrow 
region of plastic behaviour before the crust fails \citep{2009PhRvL.102s1102H}, 
but one should perhaps keep in mind that the levels of shear involved in the 
simulation may not lead to a true representation of matter that is deformed more 
gently. Let us, for the sake of the argument, suppose that this is the case and 
that the crust exhibits ideal plasticity above the chosen yield strain. If this 
were to happen, the strain would locally saturate at the yield limit even if the 
imposed force increased. One can then imagine applying a deforming force to 
source a neutron-star mountain and then increasing it until some point in the 
crust reaches yield strain. This is essentially the calculation we have done, as 
we did not model the behaviour beyond this point. Allowing for (ideal) plastic 
flow as the force is further increased, one may envisage that the entire crust 
may saturate at the yield strain. This is, of course, pure speculation 
\citep[although there have been several notable discussions about the relevance 
of plastic deformations of the neutron-star crust; see][]
{1970PhRvL..24.1191S, 2003ApJ...595..342J, 2010MNRAS.407L..54C}, but it 
might explain how a real system could reach the maximum strain configuration 
imposed in the \citet{2000MNRAS.319..902U} argument. As we already suggested, 
detailed evolutionary calculations that take into account the physical 
processes that produce the mountain will be required to make progress on the 
problem. 

Another natural avenue for future research is to generalise our calculation to 
relativity. This would be an important step as it brings realistic equations of 
state into play. Indeed, this is what we shall explore in the next chapter.


%% file: sections/chapter-5.tex
\chapter{Mountains in relativity}
\label{ch:RelativisticMountains}

In the previous chapter (Chap.~\ref{ch:Mountains}), we discussed how one 
computes mountains on neutron stars. Due to issues with previous calculations, 
we introduced a novel scheme that enables one to satisfy all the relevant 
boundary conditions of the problem. In this chapter, we generalise this work 
to relativistic gravity, as was done in \citet{2021MNRAS.507..116G}.

We dedicate Sec.~\ref{sec:RelativisticPerturbations} to deriving the fluid 
and elastic perturbation equations in relativity in order to construct fluid 
stars and stars with elastic crusts. As with the Newtonian calculation, the 
elastic crust presents a boundary-value problem, so we detail the relevant 
interface conditions. In Sec.~\ref{sec:RelativisticSources}, we consider three 
sources for the fiducial force that source the mountains. Because we conduct 
this work in relativity, we are able to consider the effect of the equation of 
state on the size of the deformations in Sec.~\ref{sec:Dependence}. We summarise 
in Sec.~\ref{sec:Summary5}.

\section{Relativistic perturbations}
\label{sec:RelativisticPerturbations}

Conceptually, the analysis of the mountain problem proceeds as in the Newtonian 
case (Sec.~\ref{sec:NewtonianPerturbations}). We need to generate stellar models 
with and without elastic crusts (stars C and A, respectively; 
Fig.~\ref{fig:StarsScheme}). As before, we build static perturbations on top of 
spherical backgrounds. In relativity, we use the relativistic structure 
equations~(\ref{eqs:TOV}) for the background.

\subsection{The fluid}
\label{sec:RelativisticFluidPerturbations}

The standard approach to computing non-radial stellar perturbations in general 
relativity is to follow \citet{1967ApJ...149..591T}. We use the Regge-Wheeler 
gauge \citep{1957PhRv..108.1063R} and focus on static, polar $\ell \geq 2$ 
perturbations, which leads to the Eulerian perturbation of the metric 
\begin{equation}
	h_{a b} dx^a dx^b = e^\nu H_0 Y_{\ell m} dt^2 + 2 H_1 Y_{\ell m} dt dr 
		+ e^\lambda H_2 Y_{\ell m} dr^2 
		+ r^2 K Y_{\ell m} (d\theta^2 + \sin^2 \theta \, d\phi^2),
	\label{eq:PerturbedMetric}
\end{equation}
where $H_0(r)$, $H_1(r)$, $H_2(r)$ and $K(r)$ describe the response of the 
spacetime to the perturbations. In the Newtonian limit, the terms in the 
perturbed metric -- specifically $H_0$ -- can be related to the perturbed 
potential $\delta \Phi$. The perturbed metric is sourced by the perturbations to 
the matter content of the spacetime, characterised by the linearised 
stress-energy tensor $\delta T_a^{\hphantom{a} b}$. This coupling is contained 
in the linearised Einstein equations [\textit{cf.} a perturbation of 
(\ref{eq:EinsteinEquations})], 
\begin{equation}
	\delta G_a^{\hphantom{a} b} = 8 \pi \, \delta T_a^{\hphantom{a} b},
	\label{eq:PerturbedEinsteinEquations}
\end{equation}
where $\delta G_a^{\hphantom{a} b}$ is the perturbed Einstein tensor. The 
derivation of the perturbed Einstein tensor is rather laborious and not 
particularly insightful, so we simply state the result 
\citep{1992PhRvD..46.4289K}:
\begin{equation}
\begin{split}
	\delta G_a^{\hphantom{a} b} = h^{b c} \left( R_{a c} 
		- \frac{1}{2} g_{a c} R \right) 
		+ \frac{1}{2} g^{b c} [ \nabla_d \nabla_a h_c^{\hphantom{c} d} 
		&+ \nabla_d \nabla_c h_a^{\hphantom{a} d} - \nabla_d \nabla^d h_{a c} 
		- \nabla_a \nabla_c h \\
		- h_{a c} R 
		&- g_{a c} ( h^{d e} R_{d e} + \nabla_d \nabla^e h_e^{\hphantom{e} d} 
		- \nabla_d \nabla^d h ) ],
\end{split}
	\label{eq:PerturbedEinstein}
\end{equation}
where $h \equiv h_a^{\hphantom{a} a}$ and the covariant derivatives are all 
associated with the background metric. Although we will not make use of them for 
this analysis, for the sake of completeness, we also have the linearised 
conservation equations from (\ref{eq:ConservationEnergyMomentum}), 
\begin{equation}
	\delta (\nabla_b T_a^{\hphantom{a} b}) = 0.
	\label{eq:PerturbedConservation}
\end{equation}
One should note that the Eulerian perturbation and the covariant derivative do 
not, in general, commute. In order to evaluate (\ref{eq:PerturbedConservation}), 
one needs to obtain the variation of the connection coefficients, 
$\delta \Gamma^a_{\hphantom{a} b c}$.

As noted in Sec.~\ref{sec:Perturbations}, in relativity, we have the Lagrangian 
perturbations to the four-velocity (\ref{eq:LagrangianFourVelocity}), 
\begin{equation}
	\Delta u^a = \frac{1}{2} u^a u^b u^c \Delta g_{b c},
	\label{eq:LagrangianFourVelocity2}
\end{equation}
and the baryon number density (\ref{eq:BaryonNumberConservation}), 
\begin{equation}
	\Delta n = - \frac{1}{2} n \bot^{a b} \Delta g_{a b},
	\label{eq:BaryonNumberConservation2}
\end{equation}
where we have the Lagrangian variation of the metric 
$\Delta g_{a b} = h_{a b} + \nabla_a \xi_b + \nabla_b \xi_a$ and the projection 
operator orthogonal to the flow 
$\bot_a^{\hphantom{a} b} = u_a u^b + \delta_a^{\hphantom{a} b}$. At this point, 
we need to define the static displacement vector \citep{2011PhRvD..84j3006P}, 
\begin{equation}
	\xi^a = 
	\begin{bmatrix}
		0 \\
		r^{-1} W \\
		r^{-2} V \partial_\theta \\
		(r \sin \theta)^{-2} V \partial_\phi \\
	\end{bmatrix}
	Y_{\ell m}, 
\end{equation}
with functions $W(r)$ and $V(r)$ describing the radial and tangential 
displacements, respectively.%
\footnote{Note that this form of the displacement vector is slightly different 
to the definition used in the Newtonian case (\ref{eq:NewtonianDisplacement}).}
We can obtain the Eulerian perturbation to the four-velocity from 
(\ref{eq:EulerianLagrangian}) and (\ref{eq:LagrangianFourVelocity2}), 
\begin{equation}
	\delta u^a = \frac{1}{2} u^a u^b u^c h_{b c}
		+ \bot_b^{\hphantom{b} a} \mathcal{L}_u \xi^b,
\end{equation}
which has components
\begin{equation}
	\delta u^t = \frac{1}{2} e^{-\nu/2} H_0 Y_{\ell m}, \qquad \delta u^i = 0.
\end{equation}
Similar to the Newtonian case, for static perturbations, the displacement vector 
does not appear in the linearised four-velocity.

Equation~(\ref{eq:BaryonNumberConservation2}) guarantees that the total number 
of baryons in the star is conserved. Computing 
Eq.~(\ref{eq:BaryonNumberConservation2}) explicitly gives 
\begin{equation}
	\Delta n = - \frac{n}{r^2} \left[ r^2 \left( K + \frac{1}{2} H_2 \right) 
		- \ell (\ell + 1) V + r \frac{dW}{dr} + \left( 1 
		+ \frac{1}{2} r \frac{d\lambda}{dr} \right) W \right] Y_{\ell m}.
	\label{eq:BaryonNumberConservation3}
\end{equation}
We continue to focus on barotropic matter, $\varepsilon = \varepsilon(n)$, 
(see discussion in Sec.~\ref{sec:EOS}) which means that 
\begin{equation}
	\Delta \varepsilon = \frac{d\varepsilon}{dn} \Delta n = \mu \Delta n,
	\label{eq:Deltavarepsilon}
\end{equation}
where $\mu$ is the chemical potential. Therefore, we can also write 
$p = p(\varepsilon)$, that leads to the relations 
\begin{equation}
	\Delta p = \frac{dp}{d\varepsilon} \Delta \varepsilon 
		= c_\text{s}^2 \Delta \varepsilon, \qquad 
		\delta p = c_\text{s}^2 \delta \varepsilon,
	\label{eq:Deltap}
\end{equation}
where we have identified the relativistic speed of sound $c_\text{s}$. We will 
use (\ref{eq:Deltap}) to connect $\delta p$ and $\delta \varepsilon$. By the 
first law of thermodynamics~(\ref{eq:FirstLawSimple}) for a single-species fluid 
composed of baryons, the baryon chemical potential is given by 
\begin{equation}
	\mu = \frac{\varepsilon + p}{n}.
\end{equation}
Hence, (\ref{eq:BaryonNumberConservation3}) and (\ref{eq:Deltavarepsilon}) 
combine to give 
\begin{equation}
	\Delta \varepsilon = - \frac{1}{r^2} (\varepsilon + p) \left[ r^2 \left( K 
		+ \frac{1}{2} H_2 \right) - \ell (\ell + 1) V + r \frac{dW}{dr} 
		+ \left( 1 + \frac{1}{2} r \frac{d\lambda}{dr} \right) W \right] 
		Y_{\ell m}.
\end{equation}
Therefore, by (\ref{eq:Deltap}), 
\begin{equation}
	\Delta p = - \frac{1}{r^2} c_\text{s}^2 (\varepsilon + p) 
		\left[ r^2 \left( K + \frac{1}{2} H_2 \right) - \ell (\ell + 1) V 
		+ r \frac{dW}{dr} + \left( 1 
		+ \frac{1}{2} r \frac{d\lambda}{dr} \right) W \right] Y_{\ell m}.
	\label{eq:Deltap1}
\end{equation}
We also have, from the relation between Lagrangian and Eulerian 
variations~(\ref{eq:EulerianLagrangian}), 
\begin{equation}
	\Delta p = \delta p + \xi^r \frac{dp}{dr} = \delta p 
		- \frac{1}{2 r} (\varepsilon + p) \frac{d\nu}{dr} W Y_{\ell m}.
	\label{eq:Deltap2}
\end{equation}
We will use Eqs.~(\ref{eq:Deltap1}) and (\ref{eq:Deltap2}) to close our system 
of equations for the crustal perturbations. Indeed, Eqs.~(\ref{eq:Deltap1}) and 
(\ref{eq:Deltap2}) hold throughout also in the fluid, but, because we are unable 
to determine the displacement in the fluid, we currently have no use for these 
expressions.

The matter content of the spacetime is encoded in the stress-energy tensor. To 
complete the specification of the linearised field equations, we use the 
stress-energy tensor for a perfect fluid~(\ref{eq:StressEnergyTensor}) to 
obtain 
\begin{equation}
	\delta T_a^{\hphantom{a} b} = (\delta \varepsilon + \delta p) u_a u^b 
		+ \delta p \, \delta_a^{\hphantom{a} b} 
		+ (\varepsilon + p) (\delta u_a u^b + u_a \delta u^b).
	\label{eq:PerturbedStressEnergyFluid}
\end{equation}
Recall that, as we did in Chap.~\ref{ch:Mountains}, we expand the perturbed 
quantities as shown in (\ref{eq:SphericalHarmonicDecomposition}) and specialise 
to a particular $(\ell, m)$ mode. In this calculation, we have 
$(\ell, m) = (2, 2)$.

One obtains the following second-order ordinary differential equation that 
governs the fluid perturbations by inserting 
Eqs.~(\ref{eq:PerturbedStressEnergyFluid}) and (\ref{eq:PerturbedEinstein}) into 
(\ref{eq:PerturbedEinsteinEquations}): 
\begin{subequations}\label{eqs:RelativisticFluidPerturbations}
\begin{equation}
\begin{split}
	\frac{d^2H_0}{dr^2} + \left[ \frac{2}{r} 
		+ \frac{1}{2} \left( \frac{d\nu}{dr} - \frac{d\lambda}{dr} \right) 
		\right] \frac{dH_0}{dr} 
		+ \Bigg\{ \frac{2}{r^2} - [2 + \ell (\ell + 1)] \frac{e^\lambda}{r^2}& \\
		+ \frac{2}{r} \left( 2 \frac{d\nu}{dr} + \frac{d\lambda}{dr} \right) 
		- \left( \frac{d\nu}{dr} \right)^2 \Bigg\} H_0 
		= - 8 \pi e^\lambda (\delta \varepsilon + \delta p)& 
	\label{eq:H_0Fluid} 
\end{split}
\end{equation}
and 
\begin{equation}
	\delta p = \frac{e^{-\lambda}}{16 \pi r} 
		\left( \frac{d\nu}{dr} + \frac{d\lambda}{dr}\right) H_0. 
	\label{eq:deltapFluid}
\end{equation}
For completeness, the other metric quantities are given by 
\begin{gather}
	H_1 = 0, \\ 
	H_2 = H_0 \label{eq:H_2Fluid}
\end{gather}
and 
\begin{equation}
	[\ell (\ell + 1) - 2] e^\lambda K = r^2 \frac{d\nu}{dr} \frac{dH_0}{dr} 
		+ \left[ \ell (\ell + 1) e^\lambda - 2 
		- r \left( \frac{d\nu}{dr} + \frac{d\lambda}{dr} \right) 
		+ r^2 \left( \frac{d\nu}{dr} \right)^2 \right] H_0.
	\label{eq:KFluid}
\end{equation}
\end{subequations}
Equations~(\ref{eqs:RelativisticFluidPerturbations}) are the relativistic 
analogue to the Newtonian fluid perturbation 
equations~(\ref{eqs:NewtonianFluidPerturbations}). Indeed, by combining 
Eqs.~(\ref{eq:H_0Fluid}) and (\ref{eq:deltapFluid}), one has a second-order 
differential equation for $H_0$, that is analogous to 
Eqs.~(\ref{eqs:NewtonianFluidPerturbations}). We should note that the 
perturbation equations in this form do not necessarily include the force that 
sources the deformation -- unless the force satisfies the relativistic analogue 
to the perturbed Laplace's equation, $\delta G_a^{\hphantom{a} b} = 0$. In this 
situation, which is effectively the tidal problem (see Chap.~\ref{ch:Tides}), 
the perturbed metric function $H_0$ contains both the gravitational and tidal 
potentials. However, in some sense, we can absorb the deforming force into a 
potential in $H_0$, like we did with $U$ in the Newtonian case [see 
(\ref{eq:Force}) and supporting text]. When we consider examples, we will 
demonstrate how one may explicitly include the force.

As remarked above, given the static nature of the problem, one is unable to 
calculate the displacement vector in the fluid, since the functions $W$ and $V$ 
do not appear in the perturbed stress-energy tensor.

As in the Newtonian problem, there are two boundary conditions that the 
perturbed metric potential must satisfy. At the origin, the solution must be 
regular, 
\begin{equation}
	H_0(0) = 0.
	\label{eq:PerturbedMetricPotentialCentre}
\end{equation}
The surface boundary condition is a little more involved. As we show in 
Appendix~\ref{app:Interface}, the function $H_0$ must be continuous across all 
boundaries. In the vacuum exterior, $r \geq R$, one should note that 
$\nu = - \lambda$, therefore, Eqs.~(\ref{eq:H_0Fluid}) and 
(\ref{eq:deltapFluid}) reduce to 
\begin{equation}
	\frac{d^2H_0}{dr^2} + \left( \frac{2}{r} - \frac{d\lambda}{dr} \right)
		\frac{dH_0}{dr} - \left[ \ell (\ell + 1) \frac{e^\lambda}{r} 
		+ \left( \frac{d\lambda}{dr} \right)^2 \right] H_0 = 0.
	\label{eq:H_0ExteriorODE}
\end{equation}
This is the relativistic analogue to Laplace's equation for polar perturbations. 
By a simple change of variables to $x = r / M - 1$, one can transform 
(\ref{eq:H_0ExteriorODE}) to 
\begin{equation}
	\frac{d}{dx} \left[ (1 - x^2) \frac{dH_0}{dx} \right] 
		+ \left[ \ell (\ell + 1) - \frac{4}{1 - x^2} \right] H_0.
	\label{eq:AssociatedLegendre}
\end{equation}
Equation~(\ref{eq:AssociatedLegendre}) is written in the form of the associated 
Legendre equation and has the general solution in terms of the associated 
Legendre polynomials $\mathcal{Q}_{\alpha \beta}(x)$ and 
$\mathcal{P}_{\alpha \beta}(x)$ with $\alpha = \ell$ and $\beta = 2$,%
\footnote{Usually Legendre polynomials are defined on the unit complex disc. We 
work with $x = r / M - 1 > 1$ and so we must change the sign in the argument of 
the logarithmic terms in $\mathcal{Q}_{\ell 2}(x)$, or, equivalently, take their 
real part.}
\begin{equation}
	H_0(r) = c_1 \mathcal{Q}_{\ell 2}(r / M - 1) 
		+ c_2 \mathcal{P}_{\ell 2}(r / M - 1),
\end{equation}
where $c_1$ and $c_2$ are constants that determine the amplitude of the 
perturbations. When we specialise to quadrupolar ($\ell = 2$) perturbations, the 
general solution becomes 
\begin{equation}
\begin{split}
	H_0(r) = c_1 \left( \frac{r}{M} \right)^2 \left( 1 - \frac{2 M}{r} \right) 
        \Bigg[&- \frac{M (M - r) (2 M^2 + 6 M r - 3 r^2)}{r^2 (2 M - r)^2} \\
        &+ \frac{3}{2} \ln \left( \frac{r}{r - 2 M} \right) \Bigg] 
        + 3 c_2 \left( \frac{r}{M} \right)^2 \left( 1 - \frac{2 M}{r} \right). 
\end{split}
	\label{eq:H_0Exterior}
\end{equation}
From (\ref{eq:H_0Exterior}), we observe two solutions to the relativistic 
Laplace's equation: a decreasing solution with $c_1$, that is associated with 
the gravitational potential of the star, and an increasing solution with $c_2$, 
that may be associated with an external tidal potential (should that be the 
physical source of the perturbations). For applications other than tidal 
deformations, one can assume $c_2 = 0$ and we have the boundary condition at the 
surface 
\begin{equation}
\begin{split}
    - \frac{1}{2} (R - 2 M) \frac{dH_0}{dr}(R) = \Bigg\{& - 1 + \frac{M}{R} 
        - 8 M^5 \bigg[ 2 M R (M - R) (2 M^2 + 6 M R - 3 R^2) \\ 
		&+ 3 R^3 (R - 2 M)^2 \ln \left( \frac{R - 2 M}{R} \right) \bigg]^{-1} 
		\Bigg\} H_0(R),
\end{split}
	\label{eq:PerturbedMetricPotentialSurface}
\end{equation}
which is the relativistic analogue (for $\ell = 2$ perturbations) of the 
Newtonian boundary condition~(\ref{eq:PerturbedPotentialSurface}). Thus, 
Eqs.~(\ref{eq:PerturbedMetricPotentialCentre}) and 
(\ref{eq:PerturbedMetricPotentialSurface}) are the boundary conditions on the 
perturbed metric potential $H_0$.

Using regularity (\ref{eq:PerturbedMetricPotentialCentre}), we find, for small 
$r$,
\begin{equation}
	H_0(r) = a_0 r^\ell [1 + \order(r^2)],
	\label{eq:Initial}
\end{equation}
where $a_0$ is a constant. With the intuition we developed from the Newtonian 
problem in Sec.~\ref{subsec:NewtonianFluid}, we note that this initial condition 
enables one to solve the coupled ordinary differential 
equations~(\ref{eqs:RelativisticFluidPerturbations}) with no freedom to impose 
(\ref{eq:PerturbedMetricPotentialSurface}). Hence, one needs a perturbing force 
in order to construct non-spherical equilibria. Indeed, suppose one considers 
the case where the star is perturbed by a presence of a companion that exerts a 
tidal force on it. In this situation, $H_0$ includes the tidal potential and the 
gravitational potential part of $H_0$ automatically satisfies 
(\ref{eq:PerturbedMetricPotentialSurface}).

\subsection{The elastic crust}
\label{subsec:RelativisticElasticPerturbations}

For an elastic material with shear modulus $\check{\mu}$, the Lagrangian 
perturbation to the anisotropic stress tensor is \citep{2019CQGra..36j5004A}%
\footnote{Note that we use the opposite sign for our definition here to that 
used in Eq.~(\ref{eq:ShearStressTensor}).}
\begin{equation}
	\Delta \pi_{a b} = - 2 \check{\mu} \Delta s_{a b},
\end{equation}
where the perturbed strain tensor $\Delta s_{a b}$ is given by 
\begin{equation}
	2 \Delta s_{a b} = \left( \bot^c_{\hphantom{c} a} \bot^d_{\hphantom{d} b} 
		- \frac{1}{3} \bot_{a b} \bot^{c d} \right) \Delta g_{c d}.
	\label{eq:Strain}
\end{equation}
The anisotropic stress tensor is symmetric and trace-free. Because star C 
(Fig.~\ref{fig:StarsScheme}) is relaxed in a spherical shape, we find 
\begin{equation}
	\delta \pi_a^{\hphantom{a} b} = - \check{\mu} \left( \bot^c_{\hphantom{c} a} 
		\bot^{d b} - \frac{1}{3} \bot_a^{\hphantom{a} b} \bot^{c d} \right) 
		\Delta g_{c d}.
	\label{eq:StressElastic}
\end{equation}
Summing the elastic stress tensor~(\ref{eq:StressElastic}) and the fluid 
stress-energy tensor~(\ref{eq:PerturbedStressEnergyFluid}) and inserting these 
expressions into the linearised field 
equations~(\ref{eq:PerturbedEinsteinEquations}) provides the information needed 
to solve the perturbations in the elastic crust.

For the application of the boundary conditions, it is convenient to define the 
following variables that are related to the traction components:%
\footnote{Note that here $T_1$ and $T_2$ are dimensionless, whereas, in the 
Newtonian calculation, their analogues have dimensions of stress 
(\ref{eqs:TractionVariables}).}
\begin{subequations}\label{eqs:RelativisticTractionVariables}
\begin{equation}
	T_1 Y_{\ell m} \equiv r^2 \delta \pi_r^{\hphantom{r} r} 
		= \frac{2 \check{\mu}}{3} \left[ r^2 (K - H_2) - \ell (\ell + 1) V 
		- 2 r \frac{dW}{dr} + \left( 4 - r \frac{d\lambda}{dr} \right) W \right] 
		Y_{\ell m}
\end{equation}
and 
\begin{equation}
	T_2 \partial_\theta Y_{\ell m} \equiv r^3 \delta \pi_r^{\hphantom{r} \theta} 
		= - \check{\mu} \left( r \frac{dV}{dr} - 2 V + e^\lambda W \right) 
		\partial_\theta Y_{\ell m}.
\end{equation}
\end{subequations}

Due to the introduction of the elastic crust, the perturbation equations become 
more complicated when compared to the fluid case. However, some of the 
perturbed Einstein equations remain unchanged. Since 
$\delta \pi_t^{\hphantom{t} b} = 0$, the [$tt$] component provides 
\begin{subequations}
\begin{equation}
\begin{split}
	e^{-\lambda} r^2 \frac{d^2K}{dr^2} 
		+ e^{-\lambda} \left( 3 
		- \frac{r}{2} \frac{d\lambda}{dr} \right) r \frac{dK}{dr} 
		- \left[ \frac{1}{2} \ell (\ell + 1) - 1 \right]& K \\
		- e^{-\lambda} r \frac{dH_2}{dr} - \left[ \frac{1}{2} \ell (\ell + 1) 
	+	 e^{-\lambda} \left( 1 - r \frac{d\lambda}{dr} \right) \right]& H_2 
		= - 8 \pi r^2 \delta \varepsilon.
\end{split}
	\label{eq:PEEtt}
\end{equation}
Because $\delta \pi_a^{\hphantom{a} b}$ is traceless, we can take the trace of 
the perturbed Einstein equations to obtain another equation that has no explicit 
dependence on the elasticity. We combine the trace with (\ref{eq:PEEtt}) to 
obtain 
\begin{equation}
\begin{split}
	- r^2 \frac{d^2H_0}{dr^2} 
		+ \left[ r \left( \frac{1}{2} \frac{d\lambda}{dr} 
		- \frac{d\nu}{dr} \right) - 2 \right]& r \frac{dH_0}{dr} 
		+ \ell (\ell + 1) e^\lambda H_0 
		- \frac{1}{2} r^2 \frac{d\nu}{dr} \frac{dH_2}{dr} \\
		+ \left[ 2 (e^\lambda - 1) 
		- r \left( 3 \frac{d\nu}{dr} + \frac{d\lambda}{dr} \right) \right]& H_2 
		+ r^2 \frac{d\nu}{dr} \frac{dK}{dr} 
		= 8 \pi e^\lambda r^2 (\delta \varepsilon + 3 \delta p).
\end{split}
	\label{eq:PEEtrace}
\end{equation}
Furthermore, we find from the [$tr$] component that, as in the fluid case, 
\begin{equation}
	H_1 = 0.
\end{equation}

Now, we consider the non-zero components of $\delta \pi_a^{\hphantom{a} b}$ to 
include elasticity into the system. The difference between the [$\theta\theta$] 
and [$\phi\phi$] components leads to the algebraic relation 
\begin{equation}
    H_2 - H_0 = 32 \pi \check{\mu} V,
	\label{eq:PEEdifference}
\end{equation}
which will be useful to eliminate $H_2$ from our equations. This expression 
shows how the equality~(\ref{eq:H_2Fluid}) from the fluid perturbations is 
spoiled with the introduction of elasticity. We can use the [$r\theta$] 
component and (\ref{eq:PEEdifference}) to provide 
\begin{equation}
	\frac{dK}{dr} = \frac{dH_0}{dr} + \frac{d\nu}{dr} H_0 
		+ \frac{16 \pi}{r} \left( 2 + r \frac{d\nu}{dr} \right) \check{\mu} V 
		- \frac{16 \pi}{r} T_2.
	\label{eq:PEErthetaa}
\end{equation}
The sum of the [$\theta\theta$] and [$\phi\phi$] components gives 
\begin{equation}
\begin{split}
	\delta p = \frac{e^{-\lambda}}{16 \pi r} \left( \frac{d\nu}{dr} 
		+ \frac{d\lambda}{dr} \right) H_0 
		+ \frac{e^{-\lambda}}{r^2} \bigg\{ e^\lambda [2 - \ell (\ell + 1)] 
		\check{\mu} V& \\
		+ \frac{e^\lambda}{2} T_1 - r \frac{dT_2}{dr} 
		- \left[ \frac{1}{2} r \left( \frac{d\nu}{dr} 
		- \frac{d\lambda}{dr} \right) + 1 \right] T_2& \bigg\},
\end{split}
	\label{eq:PEEsum}
\end{equation}
where we have simplified using (\ref{eq:PEEdifference}) and 
(\ref{eq:PEErthetaa}). The final equation we will use from the perturbed 
Einstein equations is the [$rr$] component combined with 
Eqs.~(\ref{eq:PEEdifference})--(\ref{eq:PEEsum}), 
\begin{equation}
\begin{split}
	[\ell (\ell + 1) - 2] e^\lambda K = r^2 \frac{d\nu}{dr} &\frac{dH_0}{dr} 
		+ \Bigg[ \ell (\ell + 1) e^\lambda - 2 - r \left( \frac{d\nu}{dr} 
		+ \frac{d\lambda}{dr} \right) \\
		+ r^2 &\left( \frac{d\nu}{dr} \right)^2 \Bigg] H_0 
		+ 16 \pi \left\{ [\ell (\ell + 1) - 2] e^\lambda 
		+ r^2 \left( \frac{d\nu}{dr} \right)^2 \right\} \check{\mu} V \\ 
		- 24 &\pi e^\lambda T_1 + 16 \pi r \frac{dT_2}{dr} 
		- 8 \pi \left[ 2 + r \left( \frac{d\nu}{dr} 
		+ \frac{d\lambda}{dr} \right) \right] T_2.
\end{split}
	\label{eq:PEErr}
\end{equation}
\end{subequations}
When $\check{\mu} = 0$, this reduces to (\ref{eq:KFluid}), as expected.

The next step is to formulate the system of equations in a way that is 
straightforward to integrate numerically. Clearly, there is a lot of freedom 
in how one can do this. We choose to work with the functions 
$(dH_0/dr, H_0, K, W, V, T_2)$ as our integration variables. It is useful to 
observe that through (\ref{eq:PEEdifference}) one can reduce the order of the 
system to eliminate $H_2$. In contrast to the fluid case, we are able to solve 
for the components of the displacement vector by using the definitions of the 
traction variables (\ref{eqs:RelativisticTractionVariables}). To be precise, we 
can integrate 
\begin{subequations}\label{eqs:RelativisticElasticPerturbations}
\begin{equation}
	\frac{dW}{dr} - \left( \frac{2}{r} 
		- \frac{1}{2} \frac{d\lambda}{dr} \right) W 
		= \frac{1}{2} r (K - H_0) - \left[ 16 \pi r \check{\mu} 
		+ \frac{\ell (\ell + 1)}{2 r} \right] V - \frac{3}{4 r \check{\mu}} T_1 
	\label{eq:WTraction}
\end{equation}
and 
\begin{equation}
	\frac{dV}{dr} - \frac{2}{r} V = - \frac{e^\lambda}{r} W 
		- \frac{1}{r \check{\mu}} T_2.
\end{equation}
We obtain an algebraic relation by combining (\ref{eq:PEEsum}) and 
(\ref{eq:PEErr}) in such a way as to remove $dT_2/dr$. This gives us an equation 
that involves $\delta p$ and $T_1$, 
\begin{equation}
\begin{split}
	16 \pi r^2 e^\lambda \delta p = r^2 \frac{d\nu}{dr} &\frac{dH_0}{dr} 
		+ \left[ \ell (\ell + 1) e^\lambda - 2 
		+ r^2 \left( \frac{d\nu}{dr} \right)^2 \right] H_0 
		+ [2 - \ell (\ell + 1)] e^\lambda K \\
		+ 16 &\pi r^2 \left( \frac{d\nu}{dr} \right)^2 \check{\mu} V 
		- 16 \pi e^\lambda T_1 
		- 16 \pi \left( 2 + r \frac{d\nu}{dr} \right) T_2.
\end{split}
	\label{eq:AlgebraicRelationa}
\end{equation}
From (\ref{eq:PEEsum}), we can obtain an equation to integrate for $T_2$, 
\begin{equation}
\begin{split}
	\frac{dT_2}{dr} + \left[ \frac{1}{2} \left( \frac{d\nu}{dr} 
		- \frac{d\lambda}{dr} \right) + \frac{1}{r} \right] T_2 
		= - &e^\lambda r \delta p + \frac{1}{16 \pi} \left( \frac{d\nu}{dr} 
		+ \frac{d\lambda}{dr} \right) H_0 \\ 
		+ &\frac{e^\lambda}{r} [2 - \ell (\ell + 1)] \check{\mu} V 
		+ \frac{e^\lambda}{2 r} T_1.
\end{split}
	\label{eq:T_2Elastic}
\end{equation}
It is easy to see that (\ref{eq:T_2Elastic}), when $\check{\mu} = 0$, reduces to 
give (\ref{eq:deltapFluid}). We combine 
Eqs.~(\ref{eq:PEEtrace})--(\ref{eq:PEErthetaa}) to get 
\begin{equation}
\begin{split}
	\frac{d^2H_0}{dr^2} + \left[ \frac{2}{r} 
		+ \frac{1}{2} \left( \frac{d\nu}{dr} 
		- \frac{d\lambda}{dr} \right) \right] \frac{dH_0}{dr} 
		+ &\Bigg\{ \frac{2}{r^2} - [2 + \ell (\ell + 1)] \frac{e^\lambda}{r^2} \\
		+ \frac{1}{r} \left( 3 \frac{d\nu}{dr} + \frac{d\lambda}{dr} \right) 
		- \left( \frac{d\nu}{dr} \right)^2 \Bigg\} H_0 
		= - 8 \pi &\Bigg\{ e^\lambda (3 \delta p + \delta \varepsilon) 
		+ 2 \frac{d\nu}{dr} \frac{d(\check{\mu} V)}{dr} \\
		+ 8 \bigg[ \frac{1 - e^\lambda}{r^2} 
		+ \frac{1}{2 r} \left( 2 \frac{d\nu}{dr} + \frac{d\lambda}{dr} \right)&
		- \frac{1}{4} \left( \frac{d\nu}{dr} \right)^2 \bigg] \check{\mu} V 
		+ \frac{2}{r} \frac{d\nu}{dr} T_2 \Bigg\}.
\end{split}
	\label{eq:H_0Elastic}
\end{equation}
In the fluid, where the shear modulus vanishes, one can verify that 
(\ref{eq:H_0Elastic}) reduce to give (\ref{eq:H_0Fluid}). The final equation we 
need from the perturbed Einstein equations (\ref{eq:PEErthetaa}), usefully, 
needs no further alteration, 
\begin{equation}
	\frac{dK}{dr} = \frac{dH_0}{dr} + \frac{d\nu}{dr} H_0 
		+ \frac{16 \pi}{r} \left( 2 + r \frac{d\nu}{dr} \right) \check{\mu} V 
		- \frac{16 \pi}{r} T_2.
	\label{eq:PEErthetab}
\end{equation}
To close this system of equations, we need to consider the thermodynamics. We 
can use the expressions we obtained from baryon-number conservation 
(\ref{eq:Deltap1}) and (\ref{eq:Deltap2}) to obtain a second algebraic relation 
involving $\delta p$ and $T_1$ along with (\ref{eq:WTraction}), 
\begin{equation}
	\frac{3}{4 \check{\mu}} T_1 = \frac{r^2}{(\rho + p) c_\text{s}^2} \delta p 
		+ \frac{3}{2} r^2 K - \frac{3}{2} l (l + 1) V 
		+ \left( 3 - \frac{r}{2 c_\text{s}^2} \frac{d\nu}{dr} \right) W.
	\label{eq:AlgebraicRelationb}
\end{equation}
\end{subequations}
We use (\ref{eq:AlgebraicRelationa}) and (\ref{eq:AlgebraicRelationb}) to 
determine $\delta p$ and $T_1$. 
Equations~(\ref{eqs:RelativisticElasticPerturbations}) fully specify the elastic 
perturbation problem.

\subsection{Interface conditions}
\label{subsec:RelativisticInterface}

In order to connect the fluid core and ocean to the elastic crust of star C, we 
must consider the interface conditions. By treating the crust as relaxed in a 
spherical configuration, we know that the background quantities will be 
continuous.%
\footnote{Of course, should one use an equation of state that involves 
discontinuities at such an interface, that would need to be taken into account. 
For a treatment that involves discontinuities in the equation of state see 
\citet{2020ApJ...895...28P}.}
To determine how the perturbed quantities behave at an interface, we must 
calculate the first and second fundamental forms (also known as the intrinsic 
and extrinsic curvatures), which must be continuous throughout the spacetime. We 
describe this calculation in detail in Appendix~\ref{app:Interface}. 

The first fundamental form implies that the functions $H_0$, $K$ and $W$ are 
continuous. From the second fundamental form, we find that the traction must be 
continuous: thus, $(T_1 / r^2 + \delta p)$ and $T_2$ are continuous. We will 
continue to assume that the shear modulus is non-zero throughout the crust and, 
therefore, must be discontinuous at a fluid-elastic boundary. We can use 
continuity of the radial traction~(\ref{appeq:H_0primeb}) along with 
(\ref{eq:KFluid}) to obtain an expression which is true in the elastic crust at 
an interface, 
\begin{equation}
\begin{split}
	r^2 \frac{d\nu}{dr} \frac{dH_{0 \text{E}}}{dr} 
		= - \Bigg[& \ell (\ell + 1) e^\lambda - 2 
		- r \left( \frac{d\nu}{dr} + \frac{d\lambda}{dr} \right) 
		+ r^2 \left( \frac{d\nu}{dr} \right)^2 \Bigg] H_{0 \text{F}} \\
		+ [& \ell (\ell + 1) - 2 ] e^\lambda K_\text{F} 
		- 16 \pi r^2 \left( \frac{d\nu}{dr} \right)^2 \check{\mu} V_\text{E},
\end{split}
	\label{eq:H_0primeCondition}
\end{equation}
where we re-introduce the subscripts $\text{F}$ and $\text{E}$ to denote fluid 
and elastic, respectively, to make explicit which side of the interface the 
perturbed quantities are on. From (\ref{appeq:H_0primeb}), we observe that, 
while the radial displacement must be continuous, this does not necessarily have 
to be the case for the tangential piece. From continuity of the tangential 
traction, we know that $T_2 = 0$ at the crustal boundaries.

In the core, we calculate $(dH_0/dr, H_0)$ using 
Eqs.~(\ref{eqs:RelativisticFluidPerturbations}) and, thus, obtain $K$ via the 
algebraic equation~(\ref{eq:KFluid}). In the crust, the order of the system 
increases, as we need to determine the additional functions $W$, $V$ and $T_2$. 
The system that describes the crust is $(dH_0/dr, H_0, K, W, V, T_2)$ with 
Eqs.~(\ref{eqs:RelativisticElasticPerturbations}). With the six boundary 
conditions at the fluid-elastic interfaces -- continuity of $H_0$ and $K$, 
$T_2 = 0$ and (\ref{eq:H_0primeCondition}) -- the system is well posed as a 
boundary-value problem.

\section{The deforming force}
\label{sec:RelativisticSources}

Now, we can turn our attention to considering some example forces in relativity. 
Because we can generate fully relativistic stellar models with this formalism, 
it is appropriate to consider a more realistic description of the nuclear 
matter. There are two quantities of the matter that we need to determine for our 
models: the fluid pressure-energy density relationship, $p(\varepsilon)$, and 
the shear modulus of the crust, $\check{\mu}$.

For the pressure-density relation, we first use the analytic BSk24 equation of 
state \citep{2018MNRAS.481.2994P} for the high-density regions 
($\varepsilon > \SI{e6}{\gram\per\centi\metre\cubed}$) of the star and the 
\citet{2001A&A...380..151D} table for the low-density regions 
($\varepsilon \leq \SI{e6}{\gram\per\centi\metre\cubed}$). To prescribe the 
shear-modulus profile, we use the \citet{1990PhRvA..42.4867O} result
\begin{equation}
    \check{\mu} = 0.1994 \left( \frac{4 \pi}{3} \right)^{1/3} 
        \left( \frac{1 - X_\text{n}}{A} n \right)^{4/3} (e Z)^2,
\end{equation}
where $X_\text{n}$ is the free-neutron fraction, $A$ and $Z$ are the atomic and 
proton numbers, respectively, $n$ is the baryon-number density and $e$ is the 
fundamental electric charge. We neglect any phase transitions in the crust, as 
they would significantly complicate the calculation. Should phase transitions 
exist at the boundaries of the crust, one would need to take into account the 
discontinuities in the background quantities. For the nuclear-matter parameters, 
we use the BSk24 results for the inner crust and the HFB-24 model 
\citep{2013PhRvC..88b4308G} along with the experimental data from the 2016 
Atomic Mass Evaluation \citep{2017ChPhC..41c0003W} for the outer crust 
\citep[see Table~4 in][]{2018MNRAS.481.2994P}. The location of the core-crust 
transition is given by the BSk24 results and the crust-ocean transition is 
taken to be the lowest density in the outer-crust model. The complexity of this 
prescription is due to our attempt to use a consistent model for the 
neutron-star physics.

\begin{figure}[h]
    \centering
	\includegraphics[width=0.8\textwidth]{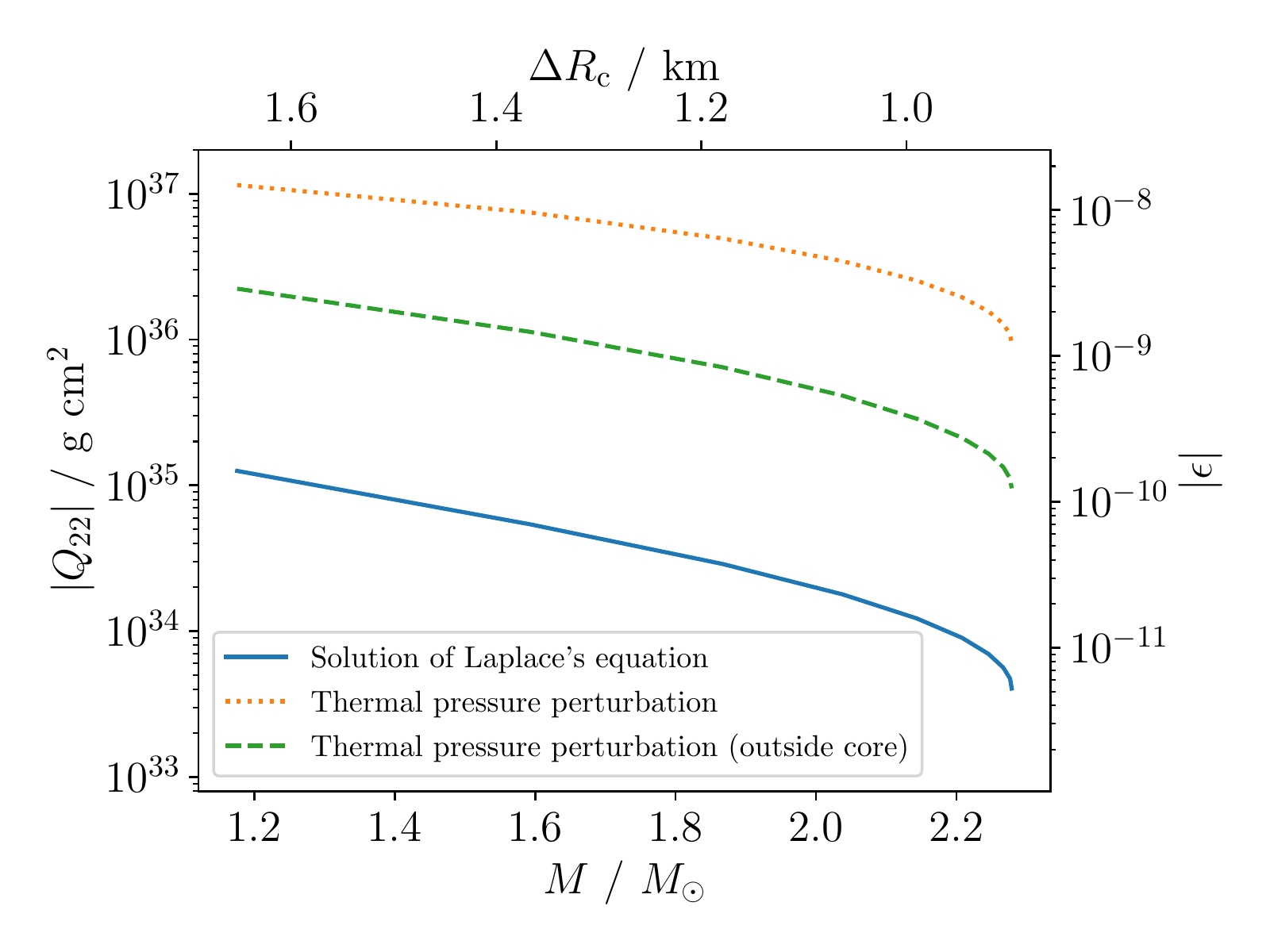}
    \caption[The maximum quadrupole and ellipticity due to the different forces]
			{The maximum quadrupole and ellipticity due to the different forces 
			 as functions of stellar mass and crustal thickness, 
			 $\Delta R_\text{c}$. We show results for the force corresponding to 
			 the solution of the relativistic Laplace's equation (solid blue 
			 line), the thermal pressure perturbation (dotted orange line) and 
			 the thermal pressure perturbation that acts outside the core (dashed 
			 green line).}
    \label{fig:Forces}
\end{figure}

We consider three sources for the perturbations: (i) a deforming potential that 
is a solution to the relativistic Laplace's equation, (ii) a thermal pressure 
perturbation and (iii) a thermal pressure perturbation that only acts outside 
the core. Sources (i) and (ii) are the relativistic equivalents of two forces we 
considered in our Newtonian calculation (Chap.~\ref{ch:Mountains}). As 
emphasised in the previous chapter, we note that the force in this problem will 
be related to the evolutionary history of the star. The examples we consider 
here have not been explicitly connected to any formation scenario. They should 
merely provide an illustration of how this calculation is carried out. To 
calculate the mountains for each example, we follow our scheme outlined in 
Section~\ref{sec:Mountains}. The perturbations are normalised by ensuring star C 
reaches breaking strain at a point in its crust, according to the von Mises 
criterion (see the relativistic version below). In practice, this means we take 
the point in the crust where the strain is greatest and set that to breaking 
strain. The results for the different forces are summarised in 
Fig.~\ref{fig:Forces}.

We calculate the structure of the fluid core of stars A and C using 
Eqs.~(\ref{eqs:RelativisticFluidPerturbations}) with the boundary condition at 
the centre~(\ref{eq:PerturbedMetricPotentialCentre}). For the crust of star C, 
we apply the interface conditions outlined in 
Sec.~\ref{subsec:RelativisticInterface} to 
Eqs.~(\ref{eqs:RelativisticElasticPerturbations}). For the fluid ocean of both 
stars, we evaluate the fluid perturbation 
equations~(\ref{eqs:RelativisticFluidPerturbations}) up to the surface, where 
the gravitational potential part of $H_0$ must satisfy 
(\ref{eq:PerturbedMetricPotentialSurface}). The numerical scheme is explained in 
Appendix~\ref{app:NumericalScheme}.

\subsection{A solution to the relativistic Laplace's equation}

The relativistic Laplace's equation is given by the vacuum Einstein equations 
(\ref{eq:H_0ExteriorODE}). By comparing (\ref{eq:PerturbedMetricPotential}) and 
(\ref{eq:H_0Exterior}), we identify 
\begin{equation}
    Q_{2 m} = \frac{M^3 c_1}{\pi},
	\label{eq:RelativisticQuadrupole}
\end{equation}
in agreement with Eq.~(41) in \citet{2013PhRvD..88d4004J}. Thus, as in the 
Newtonian calculation, the quadrupole can be obtained by examining the 
exterior gravitational potential of the star. The challenge is to disentangle 
the other potentials acting on the star that will be contained in $H_0$.

From (\ref{eq:H_0Exterior}), we can calculate the constants $c_1$ and $c_2$ 
using $H_0$ and $dH_0/dr$ at the surface:
\begin{subequations}
\begin{equation}
	c_1 = \frac{R (R - 2 M)}{8 M^3} \left[ R (2 M - R) \frac{dH_0}{dr}(R) 
		+ 2 (R - M) H_0(R) \right]
	\label{eq:PotentialAmplitude}
\end{equation}
and
\begin{equation}
\begin{split}
	c_2 = \frac{1}{48 M^3 R (2 M - R)} \Bigg\{
		- 2 M \bigg[ R (4 M^4 + 6 M^3 R - 22 M^2 R^2 
		+ 15 M R^3 - 3 R^4) \frac{dH_0}{dr}(R)& \\
		+ 2 (2 M^4 - 2 M^3 R + 13 M^2 R^2 
		- 12 M R^3 + 3 R^4) H_0(R) &\bigg] \\
		+ 3 R^2 (R - 2 M)^2 \left[ R (2 M - R) \frac{dH_0}{dr}(R) 
		+ 2 (R - M) H_0(R) \right] \ln\left( \frac{R}{R - 2 M} \right) &\Bigg\}.
\end{split}
	\label{eq:ForceAmplitude}
\end{equation}
\end{subequations}
At the surface, $H_0$ and its derivative are continuous, so we can use their 
values at this point to calculate the force amplitude (\ref{eq:ForceAmplitude}) 
and the quadrupole (\ref{eq:RelativisticQuadrupole}).

In the formalism of \citet{2019CQGra..36j5004A}, the von Mises strain for a star 
with an unstrained background is
\begin{equation}
    \Theta = \sqrt{\frac{3}{2} \Delta s_{a b} \Delta s^{a b}}.
	\label{eq:RelativisticvonMisesStrain}
\end{equation}
The von Mises criterion states that a point in the crust fractures when the 
von Mises strain reaches the threshold, $\Theta \geq \Theta^\text{break}$, at 
some point. For $(\ell, m) = (2, 2)$ perturbations, we have the explicit expression
\begin{equation}
\begin{split}
    \Theta^2 = \frac{45}{512 \pi} \Bigg\{ &\sin^2 \theta \Bigg[ 
        3 \sin^2 \theta \cos^2 2 \phi \left( \frac{T_1}{r^2 \mu} \right)^2 \\ 
        &\qquad\quad+ 4 e^{-\lambda} (3 + \cos 2 \theta 
		- 2 \sin^2 \theta \cos 4 \phi) 
        \left( \frac{T_2}{r^2 \mu} \right)^2 \Bigg] \\
        &+ (35 + 28 \cos 2 \theta + \cos 4 \theta + 8 \sin^4 \theta \cos 4 \phi) 
        \left( \frac{V}{r^2} \right)^2 \Bigg\}.
\end{split}
	\label{eq:RelativisticvonMisesStrain22}
\end{equation}
Following the results of molecular-dynamics simulations by 
\citet{2009PhRvL.102s1102H}, we take the breaking strain to be 
$\Theta^\text{break} = 0.1$.

Therefore, we compute stars A and C using the 
fluid~(\ref{eqs:RelativisticFluidPerturbations}) and elastic perturbation 
equations~(\ref{eqs:RelativisticElasticPerturbations}) for $(\ell, m) = (2, 2)$ 
perturbations. At the surface, we obtain the amplitude of the force from 
(\ref{eq:ForceAmplitude}) and increase its amplitude until a point in the crust 
breaks according to (\ref{eq:RelativisticvonMisesStrain22}). Then, once the two 
stars are normalised to the same force, we calculate the quadrupole moment with 
(\ref{eq:RelativisticQuadrupole}) and (\ref{eq:PotentialAmplitude}).

Figure~\ref{fig:Forces} shows the maximum deformations with such a force for 
varying stellar mass. In the equivalent Newtonian case, for a star with 
$M = \SI{1.4}{\solarMass}$, $R = \SI{10}{\kilo\metre}$, we found 
$Q_{2 2} = \SI{1.7e37}{\gram\centi\metre\squared}$, 
$\epsilon = \num{2.2e-8}$. We see that the corresponding maximum deformation for 
a relativistic $M = \SI{1.4}{\solarMass}$ star is two orders of magnitude lower. 
This suppression has two contributions. In going from Newtonian to relativistic 
gravity, the maximum size of the quadrupole that a crust can support decreases. 
This suppression was observed by \citet{2013PhRvD..88d4004J} in their 
relativistic calculation and has also been seen in tidal- 
\citep{2008ApJ...677.1216H, 2010PhRvD..81l3016H, 2009PhRvD..80h4035D, 
2009PhRvD..80h4018B} 
and magnetic-deformation calculations 
\citep{2004ApJ...600..296I, 2010MNRAS.406.2540C, 2012PhRvD..86d4012Y, 
2012MNRAS.427.3406F}. 
This behaviour has been attributed to the stiffness of the external, vacuum 
spacetime, that suppresses the quadrupole in the matching at the stellar surface 
\citep{2013PhRvD..88d4004J}.

\begin{figure}[h]
    \centering
	\includegraphics[width=0.8\textwidth]{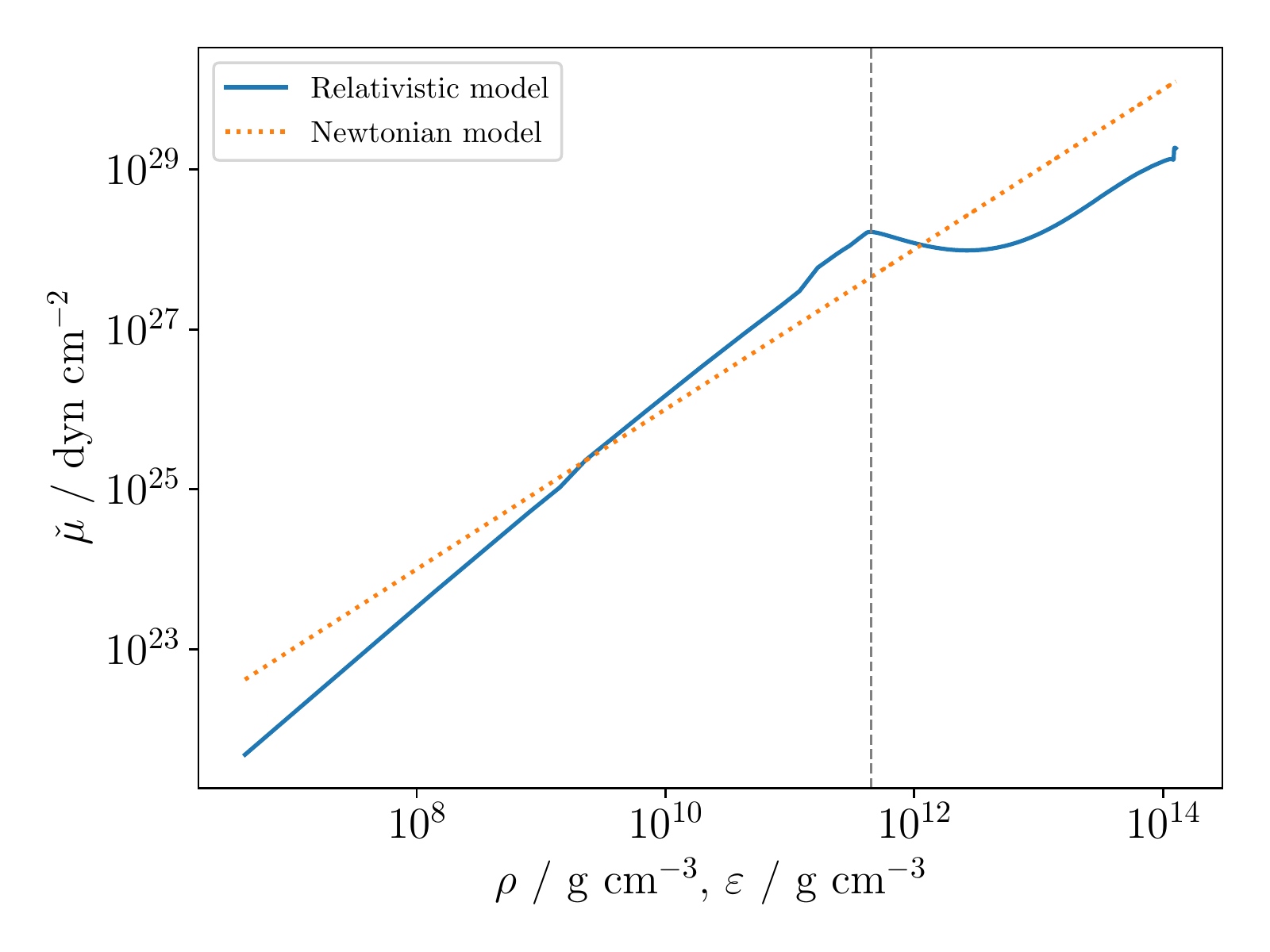}
    \caption[The shear-modulus profiles]{The shear modulus as a function of 
			 energy density, $\varepsilon$, for the model of the crust in this 
             relativistic calculation (solid blue line) and as a function of 
             mass density, $\rho$, for the linear model used in the Newtonian 
             calculation (dotted orange line). We show the density of neutron 
			 drip, that corresponds to the inner-outer crust transition 
			 (dashed grey line).}
	\label{fig:Models}
\end{figure}

The second effect comes from the equation of state. Focusing on the role of the 
matter model in our calculations, we found that the point where the strain was 
the largest for all the forces we considered was the top of the crust. This is 
where the crust yields and this behaviour is consistent with previous 
calculations of neutron-star crusts 
\citep{2020PhRvD.101j3025G, 2021MNRAS.500.5570G}. In Fig.~\ref{fig:Models}, we 
compare the shear-modulus profile used in this work with the linear model used 
in the Newtonian calculation. Although the linear model appears to be a 
reasonable approximation to the more realistic model, there are key areas where 
the two differ. Of particular importance to the maximum-mountain calculation, 
the realistic shear-modulus profile is significantly weaker in the lower-density 
outer crust (approximately an order of magnitude smaller at the top of the 
crust). This plays a pivotal role in determining the size of the mountains that 
the crust can support as the breaking strain scales with the shear modulus.

Our results are notably at least two to three orders of magnitude smaller than 
the mountains obtained by \citet{2013PhRvD..88d4004J}. As was noted in 
Chap.~\ref{ch:Mountains}, the fact that our scheme produces smaller mountains 
than calculated in previous work is not particularly surprising. Indeed, the 
very nature of the \citet{2000MNRAS.319..902U} approach [that 
\citet{2013PhRvD..88d4004J} follow] is to ensure that the \textit{entire} crust 
is at breaking strain, whereas in our scheme (in order to correctly satisfy the 
boundary conditions of the problem) breaking strain is reached at a point. 
Clearly, the size of the mountains depend on the force, so it is natural to 
explore other choices to see if we can obtain larger deformations. We will go on 
to consider a couple of other examples.

\subsection{A thermal pressure perturbation}
\label{subsec:Thermal}

The next source for the perturbations we consider is a thermal pressure. For 
this case, we assume that the pressure has a thermal component of the form 
reminiscent of (\ref{eq:ThermalPressure}),
\begin{equation}
    \delta p_\text{th} = \frac{k_\text{B} \varepsilon}{m_\text{b}} \delta T.
\end{equation}
We will assume the temperature is given by (\ref{eq:Temperature}). To 
incorporate this force into the perturbation 
equations~(\ref{eqs:RelativisticFluidPerturbations}) and 
(\ref{eqs:RelativisticElasticPerturbations}), they must be adjusted by 
$\delta p \rightarrow \delta p + \delta p_\text{th}$. In this case, $H_0$ 
corresponds solely to the gravitational potential of the star.

For this example, the amplitude $a_0$ in (\ref{eq:Initial}) is connected to the 
magnitude of the temperature perturbation. It is constrained by ensuring the 
function $H_0$ matches the exterior solution (\ref{eq:H_0Exterior}) with 
$c_2 = 0$. At the surface, the solution must satisfy 
(\ref{eq:PerturbedMetricPotentialSurface}). The temperature perturbation, 
$\delta T(R)$, is then increased until a point in the crust breaks.%
\footnote{For a canonical $M = \SI{1.4}{\solarMass}$ star, we found the 
crust breaks when $\delta T(R) = \SI{9.4e4}{\kelvin}$. As before, this 
temperature perturbation is simply a source term for the pressure 
perturbation~(\ref{eq:ThermalPressure}) and not a specific physical mechanism.}

The maximum mountains built using this force are shown in Fig.~\ref{fig:Forces}. 
Compared to the solution to the relativistic Laplace's equation, the thermal 
pressure perturbation produces mountains approximately two orders of magnitude 
larger. As we saw in the Newtonian calculation, this illustrates that the size 
of the mountains that can be built are highly dependent on their formation 
history.

\sloppy
For comparison, the corresponding force in our Newtonian calculation gave 
$Q_{2 2} = \SI{4.0e38}{\gram\centi\metre\squared}$, $\epsilon = \num{5.2e-7}$. 
Hence, the suppression for the thermal pressure is weaker than in the previous 
example.

\fussy
We also considered the situation where the top of the crust was moved to neutron 
drip in order to assess the impact of removing the weaker regions of the crust. 
In this case, for an $M = \SI{1.4}{\solarMass}$ star, we obtained 
$Q_{2 2} = \SI{5.9e38}{\gram\centi\metre\squared}$, $\epsilon = \num{7.7e-7}$, 
which is even larger than what was obtained in the Newtonian calculation. This 
is not particularly surprising since we only focus on the inner crust, which is 
orders of magnitude stronger in the shear modulus than most of the outer crust 
(see Fig.~\ref{fig:Models}). We also found that the crust no longer yielded at 
the top. This illustrates the role the shear modulus plays in supporting the 
mountains.

\subsection{A thermal pressure perturbation outside the core}

We also consider the case where the thermal pressure perturbation does not reach 
into the core. One could imagine a scenario where the surface of the neutron 
star is heated and this heating does not penetrate to the core. We will assume 
that the thermal pressure has a finite value at the base of the crust and exists 
in the crust and ocean. We use the same form for the temperature perturbation 
(\ref{eq:Temperature}).

Due to the specific nature of this force, one must be careful in setting up the 
calculation. Since the core is unperturbed, we will assume 
$H_0(r_\text{base}) = W(r_\text{base}) = 0$. Because the force suddenly appears 
at the base, we assume that $dH_0(r_\text{base})/dr$ is non-zero. For the fluid 
star (star A), this is sufficient to calculate the mountains. The precise value 
of $dH_0/dr$ at the base is determined by ensuring the surface boundary 
condition (\ref{eq:PerturbedMetricPotentialSurface}) is satisfied.

For the star with a crust (star C), we need to pay attention to the traction 
conditions. Since $dH_0 / dr$ needed to have a finite value at the base of the 
crust in the fluid star, we effectively violated the radial traction 
condition~(\ref{appeq:H_0primeb}). However, we can still use the tangential 
traction condition to demand $T_2(r_\text{base}) = 0$. This leaves two free 
values, $K(r_\text{base})$ and $V(r_\text{base})$. At the top of the crust we 
still impose both traction conditions, which constrains $K(r_\text{base})$ and 
$V(r_\text{base})$ and, thus, the problem is well posed. As was the case for the 
fluid star, $dH_0(r_\text{base})/dr$ is constrained via 
(\ref{eq:PerturbedMetricPotentialSurface}).

We plot the maximum mountains for this case in Fig.~\ref{fig:Forces}. As 
compared to the thermal pressure that acts throughout the star, the mountains 
produced in this example are approximately an order of magnitude smaller.

In summary, we have seen with all our examples in Fig.~\ref{fig:Forces} that the 
maximum deformations the crust can sustain are small. Nevertheless, it is 
interesting to note that our results are, in principle, large enough not to 
contradict the minimum deformation argument of \citet{2018ApJ...863L..40W} and 
the quadrupoles that can describe the accreting millisecond pulsar population 
(Chap.~\ref{ch:PopulationSynthesis}).

\section{Dependence on the equation of state}
\label{sec:Dependence}

As the calculation is done in full general relativity, we have the opportunity 
to assess the impact of the equation of state. We explore a subset of the chiral 
effective-field-theory models combined with a speed-of-sound parametrisation 
\citep[see][]{2018ApJ...860..149T}. [These models were recently used by 
\citet{2020NatAs...4..625C} to obtain constraints on neutron-star radii from 
observational data.] Chiral effective field theory is a systematic framework for 
low-energy hadronic interactions. For low densities, the theory describes matter 
using nucleons and pions, where the interactions are expanded in powers of 
momenta and all the relevant operators in strong interactions are included 
\citep{1990PhLB..251..288W, 1991NuPhB.363....3W, 1994PhRvC..49.2932V, 
2009RvMP...81.1773E, 2011PhR...503....1M}. 
One then uses quantum Monte Carlo methods to solve the many-body Schr{\"o}dinger 
equation to obtain an equation of state 
\citep{2010PhRvC..82a4314H, 2013PhRvC..88b5802K, 2015RvMP...87.1067C, 
2018ApJ...860..149T}. 
Chiral effective field theory is expected to describe matter well up to between 
one to two times nuclear saturation density. \citet{2018ApJ...860..149T} 
extended the equations of state to higher densities, outside the low-energy 
regime of the chiral effective field theory, using a speed-of-sound 
parametrisation to ensure that causality was not violated.

\begin{figure}[h]
    \includegraphics[width=0.49\textwidth]{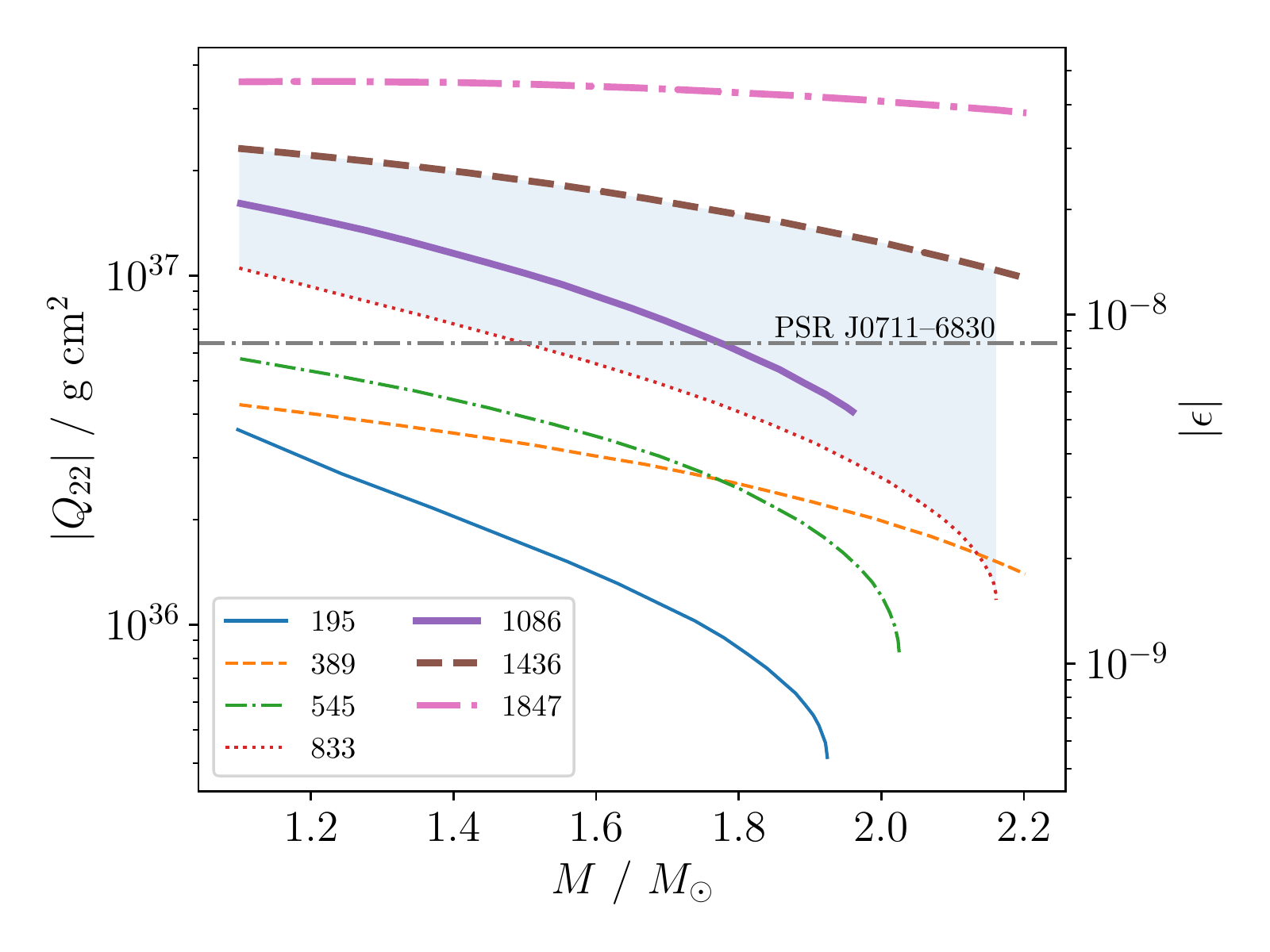}
	\includegraphics[width=0.49\textwidth]{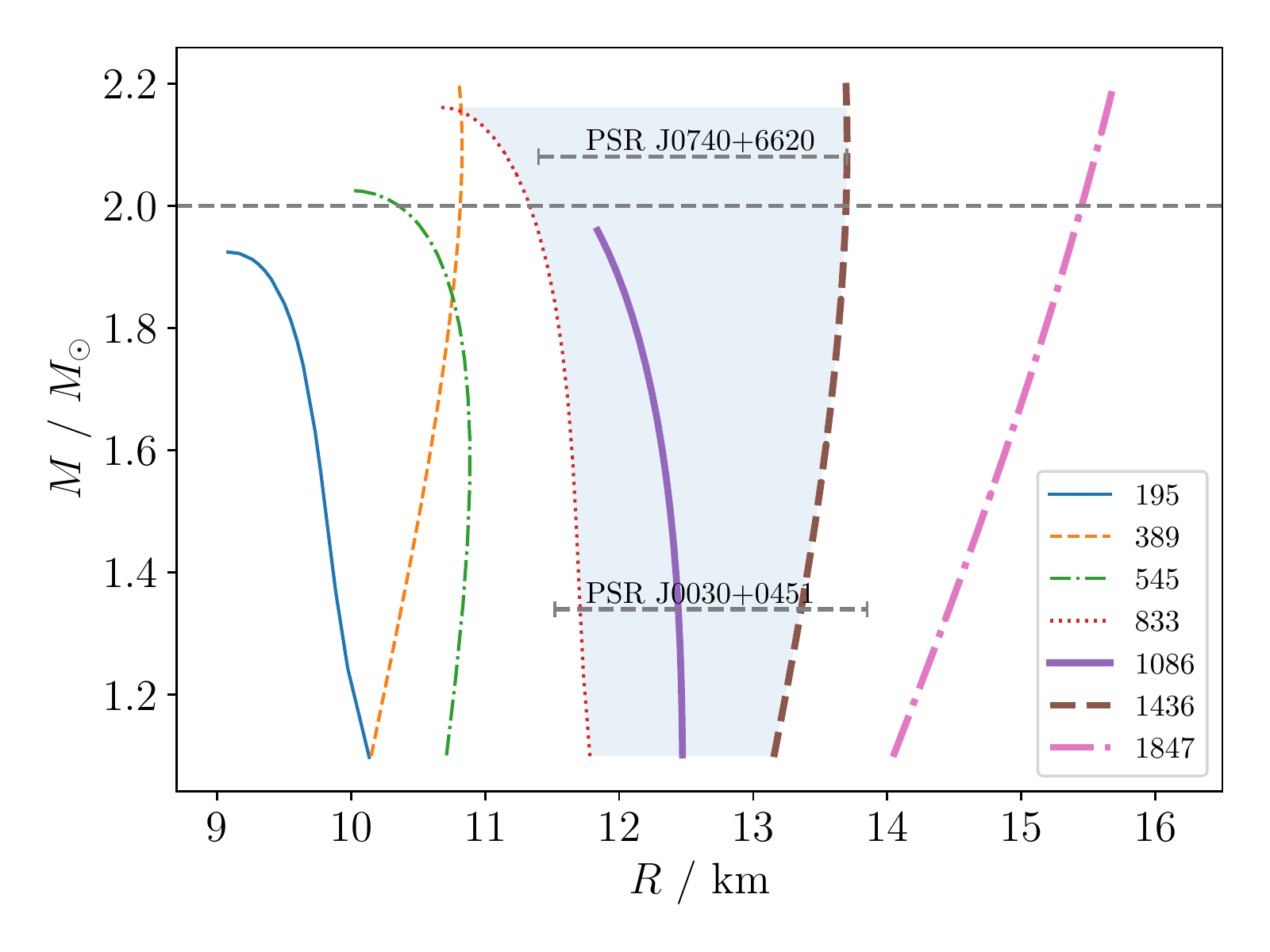}
    \caption[The maximum quadrupole and ellipticity due to the different chiral 
			 effective-field-theory equations of state]
			{The maximum quadrupole and ellipticity due to thermal pressure 
			 perturbations as functions of stellar mass for different chiral 
			 effective-field-theory equation-of-state models (left panel) and 
			 the corresponding mass-radius diagram traced out by the background 
			 stellar models (right panel). To compare with observational 
			 constraints, we indicate $M = \SI{2}{\solarMass}$ (middle dashed 
			 grey line) and the range of radii 
			 $11.52 \leq R \ / \ \si{\kilo\metre} \leq 13.85$ 
			 (bottom dashed grey line) and 
			 $11.41 \leq R \ / \ \si{\kilo\metre} \leq 13.69$ 
			 (top dashed grey line) measured by NICER for PSR J0030+0451 and PSR 
			 J0740+6620, respectively, in the right panel. We shade the region 
			 between the two models 833 and 1436 that (roughly) satisfy the 
			 constraints to give an indication of the range of possible maximum 
			 deformations for this force. In the left panel, we indicate the 
			 upper limit on the ellipticity of PSR J0711--6830 from 
			 gravitational-wave searches, $\epsilon < \num{8.3e-9}$ (dash-dotted 
			 grey line). The equations of state are indexed according to their 
			 radii for an $M = \SI{1.4}{\solarMass}$ star. All stellar models 
			 considered are stable to radial perturbations.}
	\label{fig:Chiral}
\end{figure}

We consider a selection of models for the pressure-density relation 
[supplemented by the \citet{2001A&A...380..151D} table for the low-density 
region ($\varepsilon \leq \SI{e6}{\gram\per\centi\metre\cubed}$)] and subject 
the stars to thermal pressure perturbations (as described in 
Sec.~\ref{subsec:Thermal}). We choose this mechanism since it produced the 
largest mountains from the examples we considered. The results are shown in 
Fig.~\ref{fig:Chiral}.

It should be noted that there are observational and theoretical constraints on 
the mass and radius of neutron stars. From observations, it is apparent that 
the true nuclear-matter equation of state must be able to support 
\SI{2}{\solarMass} neutron stars \citep{2013Sci...340..448A}. 
Additionally, there are recent constraints on the radius from NICER: 
PSR J0030+0451, with mass $M = \SI{1.34}{\solarMass}$, was measured to have 
$11.52 \lesssim R \ / \ \si{\kilo\metre} \lesssim 13.85$ 
\citep{2019ApJ...887L..21R} and PSR J0740+6620, with mass 
$M = \SI{2.08}{\solarMass}$, was measured to have 
$11.41 \lesssim R \ / \ \si{\kilo\metre} \lesssim 13.69$ 
\citep{2021arXiv210506980R}. There have been studies combining observations and 
theory to constrain the radius of a canonical $M = \SI{1.4}{\solarMass}$ neutron 
star. \citet{2020ApJ...893L..21R} used the NICER observation of PSR J0030+0451 
along with GW170817 to obtain a constraint of 
$11.75 \lesssim R \ / \ \si{\kilo\metre} \lesssim 13.5$. However, there is some 
degree of statistical uncertainty associated with this range and such 
constraints will need to be updated with future detections. Indeed, the study of 
\citet{2020NatAs...4..625C}, that combined nuclear theory with observations of 
GW170817, found the contrasting (and more stringent) range 
$10.4 \lesssim R \ / \ \si{\kilo\metre} \lesssim 11.9$ for an 
$M = \SI{1.4}{\solarMass}$ neutron star. We indicate $M = \SI{2}{\solarMass}$ 
and the radius ranges from the NICER measurements to the right panel of 
Fig.~\ref{fig:Chiral}. The majority of the equations of state that we consider 
support stars with $M = \SI{2}{\solarMass}$. To give some indication of the 
accepted range of equations of state in the mass-radius diagram, we shade the 
region between models 833 and 1436, which (roughly) satisfy the observational 
constraints. In order to give context for some of the most constrained upper 
limits from gravitational-wave data, we show the deformation constraint for PSR 
J0711--6830 of $\epsilon < \num{8.3e-9}$ in the left panel 
\citep{2020ApJ...902L..21A}.

In the left panel of Fig.~\ref{fig:Chiral}, we show the maximum deformations 
due to thermal pressure perturbations for the different equations of state. 
It is perhaps surprising to observe that there is a range of approximately two 
orders of magnitude across all the equations. Thus, these mountain calculations 
are quite sensitive to the equation of state. However, as indicated by the 
shaded region of Fig.~\ref{fig:Chiral}, the range of maximum deformations is 
narrower when we consider recent observational constraints on the equation of 
state.

One can also see that there is a relationship between the radius, $R$, of a 
star with a given mass, $M$, and the maximum deformation it supports. This 
should not be surprising, since equations of state that produce stars with 
larger radii will also have thicker crusts that can support larger deformations.

\section{Summary}
\label{sec:Summary5}

There is hope on the horizon that we will soon detect gravitational waves from 
rotating neutron stars for the very first time. Indeed, there are continued 
efforts to improve the data-analysis techniques 
\citep[see, \textit{e.g.},][]{2021PhRvD.103f4027B, 2020PhRvL.125q1101D, 
2021PhRvD.103f3019D, 2021ApJ...909...79S, 2021ApJ...906L..14Z}, 
along with plans in development for third-generation gravitational-wave 
detectors to be constructed in the (hopefully) not-too-distant future 
\citep{2020JCAP...03..050M}.

In Chap.~\ref{ch:Mountains}, we surveyed previous maximum-mountain calculations 
and found that there were issues relating to boundary conditions that must be 
satisfied for realistic neutron stars. In particular, the usual approach to 
calculating maximum neutron-star mountains of \citet{2000MNRAS.319..902U} 
assumes a strain field that violates continuity of the traction. We introduced a 
new scheme for calculating mountains that gives one full control of the boundary 
conditions at the cost of requiring a knowledge of the deforming force that 
sources the mountains. However, it is unclear what this force should be. In 
reality, the force is related to the formation history of the star, that may 
involve complex mechanisms like quakes and accretion from a companion. 
Therefore, in order to get a handle on the force, evolutionary calculations that 
consider the history of neutron stars will be necessary. It should be noted that 
such calculations would need to be ambitious in order to take into account the 
physics that may be important in the evolution of the crust, such as cooling, 
freezing, spin down, magnetic fields and cracking (to name but a few 
mechanisms).

In this study, we generalised our scheme from Chap.~\ref{ch:Mountains} to 
relativity. We considered three examples for the deforming force and found the 
most promising results for thermal pressure perturbations. In constructing 
relativistic stellar models with a realistic equation of state, we noted that 
the size of the mountains was suppressed (in some cases, quite significantly). 
This was generally due to two factors: (i) using relativistic gravity and (ii) 
the shear modulus of the crust was weaker at points than the simple model used 
in our Newtonian study. For most of the examples we examined, the crust yielded 
first at the top, where the shear modulus is the weakest. These results also 
point towards the necessity of evolutionary calculations in making progress on 
mountain calculations. This is evident from the role the deforming force plays 
in how large the mountains can be.

We have also demonstrated how the mountains are sensitive to the 
fluid pressure-density relationship for nuclear matter, suggesting a range of 
uncertainty larger than an order of magnitude for a range of equations of state 
that satisfy current observational constraints on the mass and radius. For this 
analysis we considered a subset of equations of state from chiral effective 
field theory, that all obey causality, and subjected the stars to thermal 
pressure perturbations.

We made an effort to build a consistent neutron-star model, based on the BSk24 
equation of state \citep{2018MNRAS.481.2994P}. However, our treatment of the 
crust is still somewhat simplistic: we assume that the crust behaves like an 
elastic solid up to some breaking strain, at which it yields and all the strain 
is subsequently released. As discussed in Sec.~\ref{sec:Summary4}, a possible 
solution to constructing larger mountains than the ones we have been able to 
obtain perhaps lies in plasticity. Plastic solids behave elastically up to some 
strain and, beyond that strain, deform in such a way that they retain some of 
the strain. It is perhaps likely that neutron-star crusts exhibits some of this 
behaviour. Suppose a neutron-star crust is modelled as an ideal plastic. It is 
then deformed up to its elastic yield limit at a point in the crust and the 
strain saturates. Even if the force is increased, the strain stays the same. One 
could then continue to apply forces to the crust in order to build as large a 
mountain as possible. This could connect real neutron stars to the maximally 
strained configuration imposed in \citet{2000MNRAS.319..902U}. At present, this 
is speculation, even though there have been some interesting discussions of 
plasticity in neutron-star crusts 
\citep{1970PhRvL..24.1191S, 2003ApJ...595..342J, 2010MNRAS.407L..54C}. This idea 
certainly seems worthy of future studies.


%% file: sections/chapter-6.tex
\chapter{Tidal deformations}
\label{ch:Tides}

We have discussed how one calculates mountains in a self-consistent manner in 
Chaps.~\ref{ch:Mountains} and \ref{ch:RelativisticMountains}. We have seen that, 
in order to obtain the mountains, one must specify the fiducial force that gives 
the star its non-spherical shape. At present, we do not have a good 
understanding of what such a force should be (at least not at a level that would 
be calculable) and, indeed, progress in this direction may have to rely on 
detailed evolutionary calculations of physical scenarios that lead to mountains. 
However, there are neutron-star gravitational-wave scenarios where we do know 
what the deforming mechanism will be: binary neutron stars experiencing tidal 
perturbations. In this chapter, we will study static tidal deformations of 
neutron stars with elastic crusts, following \citet{2020PhRvD.101j3025G}.

We set the scene in Sec.~\ref{sec:k_2Def} by introducing the tidal Love numbers 
in general relativity and showing how they can be extracted from the metric. In 
Sec.~\ref{sec:Impact}, we apply the relativistic perturbation formalism 
outlined in Sec.~\ref{sec:RelativisticPerturbations} to compute the tidal 
deformations of a neutron star with an elastic crust using a realistic equation 
of state. In Sec.~\ref{sec:CrustFailure}, we explore whether the crust will 
break during a binary inspiral. At the end of the chapter, we summarise in 
Sec.~\ref{sec:Summary6}.

\section{Definition of the tidal Love number}
\label{sec:k_2Def}

Neutron stars undergoing binary inspiral and merger are confirmed (and highly 
celebrated) gravitational-wave sources. In 2017, the LIGO and Virgo detectors 
observed a gravitational-wave signal from a binary neutron star, GW170817 
\citep{2017PhRvL.119p1101A}. More recently, in 2019, another binary merger was 
detected, GW190425 \citep{2020ApJ...892L...3A} and, in 2020, two neutron 
star-black hole binaries were observed \citep{2021ApJ...915L...5A}. Prior to 
these detections, it had long been considered the case that binary-neutron-star 
systems would be promising candidates for gravitational-wave emission 
\citep{1993PhRvL..70.2984C, 2002gr.qc.....4090C}. 

One of the exciting prospects of gravitational-wave observations is that 
they can provide model-independent constraints on the equation of state and, 
indeed, have done in the case of GW170817 
\citep{2017PhRvL.119p1101A, 2017ApJ...850L..34B, 2018PhRvL.120q2703A, 
2018ApJ...857L..23R, 2018PhRvL.120z1103M, 2018PhRvL.121i1102D, 
2018PhRvL.121p1101A, 2019PhRvX...9a1001A}. 
The gravitational-wave signal emitted from inspiralling neutron stars differs 
slightly from that of inspiralling black holes. The very fact that neutron stars 
are extended bodies introduces finite-size corrections to the gravitational-wave 
signal.%
\footnote{It is important to accurately track the orbital dynamics of each 
neutron star during the inspiral in order to build up a large enough 
signal-to-noise ratio for a confident detection.}
The dominant finite-size effect comes from the tidal deformation that each 
star's gravitational field induces on the other. Since this effect depends on 
the density distribution of the star, it may be used as a diagnostic to probe 
the neutron-star interior. However, neutron stars are believed to have solid 
crusts close to their surfaces, which introduce further complexity into 
prospective descriptions of the interior \citep{2008LRR....11...10C}. In recent 
years, there have been a few efforts in the direction of understanding the 
impact that the inclusion of an elastic crust makes on tidal deformations in 
neutron stars \citep{2011PhRvD..84j3006P, 2019PhRvD.100d4056B}, including our 
own work \citep{2020PhRvD.101j3025G}. 

For tidal deformations, it is appropriate to work in the 
\textit{adiabatic limit}, where the variations in the tidal field are assumed to 
be slow compared to the timescale associated with the star's internal response 
\citep{2019MNRAS.489.4043A}. Therefore, one uses static perturbations. 

A star immersed in a time-independent, external, quadrupolar tidal field 
$\mathcal{E}_{i j}$ will develop a quadrupole moment $Q_{i j}$ in response. In 
the Newtonian limit, $Q_{i j}$ is given by [see (\ref{eq:QuadrupoleTensor})]%
\footnote{It is assumed that the tidal field is weak and thus affects the 
mass-density distribution of the star in a perturbative way.}
\begin{equation}
    Q_{i j} = \int_V \delta \rho(x^i) \left( x_i x_j 
        - \frac{1}{3} r^2 g_{i j}\right) dV.
        \label{eq:QuadrupoleTensorPerturbed}
\end{equation}
To linear order, one can relate the quadrupole moment to the tidal field by 
\citep{2008ApJ...677.1216H}
\begin{equation}
    Q_{i j} = - \frac{2}{3} \frac{R^5}{G} k_2 \mathcal{E}_{i j}, 
    \label{eq:QErelation}
\end{equation}
where we meet the $\ell = 2$ tidal Love number $k_2$ (see 
Sec.~\ref{subsubsec:Tidal} for a brief introduction to the tidal Love numbers in 
Newtonian gravity). Since much of this analysis will make use of the Newtonian 
limit, we will reinstate factors of $G$ and $c$ in relevant expressions. We 
briefly review the procedure of calculating the tidal Love number below. For 
other detailed discussions on the subject of tidal deformations in general 
relativity, see \citet{2008ApJ...677.1216H}, \citet{2009PhRvD..80h4018B}, 
\citet{2009PhRvD..80h4035D} and \citet{2019arXiv190500408A}.

The Love number can be extracted from the exterior metric. We specialise the 
perturbing potential in (\ref{eq:PerturbedMetricPotential}) to the tidal field 
\citep[see, \textit{e.g.},][]{2009PhRvD..80h4018B}, 
\begin{equation}
    - \frac{1 + g_{t t} + h_{t t}}{2} = - \frac{M}{r} 
        + \sum_{m = -\ell}^\ell \left[ - \frac{4 \pi}{2 \ell + 1} B_\ell(r) 
        \frac{Q_{\ell m}}{r^{\ell + 1}} 
        + \frac{4 \pi (\ell - 2)!}{(2 \ell + 1)!!} A_\ell(r) r^\ell 
        \mathcal{E}_{\ell m} \right] Y_{\ell m}(\theta, \phi),
    \label{eq:PerturbedMetricTidal}
\end{equation}
where $A_\ell(r)$ is a function that goes to unity in the Newtonian limit 
\citep[this is given as $A_1$ in Table I of][]{2009PhRvD..80h4018B}. Because we 
work with quadrupolar deformations, we have $\ell = 2$.

We are free to work with either the tensors $Q_{i j}$ and $\mathcal{E}_{i j}$ or 
their spherical-harmonic counterparts $Q_{2 m}$ and $\mathcal{E}_{2 m}$ (see 
Sec.~\ref{subsec:NewtonianMultipoles}). By comparing 
(\ref{eq:PotentialMultipolel}) and (\ref{eq:PotentialMultipolelm}), we find%
\footnote{One could also obtain this expression using the decomposition in 
\citet{2008ApJ...677.1216H} based on the formalism developed in 
\citet{1980RvMP...52..299T}. However, one would need to take into account the 
different conventions for the multipole moments (see 
earlier footnote~\ref{foot:Multipole} in Chap.~\ref{ch:Structure}). 
Additionally, one could brute force this calculation by inserting pure 
mass-density perturbations of order $\ell$ into the quadrupole moment 
tensor~(\ref{eq:QuadrupoleTensorPerturbed}) and contracting the free indices 
with $\hat{x}^i$.} 
\begin{equation}
    Q_{i j \ldots q}^\ell \hat{x}^i \hat{x}^j \ldots \hat{x}^q 
        = \frac{4 \pi \ell!}{(2 \ell + 1)!!} 
        \sum_{m = - \ell}^\ell Q_{\ell m} Y_{\ell m},
\end{equation}
where $\hat{x}^i = x^i / r$ is the unit position vector in the exterior of the 
star and $Q_{i j \ldots q}^\ell$ is the multipole moment tensor of order $\ell$ 
given by (\ref{eq:MultipoleMomentTensor}). For quadrupolar deformations, 
\begin{subequations}\label{eqs:Decomposition}
\begin{equation}
    Q_{i j} \hat{x}^i \hat{x}^j = \frac{8 \pi}{15} 
        \sum_{m = - 2}^2 Q_{2 m} Y_{2 m}.
\end{equation}
Thus, we also have 
\begin{equation}
    \mathcal{E}_{i j} \hat{x}^i \hat{x}^j = \frac{8 \pi}{15} 
        \sum_{m = - 2}^2 \mathcal{E}_{2 m} Y_{2 m}.
\end{equation}
\end{subequations}
Furthermore, one can verify that $Q_{i j}$ vanishes for any perturbation with 
$\ell \neq 2$. Hence, by (\ref{eq:QErelation}), we have 
\begin{equation}
    Q_{2 m} = - \frac{2}{3} \frac{R^5}{G} k_2 \mathcal{E}_{2 m}.
    \label{eq:QErelation2}
\end{equation}
We can, without any loss of generality, assume that only one $\mathcal{E}_{2 m}$ 
is non-vanishing \citep{2008ApJ...677.1216H}.

One can identify the coefficients $c_1$ and $c_2$ with the quadrupole and the 
tidal field, respectively, by comparing the exterior solution to the vacuum 
field equations~(\ref{eq:H_0Exterior}) to the form of the exterior 
metric~(\ref{eq:PerturbedMetricTidal}): 
\begin{subequations}\label{eqs:Coefficients}
\begin{gather}
	c_1 = \pi \frac{c^4}{G^2 M^3} Q_{2 m}, \\
	c_2 = - \frac{8 \pi}{45} \frac{G^2 M^2}{c^6} \mathcal{E}_{2 m},
    \label{eq:c_2a}
\end{gather}
\end{subequations}
and, thus, using (\ref{eq:QErelation2}), obtain 
\begin{equation}
	\frac{c_1}{c_2} = \frac{15}{4} \frac{k_2}{C^5}, 
    \label{eq:Ratio}
\end{equation}
where $C \equiv G M / (c^2 R)$ is the star's compactness. Because $H_0$ and 
$dH_0/dr$ are continuous between the interior and the vacuum at the surface, we 
can use (\ref{eq:H_0Exterior}) to determine the ratio $c_1 / c_2$ in terms of 
the interior solutions at $r = R$. This gives the result 
\citep{2008ApJ...677.1216H} 
\begin{equation}
\begin{split}
	k_2 = \frac{8 C^5}{5} (1& - 2 C)^2 [2 + 2 C (y - 1) - y] 
	    \Big\{ 2 C [6 - 3 y + 3 C (5 y - 8)] \\
	    &+ 4 C^3 [13 - 11 y + C (3 y - 2) + 2 C^2 (1 + y)] \\
	    &+ 3 (1 - 2 C)^2 [2 - y + 2 C (y - 1)] \ln(1 - 2 C) \Big\}^{-1},
\end{split}
    \label{eq:k_2}
\end{equation}
where we have introduced the parameter $y \equiv R [dH_0(R)/dr] / H_0(R)$. It is 
interesting to note that for the computation of the Love number the amplitude 
$a_0$ in the initial condition (\ref{eq:Initial}) may be chosen freely. The 
reason for this is intuitive. Since the tidal Love number is a measure of how 
deformable a star is in the presence of a quadrupolar field, it is independent 
of the exact details of the external field and, therefore, the calculation of 
this quantity is insensitive to the magnitude. We see this in (\ref{eq:k_2}) as 
the ratio $y$ means that dependence on $a_0$ exactly cancels. For our analysis, 
we will focus on the dimensionless tidal deformability parameter, 
\begin{equation}
	\Lambda = \frac{2}{3} \frac{k_2}{C^5},
\end{equation}
to enable direct comparison with gravitational-wave constraints 
\citep[see, \textit{e.g.},][]{2018PhRvL.121p1101A}. 

\section{The impact of an elastic crust}
\label{sec:Impact}

We assume the background of the star to be relaxed. Thus, the elastic 
perturbation formalism we developed in 
Sec.~\ref{subsec:RelativisticElasticPerturbations} is valid. The star is 
partitioned into three regions: a fluid core, a solid crust and a fluid ocean.

To accurately prescribe the crust-ocean transition, we consider the melting 
point of the crust. The Coulomb lattice melts when the thermal energy, 
\begin{equation}
	E_\text{th} = k_\text{B} T,
\end{equation}
exceeds the interaction energy of the lattice, 
\begin{equation}
	E_\text{Coul} = \frac{Z^2 e^2}{a}, 
\end{equation}
where $a$ is the mean spacing between nuclei, by a critical factor $1/\Gamma$, 
\begin{equation}
	E_\text{th} \geq \frac{1}{\Gamma} E_\text{Coul}, 
\end{equation}
where $\Gamma \approx 173$. We assume that the crust forms a body-centred cubic 
lattice, which effectively has two nuclei per unit cube, so given the number 
density of nuclei, $n_\text{n}$, we have 
\begin{equation}
	n_\text{n} a^3 = 2.
\end{equation}
The mass density at which the crust begins to melt is, therefore, obtained from 
\begin{equation}
	\rho_\text{top} = A m_\text{b} n_\text{n} 
	    = 2 A m_\text{b} \left( \frac{\Gamma k_\text{B} T}{Z^2 e^2} \right)^3 
	    \approx \num{6.72e5} \ \left( \frac{A}{56} \right) 
		\left( \frac{Z}{26} \right)^{-2/3} 
		\left( \frac{T}{\SI{e7}{\kelvin}} \right)^3 
		\ \si{\gram\per\centi\metre\cubed}.
\end{equation}
For our prescription, we assume that the outer parts of the crust are composed 
of iron, $Z = 26$ and $A = 56$, and a temperature of $T = \SI{e7}{\kelvin}$. 
We assume, towards the surface of the star, that 
$\varepsilon_\text{top} \approx \rho_\text{top}$.

\begin{figure}[h]
    \centering
	\includegraphics[width=0.7\textwidth]{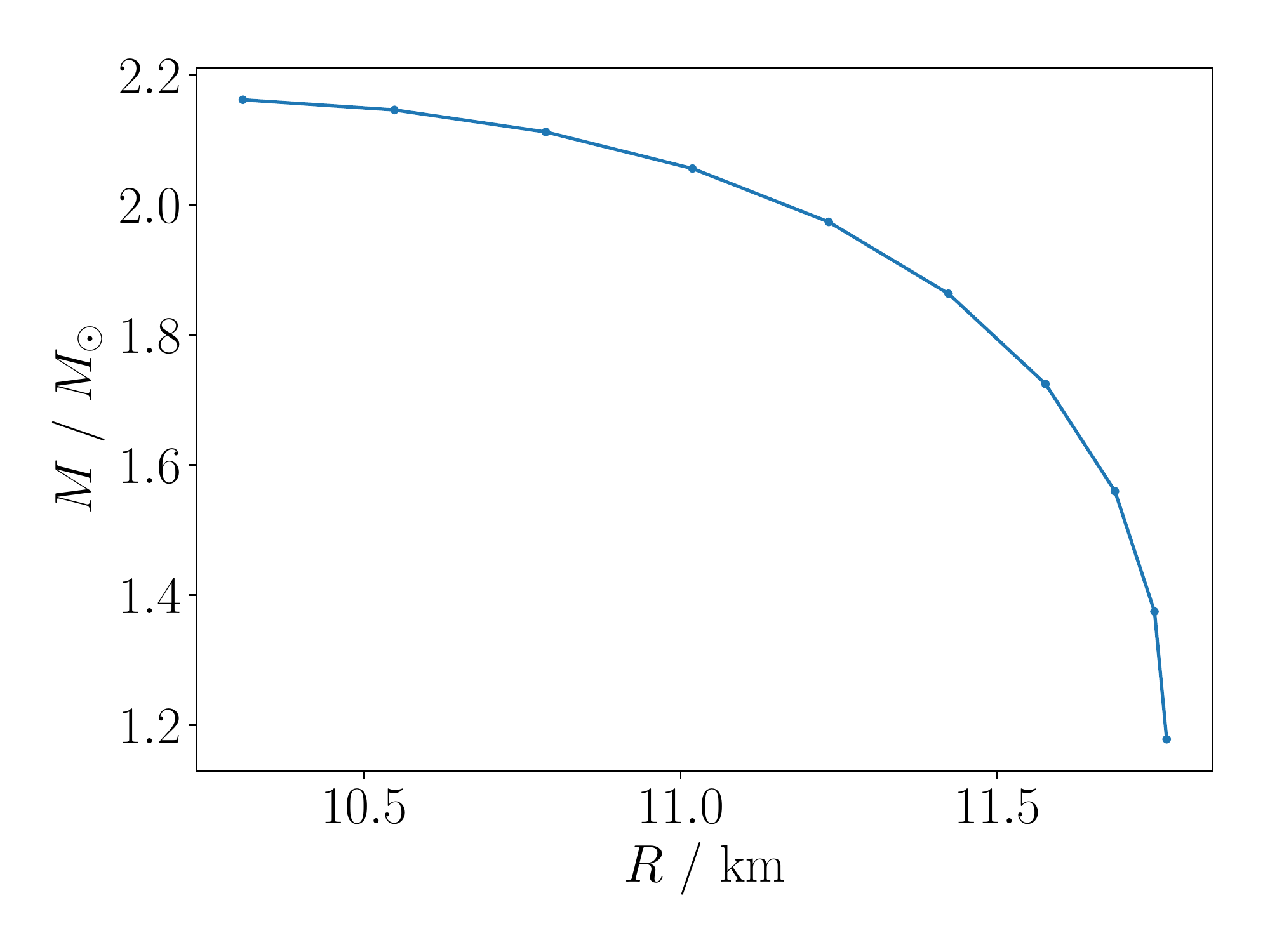}
    \caption[Mass-radius diagram for equation of state]{The mass-radius diagram 
             for the central densities considered, showing that the stellar 
             models considered are stable to radial perturbations.}
    \label{fig:MassRadius}
\end{figure}

We use the BSk19 analytic equation of state \citep{2013A&A...560A..48P} for the 
high-density parts ($\varepsilon > \SI{5e5}{\gram\per\centi\metre\cubed}$) of 
the star and the equation-of-state table from \citet{2001A&A...380..151D} for 
the low-density regions 
($\varepsilon \leq \SI{5e5}{\gram\per\centi\metre\cubed}$). We parametrise each 
stellar model according to its central density and integrate 
Eqs.~(\ref{eqs:TOV}) for the background. The background is solved along with 
Eqs.~(\ref{eqs:RelativisticFluidPerturbations}) in the fluid regions of the star 
and Eqs.~(\ref{eqs:RelativisticElasticPerturbations}) in the crust. The results 
of the integrations are summarised in Table~\ref{tab:Results}. The mass and 
radius of each stellar model is presented in Fig.~\ref{fig:MassRadius} to show 
that they are all stable to radial perturbations.

\begin{table}[h]
    \caption[Tidal deformabilities for stars with elastic crusts]{Results of the 
             numerical integrations of the perturbation equations using the 
             BSk19 equation of state for the high-density regions 
             \citep{2013A&A...560A..48P} and the equation of state from 
             \citet{2001A&A...380..151D} for the low-density layers of the star. 
             Each stellar model is determined by the central density 
             $\varepsilon_\text{c}$. We provide the radius $R$, mass 
             $M$, compactness $C$ and crustal thickness $\Delta R_\text{c}$ for 
             each star. The tidal deformability for the fluid stars, 
             $\Lambda_\text{fluid}$, and those with elastic crusts, 
             $\Lambda_\text{crust}$, are shown, along with the relative 
             difference between them, where 
             $\Delta\Lambda \equiv \Lambda_\text{crust} - \Lambda_\text{fluid}$. 
             From the differences between the tidal deformabilities, we see that 
             the correction due to the presence of a crust is very small.}
\resizebox{\textwidth}{!}{
\begin{tabular}{ c c c c r r c c }
	$\varepsilon_\text{c}$ / \SI{e15}{\gram\per\centi\metre\cubed} & $R$ / \si{\kilo\metre} & 
	$M$ / \si{\solarMass} & $C$ & 
	\multicolumn{1}{c}{$\Lambda_\text{crust}$} & 
	\multicolumn{1}{c}{$\Lambda_\text{fluid}$} & 
	$\Delta\Lambda / \Lambda_\text{fluid}$ & 
	$\Delta R_\text{c}$ / \si{\kilo\metre} \\
	\hline 
	\num{2.500} & \num{10.309} & \num{2.162} & \num{0.310} & 
	\num{3.95452368861} & \num{3.95452375613} & \num{-1.707e-08} & \num{0.278} \\
	\num{2.203} & \num{10.548} & \num{2.146} & \num{0.301} & 
	\num{5.47304743630} & \num{5.47304753879} & \num{-1.873e-08} & \num{0.307} \\
	\num{1.941} & \num{10.787} & \num{2.112} & \num{0.289} & 
	\num{8.01552675384} & \num{8.01552692367} & \num{-2.119e-08} & \num{0.343} \\
	\num{1.710} & \num{11.019} & \num{2.056} & \num{0.276} & 
	\num{12.49881579778} & \num{12.49881610913} & \num{-2.491e-08} & \num{0.391} \\
	\num{1.507} & \num{11.234} & \num{1.974} & \num{0.260} & 
	\num{20.87848456088} & \num{20.87848520087} & \num{-3.065e-08} & \num{0.451} \\
	\num{1.327} & \num{11.423} & \num{1.864} & \num{0.241} & 
	\num{37.57928548491} & \num{37.57928697864} & \num{-3.975e-08} & \num{0.529} \\
	\num{1.170} & \num{11.576} & \num{1.725} & \num{0.220} & 
	\num{73.25641923536} & \num{73.25642323791} & \num{-5.464e-08} & \num{0.629} \\
	\num{1.031} & \num{11.686} & \num{1.560} & \num{0.197} & 
	\num{155.28383741339} & \num{155.28384983242} & \num{-7.998e-08} & \num{0.758} \\
	\num{0.908} & \num{11.748} & \num{1.375} & \num{0.173} & 
	\num{358.77773415782} & \num{358.77777902699} & \num{-1.251e-07} & \num{0.926} \\
	\num{0.800} & \num{11.768} & \num{1.178} & \num{0.148} & 
	\num{903.80359991409} & \num{903.80378919034} & \num{-2.094e-07} & \num{1.144} \\
\end{tabular}
}
    \label{tab:Results}
\end{table}

\begin{figure}[h]
	\includegraphics[width=0.49\textwidth]{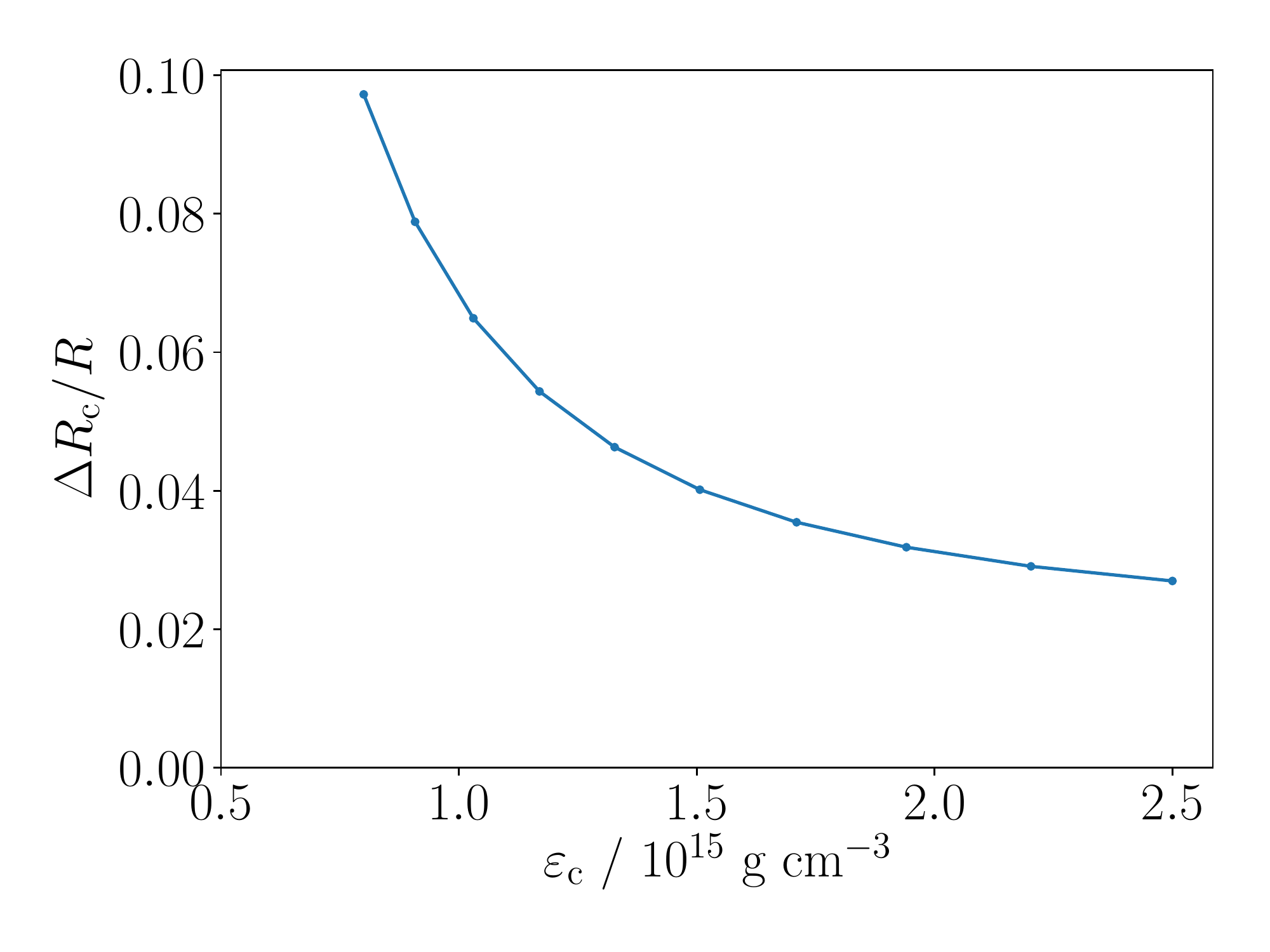}
    \includegraphics[width=0.49\textwidth]{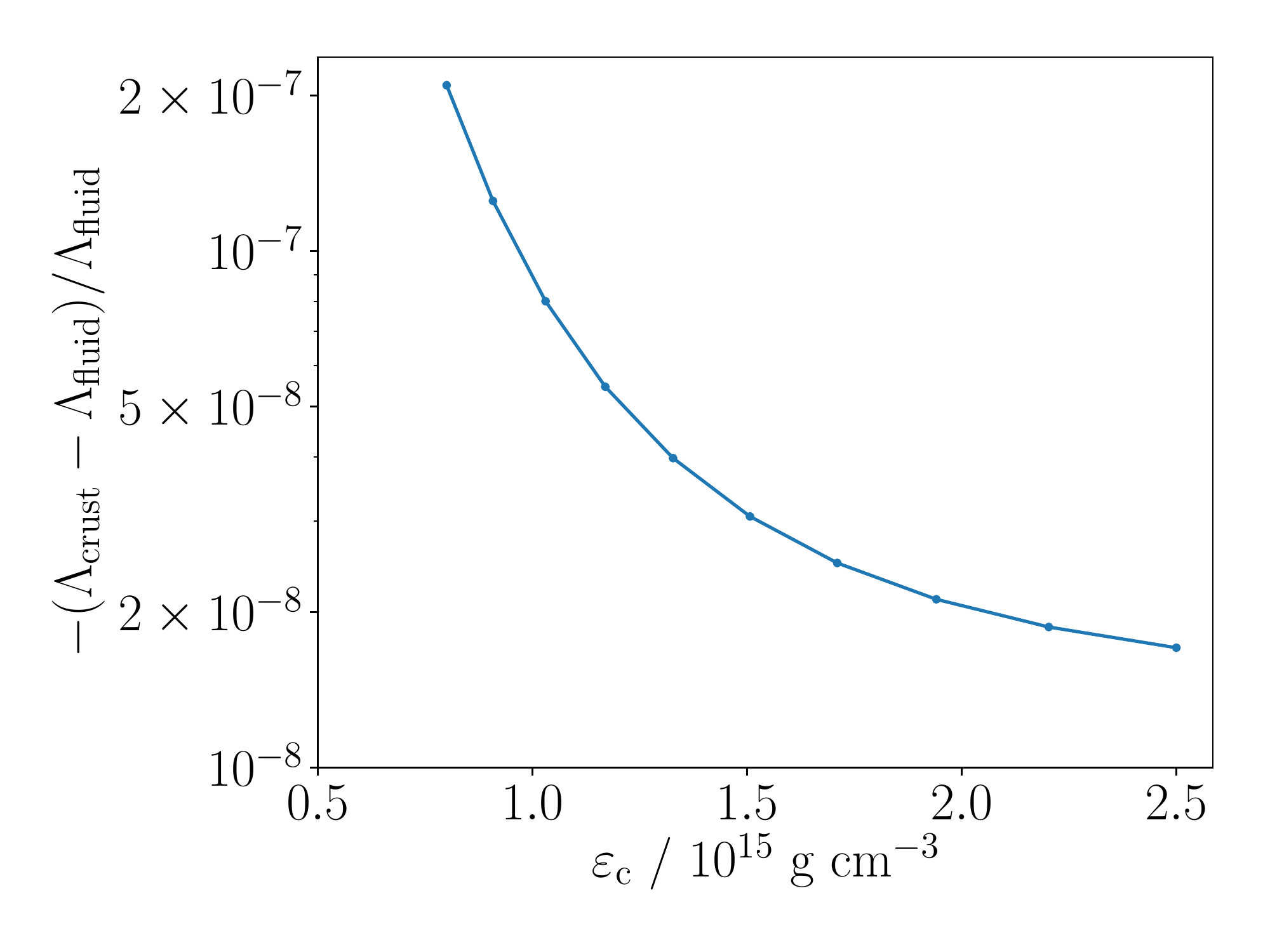}
	\caption[Crustal thickness and tidal deformability]{The ratio of crustal 
             thickness to stellar radius (left panel) 
             and the relative change in tidal deformability due to the 
             presence of a crust (right panel) as functions of central 
             density. As the central density approaches the core-crust 
			 transition (which occurs at 
			 $\varepsilon_\text{base} = \SI{1.3e14}{\gram\per\centi\metre\cubed}$) 
             the crust occupies a much larger fraction of the star, so both 
             quantities become more significant.}
    \label{fig:ThicknessLove}
\end{figure}

For each stellar model, we compute the tidal deformability in the presence of an 
elastic crust, $\Lambda_\text{crust}$, as well as when a crust is not present, 
$\Lambda_\text{fluid}$, for comparison. We also calculate the thickness of the 
crust, $\Delta R_\text{c}$. We show these quantities in 
Fig.~\ref{fig:ThicknessLove} against the central density. In agreement with 
\citet{2011PhRvD..84j3006P}, we find that the inclusion of an elastic crust has 
an almost negligible impact on the tidal deformability -- the correction is the 
largest for the least compact stars at around two parts in $\num{e7}$. This is 
because, as the compactness decreases, the crust takes up a much larger fraction 
of the star. Moreover, as one would expect, the crust works to resist the star's 
deformation which is why the tidal deformabilities computed with a crust are 
smaller.

\begin{figure}[h]
    \centering
    \includegraphics[width=0.7\textwidth]{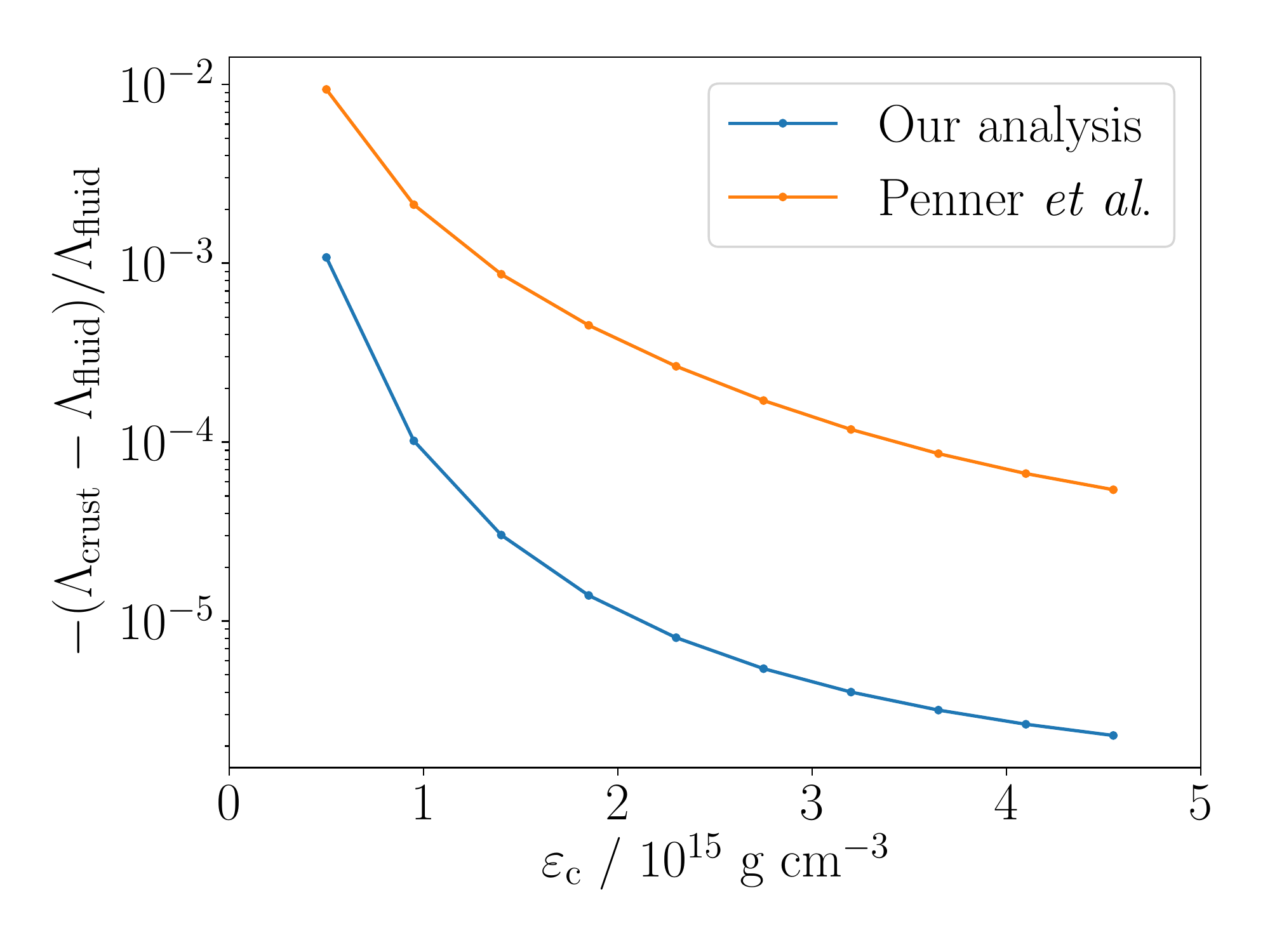}
    \caption[Tidal deformability comparison]{The relative change in the tidal 
             deformability due to the presence 
             of a crust against the central density for a polytropic equation of 
             state with a linear shear modulus. We compare our results (blue) 
             with those of \citet{2011PhRvD..84j3006P} (orange).}
    \label{fig:LoveCompare}
\end{figure}

To facilitate direct comparison with \citet{2011PhRvD..84j3006P} we also 
integrated the perturbation equations with a polytropic equation of state and a 
shear modulus that scales linearly with the pressure. We used the same 
parameters as \citet{2011PhRvD..84j3006P} and moved the core-crust transition to 
$\varepsilon_\text{base} = \SI{2e14}{\gram\per\centi\metre\cubed}$ and the 
crust-ocean transition to 
$\varepsilon_\text{top} = \SI{e7}{\gram\per\centi\metre\cubed}$. The 
result is shown in Fig.~\ref{fig:LoveCompare}. In our calculation, we find that 
the tidal deformability is approximately an order of magnitude less sensitive to 
the inclusion of an elastic crust than reported by \citet{2011PhRvD..84j3006P}. 
This quantifies the effect of the error we corrected in (\ref{eq:PEErr}). (It is 
interesting to note that the crust has a more significant effect in this simple 
model as compared to the results from the realistic equation of state.)

Furthermore, we note that our results are in stark contrast with those of 
\citet{2019PhRvD.100d4056B} who find that the crust can make corrections to the 
tidal deformability of the order of $\sim 1 \%$. The reason for this 
disagreement is twofold. Firstly, \citet{2019PhRvD.100d4056B} write down 
expressions for the components of the displacement vector in the fluid and, 
thus, supposedly compute them in the fluid. This enables them to treat the 
system of coupled ordinary differential equations as an initial-value problem 
for the entire star and they use the continuity of the traction in order to 
match the fluid and elastic regions. As we noted in 
Sec.~\ref{sec:RelativisticFluidPerturbations}, due to the static nature of the 
problem, extracting equations for the components of the displacement vector in 
the fluid is impossible. Additionally, by computing the perturbations as an 
initial-value problem means that one does not have the necessary freedom to 
enforce the traction conditions to be satisfied at the top of the crust, since 
the boundary conditions at the centre and the continuity conditions at the 
core-crust interface are sufficient to carry out the integrations. The second 
reason is due to the fact that \citet{2019PhRvD.100d4056B} do not have an outer 
fluid ocean in their stellar model, but instead have an exposed crust. In such a 
model, $dH_0/dr$ is discontinuous [see (\ref{appeq:H_0primeb})] and, therefore, 
one cannot use (\ref{eq:k_2}) as they do in order to compute the Love number. 
However, we note that the shear modulus at the top of the crust is expected to 
be small and so the discontinuity in $dH_0/dr$ will be small. The difference in 
these results is important. If one assumes that third-generation 
gravitational-wave detectors will be able to constrain $\Lambda$ to within a few 
percent \citep{2020JCAP...03..050M}, then our results show that the effect of 
the crust will not be measurable, which is at odds with the results of 
\citet{2019PhRvD.100d4056B}. 

\section{Crustal failure during inspiral}
\label{sec:CrustFailure}

The formalism described in Sec.~\ref{sec:RelativisticPerturbations} allows us to 
calculate the interior structure of a neutron star with an elastic crust that is 
experiencing static, polar perturbations. We can apply this formalism to 
determine when and where the crust will begin to fracture during a 
binary-neutron-star inspiral, as was done in \citet{2012ApJ...749L..36P}. In 
contrast to the computation of the tidal Love number, the amplitude of the 
perturbations is important for this calculation. Therefore, we must normalise 
our perturbations by matching the interior solution to the exterior at the 
surface and, thus, constrain the amplitude.

We consider a binary separated by distance $d$ where the companion star is of 
mass $M_\text{comp}$. We assume $d \gg r$, as is appropriate in the adiabatic 
regime, and work in the Newtonian limit for the normalisation. By Kepler's third 
law, the angular frequency of the binary $\Omega_\text{orbit}$ is given by 
\begin{equation}
	\Omega_\text{orbit}^2 = \frac{G (M + M_\text{comp})}{d^3}.
    \label{eq:KeplersThirdLaw}
\end{equation}
This is related to the orbital frequency of the binary, $f_\text{orbit}$, by 
$\Omega_\text{orbit} = 2 \pi f_\text{orbit}$. First, let us estimate the 
gravitational-wave frequency at merger for an equal-mass binary, 
$M = M_\text{comp}$. We assume that the point of merger corresponds to when the 
two stars touch, $d = 2 R$, and since gravitational waves radiate at twice the 
orbital frequency, $f_\text{GW} = 2 f_\text{orbit}$, we find 
\begin{equation}
	f_\text{GW}^\text{merger} = \frac{1}{2 \pi} \sqrt{\frac{G M}{R^3}} 
	    \approx \num{2170} \, \left( \frac{M}{\SI{1.4}{\solarMass}} \right)^{1/2} 
	    \left( \frac{R}{\SI{10}{\kilo\metre}} \right)^{-3/2} \, \si{\hertz}. 
    \label{eq:MergerFrequency}
\end{equation}

Expanding around $r = 0$, one can show that the external field due to the 
presence of the companion is 
\begin{equation}
    \Phi_\text{ext}(x^i) = - \frac{G M_\text{comp}}{d} 
        - \frac{G M_\text{comp}}{d^2} r \hat{m}_i \hat{x}^i 
        - \frac{3}{2} \frac{G M_\text{comp}}{d^3} r^2 
        \left( \hat{m}_i \hat{m}_j - \frac{1}{3} g_{i j} \right) 
        \hat{x}^i \hat{x}^j + \order(r^3), 
    \label{eq:ExternalField}
\end{equation}
where $\hat{m}^i$ is the unit vector that points from the centre of the star to 
the centre of the companion. The tidal piece can be expressed using the 
$(\ell, m) = (2, 0)$ spherical harmonic, 
\begin{equation}
    \Phi_\text{tidal}(x^i) 
        = - \sqrt{\frac{4 \pi}{5}} \frac{G M_\text{comp}}{d^3} r^2 Y_{2 0}.
\end{equation}
This means that the binary is orientated such that $\theta = 0$ points in the 
direction of $\hat{m}^i$. The tidal multipole in the Newtonian limit is given by, 
using (\ref{eq:ExternalField}),
\begin{equation}
    \mathcal{E}_{i j} 
        = \frac{\partial^2 \Phi_\text{ext}}{\partial x^i \partial x^j} 
        = - 3 \frac{G M_\text{comp}}{d^3} 
        \left( \hat{m}_i \hat{m}_j - \frac{1}{3} g_{i j} \right).
\end{equation}
Using the decomposition of (\ref{eqs:Decomposition}) one can show that the 
non-vanishing $\mathcal{E}_{2 m}$ is 
\begin{equation}
	\mathcal{E}_{2 0} = - \frac{3}{2} \sqrt{\frac{5}{\pi}} 
        \frac{G M_\text{comp}}{d^3},
\end{equation}
and, therefore, by (\ref{eq:c_2a}) we find 
\begin{equation}
	c_2 = \frac{4}{3} \sqrt{\frac{\pi}{5}} 
        \frac{G^3 M^2 M_\text{comp}}{c^6 d^3} 
	    = \frac{4 \pi^2}{3} \sqrt{\frac{\pi}{5}} 
	    \frac{G^2 M^2 M_\text{comp}}{c^6 (M + M_\text{comp})} f_\text{GW}^2.
    \label{eq:c_2b}
\end{equation}
Here we have chosen to parametrise the point in the inspiral by the 
gravitational-wave frequency over the separation by using 
(\ref{eq:KeplersThirdLaw}). Equations~(\ref{eq:Ratio}) and (\ref{eq:c_2b}) 
provide the necessary information to normalise the perturbations to a binary 
that is emitting gravitational waves with frequency $f_\text{GW}$.

We use the von Mises criterion to determine when the crust begins to break. 
We obtain the von Mises strain, where the unperturbed configuration is 
unstrained, through (\ref{eq:RelativisticvonMisesStrain}). The crust fractures 
when the von Mises strain reaches the threshold yield point. Using the 
definition of the traction variables (\ref{eqs:RelativisticTractionVariables}) 
with (\ref{eq:Strain}) and specialising to $(\ell, m) = (2, 0)$ perturbations, 
one finds 
\begin{equation}
    \Theta^2 = \frac{45}{256 \pi} \frac{1}{r^4} 
    \left[ (3 \cos^2 \theta - 1)^2 \left( \frac{T_1}{\check{\mu}} \right)^2 
    + 12 e^{-\lambda} \sin^2(2 \theta) \left( \frac{T_2}{\check{\mu}} \right)^2 
    + 48 \sin^4 \theta V^2 \right].
    \label{eq:RelativisticvonMises}
\end{equation}
The advantage of using the von Mises strain is that it is a function of 
position, and so we can identify where the crust is the weakest as well as when 
it breaks. Taking the breaking strain to be $\Theta^\text{break} = 0.1$ 
\citep{2009PhRvL.102s1102H}, we can calculate when the crust will break, at each 
point, by imposing that the strain in (\ref{eq:RelativisticvonMises}) is equal 
to $\Theta^\text{break}$ to normalise the perturbations and then determining the 
gravitational-wave frequency $f_\text{GW}^\text{break}$ which corresponds to 
that amplitude using (\ref{eq:Ratio}) and (\ref{eq:c_2b}).

\begin{figure}[h]
    \centering
	\includegraphics[width=0.9\textwidth]{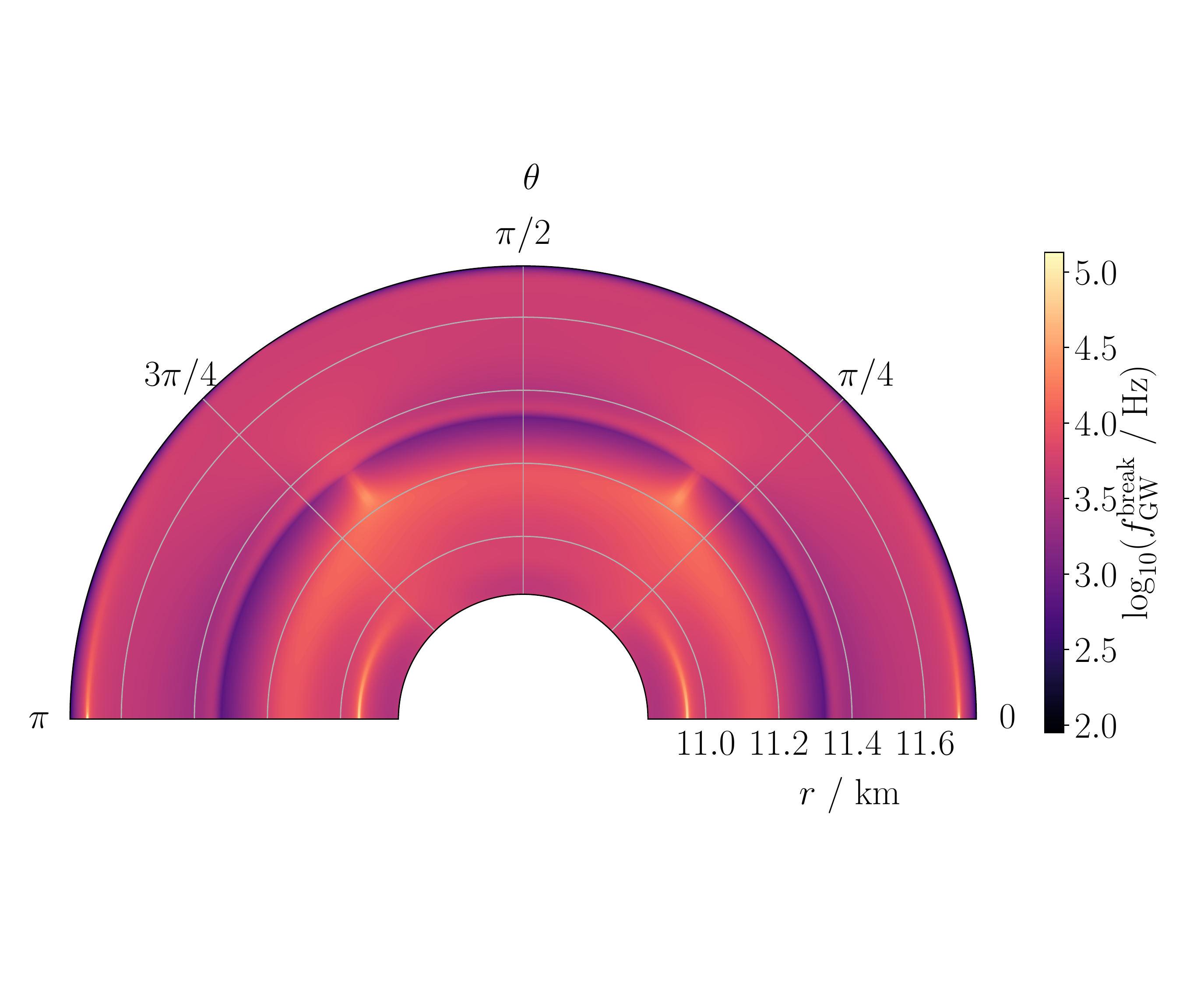}
    \caption[Crustal failure]{The gravitational-wave frequency at failure across 
             the elastic 
             crust. We can see that the majority of the crust will not fail 
             before merger.}
    \label{fig:BreakingFrequency}
\end{figure}

\begin{figure}[h]
    \centering
	\includegraphics[width=0.9\textwidth]{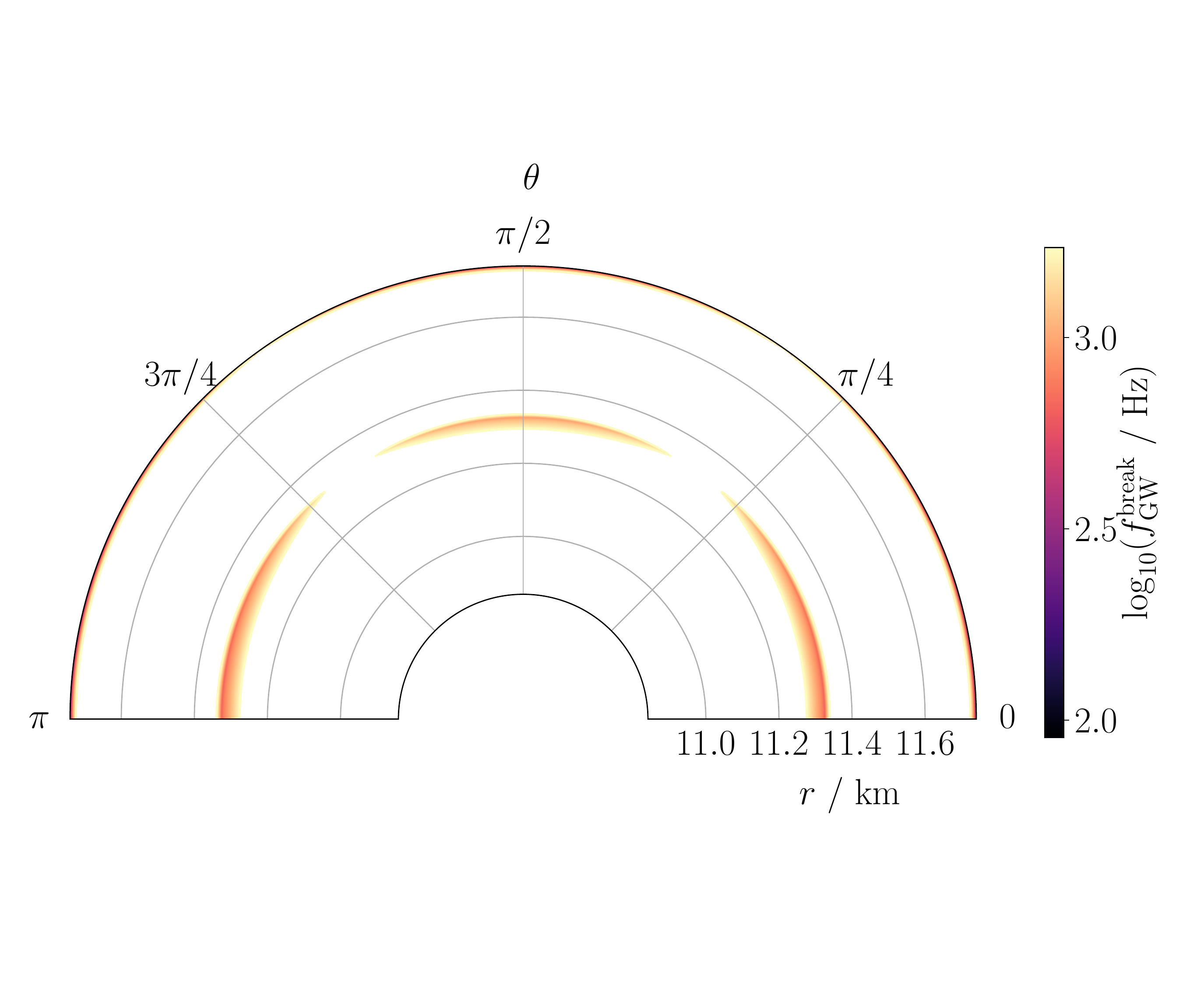}
    \caption[Saturated crustal failure]{The gravitational-wave frequency at 
             failure for the locations in 
             the elastic crust that yield before merger. This shows that the 
             stars will merge with the crust largely intact.}
    \label{fig:BreakingFrequencySaturate}
\end{figure}

\begin{figure}[h]
    \centering
	\includegraphics[width=0.7\textwidth]{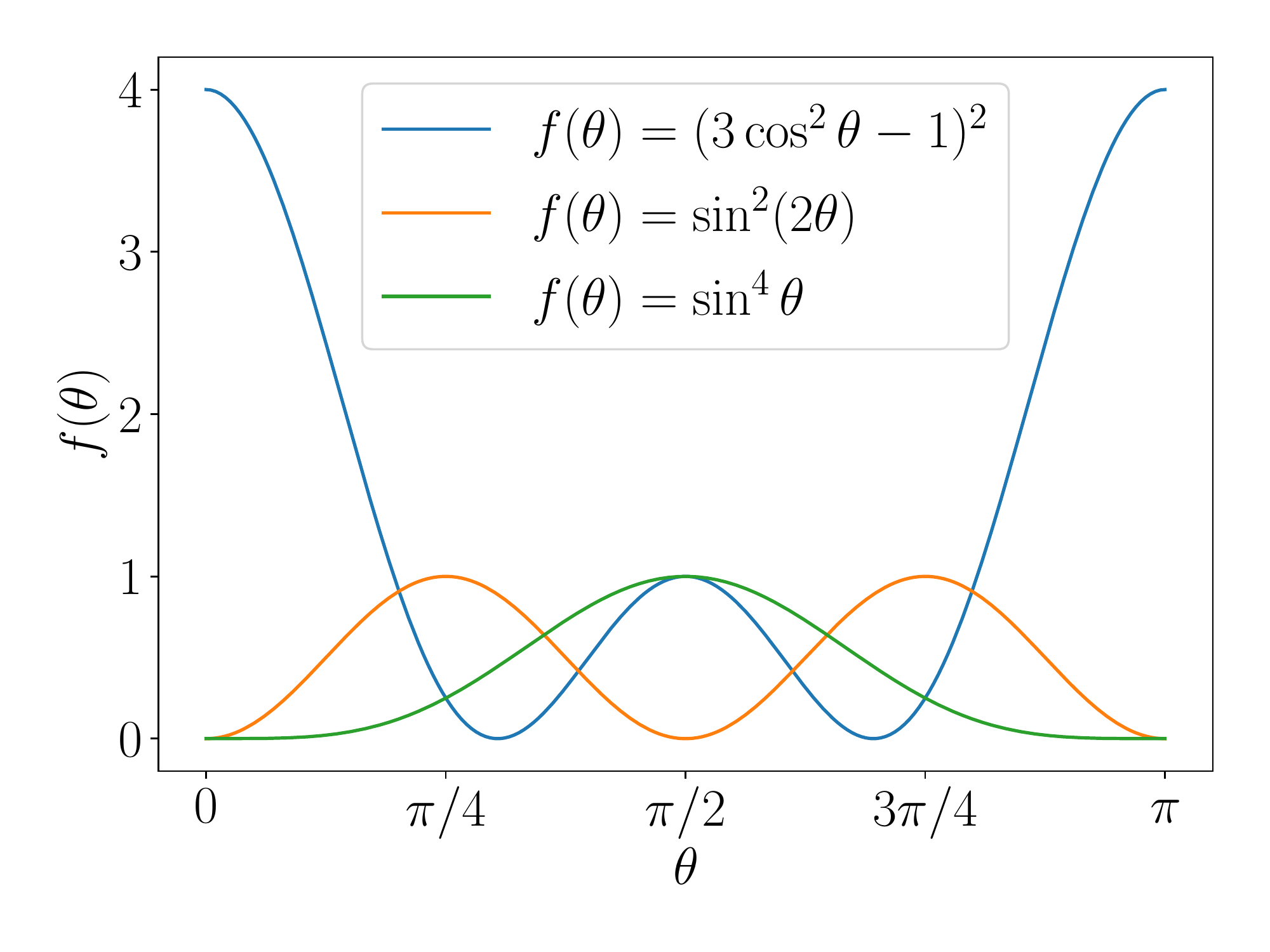}
    \caption[Angular basis for the von Mises strain]{The angular basis of the 
             von Mises strain for $(\ell, m) = (2, 0)$ perturbations.}
    \label{fig:AngularBasis}
\end{figure}

\begin{figure}[h]
    \centering
	\includegraphics[width=0.7\textwidth]{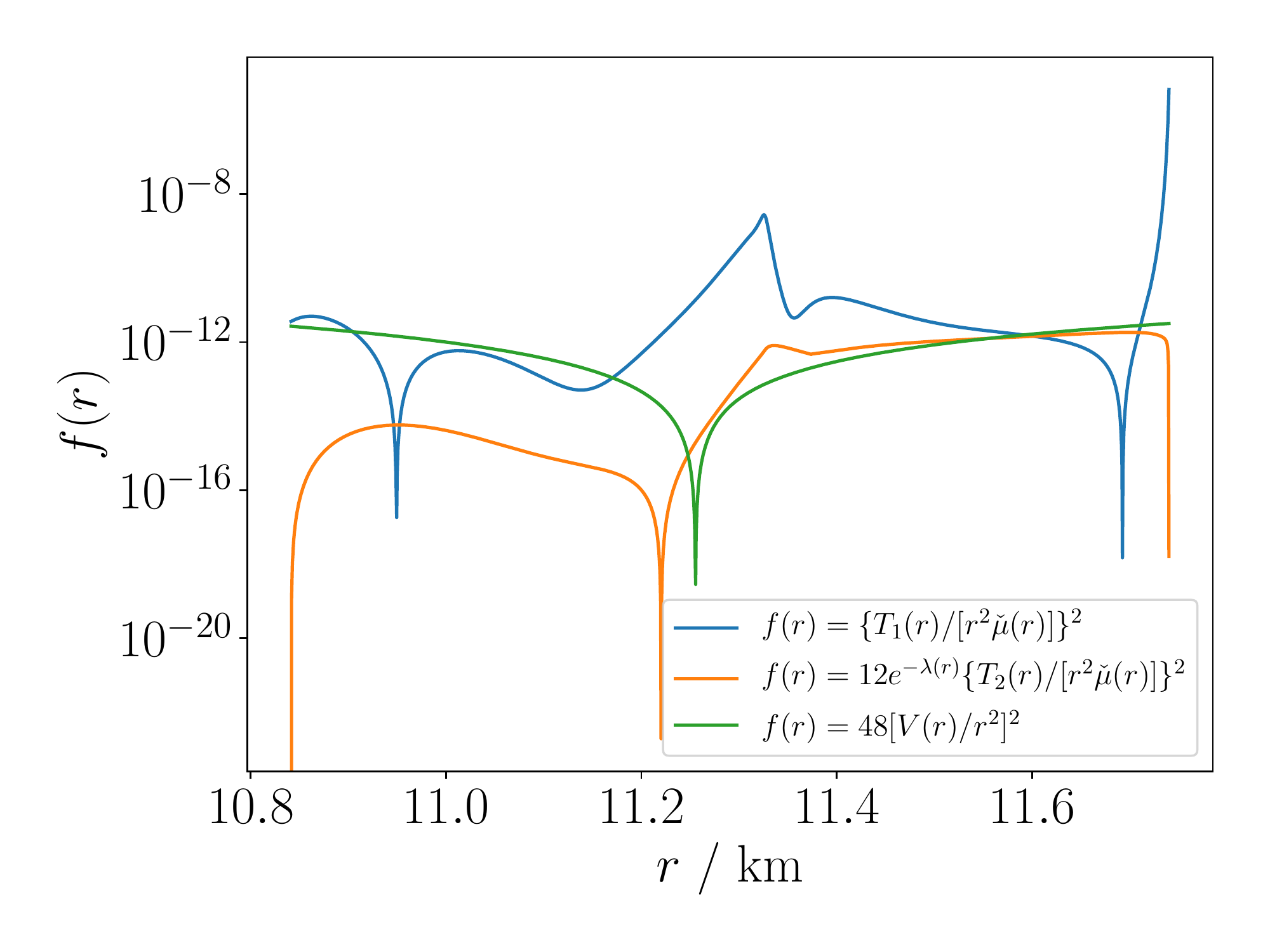}
    \caption[Traction and tangential displacement in the crust]{The radial 
             dependence of the radial and tangential traction 
             variables and the tangential displacement function normalised to a 
             binary radiating gravitational waves with 
             $f_\text{GW} = \SI{10}{\hertz}$. At neutron drip, around 
             $r = \SI{11.3}{\kilo\metre}$, we find that the tangential functions 
             combine to give local minima for most values of $\theta$. Note that 
             the vertical axis is in logarithmic scale -- the cusps correspond to 
             when the functions cross zero and change sign.}
    \label{fig:RadialBasis}
\end{figure}

As an illustration, we use the same equation of state as in 
Sec.~\ref{sec:Impact}. We assume the binary is equal mass with 
$M = M_\text{comp} = \SI{1.4}{\solarMass}$, for which we obtain a star with 
radius $R = \SI{11.74}{\kilo\metre}$. In Fig.~\ref{fig:BreakingFrequency} we 
show the gravitational-wave frequency when the crust breaks at each point. 
Figure~\ref{fig:BreakingFrequencySaturate} focuses on the regions of the star 
that break before merger. There is a clear phase transition at neutron drip 
(around $r = \SI{11.3}{\kilo\metre}$), where the inner crust is, on average, 
stronger than the outer crust. The crust is notably strong at neutron drip 
close to $\theta \approx \pi / 4$ and $3 \pi / 4$. The reason for this is, as 
the star becomes more oblate, the parts closest to the poles and equator are 
stretched the most. The region that stretches the least is in between these two 
regions at $\theta \approx \pi / 4$ and $3 \pi / 4$. This effect can be seen in 
Figs.~\ref{fig:AngularBasis} and \ref{fig:RadialBasis} where the tangential 
functions $T_2 / \check{\mu}$ and $V$ combine to give a local minimum in the von 
Mises strain at these angles and, thus, a local maximum in the breaking 
frequency. The maxima along the equator, $\theta = 0$ and $\pi$, are where 
the magnitude of the radial traction function $T_1 / \check{\mu}$ reaches a 
local minimum (as shown in Fig.~\ref{fig:RadialBasis}).

We note that, as compared to typical merger frequencies 
(\ref{eq:MergerFrequency}), our results suggest that the vast majority of the 
crust will not fracture before merger. In fact, the crust will only fail at 
neutron drip and in the very outermost part of the crust before coalescence. 
This is in contrast to the results of \citet{2012ApJ...749L..36P}, who obtain 
significantly lower breaking frequencies throughout the crust. This is likely 
related to the errors in the analysis of \citet{2011PhRvD..84j3006P} that we 
have pointed out above. Since the crust will be mostly intact by the point of 
merger, there is unlikely to be a significant amount of strain energy released 
available for an associated electromagnetic signal. However, it has been 
suggested that the resonant excitation of oscillation modes may lead to 
electromagnetic flares, although this is a different mechanism 
\citep{2012PhRvL.108a1102T}.

\section{Summary}
\label{sec:Summary6}

With the advent of gravitational-wave detections of binary-neutron-star mergers, 
we have a promising new method of constraining the equation of state of nuclear 
matter. The gravitational waveforms from these events are sensitive to tidal 
effects in the binaries which carry model-independent information on the 
equation of state.

In this chapter, we have explored the impact of an elastic crust on tidal 
deformations of neutron stars. We used the formalism detailed in 
Sec.~\ref{sec:RelativisticPerturbations} which enables one to compute static, 
polar perturbations of a neutron star with an elastic component. This was 
necessary to resolve discrepancies between previous studies 
\citep{2011PhRvD..84j3006P, 2019PhRvD.100d4056B}. There are mistakes in the 
crustal perturbation equations presented by \citet{2011PhRvD..84j3006P}, in 
particular arising from the analogous equation to (\ref{eq:PEErr}). This meant 
they marginally overestimated the impact of a crust on tidal deformations and 
consequently this affected their analysis on when the crust will break in a 
binary inspiral \citep{2012ApJ...749L..36P}. Meanwhile, the work of 
\citet{2019PhRvD.100d4056B} calculates the static displacement vector in the 
fluid regions of the star. However, such a calculation should not be possible 
due to the static nature of the problem. This means that they cannot correctly 
impose continuity of the traction at the top of the crust. Moreover, 
\citet{2019PhRvD.100d4056B} do not correctly calculate the tidal Love number 
for the assumed stellar model with an exposed crust.

We have applied our formalism to the computation of static, quadrupolar 
perturbations of a neutron star sourced by an external tidal field. We 
calculated the quadrupolar perturbations for a realistic equation of state that 
includes an elastic crust. We have shown that the inclusion of an elastic crust 
has a very small effect on the tidal deformability of a star, in the range of 
$\sim \numrange{e-8}{e-7}$ for realistic models -- even smaller than what one 
would calculate using simplistic equations of state. We found that our results 
are an order of magnitude smaller than what was reported by 
\citet{2011PhRvD..84j3006P} and significantly smaller than the results of 
\citet{2019PhRvD.100d4056B}. This means the impact of a crust on 
binary-neutron-star mergers is not expected to be detectable for current and 
next-generation gravitational-wave detectors.

We used our integrations to calculate when and where the crust would fail during 
a binary inspiral with component masses 
$M = M_\text{comp} = \SI{1.4}{\solarMass}$. We found that the crust is much 
stronger than estimated in previous work \citep{2012ApJ...749L..36P}. Only the 
small regions close to neutron drip and the outer layers of the crust will 
fracture before merger.


%% file: sections/chapter-7.tex
\chapter{Conclusions}
\label{ch:Conclusions}

Neutron stars are weird and wonderful objects. We have been aware of their 
existence for over half a century and yet there is still much more we can learn 
about them. Their status as the most compact state of matter (that we know of) 
means they provide a remarkable opportunity to study various aspects of 
cutting-edge physics. Perhaps the most exciting prospect is to uncover details 
about the ultra-dense equation of state.

With the aid of kilometre-long gravitational-wave interferometers, we have 
witnessed some of the Universe's most impressive sights: the coalescence of 
black holes and neutron stars. These events had never been seen before. The 
first observation of a binary-neutron-star inspiral and merger presented the 
opportunity to constrain the equation of state in a uniquely model-independent 
fashion. The hope is that we will continue to detect binary neutron stars and 
eventually start observing other neutron-star scenarios with gravitational-wave 
instruments.

In this thesis, we have studied how neutron stars can be deformed with the 
presence of a mountain and by the tidal field of a companion star. Both these 
situations can give rise to gravitational radiation.

Although we are yet to detect rotating neutron stars with gravitational waves, 
there are other signatures we can look for to get an indication of whether we 
should expect such radiation to play a significant role in their dynamics. An 
unresolved problem in the study of accreting neutron stars is the fact that none 
of the observed systems spin close to the centrifugal break-up frequency. This 
suggests that there must be some (as yet unaccounted for) mechanism acting in 
these systems that takes angular momentum away from the star. 

We considered whether gravitational-wave emission could explain the features of 
the observed spin distribution of accreting neutron stars in 
Chap.~\ref{ch:PopulationSynthesis}. To answer this question, we simulated 
populations of accreting neutron stars. We used a model for the torques acting 
on the neutron star that accounted for accretion and magnetic-field effects. We 
also worked with persistent and transient accretion. Promisingly, we found that 
one could recreate the observed spin distribution with the inclusion of 
gravitational-wave torques. In addition to a permanent deformation, we showed 
that one could also obtain this behaviour from thermal mountains and unstable 
\textit{r}-modes. However, based on the distributions alone, it is difficult to 
distinguish between the competing mechanisms.

A natural extension of this study would be to try to connect the 
accreting-neutron-star population with the population of rapidly rotating radio 
pulsars. In particular, it would be interesting to see whether gravitational 
waves are relevant in this process and can obtain the spin distribution of 
radio pulsars. Towards building a realistic model for the spin evolution, one 
could consider a more sophisticated treatment for transient accretion that takes 
into account variable outburst/quiescent cycles.

In this work, we obtained the observed spin distribution for accreting neutron 
stars with a permanent quadrupole moment of 
$Q_{2 2} = \SI{e36}{\gram\centi\metre\squared}$ (corresponding to an 
ellipticity of $\epsilon \approx \num{e-9}$). We found similar results for 
thermal mountains with the fractional quadrupolar heating of 
$\delta T_\text{q} / \delta T = \num{4e-4}$ and unstable \textit{r}-modes with 
amplitude $\alpha = \num{e-7}$. Understanding why there is a preference for 
these values will require further investigation.

In light of this result, we went on to consider how neutron-star mountains are 
calculated in Newtonian gravity in Chap.~\ref{ch:Mountains}. Over the past two 
decades, there have been a number of studies on the maximum mountain that a 
neutron star can support before the elastic crust fractures. This is a relevant 
question since such a result would provide interesting context to the upper 
limits on these deformations that have been obtained thus far, as well as give 
an indication of the maximum spin-frequency pulsars should be able to obtain. 
However, previous maximum-mountain studies had issues with satisfying the 
boundary conditions of the problem. 

In order to resolve these issues, we developed a new scheme for calculating 
mountains on neutron stars that explicitly includes the fiducial force which 
gives the star its non-spherical shape. By introducing the force, we have full 
control over the boundary conditions. One of the advantages of using our scheme 
is that one can compute all the relevant quantities. But, one must prescribe the 
deforming force. In reality, the force that gives rise to the mountains is 
(presently) unknown. Indeed, to make progress on this, future evolutionary 
calculations will be necessary to motivate the form of the force. As a proof of 
principle, we calculated the mountains produced by a few example forces. We 
found that the maximum mountains the considered forces produced were in the 
range of a factor of a few to two orders of magnitude below previous estimates, 
$\num{1.7e37} \leq Q_{2 2} \ / \ \si{\gram\per\centi\metre\squared} \leq 
\num{4.4e38}$ ($\num{2.2e-8} \leq \epsilon \leq \num{5.7e-7}$). 
This was not a surprising result since, in prescribing the force, the breaking 
strain was reached at a point -- rather than throughout the entire crust, which 
previous studies assumed.

In Chap.~\ref{ch:RelativisticMountains}, to construct more realistic models, we 
generalised the Newtonian scheme for neutron-star mountains to relativistic 
gravity. In addition to considering how the mountains depend on the deforming 
force in relativity, we also were able to explore how the equation of state 
impacted the size of the mountains. As has been noted for tidal- and 
magnetic-deformation calculations, we observed a suppression in the magnitude of 
the mountains in going from Newtonian to relativistic gravity. We also found 
that the shear modulus of the crust is important in supporting the mountains: in 
particular, for the sources we looked at, the shear modulus at the top of the 
crust was a limiting factor. Finally, we demonstrated how the mountains are 
sensitive to the fluid pressure-density relationship. For equations of state 
that satisfy current observational constraints, the deformations spanned a range 
of approximately an order of magnitude for a given deforming force.

In studying mountains in both Newtonian gravity and full general relativity, we 
have seen that there is a great deal of uncertainty in what the \textit{maximum} 
mountain is that the neutron-star crust can sustain. It is clear that the 
formation history of the star plays a crucial role in how large the mountains 
can be. Additionally, the equation of state of the neutron-star matter is also 
important.

A potentially promising direction for future work is what role plasticity may 
play in maximally strained neutron-star mountains. Plastic materials hold on to 
some of the strain when the yield limit is reached. It is possible that plastic 
flow in the crust could enable the star to attain a higher strain configuration 
and perhaps a larger quadrupole moment.

In Chap.~\ref{ch:Tides}, we went on to study a scenario in which the deforming 
force is well known. We explored the impact of an elastic crust in tidal 
deformations of neutron stars in general relativity. With the recent detections 
of binary neutron stars, it proves timely to go beyond the usual approximation 
of neutron stars as perfect-fluid bodies and consider the role of the crust. 

We found that the crust has a very small effect on the tidal deformability of a 
star, in the range of $\sim \numrange{e-8}{e-7}$. Such an effect is beyond the 
expected sensitivity of third-generation gravitational-wave detectors. We also 
computed when and where the crust would fail during a equal-mass binary inspiral 
with component masses of \SI{1.4}{\solarMass}. In comparison to previous work, 
we found that the crust is (perhaps surprisingly) robust and the majority of it 
will remain intact up until merger.

With the promise of more detections of binary-neutron-star inspirals, similar 
studies -- extending the description of neutron stars beyond the standard 
perfect-fluid approximation -- would be worthwhile. It would be interesting to 
study whether superfluidity and magnetic fields could leave measurable imprints 
on the waveform.

With the continued improvements of current-generation detectors and the 
developments of new, more-sensitive interferometers, alongside our improving 
theoretical understanding, there are good reasons to be optimistic about 
observing novel neutron-star scenarios using gravitational waves. However, in 
the meantime, there is plenty of work for us to do. The future is bright in 
gravitational waves.


%% file: sections/appendix-A.tex
\chapter{Solving the equations of stellar structure}
\label{app:Numerical}

This appendix provides additional information on solving the equations of 
structure for non-rotating, spherical stellar models.

\section{Polytropes}
\label{appsec:Polytropes}

The polytropic relation is a popular idealisation for barotropic equations of 
state:
\begin{equation}
	p(\rho) = K \rho^{1 + 1/n},
	\label{appeq:Polytrope}
\end{equation}
where $K$ is the polytropic constant and $n$ is the polytropic index.

One can take the divergence of the equation of hydrostatic equilibrium 
(\ref{eq:HydrostaticEquilibrium}) to write 
\begin{equation}
	\frac{1}{r^2} \frac{d}{dr}\left( \frac{r^2}{\rho} \frac{dp}{dr} \right) 
		= - 4 \pi G \rho.
	\label{appeq:HydrostaticEquilibrium}
\end{equation}
This equation may be written in dimensionless form by introducing the following 
variables:
\begin{equation}
	\rho = \rho_\text{c} \theta^n, \quad r = a \xi,
\end{equation}
where $\theta$ is the polytropic temperature, $\xi$ is the dimensionless length 
and $a$ is a constant with dimensions of length. From the equation of 
state~(\ref{appeq:Polytrope}), 
\begin{equation}
	p = p_\text{c} \theta^{n + 1},
\end{equation}
where $p_\text{c} = K \rho_\text{c}^{1 + 1/n}$. One chooses $a$ in order to 
simplify the differential equation,
\begin{equation}
	a = \sqrt{\frac{(n + 1) K \rho_\text{c}^{1/n - 1}}{4 \pi G}}.
\end{equation}
Thus, (\ref{appeq:HydrostaticEquilibrium}) becomes 
\begin{equation}
	\frac{1}{\xi^2} \frac{d}{d\xi} \left( \xi^2 \frac{d\theta}{d\xi} \right) 
		= - \theta^n.
	\label{appeq:LaneEmden}
\end{equation}
Equation~(\ref{appeq:LaneEmden}) is called the \textit{Lane-Emden equation}. The 
Lane-Emden equation has the boundary conditions at the centre, $\xi = 0$, 
\begin{subequations}
\begin{equation}
	\theta(0) = 1
	\label{appeq:thetaCentre}
\end{equation}
and 
\begin{equation}
	\frac{d\theta}{d\xi}(0) = 0.
	\label{appeq:dtheta_dxiCentre}
\end{equation}
\end{subequations}
Equation~(\ref{appeq:thetaCentre}) follows from the fact 
$\rho = \rho_\text{c} \theta^n$. Equation~(\ref{appeq:dtheta_dxiCentre}) is 
obtained by examining (\ref{eq:dp_dr}) near the centre, noting that 
$m(r) \approx 4 \pi \rho_\text{c} r^3 / 3$ in this region.

A solution of (\ref{appeq:LaneEmden}) depends on one parameter only: the 
polytropic index $n$. The other two constants $K$ and $\rho_\text{c}$ are 
scaling parameters that provide dimensions to the physical star. The Lane-Emden 
equation can be integrated outwards from $\xi = 0$ until $\theta$ goes to zero at 
some $\xi_1$. This defines the surface of the polytrope. The physical radius is, 
thus, $R = a \xi_1$ and the total mass is $M = \int_0^R 4 \pi r^2 \rho dr$. 
Therefore, one can readily show that 
\begin{equation}
	R = \left[ \frac{(n + 1) K}{4 \pi G} \right]^{1/2} 
		\rho_\text{c}^{(1 - n)/(2 n)} \xi_1
\end{equation}
and 
\begin{equation}
	M = 4 \pi \left[ \frac{(n + 1) K}{4 \pi G} \right]^{3/2} 
		\rho_\text{c}^{(3 - n)/(2 n)} \xi_1^2 
		\left| \frac{d\theta}{d\xi}(\xi_1) \right|.
\end{equation}
For a solution to a given polytropic index, one can choose $K$ and 
$\rho_\text{c}$ to give a specific $R$ and $M$.

The Lane-Emden equation admits three analytic solutions: $n = 0$, which 
corresponds to an incompressible fluid with $\rho = \text{const}$; $n = 1$, 
where $p \propto \rho^2$; and $n = 5$, known as the Roche model, which has an 
infinite extent. For $n = 1$ polytropes, which are useful for approximating 
Newtonian neutron stars, the polytropic temperature is given by 
\begin{equation}
	\theta(\xi) = \frac{\sin \xi}{\xi}
\end{equation}
and the polytrope has its first root at $\xi_1 = \pi$.

\section{Numerical integration of relativistic stars}
\label{appsec:NumericalTOV}

We have the equations of stellar structure (\ref{eqs:TOV}), 
\begin{subequations}\label{appeqs:TOV}
\begin{gather}
	\frac{dm}{dr} = 4 \pi r^2 \varepsilon, \\
	\frac{dp}{dr} = - \frac{(\varepsilon + p) (m + 4 \pi r^3 p)}{r (r - 2 m)}
\end{gather}
and
\begin{equation}
	\frac{d\nu}{dr} = \frac{2(m + 4 \pi r^3 p)}{r (r - 2 m)},
\end{equation}
\end{subequations}
along with a barotropic equation of state, $p = p(\varepsilon)$. 

In order to solve the coupled system of differential equations 
(\ref{appeqs:TOV}), we must provide boundary conditions. 
We choose some central value for the energy density $\varepsilon_\text{c}$ 
which has the corresponding pressure $p_\text{c} = p(\varepsilon_\text{c})$ at 
$r = 0$. At the stellar centre, the mass is zero, $m(0) = 0$. The surface of the 
star is defined as when the pressure vanishes, $p(R) = 0$. At this point, the 
interior metric must match smoothly to the exterior, 
$e^{\nu(R)} = 1 - 2 M / R$. Usefully, because the metric potential neatly 
decouples from the other two equations, one is free to assign a constant to 
$\nu(0)$ and then correct this constant at the surface by satisfying the 
boundary condition.

The equations of stellar structure (\ref{appeqs:TOV}) 
present an initial-value problem. The numerical integration of these equations 
should begin at the centre, $r = 0$, with initial values for $m$, $p$ and 
$\nu$ specified, and integrate through the star until $p = 0$. This determines 
the surface, $r = R$.

However, one should note that Eqs.~(\ref{appeqs:TOV}) 
are singular at the origin. To begin integrating, one must start at a small 
value $r$ away from the origin. The initial values for $m$ and $p$ are obtained 
by considering power-series expansions around $r = 0$, \textit{e.g.},
\begin{equation}
	m(r) = \sum_{\alpha = 0}^\infty \frac{1}{\alpha!} m_\alpha r^\alpha,
\end{equation}
inserting these expansions into Eqs.~(\ref{appeqs:TOV}) 
and comparing powers of $r$. One finds that the even powers of $r$ in $m(r)$ 
vanish, as do the odd powers in $\varepsilon(r)$ and $p(r)$. For the non-zero 
components, one finds 
\begin{subequations}
\begin{align}
	p_0 &= p_\text{c}, \\
	\varepsilon_0 &= \varepsilon_\text{c}, \\
	p_2 &= - \frac{4 \pi}{3} (\varepsilon_\text{c} + p_\text{c}) 
		(\varepsilon_\text{c} + 3 p_\text{c}), \\
	\varepsilon_2 &= \frac{p_2}{c_\text{s}^2(0)}, \\
	m_3 &= 8 \pi \varepsilon_\text{c}, \\
	m_5 &= 48 \pi \varepsilon_2, 
\end{align}
\end{subequations}
where $c_\text{s}^2 = dp / d\varepsilon$ is the squared sound speed and 
$c_\text{s}^2(0)$ is its central value. Therefore, 
\begin{subequations}
\begin{align}
	m(r) &= \frac{4 \pi}{3} \varepsilon_\text{c} r^3 
		- \frac{8 \pi^2}{15 c_\text{s}^2(0)} 
		(\varepsilon_\text{c} + p_\text{c}) 
		(\varepsilon_\text{c} + 3 p_\text{c}) r^5 + \order(r^7), \\
	p(r) &= p_\text{c} - \frac{2 \pi}{3} (\varepsilon_\text{c} + p_\text{c}) 
		(\varepsilon_\text{c} + 3 p_\text{c}) r^2 + \order(r^4), \\
	\varepsilon(r) &= \varepsilon_\text{c} 
		- \frac{2 \pi}{3 c_\text{s}^2(0)} 
		(\varepsilon_\text{c} + p_\text{c}) 
		(\varepsilon_\text{c} + 3 p_\text{c}) r^2 + \order(r^4).
\end{align}
\end{subequations}
The value $\nu(r)$ near the origin is arbitrary because it will be corrected at 
the surface.


%% file: sections/appendix-B.tex
\chapter{The spherical harmonics}
\label{app:Harmonics}

This appendix provides a brief introduction to the well-known (and very useful) 
spherical harmonics. Traditionally, one meets the spherical harmonics by 
examining the solutions of Laplace's equation. In Newtonian flat space, the 
Laplacian $\nabla^2$ of a scalar quantity $f$ is given by 
\begin{equation}
	\nabla^2 f = \frac{1}{r^2} \frac{\partial}{\partial r} 
		\left( r^2 \frac{\partial f}{\partial r} \right) 
		+ \frac{1}{r^2 \sin \theta} \frac{\partial}{\partial \theta} 
		\left( \sin \theta \frac{\partial f}{\partial \theta} \right) 
		+ \frac{1}{r^2 \sin^2 \theta} \frac{\partial^2 f}{\partial \phi^2}.
\end{equation}
We look for solutions of Laplace's equation, 
\begin{equation}
	\nabla^2 f = 0,
\end{equation}
using separation of variables, 
$f(r, \theta, \phi) = R(r) \Theta(\theta) \Phi(\phi)$. Thus, we obtain 
\begin{equation}
	\frac{1}{R} \frac{d}{dr} \left( r^2 \frac{dR}{dr} \right) 
		+ \frac{1}{\Theta} \frac{1}{\sin \theta} \frac{d}{d \theta} 
		\left( \sin \theta \frac{d \Theta}{d \theta} \right) 
		+ \frac{1}{\Phi} \frac{1}{\sin^2 \theta} \frac{d^2 \Phi}{d \phi^2} = 0.
	\label{appeq:Laplaces}
\end{equation}

Suppose we rearrange (\ref{appeq:Laplaces}) so that all terms with a $\phi$ 
dependence sit on the right-hand side. Then, we note that the left-hand side 
depends on $r$ and $\theta$. This implies that we have a constant $\alpha$ so 
that 
\begin{equation}
	\frac{\sin^2 \theta}{R} \frac{d}{dr} \left( r^2 \frac{dR}{dr} \right) 
		+ \frac{\sin \theta}{\Theta} \frac{d}{d \theta} 
		\left( \sin \theta \frac{d \Theta}{d \theta} \right) 
		= - \frac{1}{\Phi} \frac{d^2 \Phi}{d \phi^2} 
		= \alpha.
	\label{appeq:alpha}
\end{equation}
We can solve (\ref{appeq:alpha}) for $\Phi$ to find 
\begin{equation}
	\Phi(\phi) \propto e^{\sqrt{- \alpha} \phi}.
\end{equation}
Because $\phi$ is a periodic variable, $\Phi(\phi + 2 \pi) = \Phi(\phi)$, this 
implies that $\alpha = m^2$, where $m$ is an integer. If $m = 0$, then the 
general solution is $\Phi(\phi) = a \phi + b$, with $a$ and $b$ as integration 
constants. Periodicity means that $a = 0$, therefore, only one solution is 
allowed. 

Using similar logic as above, we put terms that depend on $r$ on one side and 
terms that depend on $\theta$ on the other, 
\begin{equation}
	\frac{1}{R} \frac{d}{dr} \left( r^2 \frac{dR}{dr} \right) 
		= - \frac{1}{\Theta} \frac{1}{\sin \theta} \frac{d}{d \theta} 
		\left( \sin \theta \frac{d \Theta}{d \theta} \right) 
		+ \frac{m^2}{\sin^2 \theta} = \beta,
	\label{appeq:beta}
\end{equation}
where we introduce the constant $\beta$. Therefore, we have the differential 
equation for the polar-angle function 
\begin{equation}
	\frac{1}{\sin \theta} \frac{d}{d \theta} 
		\left( \sin \theta \frac{d \Theta}{d \theta} \right) 
		+ \left( \beta - \frac{m^2}{\sin^2 \theta} \right) \Theta = 0.
	\label{appeq:Polar}
\end{equation}
By changing variables to $x = \cos \theta$, the differential equation 
(\ref{appeq:Polar}) becomes 
\begin{equation}
	\frac{d}{dx} \left[ (1 - x^2) \frac{d \Theta}{dx} \right] 
		+ \left( \beta - \frac{m^2}{1 - x^2} \right) \Theta = 0.
	\label{appeq:PolarLegendre}
\end{equation}
Equation~(\ref{appeq:PolarLegendre}) is in the form of the associated Legendre 
equation. It has a regular solution when $\beta = \ell (\ell + 1)$, where $\ell$ 
is a positive integer and $|m| \leq \ell$. The solution is 
\begin{equation}
	\Theta(\theta) \propto \mathcal{P}_{\ell m} (\cos \theta),
\end{equation}
where $\mathcal{P}_{\ell m} (x)$ is an associated Legendre polynomial, 
\begin{equation}
	\mathcal{P}_{\ell m} (x) = (- 1)^m (1 - x^2)^{m / 2} \frac{d^m}{dx^m} 
		\mathcal{P}_\ell (x),
\end{equation}
and $\mathcal{P}_\ell (x)$ is a Legendre polynomial, 
\begin{equation}
	\mathcal{P}_\ell (x) = \frac{1}{2^\ell \ell!} \frac{d^\ell}{dx^\ell} 
		(x^2 - 1)^\ell.
\end{equation}

From (\ref{appeq:beta}), we obtain an equation for $R(r)$, 
\begin{equation}
	\frac{d}{dr} \left( r^2 \frac{dR}{dr} \right) = \ell (\ell + 1) R,
\end{equation}
that has solutions of the form 
\begin{equation}
	R(r) \propto r^\ell, \qquad R(r) \propto 1 / r^{\ell + 1}.
\end{equation}

Combining the above results, the general solution to Laplace's equation is 
\begin{equation}
	f(r, \theta, \phi) = (A r^\ell + B / r^{\ell + 1}) 
		\mathcal{P}_{\ell m} (\cos \theta) e^{i m \phi},
\end{equation}
where $A$ and $B$ are constants.

At this point, we focus on the angular part of $f$, 
$g_{\ell m} (\theta, \phi) \propto \Theta(\theta) \Phi(\phi)$, which depends on 
the numbers $\ell$ and $m$. We can normalise $g_{\ell m} (\theta, \phi)$ so 
that 
\begin{equation}
	\int_{\theta = 0}^\pi \int_{\phi = 0}^{2 \pi} g_{\ell' m'}^* (\theta, \phi) 
		g_{\ell m} (\theta, \phi)  \sin \theta d \theta d \phi 
		= \delta_{\ell \ell'} \delta_{m m'},
	\label{appeq:Normalisation}
\end{equation}
where $\delta_{\ell m}$ is the Kronecker delta and the star denotes complex 
conjugation. Here, we integrate over the differential solid angle on a 
two-sphere, $d\Omega = \sin \theta d\theta d\phi$. Therefore, we find 
\begin{equation}
	g_{\ell m} (\theta, \phi) = (- 1)^m \sqrt{\frac{2 \ell + 1}{4 \pi} 
		\frac{(\ell - m)!}{(\ell + m)!}} \mathcal{P}_{\ell m} (\cos \theta) 
		e^{i m \phi} \equiv Y_{\ell m} (\theta, \phi),
	\label{appeq:SphericalHarmonics}
\end{equation}
which are the famous spherical harmonics $Y_{\ell m} (\theta, \phi)$. The phase 
factor $(- 1)^m$ is a convention from quantum mechanics.

Let us examine some of the properties of the spherical harmonics. We observe 
from (\ref{appeq:SphericalHarmonics}) that 
\begin{equation}
	Y_{\ell m} = (- 1)^m Y_{\ell -m}^*.
\end{equation}
We note through (\ref{appeq:Normalisation}) that the spherical harmonics are 
orthonormal. Indeed, we also have 
\begin{equation}
	\sum_{\ell = 0}^\infty \sum_{m = - \ell}^\ell 
		Y_{\ell m}^* (\vartheta, \varphi) Y_{\ell m} (\theta, \phi) 
		= \frac{\delta(\theta - \vartheta) \delta(\phi - \varphi)}{\sin \theta},
\end{equation}
where $\delta$ is the Dirac delta function. Thus, the spherical harmonics form 
an orthonormal basis on the two-sphere. Additionally, the spherical harmonics are 
complete. This means that they are linearly independent and no function of 
$(\theta, \phi)$ exists that is orthogonal to all $Y_{\ell m} (\theta, \phi)$. 
Because they are complete, any well-behaved function of $(\theta, \phi)$ can be 
expressed as 
\begin{equation}
	F(\theta, \phi) = \sum_{\ell = 0}^\infty \sum_{m = - \ell}^\ell 
		F_{\ell m} Y_{\ell m} (\theta, \phi),
\end{equation}
with coefficients $F_{\ell m}$ given by 
\begin{equation}
	F_{\ell m} = \int_{\theta = 0}^\pi \int_{\phi = 0}^{2 \pi} 
		Y_{\ell m}^* (\theta, \phi) F(\theta, \phi) \sin \theta d\theta d\phi.
\end{equation}
One can obtain the useful relation, using some of the above equations, 
\begin{equation}
	\nabla^2 Y_{\ell m} = - \frac{\ell (\ell + 1)}{r^2} Y_{\ell m}.
	\label{eq:HarmonicsRelation}
\end{equation}
It is worthwhile noting that (\ref{eq:HarmonicsRelation}) holds for all 
spacetimes in relativity that include the two-sphere [\textit{e.g.}, the 
Schwarzschild metric (\ref{eq:StaticMetric})]. We also have the addition 
formula, which we show without proof, 
\begin{equation}
	\mathcal{P}_\ell (\cos \alpha) = \frac{4 \pi}{2 \ell + 1} 
		\sum_{m = - \ell}^\ell Y_{\ell m}^* (\hat{X}^i) Y_{\ell m} (\hat{x}^i),
\end{equation}
where $\hat{X}^i$ and $\hat{x}^i$ are two unit vectors, separated by angle 
$\alpha$.


%% file: sections/appendix-C.tex
\chapter{Numerically solving the perturbation equations}
\label{app:NumericalScheme}

Our approach to solving the interior perturbation equations is similar to as 
described in \citet{2008PhRvD..78h3008L} and \citet{2015PhRvD..92f3009K}. We 
divide our star with a crust into three layers: (i) a fluid core from $R_0 = 0$ 
to $R_1$, (ii) an elastic crust from $R_1$ to $R_2$ and (iii) a fluid ocean 
from $R_2$ to $R_3 = R$. We express the system of ordinary differential 
equations for a given layer $i$ in the form 
\begin{equation}
    \frac{d \mathbf{Y}^{(i)}}{d r} = \mathbf{Q}^{(i)} \cdot \mathbf{Y}^{(i)}, 
        \ \text{for} \ r \in [R_{i - 1}, R_i],
    \label{appeq:ODE}
\end{equation}
where $\mathbf{Y}^{(i)}(r) = [y_1(r), ..., y_{k_i}(r)]$ is an abstract 
$k_i$-dimensional vector field, $\mathbf{Q}^{(i)}(r)$ is a $k_i \times k_i$ 
matrix and $r = R_i$ denotes the end of layer $i$. As long as our differential 
equations are linear we are free to write the system in the above form.

Due to the linearity of the differential equations, we generate a set of $k_i$ 
linearly independent solutions $\mathbf{Y}_j^{(i)}(r)$ for layer $i$ and obtain 
the general solution using a linear combination of these solutions, 
\begin{equation}
    \mathbf{Y}^{(i)}(r) = \sum_{j = 1}^{k_i} c_j^{(i)} \mathbf{Y}_j^{(i)}(r), 
	\label{appeq:General}
\end{equation}
where the coefficients $c_j^{(i)}$ are constants to be determined from 
boundary and interface conditions. We generate these linearly independent 
solutions by choosing linearly independent start vectors 
$\mathbf{Y}_j^{(i)}(R_{i - 1})$ and integrating through the layer using 
(\ref{appeq:ODE}) up to $r = R_i$. (Note that, in theory, there is no 
reason one could not do the reverse, integrating from $r = R_i$ to $R_{i - 1}$, 
should they wish.) \textit{A priori}, we do not have any 
additional information about layer $i$ and would na{\"i}vely integrate $k_i$ 
linearly independent start vectors. However, we can reduce the computational 
effort by applying relevant boundary conditions. For example, should a variable 
vanish at an interface, one could simply set this variable to zero in the start 
vectors and reduce the number of necessary linearly independent solutions by 
one.

We will initially focus this discussion to the Newtonian variables, but the 
logic still applies to the relativistic formulation. The fluid regions of the 
star are governed by Eqs.~(\ref{eqs:NewtonianFluidPerturbations}) and so we 
have the abstract two-dimensional vector field 
\begin{equation}
    \mathbf{Y}^{(k)}(r) = [d\delta \Phi(r)/dr, \delta \Phi(r)],
\end{equation}
where $k = 1, 3$ denotes the core and ocean, respectively. The elastic region of 
the star is more complex and requires more functions 
[see Eqs.~(\ref{eqs:NewtonianElasticPerturbations})] to describe its structure. 
Thus, we use the six-dimensional vector field 
\begin{equation}
    \mathbf{Y}^{(2)}(r) = [d\delta \Phi(r)/dr, \delta \Phi(r), \xi_r(r), 
						   \xi_\bot(r), T_1(r), T_2(r)].
\end{equation}
At a fluid-elastic interface, we know the variables $d\delta \Phi/dr$, 
$\delta \Phi$, $\xi_r$ and $T_2$ are continuous. We know the values of 
$d\delta \Phi/dr$ and $\delta \Phi$ from the calculation in the fluid core and 
so we use their final values in the core to start the integration in the crust. 
Since the traction variables vanish in the fluid, we can simplify the 
integrations in the elastic by demanding that $T_2 = 0$ at an interface. For 
each of the solutions, we calculate the value for $T_1$ at the base using 
(\ref{eq:RadialTraction}). These conditions mean that we must generate two 
linearly independent solutions with the initial values for the unknown functions 
$\xi_r$ and $\xi_\bot$. [Notice that because of the 
condition~(\ref{eq:RadialTraction}), we could equivalently choose to generate 
solutions with $T_1$ instead of either $\xi_r$ or $\xi_\bot$.] At the top of 
the crust, we demand that $T_2 = 0$ and $T_1$ be equal to the result of 
(\ref{eq:RadialTraction}). We use these two constraints to solve for the 
coefficients of the general solution~(\ref{appeq:General}). At the top of the 
crust, we can straightforwardly continue the integration through the fluid 
ocean, since $d\delta \Phi/dr$ and $\delta \Phi$ are continuous.

\sloppy
In the relativistic formulation, we have 
Eqs.~(\ref{eqs:RelativisticFluidPerturbations}) for the fluid perturbations 
with variables $(dH_0/dr, H_0)$ and 
Eqs.~(\ref{eqs:RelativisticElasticPerturbations}) for the elastic 
perturbations with variables $(dH_0/dr, H_0, K, W, V, T_2)$. The idea is very 
similar to what is described above. However, $dH_0/dr$ is not continuous. The 
functions $H_0$ and $K$ are continuous across an interface and are used to begin 
integration in the crust. Again, we have $T_2 = 0$ at the base and the value for 
$dH_0/dr$ is obtained from (\ref{eq:H_0primeCondition}). These two conditions 
also hold for the top of the crust and constrain the functions $W$ and $V$. In 
the fluid ocean, the initial value for $dH_0/dr$ is calculated from $H_0$ and 
$K$ using (\ref{eq:KFluid}).

\fussy


%% file: sections/appendix-D.tex
\chapter{The interface conditions}
\label{app:Interface}

Since we consider a star with multiple layers that have phase transitions, we 
must address how the perturbation functions behave across an interface. We 
calculate the interface conditions using the geometrical approach explained in 
\citet{2002PhRvD..66j4002A}.

Let us begin by considering the level surfaces of a scalar quantity $A$. We 
assume the level surfaces to be time-like and, therefore, have the normal, 
\begin{equation}
	\mathcal{N}^a = \frac{\nabla^a A}{\sqrt{\nabla^b A \nabla_b A}},
\end{equation}
where $\mathcal{N}^a \mathcal{N}_a = 1$ is true by construction. The first 
fundamental form (also known as the intrinsic curvature or induced three-metric) 
of these level surfaces is 
\begin{equation}
	\gamma_{a b} 
	    = P_a^{\hphantom{a} c} P_b^{\hphantom{b} d} g_{c d}, 
\end{equation}
where the projection operator along the level surfaces is given by 
\begin{equation}
	P_a^{\hphantom{a} b} = \delta_a^{\hphantom{a} b} 
	    - \mathcal{N}_a \mathcal{N}^b.
\end{equation}
The second fundamental form (also known as the extrinsic curvature) of the level 
surfaces is defined as 
\begin{equation}
	K_{a b} = - P_a^{\hphantom{a} c} P_b^{\hphantom{b} d} 
        \nabla_{(c} \mathcal{N}_{d)}.
\end{equation}

Let us specialise and consider the useful decomposition of our scalar quantity 
of the form, 
\begin{equation}
    A(t, r, \theta, \phi) = A_0(r) + \delta A(t, r, \theta, \phi).
\end{equation}
Using this decomposition we obtain the following components for the normal: 
\begin{subequations}
\begin{align}
	\mathcal{N}^t &= - e^{- \nu + \lambda/2} 
	    \frac{\partial_t \delta A}{dA_0/dr} + e^{- \nu - \lambda/2} h_{t r}, \\
	\mathcal{N}^r &= e^{-\lambda/2} 
	    \left( 1 - \frac{1}{2} e^{-\lambda} h_{r r} \right), \\
	\mathcal{N}^\theta &= \frac{e^{\lambda/2}}{r^2} 
	    \frac{\partial_\theta \delta A}{dA_0/dr}, \\
	\mathcal{N}^\phi &= \frac{e^{\lambda/2}}{r^2 \sin^2 \theta} 
	    \frac{\partial_\phi \delta A}{dA_0/dr}.
\end{align}
\end{subequations}
The level surfaces of $A$, thus, have the following non-zero components of the 
first fundamental form: 
\begin{subequations}
\begin{align}
    \gamma_{t t} &= - e^\nu + h_{t t}, \\
    \gamma_{t r} &= h_{t r} - e^\lambda \frac{\partial_t \delta A}{dA_0/dr}, \\
   	 \gamma_{r \theta} &= - e^\lambda \frac{\partial_\theta \delta A}{dA_0/dr}, \\
    \gamma_{r \phi} &= - e^\lambda \frac{\partial_\phi \delta A}{dA_0/dr}, \\
    \gamma_{\theta \theta} &= r^2 + h_{\theta \theta}, \\
    \gamma_{\phi \phi} &= r^2 \sin^2 \theta + h_{\phi \phi}.
\end{align}
\end{subequations}
The non-trivial components of the second fundamental form are 
\begin{subequations}
\begin{align}
	K_{t t} &= \frac{1}{2} \frac{d\nu}{dr} e^{\nu - \lambda/2} - e^{\lambda/2} 
	    \frac{\partial_t^2 \delta A}{dA_0/dr} + e^{-\lambda/2} \partial_t h_{t r} 
	    - \frac{1}{2} e^{-\lambda/2} \partial_r h_{t t} 
	    - \frac{1}{4} \frac{d\nu}{dr} e^{\nu - 3\lambda/2} h_{r r}, \\
	K_{t r} &= \frac{1}{2} \frac{d\nu}{dr} \left( e^{\lambda/2} 
	    \frac{\partial_t \delta A}{dA_0/dr} - e^{-\lambda/2} h_{t r} \right), \\
	K_{t \theta} &= - e^{\lambda/2} 
	    \left( \frac{\partial_t \partial_\theta \delta A}{dA_0/dr} 
	    - \frac{1}{2} e^{-\lambda} \partial_\theta h_{t r} \right), \\
	K_{t \phi} &= - e^{\lambda/2} 
	    \left( \frac{\partial_t \partial_\phi \delta A}{dA_0/dr} 
	    - \frac{1}{2} e^{-\lambda} \partial_\phi h_{t r} \right), \\
	K_{r \theta} &= \frac{e^{\lambda/2}}{r} 
	    \frac{\partial_\theta \delta A}{dA_0/dr} \\
	K_{r \phi} &= \frac{e^{\lambda/2}}{r} 
	    \frac{\partial_\phi \delta A}{dA_0/dr}, \\
	K_{\theta \theta} &= - e^{-\lambda/2} r - e^{\lambda/2} 
	    \frac{\partial_\theta^2 \delta A}{dA_0/dr} 
	    - \frac{1}{2} e^{-\lambda/2} ( \partial_r h_{\theta \theta} 
	    - e^{-\lambda} r h_{r r} ), \\
	K_{\theta \phi} &= - \frac{e^{-\lambda/2}}{dA_0/dr} 
	    \left( \partial_\theta \partial_\phi \delta A 
	    - \cot\theta \partial_\phi \delta A \right), \\
\begin{split}
	K_{\phi \phi} &= - e^{-\lambda/2} r \sin^2 \theta - e^{\lambda/2} 
	    \frac{\partial_\phi^2 \delta A}{dA_0/dr} 
	    - e^{\lambda/2} \sin\theta \cos\theta \frac{\partial_\theta \delta A}{dA_0/dr} \\
	    &\quad- \frac{1}{2} e^{-\lambda/2} ( \partial_r h_{\phi \phi} 
	    - e^{-\lambda} r \sin^2 \theta h_{r r} )
\end{split}
\end{align}
\end{subequations}
Both the first and second fundamental forms must be continuous across an 
interface (in the absence of surface degrees of freedom).

As was done by \citet{1990MNRAS.245...82F}, we will consider the level surfaces 
of the radial shell, so we assign $A_0 = r$ and $\delta A = \xi^r$. We use the 
perturbed metric for polar perturbations (\ref{eq:PerturbedMetric}). Because of 
how we set up the problem by assuming the background star is in a relaxed state, 
we know that the background quantities will all be continuous across an 
interface. We further assume that there is no discontinuity in the density or 
pressure. The first fundamental form with components $\gamma_{t t}$, 
$\gamma_{t r}$, $\gamma_{\theta \theta}$ and $\gamma_{r \theta}$ show 
\begin{equation}
    [H_0]_r = 0, \quad
    [H_1]_r = 0, \quad
   	[K]_r = 0, \quad 
    \left[ \frac{\delta A}{dA_0/dr} \right]_r = 0, 
\end{equation}
where we have introduced the notation 
$[f]_r = \lim_{\epsilon \rightarrow 0} [f(r + \epsilon) - f(r - \epsilon)]$ to 
describe the continuity of a function $f(r)$ at a point $r$. The angular part of 
$\delta A$ is decomposed using spherical harmonics. For the problem we 
are analysing, $H_1$ simply vanishes. The generic condition above translates to 
$[\xi^r]_r = 0$, which is equivalent to 
\begin{equation}
    [W]_r = 0.
\end{equation}
This condition is equivalent to saying there must not be a gap in the perturbed 
material.

We have exhausted the information we can learn from continuity of the first 
fundamental form. We also notice that there is no additional information 
to be learned from the components $K_{t r}$, $K_{t \theta}$, $K_{t \phi}$,  
$K_{r \theta}$ and $K_{r \phi}$ of the second fundamental form; only components 
$K_{t t}$ and $K_{\theta \theta}$ provide more interface conditions. Continuity 
of $K_{t t}$ implies 
\begin{equation}
    \left[ \partial_r h_{t t} 
		+ \frac{1}{2} \frac{d\nu}{dr} e^{\nu - \lambda} h_{r r} \right]_r = 0.
\end{equation}
This gives 
\begin{equation}
    \left[ \frac{dH_0}{dr} \right]_r = - \frac{1}{2} \frac{d\nu}{dr} [H_2]_r.
    \label{appeq:H_0primea}
\end{equation}
Similarly, we can infer from $K_{\theta \theta}$, 
\begin{equation}
	[\partial_r h_{\theta \theta} - e^{-\lambda} r h_{r r}]_r 
	    = 0 \quad \Rightarrow \quad 
		\left[ \frac{dK}{dr} \right]_r = \frac{1}{r} [H_2]_r.
    \label{appeq:Kprime}
\end{equation}
We can combine (\ref{appeq:H_0primea}) and (\ref{appeq:Kprime}) to obtain 
\begin{equation}
    \left[ \frac{dK}{dr} - \frac{dH_0}{dr} \right]_r = \frac{1}{2 r} \left( 2 
		+ r \frac{d\nu}{dr} \right) [H_2]_r.
\end{equation}

Now, we need some information from the perturbed Einstein equations. The above 
expression can be further used along with (\ref{eq:PEErthetaa}) to provide 
\begin{equation}
	\frac{16 \pi}{r} \left( 2 + r \frac{d\nu}{dr} \right) [\check{\mu} V]_r 
		- \frac{16 \pi}{r} [T_2]_r 
		= \frac{1}{2 r} \left( 2 + r \frac{d\nu}{dr} \right) [H_2]_r.
    \label{appeq:PEErtheta2Continuity}
\end{equation}
Using continuity of $H_0$ and (\ref{eq:PEEdifference}) we find 
\begin{equation}
    [H_2]_r = 32 \pi [\check{\mu} V]_r.
    \label{appeq:PEEdifferenceContinuity}
\end{equation}
This condition states that we should expect a discontinuity in $H_2$ for two 
reasons: (i) the shear modulus vanishes in the fluid and has a finite value in 
the crust and (ii) there is no reason that the tangential displacement function 
$V$ need be continuous. This further implies through (\ref{appeq:H_0primea}) 
that $dH_0/dr$ will be discontinuous, 
\begin{equation}
	\left[ \frac{dH_0}{dr} \right]_r 
		= - 16 \pi \frac{d\nu}{dr} [\check{\mu} V]_r.
    \label{appeq:H_0primeb}
\end{equation}
Equations~(\ref{appeq:PEErtheta2Continuity}) and 
(\ref{appeq:PEEdifferenceContinuity}) imply continuity of the tangential 
traction variable, 
\begin{equation}
    [T_2]_r = 0.
    \label{appeq:T_2Continuity}
\end{equation}

Finally, we use (\ref{eq:AlgebraicRelationa}), along with the continuity 
condition (\ref{appeq:H_0primeb}), to obtain 
\begin{equation}
    \left[ \frac{T_1}{r^2} + \delta p \right]_r = 0.
	\label{appeq:T_1Continuity}
\end{equation}
Equations~(\ref{appeq:T_2Continuity}) and (\ref{appeq:T_1Continuity}) simply 
mean that the radial and tangential stresses are continuous across a 
fluid-elastic interface. These interface conditions are necessary when 
considering how the functions behave across a fluid-elastic boundary and enable 
one to carry out the integration in the crust.
